\documentclass[12pt,twoside]{book}
\usepackage{amssymb}
\usepackage{amsmath}
\usepackage[dvips]{graphicx}
\usepackage{fancyhdr}
\fancypagestyle{plain}{%
   \fancyhead{} %
    
}
\begin{document}

\setcounter{page}{1}
\newtheorem{t1}{Theorem}[section]
\newtheorem{d1}{Definition}[section]
\newtheorem{c1}{Corollary}[section]
\newtheorem{l1}{Lemma}[section]
\newtheorem{r1}{Remark}[section]
\newcommand{\cA}{{\cal A}}
\newcommand{\cB}{{\cal B}}
\newcommand{\cC}{{\cal C}}
\newcommand{\cD}{{\cal D}}
\newcommand{\cE}{{\cal E}}
\newcommand{\cF}{{\cal F}}
\newcommand{\cG}{{\cal G}}
\newcommand{\cH}{{\cal H}}
\newcommand{\cI}{{\cal I}}
\newcommand{\cJ}{{\cal J}}
\newcommand{\cK}{{\cal K}}
\newcommand{\cL}{{\cal L}}
\newcommand{\cM}{{\cal M}}
\newcommand{\cN}{{\cal N}}
\newcommand{\cO}{{\cal O}}
\newcommand{\cP}{{\cal P}}
\newcommand{\cQ}{{\cal Q}}
\newcommand{\cR}{{\cal R}}
\newcommand{\cS}{{\cal S}}
\newcommand{\cT}{{\cal T}}
\newcommand{\cU}{{\cal U}}
\newcommand{\cV}{{\cal V}}
\newcommand{\cX}{{\cal X}}
\newcommand{\cW}{{\cal W}}
\newcommand{\cY}{{\cal Y}}
\newcommand{\cZ}{{\cal Z}}
\def\cl{\centerline}
\def\bd{\begin{description}}
\def\be{\begin{enumerate}}
\def\ben{\begin{equation}}
\def\benn{\begin{equation*}}
\def\een{\end{equation}}
\def\eenn{\end{equation*}}
\def\benr{\begin{eqnarray}}
\def\eenr{\end{eqnarray}}
\def\benrr{\begin{eqnarray*}}
\def\eenrr{\end{eqnarray*}}
\def\ed{\end{description}}
\def\ee{\end{enumerate}}
\def\al{\alpha}
\def\b{\beta}
\def\bR{\bar\R}
\def\bc{\begin{center}}
\def\ec{\end{center}}
\def\d{\dot}
\def\D{\Delta}
\def\del{\delta}
\def\ep{\epsilon}
\def\g{\gamma}
\def\G{\Gamma}
\def\h{\hat}
\def\iny{\infty}
\def\La{\Longrightarrow}
\def\la{\lambda}
\def\m{\mu}
\def\n{\nu}
\def\noi{\noindent}
\def\Om{\Omega}
\def\om{\omega}
\def\p{\psi}
\def\pr{\prime}
\def\r{\ref}
\def\R{{\bf R}}
\def\ra{\rightarrow}
\def\s{\sum_{i=1}^n}
\def\si{\sigma}
\def\Si{\Sigma}
\def\t{\tau}
\def\th{\theta}
\def\Th{\Theta}

\def\vep{\varepsilon}
\def\vp{\varphi}
\def\pa{\partial}
\def\un{\underline}
\def\ov{\overline}
\def\fr{\frac}
\def\sq{\sqrt}

\def\WW{\begin{stack}{\circle \\ W}\end{stack}}
\def\ww{\begin{stack}{\circle \\ w}\end{stack}}
\def\st{\stackrel}
\def\Ra{\Rightarrow}
\def\R{{\mathbb R}}
\def\bi{\begin{itemize}}
\def\ei{\end{itemize}}
\def\i{\item}
\def\bt{\begin{tabular}}
\def\et{\end{tabular}}
\def\lf{\leftarrow}
\def\nn{\nonumber}
\def\va{\vartheta}
\def\wh{\widehat}
\def\vs{\vspace}
\def\Lam{\Lambda}
\def\sm{\setminus}
\def\ba{\begin{array}}
\def\ea{\end{array}}
\def\ds{\displaystyle}
\def\lan{\langle}
\def\ran{\rangle}
\baselineskip 15truept
\large

\thispagestyle{empty}

\frontmatter
\title{DETECTION AND QUANTIFICATION OF ENTANGLEMENT IN MULTIPARTITE QUANTUM SYSTEMS USING WEIGHTED GRAPH AND BLOCH REPRESENTATION OF STATES.},\author{By\\\\Ali Saif M. Hassan\\Department of Physics,\\University of Pune, Ganeshkhind, Pune-411007\\\\\\Under the Supervision of \\Dr. P. S. Joag \\Department of Physics \\University of Pune,\\ Pune 411007.\\\\\\Dissertation submitted in partial fulfillment of the requirements\\for the degree of\\Doctor of Philosophy in Physics}
\maketitle
\newpage
\begin{center}
\textbf{List of Publications}
\end{center}
1. \textbf{A combinatorial approach to multipartite quantum
systems: basic formulation}\\Ali Saif M. Hassan and P. S. Joag, {\it J. Phys. A} \textbf{40}, 10251 (2007).\\\\
2. \textbf{On the degree conjecture for separability of multipartite
quantum states}\\ Ali Saif M. Hassan and P. S. Joag, {\it J. Math. Phys.} \textbf{49}, 012105 (2008).\\\\
3. \textbf{Separability criterion for Multipartite Quantum States Based on The Bloch Representation of Density Matrices}\\ Ali Saif M. Hassan and P. S. Joag, {\it Quant. Infor.  Comput.} \textbf{8}, 0773-0790 (2008).\\\\
4. \textbf{Experimentally accessible geometric measure for entanglement in N-qubit pure states}\\ Ali Saif M. Hassan and P. S. Joag, {\it Phys. Rev. A} \textbf{77}, 062334 (2008).

\newpage
\begin{center}
\large {\bf Acknowledgments}
\end{center}
\begin{sloppypar}
It gives me great pleasure to express my gratitude to my supervisor Dr. Pramod S. Joag for his constant guidance, encouragement and support. He has been always more than a guide to me and remains an ideal in my life both as a physicist and as a human being.
\end{sloppypar}
\begin{sloppypar}
	I wish to thank Dr. P. Durganandini for many useful discussions and encouragement.
\end{sloppypar}
\begin{sloppypar}
	My sincere thanks go to Mr. Ali  Ahanj who worked along with me on some allied aspects of Quantum Informations theory. I enjoyed our association as members of the same research group.
\end{sloppypar}

\begin{sloppypar}
I  thank  Mr. Bhalachandra Pujari for his help with Latex and drawing figures.
\end{sloppypar}

\begin{sloppypar}
	I would like to thank the Department of Physics, University of Pune for providing necessary facilities.
\end{sloppypar}

\begin{sloppypar}
 My deepest thanks go to my family, especially my Parents, my Wife and Children, for their support and encouragement.
\end{sloppypar}
\begin{flushright}
Ali Saif M. Hassan
\end{flushright}
\begin{sloppypar}
October 2008
\end{sloppypar}

\newpage

\tableofcontents

\newpage
\begin{center}
\large {\bf Abstract}
\end{center}

This thesis is an attempt to enhance understanding of the following questions\\ A-  Given a multipartite quantum state (possibly mixed), how to find out whether it is entangled or separable? (Detection of entanglement.)\\
B- Given an entangled state, how to decide how much entangled it is? (Measure of entanglement.), in the context of multipartite quantum states.

 We have explored two approaches. In the first approach, we assign a weighted graph with multipartite quantum state and address the question of separability  in terms of these graphs and various operations involving them. In the second approach we use the so called Bloch representation of multipartite quantum states to establish new criteria for detection of multipartite entangled states. We further give a new measure for entanglement in $N$-qubit entangled pure state and formally extend it to cover $N$-qubit mixed states.

We give a method to associate a graph with an arbitrary density matrix referred to a standard orthonormal basis in the Hilbert space of a finite
dimensional quantum system. We study related issues such as classification of pure and mixed states, Von Neumann entropy, separability of multipartite quantum states and quantum operations in terms of the graphs associated with
quantum states. In order to address the separability and entanglement questions using graphs, we introduce a modified tensor product of weighted graphs, and establish its algebraic properties. In particular, we show that Werner's definition (Werner 1989 Phys. Rev. A \textbf{40} 4277) of a separable state can be written in terms of graphs, for the states in a real or complex Hilbert space. We generalize the separability criterion (degree criterion) due to Braunstein {\it et al.} (2006 Phys. Rev. A \textbf{73} 012320) to a class of weighted graphs with real weights. We have given some criteria for the Laplacian associated with a weighted graph to be positive semidefinite.
We settle the so-called degree conjecture for the separability of multipartite quantum states, which are normalized graph Laplacians, first given by Braunstein {\it et al.} [Phys. Rev. A \textbf{73}, 012320 (2006)]. The conjecture states that a multipartite quantum state is separable if and only if the degree matrix of the graph associated with the state is equal to the degree matrix of the partial transpose of this graph. We call this statement to be the strong form of the conjecture. In its weak version, the conjecture requires only the necessity, that is, if the state is separable, the corresponding degree matrices match. We prove the strong form of the conjecture for {\it pure} multipartite quantum states, using the modified tensor product of graphs defined by Ali S. M. Hassan and P. S. Joag [J. Phys. A \textbf{40}, 10251 (2007)], as both necessary and sufficient condition for separability. Based on this proof, we give a polynomial-time algorithm for completely factorizing any pure multipartite quantum state. By polynomial-time algorithm we mean that the execution time of this algorithm increases as a polynomial in $m,$ where $m$ is the number of parts of the quantum system. We give a counter-example to show that the conjecture fails, in general, even in its weak form, for multipartite mixed states. Finally, we prove this conjecture, in its weak form, for a class of multipartite mixed states, giving only a necessary condition for separability.

We give a new separability criterion, a necessary condition for separability of $N$-partite quantum states. The criterion is based on the Bloch representation of a $N$-partite quantum state and makes use of multilinear algebra, in particular, the matrization of tensors. Our criterion applies to {\it arbitrary} $N$-partite quantum states in $\mathcal{H}=\mathcal{H}^{d_1}\otimes \mathcal{H}^{d_2} \otimes \cdots \otimes \mathcal{H}^{d_N}.$ The criterion can test whether a $N$-partite state is entangled and can be applied to different partitions of the $N$-partite system.  We provide examples that show the ability of this criterion to detect entanglement. We show that this criterion can detect bound entangled states. We prove a sufficiency condition for separability of a 3-partite state, straightforwardly generalizable  to the case  $N > 3,$ under certain  condition. We also give a necessary and sufficient condition for separability of a class of $N$-qubit states which includes $N$-qubit PPT states.
We present a multipartite entanglement measure for $N$-qubit pure states, using the norm of the correlation tensor which occurs in the Bloch representation of the state. We compute this measure for  several important classes of $N$-qubit pure states such as GHZ states, W states and their superpositions. We compute this measure for interesting applications like one dimensional Heisenberg antiferromagnet.  We use this measure to follow the entanglement dynamics of Grover's algorithm. We prove that this measure possesses almost all the properties expected of a good entanglement measure, including monotonicity. Finally, we extend this measure to $N$-qubit mixed states via convex roof construction  and establish its various properties, including its monotonicity. We also introduce a related measure which has all properties of the above measure and is also additive.\\

\mainmatter

\chapter{Introduction and Overview}
\begin{figure}[!ht]
\begin{center}
\includegraphics[width=8cm,height=.5cm]{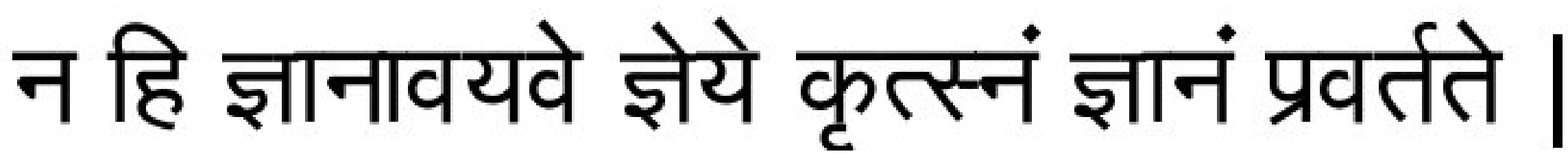}

Charak Samhita (First book of Aurveda)\\
({\it Whole is not known by knowing its parts})
\end{center}
\end{figure}
\begin{center}
\scriptsize\textsc{$\cdots$ But I can safely say that nobody understands Quantum Mechanics$\cdots$\\ Richard Feynmann}
\end{center}

Entanglement is a subtle and eluding property of quantum systems comprising many parts. Entanglement induces correlations between the measurable properties of different parts of a quantum system which cannot be reproduced by any procedure involving only the local operations (LO) on and classical communication (CC) between various parts of the system \cite{pv07}.  In consonance with this, entanglement in a quantum system cannot increase (or be created) via LOCC. This principle is connected to another intriguing property of entanglement: a multipartite quantum system can get entangled in various inequivalent ways, which cannot be transformed into each other via LOCC. However, the most challenging aspect of entanglement is that it cannot be `built in parts', that is, the entanglement of N parts is not a sum or a simple function of the entanglement of $M ( < N)$ partite subsystems \cite{eg05}.

The concept of entanglement has played a crucial role in the development of quantum physics. In the early days entanglement was mainly perceived as the qualitative
feature of quantum theory that most strikingly distinguishes it from our classical intuition. The subsequent development of Bell's inequalities made this distinction quantitative, and therefore rendered the nonlocal features of quantum theory accessible to experimental verification \cite{bel64,bel66,perb93}. Bell's inequalities may indeed
be viewed as an early attempt to quantify quantum correlations that are responsible for the counterintuitive features of quantum mechanically entangled states. At the time it was almost unimaginable that such quantum correlations between distinct quantum systems could be created in well controlled environments. However, the
technological progress of the last few decades means that we are now able to coherently prepare, manipulate, and measure individual quantum systems, as well as create controllable quantum correlations. In parallel with these developments, quantum correlations have come to be recognized as a novel resource that may be used to perform
tasks that are either impossible or very inefficient in the classical realm. These developments have provided the seed for the development of modern quantum information
science.

 Given the new found status of entanglement as a resource it is quite natural and important to discover the mathematical structures underlying its theoretical description. We will see that such a description aims to provide answers to three questions about entanglement, namely (1) its detection and classification, (2) its manipulation and, (3) its quantification.

In this thesis, we deal with the first and the third problem. We have used two approaches, the combinatorial and the geometric (Bloch representation) approaches for studying the detection problem and we give a geometric measure for quantifying the entanglement of multipartite pure states, we extend it to mixed states by convex roof construction. Our measure satisfies all properties expected of a good measure of entanglement.

In order to fathom and use entanglement and its role in various quantum phenomena, we must be able to say what entanglement is, and how we actually use it. In any quantum
communication experiment we would like to be able to distribute quantum particles across distantly separated laboratories. Perfect quantum communication is essentially
equivalent to perfect entanglement distribution. If we can transport a qubit without any decoherence \cite{gkksz}, then any entanglement shared by that qubit will also be distributed perfectly. Conversely, if we can distribute entangled states perfectly then with a small amount of classical communication, we may use teleportation \cite{ncb00} to perfectly transmit quantum states. However, in any experiment involving these processes, the effects of noise will inevitably impair our ability to send quantum states over long distances. A way of trying to overcome this problem is to distribute quantum states by using the noisy quantum channels that are available, but then to try and combat the effects of this noise using higher quality local quantum processes in the distantly separated laboratories. However, there is no reason to make the operations of separated laboratories totally independent. It turns out that the ability to perform classical communication is vital for many quantum information protocols - a prominent example being teleportation.

We have loosely described entanglement as the quantum correlations that can occur in many-party quantum states. This leads to the question  what differentiates quantum correlations from classical correlations? The distinction between `quantum' effects and 'classical' effects is frequently a cause of heated debate. However, in the context of quantum information a precise way to define classical correlations is via LOCC operations. Classical correlations can be defined as those that can be generated by LOCC operations. If we observe a quantum system and find correlations that cannot be simulated classically, then we usually attribute them to quantum effects, and hence label them quantum correlations. The entanglement is a resource because it lifts the so-called LOCC constraint, i.e. entanglement and LOCC together can perform tasks that cannot be accomplished by LOCC alone. Using LOCC-operations as the only other tool, the inherent quantum correlations of entanglement are required to implement general, and therefore nonlocal, quantum operations on two or more parts \cite{ejpp,clp01}. As LOCC-operations alone are insufficient to achieve these transformations, we conclude that entanglement may be defined as the sort of correlations that may not be created by LOCC alone.

Entanglement has proved to be a vital physical resource for various kinds of quantum-information processing, including quantum state teleportation \cite{bbcjpw,yc06}, cryptographic key distribution \cite{bbdcjmps}, classical communication over quantum channels \cite{bw92,bfsb97,bsst}, quantum error correction \cite{sho95}, quantum computational speedups \cite{deu85}, and distributed computation \cite{gro97,cb97}. Further, entanglement is expected to play a crucial role in the many particle phenomena such as quantum phase transitions, transfer of information across a spin chain \cite{on02,pv07a} etc. Therefore, quantification of entanglement of multipartite quantum  states is fundamental to the whole field of quantum information and in general, to the physics of multicomponent quantum systems.

Whereas the entanglement in pure bipartite state is well understood, the understanding of entanglement in mixed bipartite state is far from complete. In section 1.1, we review the entanglement of bipartite quantum systems. We will state the available measures and criteria for detecting entanglement for both bipartite pure and mixed states. In section 1.2, we deal with multipartite entangled states. In section 1.3, we  briefly summarize graph theory and density matrix of a graph. In section 1.4, we discuss a geometric approach i.e. Bloch representation of quantum states. In section 1.5, we close the chapter by giving some basic multilinear algebra. The material in section 1.3, 1.4 and 1.5 forms a background for chapters 2, 3, 4, and 5.

\section{ Bipartite Entanglement}

In this section, we define the entanglement in bipartite quantum states. We review the work that has been done in the bipartite systems in connection with the detection and quantification of entanglement.

Consider a system consisting of two subsystems. Quantum mechanics associates to each subsystem a Hilbert space. Let $\mathcal{H}_A$ and $\mathcal{H}_B$ denote these two Hilbert spaces, let $|i\rangle_A$ (where $i=1,2,3,\cdots$) represent a complete orthonormal basis for $\mathcal{H}_A$ and  $|j\rangle_B$ (where $j=1,2,3,\cdots$) a complete orthonormal basis for $\mathcal{H}_B$. Quantum mechanics associates with the system, i.e. the two subsystems taken together, the Hilbert space spanned by the states $|i\rangle_A \otimes |j\rangle_B$.  In  the following, we will drop the tensor product symbol $\otimes$ and write $|i\rangle_A \otimes |j\rangle_B$ as $|i\rangle_A |j\rangle_B$, and so on. Any linear combination of the basis states $|i\rangle_A  |j\rangle_B$ is a state of the system, and any state $|\psi\rangle_{AB}$ of the system can be written \cite{prb98} $$|\psi\rangle_{AB}=\sum_{ij} c_{ij}|i\rangle_A  |j\rangle_B,$$
where the $c_{ij}$ are complex coefficients, we take  $|\psi\rangle_{AB}$ to be normalized, hence $\sum_{ij} |c_{ij}|^2=1$.

If we can write $|\psi\rangle_{AB}= |\psi^{(A)}\rangle_A |\psi^{(B)}\rangle_B$, we say the $|\psi\rangle_{AB}$ is product state (separable state).
If $|\psi\rangle_{AB}$ is not a product state, we say that it is entangled.

By using local operators and classical communication (LOCC) any state $|\psi\rangle_{AB}$ of two subsystems $A$ and $B$ can be transformed to the form \cite{abhh,ncb00} $$|\psi\rangle_{AB}=\sum_{i=1}^k d_i |\phi_i\rangle_A |\phi'_i\rangle_B; \;\; k\le dim(\mathcal{H}_A \otimes \mathcal{H}_B),$$
where the positive coefficients $d_i$ are called Schimdt coefficients. The state is entangled if at least two coefficients do not vanish. Pure entangled state contains quantum correlation which can not be simulated  by any classical tools. A fundamental Theorem was proved by Bell \cite{bel64}, who showed that if the constraint of locality was imposed on the hidden variables, then there was an upper bound on the correlations of results of measurements that could be performed on the two distant systems. That upper bound, mathematically expressed by Bell's inequality \cite{bel64}, is violated by some state in quantum mechanics, thus the state contains quantum correlation which is Non-local property of quantum state \cite{per96a}.

However, in real conditions, owing to interaction with the environment, called decoherence, we encounter mixed states rather than pure ones. A mixed state is a classical mixture of pure quantum states \cite{prb98}. These mixed states can still possess some residual entanglement. A mixed state is considered to be entangled if it is not a mixture of product states \cite{wer89}. In mixed states the quantum correlations are weakened and hence the manifestations of mixed state entanglement can be very subtle \cite{pop95}. It is difficult to apply directly  the above definition of entanglement of mixed states to know whether a quantum state is entangled or not, because the mixed state contains both classical and quantum correlations, and can be prepared using infinite possible ensembles.

For pure states, it is easily shown that the CHSH inequality is violated by any nonfactorable state \cite{cfs73,gp92}, while on the other hand a factorable state trivially admits a (contextual) LHV model \cite{bel66}.

For mixed states, Werner \cite{wer89} constructed a density matrix $\rho_w$ for a pair of spin-$s$ particles. Werner's state $\rho_w$ can not be written as a sum of direct products of density matrices, $\sum_j c_j \rho^A_j\otimes \rho^B_j$, where $A$ and $B$ refer to the two distant particles and $j$ runs over the states in the ensemble. Therefore, genuine quantum correlations are involved in $\rho_w$. Nevertheless, for any pair of ideal local measurements performed on the two particles, the correlations derived from $\rho_w$ not only satisfy the CHSH inequality, but, as Werner showed \cite{wer89}, it is possible to introduce an explicit LHV model that correctly reproduces all the observable correlations for these ideal measurements \cite{per96a}. Thus for mixed states entanglement and nonlocality are two different resources.

\subsection{Quantification and Detection of Bipartite Entanglement}

 Given the wide range of tasks that exploit entanglement, one might try to define entanglement as `that property which is exploited in such protocols'. However, there is a whole range of such tasks, with a whole range of possible measures of success. This means that situations will almost certainly arise where a state $\rho_1$ is better than another state $\rho_2$ for achieving one task, but for achieving a different task $\rho_2$  is better than $\rho_1$ . Consequently using a task-based approach for quantifying entanglement will certainly not lead to a single unified perspective. However, despite this problem, it is possible to assert some general statements which are valid regardless of what your favorite use of entanglement is, as long as the key set of `allowed' operations is the LOCC class. This guides us as to how to approach the quantification of entanglement, and so we will state some of this statement  \cite{pv07} :

i) Separable states contain no entanglement

ii) All non-separable states allow some tasks to be achieved better than by LOCC alone, hence all nonseparable states are entangled.

iii) The entanglement of states does not increase under LOCC transformations.

iv) Entanglement does not change under Local Unitary operations.

v) There are maximally entangled states.

These properties give us some hints for the quantification of entanglement from the perspective of LOCC transformations in the asymptotic limit. However, one can try to salvage the  situation by taking a more axiomatic approach. One can define real valued functions that satisfy the basic properties of entanglement that we outlined above, and use these functions to attempt to quantify the amount of entanglement in a given quantum state.

We will now discuss and present a few basic axioms that any measure of entanglement should satisfy \cite{pv07,bzb06}.

 1- A bipartite entanglement measure $E(\rho)$ is a mapping from density matrices into positive real numbers. $\rho \rightarrow E(\rho) \in \mathbb{R}^+$,
 defined for states of arbitrary bipartite systems. A normalization factor is also usually included such that the maximally entangled states of two qudits has $E(|\psi^+_d\rangle) = \log d.$

2- $E(\rho)=0$ if the state is separable.

3- $E(\rho)$ does not increase on average under LOCC.

4- For pure state $|\psi\rangle \langle \psi|$ the measure reduces to the entropy of entanglement defined bellow.

 Any function $E$ satisfying the first three conditions is called an entanglement monotone. The term entanglement measure will be used for any quantity that satisfies axioms 1,2 and 4, and also does not increase under deterministic LOCC transformations. Frequently, some authors also impose additional requirements for entanglement measures: \cite{bzb06}
convexity, additivity and continuity.

The study of the LOCC transformation of pure states has so far enabled us to justify the concept of maximally entangled states and has also permitted us, in some cases,
to assert that one state is more entangled than another. However, we know that exact LOCC transformations can only induce a partial order on the set of quantum states.
The situation is even more complex for mixed states, where even the question of when it is possible to LOCC transform one state into another is a difficult problem
with no transparent solution at the time of writing.

All this means that if we want to give a definite answer as to whether one state is more entangled than another for any pair of states, it will be necessary to consider a more general setting. In this context a very natural way to compare and quantify entanglement is to study LOCC transformations of states in the so called asymptotic regime. Instead of asking whether for a single pair of particles the initial state $\rho$ may be transformed to a final state $\sigma$ by LOCC operations, we may ask whether for some large integers $m,\; n$ we can implement the `whole-sale' transformation $\rho^{\otimes n}\longrightarrow \sigma^{\otimes m}$. The largest ratio
$m/n$ for which one may achieve this would then indicate the relative entanglement content of these two states. In this setting we consider imperfect transformations between large blocks of states, such that in the limit of large block sizes the imperfections vanish.

Such an asymptotic approach will alleviate some of the problems that we encountered in the case of manipulation of single bi-partite states. It turns out that the asymptotic setting yields a unique total order on bi-partite pure states, and as a consequence, leads to a very natural measure of entanglement that is essentially unique. We will start by defining first entanglement measure - the entanglement cost $E_c(\rho)$ \cite{bbps,hht01,bdsw,woot98,woot01}. For a given state $\rho$, this measure quantifies the maximal possible rate $r$ at which one can convert blocks of two-qubit maximally entangled states into output states that approximate many copies of $\rho$, such that the approximations become vanishingly small in the limit of large block sizes. $E_c(\rho)$ measures how many maximally entangled states are required to create copies of $\rho$ by LOCC alone, we can ask about the reverse process: at what rate may we obtain maximally entangled states from an input supply of states of the form $\rho$. This process is known as entanglement distillation \cite{bbps,rain99} ( usually reserved for the pure state). $E_D(\rho)$ tells us the rate at which noisy mixed states may be converted back into the singlet state by LOCC.  Given these two entanglement measures it is natural to ask whether $E_C \stackrel{?}= E_D$, i.e. whether entanglement transformations become reversible in the asymptotic limit. This is indeed the case for pure state transformations where $E_D(\rho)$ and $E_C(\rho)$ are identical and equal to the entropy of entanglement \cite{bbps}. $E_D(\rho)$ and $E_C(\rho)$ are not equal for mixed states and also difficult to compute for mixed states, except for some simple but very special states \cite{vp98,pvp00}.

Thus, we need a related measure of entanglement, which is the entanglement of formation \cite{bdsw}. For a mixed state $\rho$ this measure is defined as $E_F(\rho)=inf\{\sum_i p_i E(|\psi_i\rangle \langle\psi_i|) : \rho=\sum_i p_i |\psi_i\rangle\langle\psi_i|\}.$
Given that this measure represents the minimal possible average entanglement over all pure state decompositions of $\rho$, where $E(|\psi\rangle\langle\psi|)= S(tr_B\{|\psi\rangle\langle\psi|\}$ is entropy of entanglement for pure states \cite{bdsw}.
The variational problem that defines $E_F$ is extremely difficult to solve in general and at present one must either resort to numerical techniques for general states \cite{avm01}, or restrict attention to cases with high symmetry (e.g. \cite{vw01,efppw,tv00}), or consider only cases of low dimensionality. Quite remarkably a closed form solution is known for bipartite qubit states \cite{woot98,woot01,avm01}. This exact formula is based on the often used two-qubit concurrence. For general bi-partite qubit states it has been shown that \cite{woot01} $E_F(\rho)=S(\frac{1+\sqrt{1-C^2(\rho)}}{2})$, with $S(x)=-x log_2 x-(1-x)log_2(1-x)$, $C$ being the concurrence. For higher dimensional systems this connection breaks down, in fact there is not even a unique definition of the concurrence \cite{uhl00,rbchm}. Another important class of measures is  entanglement measures from convex roof constructions. The entanglement of formation $E_F$ is an important example of the general concept of a convex roof construction. The convex roof $f$ of a function $f$ is defined as the largest convex function that is  bounded from above by the function $f$ for all arguments \cite{uhl00,uhl98}. The importance of the convex roof method is based on the fact that it can be used to construct entanglement monotones from any unitarily invariant and concave function of density matrices \cite{vida00}. Also various such quantities have been proposed over the years, such as the relative entropy of entanglement \cite{vprk,vp98,vpjk}, the squashed entanglement \cite{cw04} and Logarithmic Negativity \cite{vw02}. Mintert {\it et al.} \cite{mkb04} found a lower bound on I-concurrence \cite{rbchm} which is simpler to estimate than the I-concurrence itself. Another attempt of generalizing the concurrence for mixed states in higher dimensions was made by Badziag {\it et al.} in \cite{bdhhh}. Yet another proposal to deal with mixed states in higher dimensions is presented in \cite{lcok}. For more details we refer the reader to \cite{pv07}. Unfortunately, all these measures are difficult to implement experimentally and they require substantial efforts to estimate.

On the other side, there are attempts to understand the separability problem, which is to identify the states that contain classical correlations only (or no correlations at all). These states are termed separable states, and their mathematical characterization has been formulated by Werner \cite{wer89}. We call this the problem of entanglement detection.

A major step in the characterization of the separable states was done by Peres \cite{per96b} and the Horodecki family \cite{hhh96}. Peres provided very powerful necessary condition for separability. Later on, Horodecki's demonstrated that this condition is also sufficient for composite Hilbert spaces of dimension $2 \times 2$ and $2\times 3$. A density matrix that verifies Peres criterion is termed ``PPT'' for positive partial transpose. In general, there exist PPT states $\rho$ which are not separable in $\mathcal{H}_A\otimes \mathcal{H}_B$ spaces ($dim(\mathcal{H}_A) \ge 2,dim(\mathcal{H}_B)\ge 4$ or $dim(\mathcal{H}_A)\ge 3$) \cite{phor97}. The PPT entangled states have been termed ``bound entangled states'' to distinguish them from the ``free entangled states''. ``Bound entangled states'' are entangled, however, no matter how many copies of them we have, these states cannot be ``distilled'' via local operations and classical communication to the form of pure maximally entangled states \cite{hhh98}. We encounter thus new problems such as: How can one distinguish a separable state $\rho$ from a PPT entangled state $\rho$? Are all non-PPT states (NPPT states) ``free entangled'' i.e. distillable?

From the end of last century, there has been a growing effort in searching for necessary and sufficient separability criteria. Several necessary conditions for separability are known: Werner has derived a condition based on the analysis of local hidden variables (LHV) models and the mean value of the, so-called, flipping operator \cite{wer89}, the Horodecki's have proposed a necessary criterion based on the so-called $\alpha$-entropy inequalities \cite{hhh96b}, etc. A general necessary and sufficient condition for separability was discovered by the Horodecki family in terms of positive maps \cite{hhh96}. A map is defined positive if it maps positive operators into positive operators. Later on the reduction criterion of separability was introduced \cite{mphor99,cag99}. Violation of this criterion is sufficient for entanglement to be free. Sufficient conditions for separability are also known. In \cite{zhsl} it was proved that any state close enough to the completely random state $\sigma_0 = I/NM$ is separable. In \cite{phor97}, in which the first explicit examples of entangled states with PPT property were provided. Another necessary criterion of separability was formulated which is the so-called range criterion. The analysis of the range of the density matrices, initiated by P. Horodecki, turned out to be very fruitful, leading, in particular, to the algorithm of optimal decomposition of mixed states into the separable and inseparable part \cite{ls98,stv98}, and to systematic methods of constructing examples of PPT entangled states with no product vectors in their range, using either so-called unextendible product bases (UPB's) \cite{bdmsst,ter98}, or the method described in \cite{bp00}. Also, considerable progress in the study of PPT entangled states has been made \cite{kckl,hlvc}. Lewenstein {\it et al.} employ the idea of ``subtracting projectors on product vectors'' \cite{ls98,stv98}. They introduced the `edge' state, which has a property that no projection onto the product state can be subtracted from it, keeping the rest positive definite and PPT. They  mentioned a different approach to the entanglement problem, based on the so-called entanglement witnesses. An entanglement witness is an observable $W$ that reveals the entanglement of an entangled density matrix $\rho$. Rudolph \cite{rud03}; Chen and Wu \cite{cw03} discovered new criterion called computable cross norm (CCN) criterion or matrix realignment criterion. Quite remarkably the realignment criterion has been found to detect some of PPT entanglement. It also provides nice lower bound on concurrence function \cite{caf05}. General separability criteria based on local uncertainty relation valid both for discrete and continuous variables have been introduced in \cite{gmvt,ht03}. Further it has been shown \cite{hof03} that PPT entanglement can be detected by means of local uncertainty relations introduced in \cite{ht03}. This approach has been further developed and simplified by G$\ddot{u}$hne \cite{guh04} and developed also in entropic terms \cite{gl04}. Recently, Braunstein {\it et al.} \cite{bgmsw,bgs06} have initiated a new approach towards the mixed state entanglement by associating graphs with density matrices and understanding their classification using these graphs. Hildebrand {\it et al.} \cite{hms06} testified that the degree condition \cite{bgmsw} is equivalent to the PPT criterion. Sixia Yu {\it et al.} \cite{yl05} have given a new family of entanglement witnesses and corresponding positive maps that are not completely positive based on local orthogonal observables. de Vicente \cite{vic07a} has introduced a new approach to study the separability of bipartite quantum systems in arbitrary dimensions using the Bloch representation of their density matrix. He has obtained analytical lower bounds on the concurrence of bipartite quantum systems in arbitrary dimensions related to the violation of separability conditions based on local uncertainty relations and on the Bloch representation of density matrices \cite{vic07b}. Very recently G$\ddot{u}$hne {\it et al.} \cite{ghge} have proposed a unifying approach to the separability problem which uses a representation of a quantum state by a covariance matrix of locally measurable observables and they have proposed nonlinear witness \cite{gl07}. Despite many efforts and seminal results obtained in the recent years, the problem of separability of bipartite mixed states remains essentially open.

\section{ Multipartite Entanglement.}

In this section, we deal with entanglement in multipartite quantum states. We state the main difference between multipartite and bipartite entangled states. We review the work that has been done in connection with the detection and quantification of multipartite entanglement.

Multiparticle entanglement is genuinely different from entanglement in quantum systems
consisting of two parts. To understand what is so different consider, say, a quantum
system that is composed of three qubits. Each of the qubits is to be held by one of
three laboratories distantly separated. It may come as quite a surprise that states of
such composite quantum systems may contain tripartite entanglement, while at the same
time showing no bi-partite entanglement at all. In contrast to the bipartite setting, there is no longer a natural ``unit'' of entanglement, the role that was taken by the maximally entangled state  of a system of two qubits. Quite strikingly, the very concept of being maximally entangled becomes void. Instead, we will see that in two ways there are ``inequivalent kinds of entanglement''. Consider multipartite entanglement of pure quantum states. A theory of entanglement should not discriminate states that differ only by a local operation. Here, ``local operation'' can mean merely a change of local bases (LU operations) or, else, general local quantum operations assisted by classical communication, that are either required to be successful at each instance (LOCC) or just stochastically  (SLOCC). For each notion of locality, local unitary operation (LU) or Local operations and classical communication (LOCC) or just stochastic-LOCC (SLOCC), the questions that have to be addressed are how many equivalence classes exist, how are they parameterized and how can one decide whether two given states belong to the same class?

For the case of bi-partite qubit states, two quantum states are LU-equivalent if
and only if their respective Schmidt normal forms coincide. All classes are parameterized by only one real parameter. Some simple parameter counting arguments show that in the case of N-qubit systems the situation must be vastly more complex. Indeed, disregarding a global phase, it takes $2^{N+1}-2$ real parameters to fix a normalized quantum state in $\mathcal{H} = (\mathbb{C}^2)^{\otimes N}$. The group of local unitary
transformations $SU(2)\times\cdots \times SU(2)$ on the other hand has 3N real parameters \cite{eg05}. Therefore, one needs at least $2^{N+1}-3N-2$ real numbers to parameterize the sets of inequivalent pure quantum states \cite{lp98}. This lower bound turns out to be tight \cite{chs00}. It is a striking result that the ratio of non-local to local parameters grows exponentially in the number of systems. In particular, the finding rules out all hopes of a generalization of the Schmidt normal form. A general pure tripartite qubit state, say, cannot be cast into the form
$sin\theta |000\rangle + cos\theta|111\rangle$ by the action of local unitaries \cite{perb93}. Considerable effort has been undertaken to describe the structure of LU-equivalence classes by the use of invariants or normal forms \cite{lp98,chs00,grb98,rain97b,aajt}. Ac$\acute{i}$n {\it et al.} \cite{aacjlt} have proved for any pure three-qubit state the existence of local bases which allow one to build a set of five orthogonal product states in terms of which the state can be written in a unique form. This leads to a canonical form which generalizes the two-qubit Schmidt decomposition. It is uniquely characterized by the five entanglement parameters.  When one deals with SLOCC operations, the group of SLOCC,  $SL(\mathbb{C}^2)\times\cdots \times SL(\mathbb{C}^2)$ has $2^{N+1}-6N-2$ parameters that are necessary to label SLOCC equivalence classes of qubit systems. It turns out that the three-qubit pure states are partitioned into a total of Six SLOCC-equivalent classes \cite{dvc00}. The picture is complete for three-qubit : any fully entangled state is SLOCC-equivalent to either $|GHZ\rangle$ or $|W\rangle$ \cite{dvc00}. Three-qubit W-states and GHZ-states have already been experimentally realized, both purely optically using postselection  \cite{ekbkw,bpdwz} and in ion traps \cite{rrhhblbsb}. The two states behave differently, however, if a system is traced out. Specifically, tracing
out the first qubit of the GHZ state will leave the remaining systems in a complete mixture. For $|W\rangle$ will leave the remaining systems in a mixed entangled bipartite state. Thus, the entanglement of $|W\rangle$ is more robust under particle loss than the one of $|GHZ\rangle$ \cite{dvc00}. From point of view of asymptotic manipulation of multipartite quantum states, there is no longer a single essential ingredient as in bipartite the maximally entangled state or EPR-state, but many different ones. In the multi-particle case, however, it is meaningful to introduce the concept of a minimal reversible entanglement generating set $(MREGS)$. An $MREGS$ $S$ is a set of pure states such that any other state can be generated from $S$ by means of reversible asymptotic LOCC. It must be minimal in the sense that no set
of smaller cardinality possesses the same property \cite{bprst,lpsw,gpv00}. Yet, it can be shown that merely to consider maximally entangled qubit pairs is not sufficient to construct an $MREGS$ \cite{lpsw}. To find general means for constructing $MREGS$ constitutes one of the challenging open problems of the field: as long as this question is generally unresolved, the development of a ``theory of multi-particle entanglement'' in the same way as in the bi-partite setting seems infeasible.

Regardless of whether there is a unit of multipartite entanglement, researchers have tried to find some measure of multipartite entanglement of pure states. Recently , Meyer and Wallach \cite{mw02} have defined a polynomial measure which is scalable, i.e. which applies to any number of spin-1/2 particles. Wong and Christensen \cite{wc01} have proposed a potential measure of a type of entanglement of pure states of N-qubits, the N-tangle. For a system of two qubits the N-tangle is equal to the square of the concurrence, and for systems of three qubits it is equal to the ``residual entanglement''. The geometric measure of entanglement \cite{wg03} makes use of a geometric distance to the set of product state: $E_{Geo}=min|||\psi\rangle\langle\psi|-\sigma||_2$, where $||.||_2$ is the Hilbert-Schmidt norm, and the minimum is taken over all product states. The relative entropy of entanglement in the multipartite setting is defined as the minimal distance of a given state to the set of fully separable states, quantified in terms of the quantum relative entropy \cite{pv01}. There are also many measures are defined for multipartite entanglement of pure and mixed states, as in \cite{ckw00,ebheb,par04,wagm,jtfsbst,ys050706,mkbdbkm,lmbsa}. Recently Lamata {\it et al.} \cite{llss} have proposed an inductive procedure to classify $N$-partite spin-1/2 entanglement under stochastic local operations and classical communication provided such a classification is known for $N-1$ qubits. The method is based upon the analysis of the coefficient matrix of the state in an arbitrary product basis. For mixed state the classification scheme is based on separability properties \cite{dc00a}. At the lowest level there is the class of states that can be prepared using LOCC alone. Its members are called fully separable and can be written in the form $\rho=\sum_i p_i (\rho^i_1\otimes\rho^i_2\otimes\cdots\otimes\rho^i_N).$
Evidently, states of this kind do not contain entanglement. A state is referred to as k-separable, if it is fully separable considered as a state on some k-partite split. By the use of this terminology, the separability classes can be brought into a
hierarchy, where $k$-separable classes are considered to be more entangled then $l$-separable ones for $k < l$. States that are not separable with respect to any non-trivial split are fully inseparable. The number of all splits of a composite system grows exorbitantly fast with the number $N$ of its constituents. One is naturally tempted to reduce the complexity by identifying redundancies in this classification. After all, once it is established that a state is fully separable, there is no need to consider any further splits. For three systems, The five possible splits (1-2-3, 12-3, 1-23, 13-2, 123) have already been known for pure state in the above discussion. It is a counter-intuitive fact that there are mixed states that are
separable with respect to any bi-partite split but are not fully separable \cite{bdmsst}. An analogous phenomenon does not exist for pure states. The following sub-classes of the set of bi-separable, i. e. 2-separable, states are all non-empty \cite{dc00a}.
\begin{itemize}

\item 1-qubit bi-separable states with respect to the first system are separable for the
split 1-23 but not for 12-3 or 13-2.
\item 2-qubit bi-separable states with respect to the first and second system are separable for the split 1-23 and 2-13, but not for 12-3.
\item 3-qubit bi-separable states are separable with respect to any bi-partite split but are not fully separable.
\end{itemize}

Together with the inseparable states and the fully separable ones, the above sets constitute a complete classification of mixed three qubit states modulo system permutations \cite{dct99}. The quantification of entanglement of multipartite mixed states is void. The most measures are taken as convex roof constructions \cite{wg03,mkbdbkm,jtfsbst}. There is some progress in the detection of entanglement of multipartite mixed state. Kai Chen and Ling-An Wu \cite{cw02} have generalized partial transposition separability criterion for the density matrix of a multipartite quantum system. This criterion contains as special cases the famous Peres-Horodecki  criterion \cite{per96b} and the realignment criterion \cite{cw03,rud03}. Xiu-Hong Gao {\it et al.} \cite{gfw06} have derived an analytical lower bound for the concurrence of tripartite quantum mixed states \cite{mkbdbkm}. A functional relation is established relating concurrence and the generalized partial transposition \cite{cw02}. Chang-Shui  Yu and He-shan Song \cite{ys05} have presented a method to construct full separability criterion for tripartite system of qubits. Later on, they have generalized it to the higher-dimensional systems. The above criteria need a complete quantum state tomography, this can be a costly procedure. It may be desirable to detect entanglement without the need of acquiring full knowledge of the quantum state. This is where entanglement witnesses come into play. Recently, Korbicz {\it et al.} \cite{kcl05} have derived spin squeezing inequalities that generalize the concept of the spin squeezing parameter and provide necessary and sufficient conditions for genuine 2-, or 3-qubit entanglement for symmetric states, and sufficient condition for $N$-qubit states. The inequalities have a clear physical interpretation as entanglement witnesses. Also Usha Devi {\it et al.} \cite{dpr07} have shown that higher order inter-group correlations involving even number of qubits are necessarily positive semidefinite for separable symmetric $N$-qubit states. T$\acute{o}$th and G$\ddot{u}$hne \cite{tg04} have presented entanglement witnesses for detecting genuine multiqubit entanglement.  The generalized Bell-type inequality is used to characterize and detect multipartite entanglement. D$\ddot{u}$r \cite{dur01} has studied the relation between distillability of multipartite states and violation of Bell's inequality. He proved that there exist multipartite bound entangled states that
violate a multipartite Bell inequality. This implies that (i) violation of Bell's inequality is not a sufficient condition for distillability and (ii) some bound entangled states cannot be described by a local hidden  variable model.  Later on, Ac$\acute{i}$n \cite{acin02} has proved that for all the states violating this inequality there exists at least one splitting of the parties into two groups such that some pure-state entanglement can be distilled. We saw that, for bipartite systems, bound entanglement is clearly defined as it involves only two spatially separated parties and a necessary and sufficient condition for distillability of bipartite quantum states is known \cite{hhh98}. In  a multiparty setting , however, due to several distinct spatially separated configurations, the definition of bound entanglement is not unique. A multipartite quantum state is said to be bound entangled if there is no distillable entanglement between any subset as long as all the parties remain spatially separated from each other. When, however, one also allows some of the parties to group together and perform local operations collectively, two qualitatively different classes of bound entanglement arise: (a) activable bound entangled  (ABE) states. The states that are not distillable when every party is separated from every other but become distillable, if certain parties decide to group together \cite{dc00b,smo01}. This implies that there is at least one bipartite partition/cut where the state is negative under partial transposition (NPT) \cite{per96b}. Such states have been also referred to as unlockable bound entangled (UBE) states in the literature. (b) Nonactivable Bound Entangled states- states that are not distillable under any modified configuration as long as there are at least two spatially separated groups. In other words, such states are always positive under partial transposition across any bipartite partition \cite{bdmsst}. For unlockable bound entangled states, we refer the reader to \cite{dc00b,smo01,sst03}. $\dot{Z}$ukowski {\it et al.} \cite{zb02} have derived a single general Bell inequality which is a sufficient and necessary condition for the correlation function for $N$ particles to be describable in a local and realistic picture, they also derived a necessary and sufficient condition for an arbitrary $N$-qubit mixed state to violate this inequality. Later on, in \cite{zblw} it was shown that there exist pure entangled $N>2$ qubit states that do not violate any Bell inequality for $N$ particle correlation functions for experiments involving two dichotomic observables per local measuring station. Laskowski {\it et al.} \cite{lz05} have shown that the generalized Bell-type inequality, explicitly involving rotational symmetry of physical laws, is very efficient in distinguishing between true $N$-particle quantum correlations and correlations involving less particles. This applies to various types of generalized partial separabilities. Very recently, Badziag {\it et al.} \cite{bblpz} have presented an intuitive geometrical entanglement criterion. It allows formulation of simple and experimentally friendly sufficient conditions for entanglement of $N$-qubits.
Li {\it et al.} \cite{lfw08} have investigated the separability of arbitrary dimensional tripartite systems. By introducing a new operator related to transformations on the subsystems a necessary condition for the separability of tripartite
systems is presented.

\section{ Graphs and density matrix of a graph }

In this section, we give a brief summary for graphs, which  is necessary for chapter 2 and 3. We will also introduce the definition of density matrix of a graph.

A {\it graph} $G$ consists of a {\it vertex} set $V(G)$ and an {\it edge} set $E(G)$, where an edge is an unordered pair of distinct vertices of $G$. If $vu$ is an edge, then we say that $v$ and $u$ are adjacent or that $u$ is a neighbour of $v$, and denote this by writing $v\thicksim u$. A {\it loop} is an edge whose endpoints are equal. Two graph $G$ and $H$ are equal if and only if they have the same vertex set and the same edge set. Two graphs $G$ and $H$ are {\it isomorphic} if there is a bijection, $\varphi$ say, from $V(G)$ to $V(H)$ such that $v\thicksim u $ in $G$ if and only if $\varphi(v)\thicksim \varphi(u)$ in $H$. We say that $\varphi$ is an isomorphism from $G$ to $H$. Since $\varphi$ is bijection it has an inverse, which is an isomorphism from $H$ to $G$. If $G$ and $H$ are isomorphic, then we write $G \cong H$. A graph is called {\it complete} if every pair of vertices are adjacent, and the complete graph on $n$ vertices is denoted by $K_n$. A graph with no edges (but at least one vertex) is called {\it empty}. The graph with no vertices and no edges is the {\it null} graph. A {\it subgraph} of a graph $G$ is a graph $H$ such that $V(H)\subseteq V(G)$, $E(H) \subseteq E(G)$.
If $V(H)= V(G)$, we call $H$ a {\it spanning} subgraph of $G$. Any spanning subgraph of $G$ can be obtained by deleting some of the edges from $G$. The number of spanning subgraphs of $G$ is equal to the number of subsets of $E(G)$.

A subgraph $H$ of $G$ is an {\it induced} subgraph if two vertices of $V(H)$ are adjacent in $H$ if and only if they are adjacent in $G$. Any induced subgraph of $G$ can be obtained by deleting some of the vertices from $G$, along with any edges that contain a deleted vertex. The number of induced subgraphs of $G$ is equal to the number of subsets of $V(G)$.

A {\it clique } is a subgraph that is complete. It is necessarily an induced subgraph. A set of vertices that induces an empty subgraph is called an independent set. A {\it path} of length $r$ from $v$ to $u$ in a graph is a sequence of $r+1$ distinct vertices starting with $v$ and ending with $u$ such that consecutive vertices are adjacent. If there is a path between any two vertices of a graph $G$, then $G$ is {\it connected}, otherwise {\it disconnected}. Alternatively, $G$ is disconnected if we can partition its vertices into two nonempty sets, $R$ and $S$, say, such that no vertex in $R$ is adjacent to a vertex in $S$. in this case we say that $G$ is the disjoint union of two subgraphs.

A {\it cycle} is a connected graph where every vertex has exactly two neighbours. The smallest cycle is the complete graph $K_3$. An acyclic graph is a graph with no cycles. A connected acyclic graph is called a tree, and an acyclic graph is called a {\it forest}, since each component is a tree. a spanning subgraph with no cycles is called a {\it spanning tree}. A graph has a spanning tree if and only if it is connected. A {\it star} $K_{1,n}$, which consists of a single vertex with $n$ neighbours.

An isomorphism from a graph $G$ to itself is called an {\it automorphism} of $G$. An automorphism is therefore a permutation of the vertices of $G$ that maps edges to edges and nonedges to nonedges. The set of all automorphisms of $G$ forms a group, which is called the {\it automorphism group} of $G$ and denoted by $Aut(G)$. The {\it symmetric group} $Sym(V)$ is the group of all permutations of a set $V$, and so the automorphism group of $G$ is a subgroup of $Sym(V(G))$.

The {\it valency} of a vertex $v$ is the number of neighbours of $v$, and the maximum and minimum valency of a graph $G$ are the maximum and minimum values of the valencies of any vertex of $G$. A graph in which every vertex has equal valency $k$ is called {\it regular} of valency $k$ or $k$-regular. The {\it distance} $d_G(v,u)$ between two vertices $v$ and $u$ in a graph $G$ is the length of the shortest path from $v$ to $u$. A {\it leaf} is a vertex of degree (valency) 1. An {\it isolated} vertex has degree 0.

The {\it complement} $\bar{G}$ of a graph $G$ has the same vertex set as $G$, where vertices $v$ and $u$ are adjacent in $\bar{G}$ if and only if they are not adjacent in $G$. A mapping $f$ from $V(G)$ to $V(H)$ is a {\it homomorphism} if $f(v)$ and $f(u)$ are adjacent in $H$ whenever $v$ and $u$ are adjacent in $G$. A graph $G$ is called {\it bipartite} if its vertex set can be partitioned into two parts $V_1$ and $V_2$ such that every edge has one end in $V_1$ and one in$V_2$. A {\it proper colouring} of a graph $G$ is a map from $V(G)$ into some finite set of colours such that no two adjacent vertices are assigned the same colour. If $G$ can be properly coloured with a set of $k$ colours, then we say that $G$ can be properly $k$-coloured. The least value of $k$ for which $G$ can be properly $k$-coloured is the {\it chromatic number} of $G$, and is denoted by $\chi(G)$. The set of vertices with a particular colour is called a {\it colour class} of the colouring, and is an independent set.

 A homomorphism from a graph $G$ to itself is called an {\it endomorphism}, and the set of all endomorphisms of $G$ is the {\it endomorphism monoid} of $G$. A monoid is a set that has an associative binary multiplication defined on it and an identity element.

A {\it line graph} of a graph $G$ is graph $L(G)$ with the edges of $G$ as its vertices, and where two edges of $G$ are adjacent in $L(G)$ if and only if they are incident in $G$. The star $K_{1,n}$, consists of a single vertex with $n$ neighbours. For more details see \cite{bol98,godsil}.

The Cartesian product $G \square H$ of two graphs $G$ and $H$ is defined on the Cartesian product $V(G)\times V(H)$ of the vertex sets of the factors. The edge set $E(G\square H)$ is the set of all pairs $\{(u,v),(x,y)\}$ of vertices for which either $u=x$ and $\{v,y\} \in E(H)$ or $\{u,x\} \in E(G)$ and $v=y$.

The direct product (tensor product) $G\otimes H$ is defined also on the Cartesian product $V(G)\times V(H)$ of the vertex sets of the factors. Two vertices $(u_1,u_2),\; (v_1,v_2)$ are adjacent when $u_1v_1 \in E(G)$ and $u_2v_2 \in E(H).$

The adjacency matrix of a graph on $n$ vertices $G$ is an $n\times n$ matrix, denoted by $M(G)$, having rows and columns labeled by the vertices of $G$, and $ij-th$ entry defined as follows:

\begin{displaymath}
[M(G)]_{i,j} = \left\{ \begin{array}{ll}
1 & \mbox{if}\; \{v_i,v_j\} \in E(G);\\
 0 & \mbox{if}\; \{v_i,v_j\} \notin E(G). ~~~~~~~~~~~~~~~~~~~~~~(1.1)
\end{array} \right.
\end{displaymath}\\
Note that $M(G\otimes H)=M(G) \otimes M(H).$

Two distinct vertices $v_i$ and $v_j$ are said to be adjacent if $\{v_i, v_j\}\in E(G)$. The degree of a vertex $v_i \in V (G)$, denoted by $\mathfrak{d}_G(v_i)$, is the number of edges adjacent to $v_i$. Two adjacent vertices are also said to be neighbours. The degree-sum of $G$ is defined and denoted by $\mathfrak{d}_G =\sum^n_{i=1} \mathfrak{d}_G(v_i)$. Note that
$\mathfrak{d}_G = 2 |E(G)|$. The degree matrix of $G$ is an $n \times n$ matrix, denoted by $\Delta(G)$, having $ij-th$ entry defined as follows:

\begin{displaymath}
[\Delta(G)]_{i,j} = \left\{ \begin{array}{ll}
\mathfrak{d}_G(v_i) & \mbox{if}\; i=j;\\
 0 & \mbox{if}\; i\ne j.~~~~~~~~~~~~~~~~~~~~~~~~~~(1.2)
\end{array} \right.
\end{displaymath}

The {\it combinatorial Laplacian matrix} of a graph $G$ (for short, Laplacian) is the matrix $L(G)= \Delta(G)- M(G)$. Notice that $L(G)$ does not change if we add or delete loops from $G$. According to our definition of graph, $L(G)\ne 0$. Note that $L(G\otimes H)\ne L(G) \otimes L(H).$

In Standard Quantum Mechanics (that is the Hilbert space formulation of Quantum Mechanics), the state of a quantum mechanical system associated to the $n$-dimensional Hilbert space $\mathcal{H} \cong \mathbb{C}^n$ is identified with an $n\times n$ positive semidefinite, trace-one, Hermitian matrix, called a {\it density matrix}. It is easy to observe that the Laplacian of a graph is symmetric and positive semidefinite. The Laplacian of a graph $G$, scaled by the degree-sum of $G$, has trace one and it is then a density matrix. This observation leads to the following definition \cite{bgs06}.

The density matrix of a graph $G$ is the matrix

$$\sigma(G)=\frac{1}{\mathfrak{d}_G} L(G). \eqno{(1.3)}$$

\section{ Bloch Representation}

In this section, we discuss the geometric approach to a density matrix via its Bloch representation. The determination of a state on the basis of the actual measurement (experimental data) is important both for experimentalists and theoreticians. In classical physics, it is trivial because there is a one-to-one correspondence between the state and the actual measurement. On the other hand, in quantum mechanics, where a density matrix is used to describe the state, it is generally nontrivial to connect them \cite{neum32,pgsh33,des76,weig92,perb93}. the Bloch representation of the density matrix can be constructed experimently giving the required connection between the density matrix and experiments.

$N$-level quantum states are described by density operators, i.e. unit trace Hermitian positive semidefinite linear operators, which act on the Hilbert space $\mathcal{H} \simeq \mathbb{C}^N.$ The Hermitian operators acting on $\mathcal{H}$ constitute a Hilbert space themselves, the so-called Hilbert-Schmidt space denoted by $HS(\mathcal{H})$, with inner product $(\rho,\sigma)_{HS}=Tr(\rho^{\dagger}\sigma)$. Accordingly, the density operators can be expanded by any basis of this space. In particular, we can choose to expand $\rho$ in terms of the identity operator $I_N$ and the traceless Hermitian generators of $SU(N) \;\; \lambda_i \; (i=1,2,\cdots,N^2-1),$

$$\rho=\frac{1}{N}(I_N+\sum^{N^2-1}_{i=1} r_i \lambda_i). \eqno{(1.4)}$$
The generators of $SU(N)$ satisfy the orthogonality relation $$ (\lambda_i,\lambda_j)_{HS}= Tr(\lambda_i\lambda_j)=2\delta_{ij}, \eqno{(1.5)}$$
and they are characterized by the structure constants of the corresponding Lie algebra, $f_{ijk}$ and $g_{ijk}$, which are, respectively, completely antisymmetric and completely symmetric, $$\lambda_i\lambda_j = \frac{2}{N}\delta_{ij}I_N+if_{ijk}\lambda_k+g_{ijk}\lambda_k.\eqno{(1.6)}$$
The generators can be easily constructed from any orthonormal basis $\{|j\rangle\}^{N-1}_{j=0}$ in $\mathcal{H}$ \cite{he81}. The (orthogonal) generators are given by
$$\{\lambda_i\}^{N^2-1}_{i=1}=\{u_{jk},v_{jk},w_l\},\eqno{(1.7)}$$
when $i=1,\cdots,N-1$
$$\lambda_i=w_l=\sqrt{\frac{2}{l(l+1)}}\sum^l_{j=1}(|j\rangle\langle j|-l|l+1\rangle\langle l+1|), \; 1 \le l \le N-1,$$
while for $i=N,\cdots,(N+2)(N-1)/2$

$$\lambda_i=u_{jk}=|j\rangle \langle k|+|k \rangle \langle j|,$$
and for $i=N(N+1)/2,\cdots,N^2-1$
 $$\lambda_i= v_{jk}=-i(|j\rangle \langle k|- |k\rangle \langle j|),$$
 $1 \le j \le k\le N.$\\
The orthogonality relation (1.5) implies that the coefficients in (1.4) are given by
$$r_i=\frac{N}{2}Tr(\rho\lambda_i).$$
Notice that the coefficient of $I_N$ is fixed due to the unit trace condition. The vector $ \textbf{r}=(r_1r_2\cdots r_{N^2-1})^t \in \mathbb{R}^{N^2-1}$, which completely characterizes the density operator, is called Bloch vector or coherence vector. The representation (1.4) was introduced by Bloch \cite{bloc46} in the $N=2$ case and generalized to arbitrary dimensions in \cite{he81}. Any density matrix in two-level systems turns out to be characterized uniquely by a three-dimensional real vector where the length satisfies
 $$|\mathbf{\lambda}| \equiv \sqrt{\lambda_i \lambda_i}\le 1.\eqno{(1.8)}$$
 Therefore, if we define the Bloch-vector space $B(\mathbb{R}^3)$ as a ball with radius 1:
$$B(\mathbb{R}^3)=\{\mathbf{\lambda}=(\lambda_1,\lambda_2,\lambda_3) \in \mathbb{R}^3 : |\mathbf{\lambda}| \le 1\},$$
its element gives an equivalent description of the density matrix with the following bijection (one-to-one and onto) map from $B(\mathbb{R}^3)$ to the set of density matrices. $$\mathbf{\lambda} \longrightarrow \rho=\frac{1}{2}I_2+\frac{1}{2}\lambda_i \sigma_i$$
$B(\mathbb{R}^3)$ is called the Bloch ball, its surface the Bloch sphere and its element the Bloch vector. The equality in Eq.(1.8) (i.e., $|\mathbf{\lambda}|=1$), the surface of the ball (the Bloch sphere) which constitutes the set of extreme points of Bloch ball, corresponds to the set of pure states, the points interior to the Bloch ball correspond to mixed states. It has an interesting appeal from the experimentalist point of view, since in this way it becomes clear how the density operator can be constructed from the expectation values of the operators $\lambda_i$,
$$\langle \lambda_i \rangle = Tr(\rho \lambda_i)=\frac{2}{N} r_i. \eqno{(1.9)}$$
As we have seen, every density operator admits a representation as in Eq.(1.4); however, the converse is not true. A matrix of the form (1.4) is of unit trace and Hermitian, but it might not be positive semidefinite, so to guarantee this property further restrictions must be added to the coherence vector. The set of all the Bloch vectors that constitute a density operator is known as the Bloch-vector space $B(\mathbb{R}^{N^2-1}).$ from above discussion it is known that in the case $N=2$ this space equals the unit ball in $\mathbb{R}^3$ and pure states are represented by vectors on the unit sphere. The problem of determining $B(\mathbb{R}^{N^2-1})$ when $N \ge 3$ is still open and a subject of current research \cite{kk05}. However, many of its properties are known. For instance, for pure states $(\rho^2=\rho)$ it must hold $$||\mathbf{r}||_2=\sqrt{\frac{N(N-1)}{2}}, \;\;r_ir_jg_{ijk}=(N-2)r_k, \eqno{(1.10)}$$
where $||.||_2$ is the Euclidean norm on $\mathbb{R}^{N^2-1}$. In the case of mixed states, the conditions that the coherence vector must satisfy in order to represent a density operator have been recently provided in \cite{kim03,bk03}. Regretfully, their mathematical expression is rather cumbersome. It is also known \cite{har78,koss03} that $B(\mathbb{R}^{N^2-1})$ is a subset of the ball $D_R(\mathbb{R}^{N^2-1})$ of radius $R=\sqrt{\frac{N(N-1)}{2}}$, which is the minimum ball containing it, and that the ball $D_r(\mathbb{R}^{N^2-1})$ of radius $r=\sqrt{\frac{N}{2(N-1)}}$ is included in $B(\mathbb{R}^{N^2-1})$. that is,

$$D_r(\mathbb{R}^{N^2-1}) \subset B(\mathbb{R}^{N^2-1}) \subset D_R(\mathbb{R}^{N^2-1}). \eqno{(1.11)}$$
In the case of bipartite quantum systems of dimensions $M\times N\;(\mathcal{H} \simeq \mathbb{C}^M\otimes \mathbb{C}^N)$ composed of subsystems $A$ and $B$, we can analogously represent the density operators as
$$\rho=\frac{1}{MN}(I_M\otimes I_N + \sum_i r_i \lambda_i\otimes I_N +\sum_j s_j I_M \otimes  \tilde{\lambda_j} + \sum_{ij} \lambda_i \otimes \tilde{\lambda_j}), \eqno{(1.12)}$$
where $\lambda_i \; (\tilde{\lambda_j})$ are the generators of $SU(M) \; (SU(N))$. Notice that $\mathbf{r} \in \mathbb{R}^{M^2-1}$ and $\mathbf{s}\in \mathbb{R}^{N^2-1}$ are the coherence vectors of the subsystems, so that they can be determined locally,

$$\rho_A=Tr_B\rho=\frac{1}{M}(I_M+\sum_i r_i \lambda_i), \;\;\rho_B=Tr_A\rho=\frac{1}{N}(I_N+\sum_i s_i \tilde{\lambda_i}). \eqno{(1.13)}$$
The coefficients $t_{ij}$, responsible for the possible correlations, form the real matrix $T \in \mathbb{R}{(M^2-1)\times(N^2-1)}$, and, as before, they can be easily obtained by $t_{ij}=\frac{MN}{4}Tr(\rho\lambda_i\otimes  \tilde{\lambda_j})=\frac{MN}{4}\langle \lambda_i\otimes  \tilde{\lambda_j}\rangle.$

\section{ Multilinear Algebra}

In this section, we give some basic of multilinear algebra  and Higher-order tensors. Higher-order tensor decompositions are in frequent use today in a variety of fields including psychometrics \cite{tuck66,cc70,hars70}, chemometrics \cite{ad81}, image analysis \cite{vt02,savb03,wa03}, graph analysis \cite{kbk05,kb06}, signal processing \cite{cpl,mb05} and we will use it in separability problem. the two most commonly used decompositions of tensor are Tucker \cite{tuck66} and Kruskal \cite{krus77,cc70,hars70}, which can be thought of as higher-order generalizations of the matrix singular value decomposition.

We start by defining a product of matrices which are useful to us.
The Khatri-Rao product \cite{mcd80,rmb71,sbg04} is the columnwise Kronecker product (tensor product). The Khatri-Rao product of matrices $A \in \mathbb{R}^{I\times K}$ and  $B \in \mathbb{R}^{J\times K}$ is denoted by $A\odot B$ and its $(IJ)\times K$ result is defined by $$A\odot B=[a_{:1}\otimes b_{:1} \; a_{:2}\otimes b_{:2}\; \cdots \; a_{:K}\otimes b_{:K}], \eqno{(1.14)}$$
where $ a_{:j}, \; b_{:j}$ are the $jth$ columns vectors of the matricies $A$ and $B$.
Observe that the matrices in a Khatri-Rao product all have the same number of columns.\\
As an example, let $A \in \mathbb{R}^{3\times 4}$ and  $B \in \mathbb{R}^{2\times 4}$ be as follows

\begin{displaymath}
A =
\left(\begin{array}{cccc}
1 & 4 & 7 & 10 \\
2 & 5 & 8 & 11\\
3 & 6 & 9 & 12
\end{array}\right),
\end{displaymath}

\begin{displaymath}
B =
\left(\begin{array}{cccc}
1 & 2 & 3 & 8 \\
4 & 5 & 6 & 10
\end{array}\right).
\end{displaymath}

The Khatri-Rao product of $A$ and $B$ is $A\odot B \in \mathbb{R}^{6\times 4}$

\begin{displaymath}
A\odot B =
\left(\begin{array}{cccc}
1 b_{:1} & 4 b_{:2} & 7 b_{:3} & 10 b_{:4}\\
2 b_{:1}& 5 b_{:2} & 8 b_{:3}& 11 b_{:4}\\
3 b_{:1}& 6 b_{:2} & 9 b_{:3}& 12 b_{:4}
\end{array}\right),
\end{displaymath}

\begin{displaymath}
A\odot B =
\left(\begin{array}{cccc}
1  & 8 & 21 & 80 \\
4 & 20 & 42 & 100 \\
2 & 10 & 24 & 88  \\
8 & 25 & 48 & 110 \\
3 & 12 & 27 & 96  \\
12& 30 & 54 & 120
\end{array}\right).
\end{displaymath}

\subsection{Tensors}

Let $\mathcal{X}$ be an $I_1\times I_2\times \cdots\times I_N$ tensor over $\mathbb{R}$. The order of $\mathcal{X}$ is $N$. The $nth$ dimension of $\mathcal{X}$ is $I_n$. An element of $\mathcal{X}$ is specified as $\mathcal{X}_{i_1 i_2\cdots i_N},$ where $i_j \in \{1,2,\cdots,I_j\}$ for $j=1,2,\cdots,N$. The set of all tensors of size $I_1\times I_2 \times \cdots\times I_N$ is denoted by $\mathbb{S}(I_1,\cdots,I_N)$. Let $\mathcal{X},\; \mathcal{Y} \in \mathbb{S}(I_1,\cdots,I_N)$. The inner product of  $\mathcal{X}$ and $\mathcal{Y}$ is defined as $\mathcal{X}\cdot \mathcal{Y}=\sum^{I_1}_{i_1=1}\sum^{I_2}_{i_2=1}\cdots \sum^{I_N}_{i_N=1} \mathcal{X}_{i_1 i_2\cdots i_N}\mathcal{Y}_{i_1 i_2\cdots i_N}.$\\
Correspondingly, the norm of $\mathcal{X}$ , $||\mathcal{X}||$, is defined as $||\mathcal{X}||^2\equiv \mathcal{X}.\mathcal{X}=\sum^{I_1}_{i_1=1}\sum^{I_2}_{i_2=1}\cdots \sum^{I_N}_{i_N=1} \mathcal{X}^2_{i_1 i_2\cdots i_N}.$\\
We say  $\mathcal{X}$ is a unit tensor if $||\mathcal{X}||=1.$\\
A decomposed tensor is a tensor $U \in \mathbb{S}(I_1,\cdots,I_N)$ that can be written $$U=u^{(1)}\circ u^{(2)}\circ\cdots\circ u^{(N)},\eqno{(1.15)}$$
where $\circ$ denotes the outer product and each $u^{(j)}\in \mathbb{R}^{I_j}$, for $j=1,\cdots,N$. The vectors $u^{(j)}$ are called the components of $U$. $U_{i_1i_2\cdots i_N}=u^{(1)}_{i_1}u^{(2)}_{i_2}\cdots u^{(N)}_{i_N}.$
A decomposed tensor is a tensor of rank one. The set of all decomposed tensors of size $I_1\times I_2 \times \cdots\times I_N$ is denoted by $\mathcal{D}(I_1,\cdots,I_N)$.\\
\noi\textbf{Lemma 1.5.1}: \cite{kold01} Let $U,V \in \mathcal{D}$ where $U$ is defined as in Eq.(1.15)  and $V$ is defined by $$V=v^{(1)}\circ v^{(2)}\circ\cdots\circ v^{(N)}.\eqno{(1.16)}$$
Then (a) $U.V=\Pi^N_{j=1}u^{(j)}.v^{(j)}$, (b) $||U||= \Pi^N_{j=1}||u^{(j)}||_2,$
 and, (c) $U+V \in \mathcal{D}$ if and only if all but at most one of the components of $U$ and $V$ are equal (within a scalar multiple).

 Let $U,\;V \in \mathcal{D}$ be defined as Eq.(1.15) and Eq.(1.16) respectively with $||U||=||V||=1.$ We say that $U$ and $V$ are orthogonal $(U \perp V)$ if $U.V=\Pi^N_{j=1}u^{(j)}.v^{(j)}=0.$ We say that $U$ and $V$ are completely orthogonal $(U \perp_c V)$ if for every $j=1,2,\cdots,N$; $u{(j)}\perp v{(j)}$. we say that $U$ and $V$ are strongly orthogonal $(U \perp_s V)$ if $U \perp V$ and for every $j=1,\cdots,N$ $u{(j)}= \pm v{(j)}$ or $u{(j)} \perp v{(j)}$.

 Let $\mathcal{X} \in \mathbb{S}$ be a tensor $$\mathcal{X}=\sum^r_{i=1} \sigma_i U_i, \eqno{(1.17)}$$
where $\sigma > 0$ for $i=1,\cdots,r$ and each $U_i \in \mathcal{D}$ and $||U||=1$ for $i=1,\cdots,r.$

\begin{itemize}

\item The rank of  $\mathcal{X}$, denoted rank($\mathcal{X}$), is defined to be the minimal $r$ such that $\mathcal{X}$ can be expressed as in Eq.(1.17), The decomposition is called the rank decomposition.

\item The orthogonal rank of $\mathcal{X}$, denoted $rank_{\perp}(\mathcal{X})$, is defined to be the minimal $r$ such that $\mathcal{X}$ can be expressed as in Eq.(1.17) and $U_i \perp U_j$ for all $i\ne j$. The decomposition is called the orthogonal rank decomposition.
\item The complete orthogonal rank of $\mathcal{X}$, denoted $rank_{\perp_c}(\mathcal{X})$, is defined to be the minimal $r$ such that $\mathcal{X}$ can be expressed as in Eq. (1.17) and $U_i \perp_c U_j$ for all $i \ne j.$ The decomposition is called the complete orthogonal decomposition.
\item The strong orthogonal rank of  $\mathcal{X}$, denoted $rank_{\perp_s}(\mathcal{X})$, is defined to be the minimal $r$ such that $(\mathcal{X})$ can be expressed as in Eq.(1.17) and $U_i \perp_s U_j$ for all $i \ne j$. The decomposition is called the strong orthogonal rank decomposition \cite{kold01}.
\end{itemize}

 Assume an $Nth$-order tensor $\mathcal{A}\in \mathbb(C)^{I_1\times I_2\times\cdots \times I_N}.$ The matrix unfolding $A_{(n)}\in \mathbb(C)^{I_n\times (I_{n+1}I_{n+2}\cdots I_N I_1I_2\cdots I_{n-1})}$ contains the element $a_{i_1i_2\cdots i_N}$ at the position with row number $i_n$ and column number equal to
   $$(i_{n+1} - 1)I_{n+2}I_{n+3} \dots I_N I_1 I_2 \dots I_{n-1} + (i_{n+2} - 1)I_{n+3}I_{n+4}\dots I_N I_1 I_2 \dots I_{n-1} $$
 $$+ \dots +(i_N - 1)I_1 I_2 \dots I_{n-1} + (i_1 - 1)I_2I_3 \dots I_{n-1} + (i_2 - 1)I_3I_4 \dots I_{n-1} +\dots + i_{n-1}.$$
For $n=1$, we take the last term $i_{n-1}=i_{0}=i_N$.

As an example \cite{lmv00}, define a tensor $\mathcal{T}^{(3)} \in \mathbb{R}^{3\times 2 \times 3}$, by $t_{111} = t_{112} = t_{211} =-t_{212} = 1$, $t_{213} = t_{311} = t_{313} = t_{121} = t_{122} = t_{221} = -t_{222} = 2$, $t_{223} = t_{321} = t_{323} = 4$,
$t_{113} = t_{312} = t_{123} = t_{322} = 0$. The matrix unfolding $T^{(3)}_{(1)}$ is given by

\begin{displaymath}
T^{(3)}_{(1)} =
\left(\begin{array}{ccc|ccc}
1& 1 & 0 & 2 & 2 & 0 \\
1 & -1 & 2 & 2 & -2 & 4\\
2 & 0 & 2 & 4 & 0 & 4
\end{array}\right).
\end{displaymath}\\
If we refer in general to the vectors of an $Nth$-order tensor $\mathcal{X} \in \mathbb{C}^{I_1\times I_2\times \cdots \times I_N}$ as its ''$n$-mode vectors'', defined as the $I_n$-dimensional vectors obtained from $\mathcal{X}$ by
varying the index $i_n$ and keeping the other indices fixed, then we have the following
definition. The $n$-rank of $\mathcal{X}$, denoted by $R_n = rank_n(\mathcal{X})$, is the dimension of the vector space spanned by the $n$-mode vectors \cite{lmv00}. The $n$-mode vectors of $\mathcal{X}$ are the column vectors of the matrix
unfolding $A_{(n)}$ and $rank_n(\mathcal{X}) = rank(A_{(n)})$.

The $n$-mode product of a tensor $\mathcal{Y} \in \mathbb{R}^{J_1\times J_2\times\cdots\times J_N}$ by a matrix $A \in \mathbb{R}^{I\times J_n}$, is denoted by $\mathcal{Y}\times_n A$, is an $(J_1\times J_2\times\cdots J_{n-1}\times I \times J_{J+1}\cdots\times J_N$)-tensor of which the entries are given by
$(\mathcal{Y}\times_n A)_{j_1j_2\cdots j_{n-1} i j_{n+1}\cdots j_N}=\sum^{J_n}_{j_n=1} y_{j_1j_2\cdots j_{n-1} j_n j_{n+1}\cdots j_N}a_{ij_n}$ \cite{kold06}.\\

As an example \cite{kold06}, let $\mathcal{Y}$ be the following $3 \times 4 \times 2$ tensor:

\begin{displaymath}
Y_{::1} =
\left(\begin{array}{cccc}
1& 4 & 7 & 10 \\
2 & 5 & 8 & 11 \\
3 & 6 & 9 & 12
\end{array}\right),
\end{displaymath}

\begin{displaymath}
Y_{::2} =
\left(\begin{array}{cccc}
13& 16 & 19 & 22 \\
14 & 17 & 20 & 23 \\
15 & 18 & 21 & 24
\end{array}\right).
\end{displaymath}
Let $A$ be the following $2\times 3$ matrix:

\begin{displaymath}
A =
\left(\begin{array}{ccc}
1& 2 & 3 \\
4 & 5 & 6
\end{array}\right).
\end{displaymath}
Note that the number of columns in $A$ is equal to the size of mode $1$ of $\mathcal{Y}$. Thus we can compute $\mathcal{Y}\times_1 A$, which is of size $2\times 4\times 2$ and

\begin{displaymath}
(\mathcal{Y}\times_1 A)_{::1} =
\left(\begin{array}{cccc}
22& 49 & 76 & 103 \\
28 & 64 & 100 & 136
\end{array}\right),
\end{displaymath}

\begin{displaymath}
(\mathcal{Y}\times_1 A)_{::2} =
\left(\begin{array}{cccc}
130& 157 & 184 & 211 \\
172 & 208 & 244 & 280
\end{array}\right).
\end{displaymath}

(a) Given matrices $A \in \mathbb{R}^{I_m\times J_m},\; B \in \mathbb{R}^{I_n \times J_n}$
$$\mathcal{Y}\times_m A \times_n B = (\mathcal{Y}\times_m A)\times_n B =(\mathcal{Y}\times_n B) \times_m A\;\; (m \ne n)$$

(b) If  $A \in \mathbb{R}^{I\times J_n},\; B \in \mathbb{R}^{K \times I}$
$$ \mathcal{Y}\times_n A \times_n B= \mathcal{Y} \times_n(BA).$$

(c) If $A \in \mathbb{R}^{I\times J_n}$ with full rank, then

$$\mathcal{X}=\mathcal{Y}\times_n A \Rightarrow \mathcal{Y}=\mathcal{X}\times_n \mathcal{A}^{\dagger}.$$
The matrix unfolding and $n$-mode product are related via\\
\noi\textbf{Proposition 1.5.2}: \cite{kold06} Let  $\mathcal{Y} \in \mathbb{R}^{J_1\times J_2\times\cdots\times J_N}$.\\

(a) If $A \in \mathbb{R}^{I\times J_n}$. Then $$\mathcal{X}=\mathcal{Y} \times_n A \Leftrightarrow X_{(n)}= A Y_{(n)}.$$

(b) Consequently, if $A^{(n)} \in \mathbb{R}^{I_n\times J_n}$ for all $n \in {1,2,\cdots,N}$ we have
$$\mathcal{X}=\mathcal{Y}\times_1 A^{(1)}\times_2 A^{(2)}\cdots \times_N A^{(N)} \Leftrightarrow$$
$$ X_{(n)}=A^{(n)} Y_{(n)}(A^{(N)}\otimes\cdots\otimes A^{(n+1)}\otimes A^{(n-1)}\otimes\cdots \otimes A^{(1)})^T.$$
\noi\textbf{Proposition 1.5.3}: \cite{kold06} Let $\mathcal{X}, \; \mathcal{Y}  \in \mathbb{R}^{I_1\times I_2\times \cdots \times I_N}$.

(a) Then $||\mathcal{X}||=||X_{(n)}||_F$ , for $n \in \{1,2,\cdots,N\}$, where $||.||_F$ is Frobenius norm of a matrix.

(b) If $\mathcal{X}= a^{(1)}\circ a^{(2)} \circ \cdots \circ a^{(N)}$ and $\mathcal{Y}= b^{(1)}\circ b^{(2)}\circ \cdots \circ b^{(N)}$. Then $\mathcal{X}.\mathcal{Y}=\Pi^N_{n=1} a^{(n)}.b^{(n)}.$

(c)  Let $Q$ be a $J\times I_n$ orthonormal matrix. Then $||\mathcal{X}||=||\mathcal{X}\times Q||.$\\

In this thesis, we study detection and quantification of entanglement in multipartite quantum systems.
The chapters are arranged as follows :\\
\noi\textbf{Chapter 2} :  In this chapter, we give a method to associate a graph with an arbitrary density matrix referred to a standard orthonormal basis in the Hilbert space of a finite
dimensional quantum system. We study related issues such as classification of pure and mixed states, Von Neumann entropy, separability of multipartite quantum states and quantum operations in terms of the graphs associated with
quantum states. In order to address the separability and entanglement questions using graphs, we introduce a modified tensor product of weighted graphs, and establish its algebraic properties. In particular, we show that Werner's definition (Werner \cite{wer89} of a separable state can be written in terms of graphs, for the states in a real or complex Hilbert space. We generalize the separability criterion (degree criterion) due to Braunstein {\it et al.} \cite{bgmsw} to a class of weighted graphs with real weights. We have given some criteria for the Laplacian associated with a weighted graph to be positive semidefinite.\\
\noi\textbf{Chapter 3} : In this chapter, we settle the so-called degree conjecture for the separability of multipartite quantum states, which are normalized graph Laplacians, first given by Braunstein {\it et al.} \cite{bgmsw}. The conjecture states that a multipartite quantum state is separable if and only if the degree matrix of the graph associated with the state is equal to the degree matrix of the partial transpose of this graph. We call this statement to be the strong form of the conjecture. In its weak version, the conjecture requires only the necessity, that is, if the state is separable, the corresponding degree matrices match. We prove the strong form of the conjecture for {\it pure} multipartite quantum states, using the modified tensor product of graphs defined by Ali S. M. Hassan and P. S. Joag \cite{hj07}, as both necessary and sufficient condition for separability. Based on this proof, we give a polynomial-time algorithm for completely factorizing any pure multipartite quantum state. By polynomial-time algorithm we mean that the execution time of this algorithm increases as a polynomial in $m,$ where $m$ is the number of parts of the quantum system. We give a counter-example to show that the conjecture fails, in general, even in its weak form, for multipartite mixed states. Finally, we prove this conjecture, in its weak form, for a class of multipartite mixed states, giving only a necessary condition for separability.\\
 \noi\textbf{Chapter 4} : In this chapter, we give a new separability criterion, a necessary condition for separability of $N$-partite quantum states. The criterion is based on the Bloch representation of a $N$-partite quantum state and makes use of multilinear algebra, in particular, the matrization of tensors. Our criterion applies to {\it arbitrary} $N$-partite quantum states in $\mathcal{H}=\mathcal{H}^{d_1}\otimes \mathcal{H}^{d_2} \otimes \cdots \otimes \mathcal{H}^{d_N}.$ The criterion can test whether a $N$-partite state is entangled and can be applied to different partitions of the $N$-partite system.  We provide examples that show the ability of this criterion to detect entanglement. We show that this criterion can detect bound entangled states. We prove a sufficiency condition for separability of a 3-partite state, straightforwardly generalizable  to the case  $N > 3,$ under certain  condition. We also give a necessary and sufficient condition for separability of a class of $N$-qubit states which includes $N$-qubit PPT states. \\
\noi\textbf{chapter 5} : In this chapter we present a multipartite entanglement measure for $N$-qubit pure states, using the norm of the correlation tensor which occurs in the Bloch representation of the state. We compute this measure for  several important classes of $N$-qubit pure states such as GHZ states, W states and their superpositions. We compute this measure for interesting applications like one dimensional Heisenberg antiferromagnet.  We use this measure to follow the entanglement dynamics of Grover's algorithm. We prove that this measure possesses almost all the properties expected of a good entanglement measure, including monotonicity. Finally, we extend this measure to $N$-qubit mixed states via convex roof construction  and establish its various properties, including its monotonicity. We also introduce a related measure which has all properties of the above measure and is also additive.\\
\noi\textbf{Chapter 6} : In this chapter, we summarize the work presented in this thesis and give the possible ways in which this work may be developed  further.

\chapter{A combinatorial approach to multipartite quantum
systems: basic formulation}
\begin{center}
\scriptsize\textsc{Before I came here I was confused about this subject. \\ Having listened to your lecture I am still confused. But on a higher level.\\  {\it Enrico Fermi}}
\end{center}

\section{Introduction}

In chapter 1, we saw that quantum information is a rapidly expanding field of research because of its theoretical advances in fast algorithms, superdence quantum coding, quantum error correction, teleportation, cryptography and so forth \cite{abhh,bezb00,ncb00}. Most of these applications are based on entanglement in quantum states (see chapter 1).  Although entanglement in pure state systems is relatively well understood, its understanding in the so called mixed quantum states \cite{perb93}, which are statistical mixtures of pure quantum states, is at a primitive level.  Recently, Braunstein, Ghosh and Severini \cite{bgmsw,bgs06}, have initiated a new approach towards the mixed state entanglement by associating graphs with density matrices and understanding their classification using these graphs. Hildebrand, Mancini and Severini  \cite{hms06} testified that the degree condition (see chapter 3) is equivalent to the PPT-criterion. They also considered the concurrence (see chapter 1) of density matrices of graphs and pointed out that there are examples on four vertices whose concurrence is a rational number. In this chapter, we generalize the work of these authors and give a method to associate a graph with the density matrix (real or complex), of an {\it arbitrary} density operator, and also to associate a graph with the matrix representing Hermitian operator (observable) of the quantum system, both with respect to a standard orthonormal basis in Hilbert space.  We define a modified tensor product of graphs and use it to give Werner's definition for the separability of a $m$-partite quantum system, in  $\mathbb{R}^{q_1} \otimes \mathbb{R}^{q_2} \otimes \cdots \otimes \mathbb{R}^{q_m}$, as well as  $\mathbb{C}^{q_1} \otimes \mathbb{C}^{q_2} \otimes \cdots \otimes \mathbb{C}^{q_m}$ in terms of graphs. We also deal with classification of pure and mixed states and related concepts like von-Neumann entropy in terms of graphs.\\
The chapter is organized as follows.  In Section 2.2, we define weighted graphs and their generalized Laplacians which correspond to density matrices, and discuss the permutation invariance of this association.  We also deal with pure and mixed states in terms of graphs.  Section 2.3 deals with von-Neumann entropy.  Section 2.4 is concerned with separability issues as mentioned above.  In Section 2.5, we deal with graph operations which correspond to quantum operations \cite{ncb00,kraus83,preskill}.  In Section 2.6, we present a method to associate a graph with a general Hermitian matrix, having complex off-diagonal elements. We define the modified tensor product for complex weighted graphs and express the separability of mixed quantum states in a complex Hilbert space in terms of graphs, using Werner's definition. In section 2.7, we present some graphical criteria for the associated Laplacian to be positive semidefinite. Finally, we close with  a summary and some general comments. Sections 2.2 to 2.5 deal with graphs with real weights, that is, quantum states living in real Hilbert space. Graphs with  complex weights, corresponding to density operators with complex off diagonal elements are treated in section 2.6. However, a large part of the results obtained for real Hilbert space in sections 2.2 to 2.5, go over to the case of complex Hilbert space (see section 2.8 (ix)).

\section{Density matrix of a weighted graph}

\subsection{Definitions}

 A graph $G = (V, E)$ is a pair of a nonempty and finite set called vertex set $V(G)$, whose elements are called vertices and a set $E(G)\subseteq V^2(G)$ of unordered pairs of vertices called edges.  A loop is an edge of the form $\{ v_i, v_i\}$ for some vertex $v_i$.  A graph $G$ is on $n$ vertices if $|V(G)| = n.$  We call the graph as defined above a simple graph.  $|E(G)| = m+s$, where  $m$ is the number  of edges joining vertices,  $s$ is the number  of loops in $G$ \cite{west02}.

A {\it weighted graph} $(G, a)$ is a graph together with a {\it weight function} \cite{mohar91}
$$ a : V(G) \times V(G) \ra I\!\!R$$
which associates a  real number (weight) $a(\{u, v\})$ to each pair $\{u, v\}$ of vertices.  The function $a$ satisfies the following properties:

\bd
\i(i) $a(\{u, v\}) \ne 0$ if $\{u, v\} \in E(G, a)$ and $a(\{u, v\}) = 0$ if $\{u, v\} \not\in E(G, a)$.

\i(ii) $a(\{u, v\}) = a(\{v, u\})$

\i(iii) $a(v, v) \ne 0$ if $\{v,v\}\in E(G,a)$ and is zero otherwise.
\ed
If $e = \{u, v\}$ is an edge in $E(G, a)$, property (ii) allows us to write $a(e)$ or $a_{uv}$ for $a(\{u, v\})$.  A simple graph can be viewed as a weighted graph with all   nonzero weights equal to 1.

In the case of simple graphs the degree $\mathfrak{d}_v$ of a vertex $v \in V(G)$ is defined as the number of edges in $E(G)$ incident on $v$.  For a weighted graph we set

$$   \mathfrak{d}_{(G,a)} (v) = \mathfrak{d}_v = \sum_{u \in V(G,a)} a_{uv}.  \eqno{(2.1)}
 $$

The {\it adjacency matrix} of a weighted graph with $n$ vertices $M(G, a) = [a_{uv}]_{u, v \in V(G, a)}$ is a $n \times n$ matrix whose rows and columns are indexed by vertices in $V(G,a)$ and whose $uv$-th element is $a_{uv}$.  Obviously the adjacency matrix $M(G, a)$ is a real symmetric matrix with diagonal element $vv$ equal to the weight of the  loops on vertex $v$ (i.e $a_{vv})$.

The {\it degree matrix} for the weighted graph $\D(G, a)$ is a $n \times n$ diagonal matrix, whose rows and columns are labeled by vertices in $V(G, a)$ and whose diagonal elements are the degrees of the corresponding vertices.
$$\D(G, a) = diag [ \mathfrak{d}_v; v \in V(G, a)]. \eqno{(2.2)}$$
The {\it combinatorial Laplacian} of a weighted graph is defined to be
$$ L(G, a) = \D(G, a) - M(G, a). \eqno{(2.3)}$$
The {\it degree sum} of $(G, a)$ is defined as
$$\mathfrak{d}_{(G, a)} = \sum_{v \in V(G, a)}  \mathfrak{d}_v = Tr \D(G, a) . \eqno{(2.4)}$$
The Laplacian defined by Eq. (2.3) has no record of loops in the graph.  Therefore we define the {\it generalized Laplacian} of a graph $(G, a)$, which includes loops, as
$$ Q(G, a) = \D(G, a) - M(G, a) + \D_0(G, a) \eqno{(2.5)}$$
where $\D_0(G, a)$ is a $n \times n$ diagonal matrix with diagonal elements equal to the weights of the loops on the corresponding vertices
$$ [\D_0(G, a)]_{vv} = a_{vv} . \eqno{(2.6)}$$
We call $\D_0(G, a)$ the {\it loop matrix} of the graph $(G, a)$ .

For a given weighted graph $(G, a)$, the generalized Laplacian, defined by Eq. (2.5), is not necessarily a positive semidefinite matrix.  When, for a given graph $(G, a)$, the generalized Laplacian $Q(G,a)$ is positive semidefinite, we can define the density matrix of the corresponding graph $(G, a)$ as
$$ \si(G, a) = \fr{1}{\mathfrak{d}_{(G, a)}} Q(G, a) = \fr{1}{\mathfrak{d}_{(G, a)}} [L(G, a) + \D_0(G, a)] \eqno{(2.7)}$$
where $Tr(\si(G, a)) = 1$.
Note that, this definition of the density matrix of a weighted graph $(G, a)$ reduces to that of the density matrix for a simple graph without loops \cite{bgs06}.

Whenever we can define the density matrix for a graph $(G, a)$ we say that the graph $(G, a)$ has density matrix.

For any density matrix $\si$, we can obtain the corresponding graph as follows:

\bd
\i(i) Determine the number of vertices of the graph from the size $(n \times n)$ of the density matrix.  The number of vertices = $n$.  Label the vertices from 1 to $n$.

\i(ii) If the $ij$-th element of $\si$ is not zero draw an edge between vertices $v_i$ and $v_j$ with weight $- \si_{ij}$.

\i(iii) Ensure that $\mathfrak{d}_{v_i} = \si_{ii}$ by adding loop of appropriate weight to $v_i$ if necessary.
\ed

\noi{\it Example(2.1)}: For the following three matrices, we find the corresponding graphs.

\bd
\i(1) $ \si = \fr{1}{2} \left[ \ba{cc} 1 & 1 \\ 1 & 1 \ea \right] $
in $\mathbb{R}^2$.
\begin{figure}[!ht]
\begin{center}
\includegraphics[width=3cm,height=.5cm]{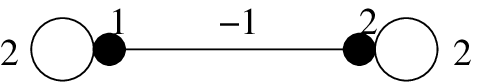}

Figure 2.1
\end{center}
\end{figure}

\i(2) $\si = \fr{1}{16} \left[ \ba{rrrr} 9 & -1 & -1 & 1 \\ -1 & 3 & -1 & -1 \\ -1 & -1 & 3 & -1 \\ 1 & -1 & -1 & 1 \ea \right] $
in  $\mathbb{R}^2 \otimes \mathbb{R}^2$.
\begin{figure}[!ht]
\begin{center}
\includegraphics[width=4cm,height=3cm]{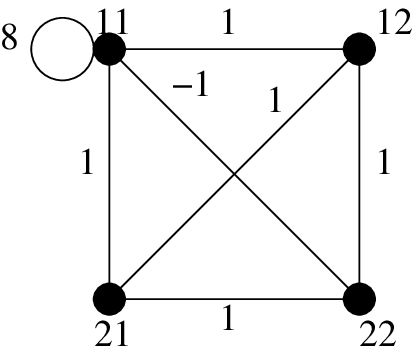}

Figure 2.2
\end{center}
\end{figure}

\i(3) $\si = \fr{1}{4} \left[ \ba{cccc} 1 & 0 & 0 & 0 \\ 0 & 1 & 0 & 0 \\ 0 & 0 & 1 & 0 \\ 0 & 0 & 0 & 1 \ea \right] $
in  $\mathbb{R}^2 \otimes \mathbb{R}^2$.
 \begin{figure}[!ht]
 \begin{center}
\includegraphics[width=4cm,height=4cm]{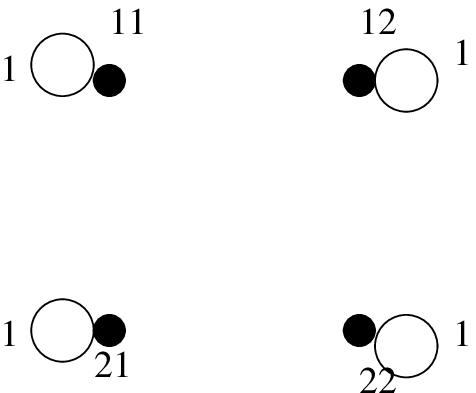}

Figure 2.3
\end{center}
\end{figure}

\ed

\subsection{Invariance under isomorphism}

Two weighted graphs $(G, a)$ and $(G', a')$ are isomorphic if there is a bijective map  \cite{godsil}
$$ \phi : V(G, a)  \longmapsto V(G', a')$$
such that
$$ \{\phi(v_i),  \phi(v_j)\} \in E(G' a') ~ \mbox{iff}~~ \{v_i, v_j\} \in E(G, a), i, j = 1, 2, \cdots, n$$
and
$$ a_{\phi(v_i) \phi(v_j)}' = a_{v_iv_j} ~~ i, j = 1, 2, \cdots, n.$$
We denote isomorphism of $(G, a)$ and $(G', a')$ by $(G, a) \cong (G' a')$.

Equivalently, two graphs $(G, a)$ and $(G', a')$ are isomorphic if there exists a permutation matrix $P$ such that
$$ P^TM(G, a) P = M(G', a').$$
Note that,
$$ P^T \D(G, a)P = \D(G', a'); ~~P^T \D_0(G, a)P = \D_0(G', a')$$
Therefore we have
$$ P^TQ(G, a) P = Q(G', a'). \eqno{(2.8)}$$
This means that $Q(G, a)$ and $Q(G', a')$ are similar and have the same eigenvalues.  Therefore, if $Q(G, a)$ is positive semidefinite then so is $Q(G', a')$.  Therefore, if $(G, a)$ has the density matrix so does $(G', a')$.  We have proved

\noi {\bf Theorem 2.2.1 :} The set of all weighted graphs having density matrix is closed under isomorphism.\hspace{\stretch{1}}$ \blacksquare$

Since isomorphism is an equivalence relation, this set is partitioned by it, mutually isomorphic graphs forming the partition.

\subsection{Correspondence with quantum mechanics}

Henceforth, we consider only the graphs having density matrix unless stated otherwise.   The basic correspondence with QM is defined by the density matrix of the graph.  For a graph with $n$ vertices the dimension of the Hilbert space of the corresponding quantum system is $n$.  To establish the required correspondence we fix an orthonormal basis in the Hilbert space $\mathbb{R}^{q_1} \otimes \mathbb{R}^{q_2} \otimes \cdots \otimes \mathbb{R}^{q_m}$ of the system, which we call the standard basis and denote it by $\{ | ijk\ell \cdots \ran\}, i, j, k, \ell \cdots = 1, 2, \cdots, n = q_1 q_2 \cdots q_m$, or by $\{ |v_i\ran\}, i = 1, \cdots, n = q_1 q_2 \cdots q_m$.  We label $n$ vertices of the graph $(G, a)$ corresponding to the given density matrix by the $n$ basis vectors in the standard basis.  We say that the graph $(G, a)$ corresponds to the quantum state (density operator) whose matrix in the standard basis is the given density matrix.  Finally, we set up a procedure, by associating appropriate projection operators with edges and loops of $(G, a)$ to reconstruct this quantum state from the graph $(G, a)$.  (See  Theorem 2.2.7). In view of Theorem 2.2.1 , if $(G,a)$ has density matrix $\si$ and $(G,a)\cong(G',a')$ with the corresponding permutation matrix $P$,then $(G',a')$ has the density matrix $P^T \si P$. All of this paragraph applies to the complex weighted graph (section 2.6).

\subsection{Pure and mixed states}

A density matrix $\rho$ is said to be pure if $Tr(\rho^2) = 1$ and mixed otherwise.  Theorem  2.2.3 gives a necessary and sufficient condition on a graph $(G, a)$ for $\si(G, a)$ to be pure. For a graph $(G, a)$  having $k$ components $(G_1, a_1) , (G_2, a_2), \cdots, (G_k, a_k)$ we write $(G, a) = (G_1, a_1) \uplus (G_2, a_2) \uplus \cdots \uplus (G_k, a_k)$ where $a_i, i = 1, \cdots, k$ are the restrictions of the weight function of the graph $(G, a)$ to the components.  We can order the vertices such that $M(G, a) = \oplus^k_{i=1} M(G_i, a_i)$. When $k = 1$, $(G, a)$ is said to be connected.  From now on we denote by $\la_1(A), \la_2(A), \cdots, \la_k(A)$ the $k$ different eigenvalues of the Hermitian matrix $A$ in the nondecreasing order.  The set of the eigenvalues of $A$ together with their multiplicities is called the spectrum of $A$ \cite{godsil,ltb85,hjb90}.

\noi {\bf Lemma 2.2.2 :} The density matrix of a graph $(G, a)$ without loops has zero eigenvalue with multiplicity  greater than or equal to the number of components of $(G, a)$ with equality applying  when weight function $a=$constant $> 0$,

\noi {\bf Proof :} Let $(G, a)$ be a graph with $n$ vertices and $m$ edges.  Since $Q(G, a)$ is positive semidefinite, for $x \in I\!\!R^n$ we must have \cite{mohar91}
$$ x^TQ(G,a)x = \sum^m_{k=1} a_{i_kj_k} (x_{i_k} - x_{j_k})^2 + \sum^s_{t=1} a_{i_ti_t} x^2_{i_t} \ge 0.$$
For the graph without loops the above inequality becomes
$$ x^TQ(G, a)x = \sum a_{i_k j_k} (x_{i_k} - x_{j_k})^2 \ge 0. \eqno{(2.9)}$$
For $x^T = (1~ 1 ~ \cdots 1)$ we can see $x^T Qx= 0$. This means that $x^T = (1~~ 1~~ 1\cdots 1)$ is an unnormalized eigenvector belonging to the eigenvalue 0 \cite{godsil}.  If there are two components $(G_1, a)$ and $(G_2, a)$ of $(G, a)$, with $n_1, m_1$ and $n_2, m_2$ being the numbers of vertices and edges in $(G_1, a)$ and $(G_2, a)$ respectively, we can decompose the sum in Eq.(2.9)  as
$$ x^TQ(G, a)x = \sum^{m_1}_{k_1=1} a_{i_{k_1}j_{k_1}} (x_{i_{k_1}} - x_{j_{k_1}})^2 + \sum^{m_2}_{k_2=1} a_{i_{k_2}j_{k_2}} (x_{i_{k_2}} - x_{j_{k_2}})^2 . \eqno{(2.10)}$$
 For $x^T = (1~1~1 \cdots 1)$ both the terms in Eq. (2.10) vanish.  Now consider two vectors $x_1^T = ( 0 ~ 0 ~ \cdots 0  1 ~ 1 \cdots 1)$ with first $n_1$ components zero and last $n_2$ components 1 and $x_2^{T} = (1~~ 1\cdots 1~~0~~0 \cdots 0)$ with first $n_1$ components 1 and last $n_2$ components zero, $(n_1 + n_2 = n)$.  Obviously the RHS of Eq. (2.10) vanishes for both $x_1$ and $x_2$.   This implies $x_1$ and $x_2$ are two orthogonal eigenvectors with eigenvalue zero.  This means that the multiplicity of zero eigen value is at least 2 (number of components in $(G, a)).$

The equality condition for $a_{uv}=$constant$ > 0 ~~ \forall ~~ \{ u, v\} \in E(G, a)$ is proved in \cite{bgs06}. \hspace{\stretch{1}}$ \blacksquare$

\noi {\bf Theorem 2.2.3 :} The necessary and sufficient condition for the state given by a graph $(G,a )$ to be pure is
$$ \sum^n_{i=1}  \mathfrak{d}^2_i + 2 \sum^m_{k=1} a^2_{i_kj_k} =  \mathfrak{d}^2_{(G, a)} \eqno{(2.11)}$$
where $ \mathfrak{d}_i$ is the degree of the vertex $v_i$, $a_{i_kj_k}$ is the weight of the edge $\{ v_{i_k}, v_{j_k}\},(v_{i_k} \ne v_{j_k})$ and $\mathfrak{d}_{(G, a)}$ is the degree sum $\mathfrak{d}_{(G, a)} = \sum\limits^n_{i=1}  \mathfrak{d}_i$.

\noi {\bf Proof :} Equation (2.11) is just the restatement of the requirement
 $Tr(\si^2(G, a))$ $=1$, which is the necessary and sufficient condition for the state $\si(G,a)$ to be pure.\hspace{\stretch{1}}$ \blacksquare$

\noi {\bf Lemma 2.2.4 :} The graph $(G,a)$ for a pure state $\si(G,a)$ has the form $(K_\ell, b) \uplus v_{\ell +1} \uplus v_{\ell +2} \uplus \cdots \uplus v_n$ for some $1 \le \ell \le n$.

\noi {\bf Proof :} Since the state is pure, it has the form
$$ | \psi \ran = \sum^\ell_{k=1} c_{i_k} |v_{i_k}\ran, 1 \le i_k \le n.$$
We can permute the basis vectors to transform this sum to $|\psi\ran = \sum\limits^\ell_{i=1} c_i|v_i\ran$.  That is, the $\ell$ basis kets contributing to the sum in the above equation become the vectors $|v_1\ran, |v_2\ran, \cdots, |v_\ell\ran$ under this permutation.  The resulting density matrix $|\psi\ran \lan \psi|$ has a block of first $\ell$ rows and first $\ell$ columns all of whose elements are nonzero, while all the other elements of density matrix are zero.  The graph corresponding to this density matrix is just the required graph. \hspace{\stretch{1}}$ \blacksquare$

\noi{\it Example (2.2)} :  We now give important cases of pure state graphs in $\mathbb{R}^2$ which we use later.

\begin{figure}[!ht]
\includegraphics[width=3cm,height=0.7cm]{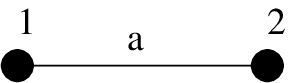}

Figure 2.4
\end{figure}

(i)~~$\si(K_2, a) = \fr{1}{2a_{12}} \left[ \ba{cc} a_{12} & - a_{12} \\ -   a_{12} & a_{12} \ea \right] = \fr{1}{2} \left[ \ba{cc} 1 & -1 \\ -1 & 1 \ea \right]   = P[ \fr{1}{\sq 2} (|v_1\ran - |v_2\ran]$, the corresponding graph is as shown in Figure 2.4.
 \begin{figure}[!ht]
\includegraphics[width=3cm,height=1cm]{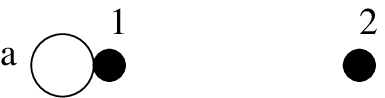}

Figure 2.5
\end{figure}

(ii)~~ $ \si(K_1, a) = \fr{1}{a} \left[ \ba{cc} a & 0 \\ 0 & 0 \ea \right] = \left[\ba{cc} 1 & 0 \\ 0 & 0 \ea \right] = P[|v_1\ran]$, the corresponding graph is as shown in Figure 2.5.
\begin{figure}[!ht]
\includegraphics[width=3cm,height=1cm]{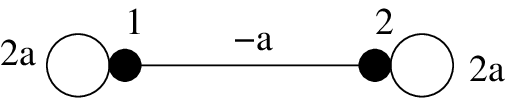}

Figure 2.6
\end{figure}

(iii)~~ $ a_{12} > 0,  \si(K_2, -a) = \fr{1}{2a_{12}} \left[ \ba{cc} a_{12} & a_{12} \\ a_{12} & a_{12} \ea \right]  = \fr{1}{2} \left[ \ba{cc} 1 & 1 \\ 1 & 1 \ea \right] = P[ \fr{1}{\sq 2} (|v_1\ran + |v_2\ran)], a > 0$. The corresponding graph is as shown in Figure 2.6.

\begin{figure}[!ht]
\includegraphics[width=3cm,height=3cm]{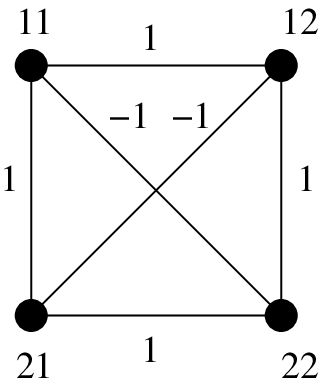}

Figure 2.7
\end{figure}
(iv) $$ \si(G, a) = \fr{1}{4} \left[ \ba{cccc} 1 & -1 & -1 & 1 \\ -1 &  1 &  1 & -1 \\ -1 & 1& 1 & -1\\ 1 & -1 & -1 & 1 \ea \right]= P[ (|-\ran  |-\ran)],$$ where $ |-\ran=\fr{1}{\sq 2}(|1\ran - |2\ran),$ the corresponding graph is as shown in Figure 2.7.
 \begin{figure}[!ht]
\includegraphics[width=4cm,height=4cm]{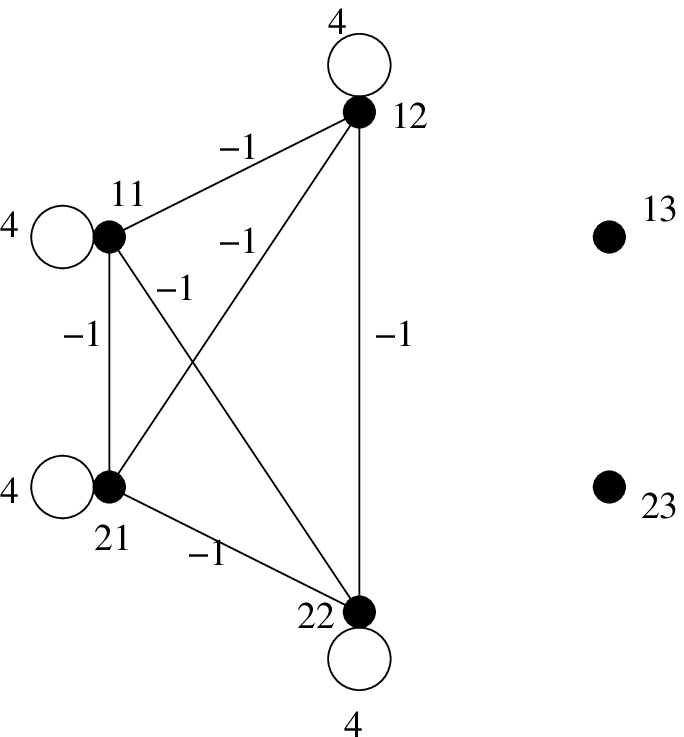}

Figure 2.8
\end{figure}

(v) $$ \si(G, a) = \fr{1}{4} \left[ \ba{cccccc} 1 & 1 & 0 & 1 & 1 & 0 \\ 1 &  1 & 0 &  1 & 1 & 0 \\ 0 & 0 & 0 & 0 & 0 &0 \\1 & 1& 0 &  1 & 1 & 0 \\ 1 & 1 & 0 & 1 & 1 & 0\\ 0 & 0 & 0 & 0 & 0 & 0 \ea \right]= P[ (|+\ran  |+\ran)],$$ where $ |+\ran=\fr{1}{\sq 2}(|1\ran + |2\ran)$
in $\mathbb{R}^2 \otimes \mathbb{R}^3$, the corresponding graph is as shown in Figure 2.8.

It may be seen that in each of the cases in example (2.2), same density matrix on the standard basis corresponds to infinite family of graphs as the  nonzero weight on  each edge or loop is multiplied by a constant.  But this is a false alarm because any weight $a \ne 1$ only changes the length of the corresponding state vector in the Hilbert space (i.e. state becomes unnormalized) which does not have any physical significance.  Another example pertaining to this situation is the random mixture (see lemma (2.3.1 )).
$$ \si(G, a) = \fr{1}{an} \left[ \ba{cccc} a \\ & a & & 0 \\ & & \ddots \\ 0 & & & a \ea \right] = \fr{1}{n} \left[ \ba{cccc} 1 & & & 0 \\ & 1 & & \\ & & 1 \\ & & \ddots \\ 0 & & & 1 \ea \right] = \fr{1}{n} I_n.$$
However, this does not lead to any contradiction because of the uniqueness of the random mixture \cite{perb93}.

All the  density matrices in (i), (ii), (iii), (iv), (v)   above represent pure states.

\noi{\bf Remark 2.2.5 :} Any graph with the weight function $a=$ constant $>0$ has the same density matrix for all $a>0$. This infinite family of graphs corresponds to the same quantum state (density operator).

\noi {\bf Definition 2.2.6 :} A graph $(H, b)$ is said to be a factor of graph $(G, a)$ if $V(H, b) = V(G, a)$ and there exists a graph $(H', b')$ such that $V(H', b') = V(G, a)$ and $M(G, a) = M(H, b) + M(H', b')$.  Thus a factor is only a spanning subgraph.  Note that
$$ a_{v_iv_j} = \left\{ \ba{lll} b_{v_iv_j} & \mbox{if} & \{v_i,v_j\} \in E(H, b) \\ b'_{v_iv_j} & \mbox{if} & \{v_i,v_j\} \in E(H', b'). \ea \right. $$
Now let  $(G, a)$ be a graph on $n$ vertices $v_1, \cdots, v_n$ having $m$ edges\\ $\{v_{i_1}, v_{j_1}\}, \cdots, \{v_{i_m}, v_{j_m}\}$ and $s$ loops $\{v_{i_1}, v_{i_1}\} \cdots \{v_{i_s}, v_{i_s}\}$ where $1 \le i_1,j_1,$ $\cdots, i_m, j_m \le n, \;\;1 \le i_1, i_2, \cdots, i_s \le n$.

Let $(H_{i_kj_k}, a_{i_kj_k})$ be the factor of $(G, a)$ such that
$$ [M(H_{i_kj_k}, a_{i_kj_k})]_{u,w} = \left\{ \ba{l} a_{i_kj_k} ~~ \mbox{if}~~ u = i_k~~ \mbox{and}~~ w = j_k ~~\mbox{or}~~ u = j_k, w = i_k \\ 0  ~~~~~~ \mbox{otherwise}. \ea \right. \eqno{(2.12)}$$
Let $(H_{i_t,i_t}, a_{i_t i_t})$ be a factor of $(G, a)$ such that
$$ [M(H_{i_ti_t}, a_{i_t i_t})]_{uw} = \left\{ \ba{l} a_{i_t i_t}~~ \mbox{when}~~ u = i_t = w \\ 0 ~~~~~ \mbox{otherwise} \ea \right. \eqno{(2.13)}$$

\noi {\bf Theorem 2.2.7 :} The density matrix of a graph $(G, a)$ as defined above with factors given by equations (2.12) and (2.13) can be decomposed as
$$ \si(G, a) = \fr{1}{\mathfrak{d}_{(G, a)}} \sum^m_{k=1} 2a_{i_kj_k} \si(H_{i_kj_k}, a_{i_kj_k}) + \fr{1}{\mathfrak{d}_{(G, a)}} \sum^s_{t=1} a_{i_ti_t} \si(H_{i_ti_t}, a_{i_ti_t}) $$
or
$$ \si(G, a) = \fr{1}{\mathfrak{d}_{(G, a)}} \sum^m_{k=1} 2a_{i_kj_k} P[\fr{1}{\sq 2}(|v_{i_k}\ran - |v_{j_k}\ran)] + \fr{1}{\mathfrak{d}_{(G, a)}}  \sum^s_{t=1} a_{i_ti_t} P[|v_{i_t}\ran]$$

\noi {\bf Proof :} From equations (2.12), (2.13) and theorem 2.2.3 and lemma 2.2.4, the density matrix
$$\si (H_{i_kj_k}, a_{i_kj_k}) = \fr{1}{2a_{i_kj_k}} [ \D(H_{i_kj_k}, a_{i_kj_k}) - M(H_{i_kj_k},a_{i_kj_k})]$$
is a pure state.  Also,
$$ \si (H_{i_ti_t}, a_{i_ti_t}) = \fr{1}{a_{i_ti_t}} [ \D_0 (H_{i_t, i_t}, a_{i_ti_t})]$$
is a pure state.  Now
$$ \D(G, a) = \sum^m_{k=1} \D(H_{i_kj_k}, a_{i_kj_k}) + \sum^s_{t=1} \D_0(H_{i_ti_t}, a_{i_ti_t})$$
$$M(G, a) = \sum^m_{k=1} M(H_{i_kj_k}, a_{i_kj_k}) + \sum^s_{t=1} \D_0(H_{i_ti_t}, a_{i_ti_t}).$$
Therefore

$$\si(G, a)  =  \fr{1}{\mathfrak{d}_{(G, a)}} \left[ \sum^m_{k=1} \D(H_{i_kj_k}, a_{i_kj_k}) - \sum^m_{k=1} M(H_{i_kj_k}, a_{i_kj_k})\right]$$
 $$+ \fr{1}{\mathfrak{d}_{(G, a)}} \left[ \sum^s_{t=1} \D_0 (H_{i_ti_t}, a_{i_ti_t})\right]$$
$$=  \fr{1}{\mathfrak{d}_{(G, a)}} \sum^m_{k=1} [\D(H_{i_kj_k}, a_{i_kj_k}) - M(H_{i_kj_k}, a_{i_kj_k})] \\
 + \fr{1}{\mathfrak{d}_{(G, a)}} \sum^s_{t=1} \D_0(H_{i_ti_t}, a_{i_ti_t})$$
$$ =  \fr{1}{\mathfrak{d}_{(G, a)}} \sum_k 2a_{i_kj_k} \si(H_{i_kj_k}, a_{i_kj_k}) \\
 + \fr{1}{\mathfrak{d}_{(G, a)}} \sum_t a_{i_ti_t} \si(H_{i_ti_t}, a_{i_ti_t}) ~~~~~~~~~~\mbox{(2.14)}$$

 In terms of the standard basis, the $uw$-th element of matrices $\si(H_{i_kj_k}, a_{i_kj_k})$ and $\si(H_{i_ti_t}, a_{i_ti_t})$ are given by $\lan v_u | \si(H_{i_kj_k} , ,a_{i_kj_k}) | v_w \ran$ and $\lan v_u | \si (H_{i_ti_t}, a_{i_ti_t}) | v_w\ran$ respectively.  In this basis
$$ \si(H_{i_kj_k}, a_{i_kj_k}) = P[ \fr{1}{\sq 2} ( | v_{i_k} \ran - | v_{j_k} \ran )]$$
$$ \si(H_{i_ti_t}, a_{i_ti_t}) = P[| v_{i_t} \ran ] .$$

Therefore equation (2.14) becomes
$$\si(G, a)  =  \fr{1}{\mathfrak{d}_{(G, a)}} \sum^m_{k=1} 2a_{i_kj_k} P[\fr{1}{\sq 2} (| v_{i_k}\ran - | v_{j_k} \ran) + \fr{1}{\mathfrak{d}_{(G, a)}} \sum ^s_{t=1} a_{i_ti_t} P[ | v_{i_t} \ran]~~~~~~~~~~\mbox{(2.15)}$$
$\hspace{\stretch{1}} \blacksquare$

\noi {\bf Remark 2.2.8 :} If all weights $a_{i_kj_k} > 0$ then equations (2.14), (2.15) give $\si(G, a)$ as a mixture of pure states.  However, in the next subsection we show that any graph $(G, a)$ having a density matrix can be decomposed into graphs (spanning subgraphs) corresponding to pure states.

\subsection{Convex combination of density matrices}

Consider two graphs $(G_1, a_1) $ and $(G_2, a_2)$ each on the same $n$ vertices, having $\si(G_1, a_1)$ and $\si(G_2, a_2)$ as their density matrices, respectively.  We give an algorithm to construct the graph $(G, a)$ whose density matrix is
$$ \si(G, a) = \la \si(G_1, a_1) + (1 - \la) \si(G_2, a_2)$$
with
$0 \le \la \le 1, \la = \al/\b, \al, \b > 0 $ being real.

We use the symbol $\sqcup$ to denote the union of the edge sets of two graphs $(G_1, a_1) $ and $(G_2, a_2)$ on the same set of vertices to give $(G,a).$ If $\{v_i,v_j\} \in E(G_1,a_1)$ and $\{v_i,v_j\} \in E(G_2,a_2)$ then $a(\{v_i,v_j\}) = a_1(\{v_i,v_j\}) + a_2(\{v_i,v_j\})$. We write $(G,a)= (G_1, a_1) \sqcup(G_2, a_2).$
 If $E(G_1,a_1)$ and $E(G_2,a_2)$ are disjoint sets, then we call the resulting graph $(G,a)$ the disjoint edge union of $(G_1,a_1)$ and $(G_2,a_2)$, we write $(G,a)= (G_1, a_1) \dotplus(G_2, a_2).$

The algorithm is as follows :

\noi {\bf Algorithm 2.2.9 :}

\bd
\i(i) Put $\la = \al/\b$ so that $(1 - \la) = \fr{\b - \al}{\b}$, where $\al>0$, $\b>0$ are real.

\i(ii) Write $\si(G, a) = \fr{1}{\b} (\al \si(G_1,a_1) + (\b - \al) \si(G_2, a_2))$.

\i(iii) Modify the weight functions of the two graphs $(G_1, a_1)$ and $(G_2, a_2)$ to get $a_1' = \al a_1$ and $a_2' = (\b - \al)a_2$.

\i(iv) The graph $(G, a)$ corresponding to $\si$ in step (ii) is
$$ (G, a) = (G_1, a_1') \sqcup (G_2, a_2') \eqno{(2.16)}$$
such that
$$ a_{v_iv_j} = (a_1')_{v_iv_j} + (a_2')_{v_iv_j} \eqno{(2.16a)}$$
$$ a_{v_iv_i} = (a_1')_{v_iv_i} + (a_2')_{v_iv_i} \eqno{(2.16b)}$$
where we take $(a_{1,2}')_{v_iv_j} = 0 = (a_{1,2}')_{v_iv_i}$  if $\{v_i,v_j\},\{v_i,v_i\} \not\in E(G_1, a_1)$ or $E(G_2, a_2)$

\ed $\hspace{\stretch{1}}$ $\blacksquare$

We can apply this algorithm to any convex combination of more than two density matrices $\si(G, a)= \sum\limits^k_{i=1} p_i \si(G_i, a_i),\; \sum\limits_i p_i = 1$, by writing $p_i = \al_i/\b, \al_i,\b >0$ and real$, i = 1, \cdots, k$.

\noi{\it Example(2.3)} : consider the density matrices

\bd
\i(i) $\si(G_1, a_1) = | + + \ran \lan + + | = \fr{1}{4} \left[ \ba{cccc} 1 & 1 & 1 & 1\\ 1 & 1 & 1 & 1 \\ 1 & 1 & 1 &1 \\ 1 & 1 & 1 & 1 \ea \right]$\\
whose graph is shown in Figure 2.9\\
\begin{figure}[!ht]
\includegraphics[width=4cm,height=3cm]{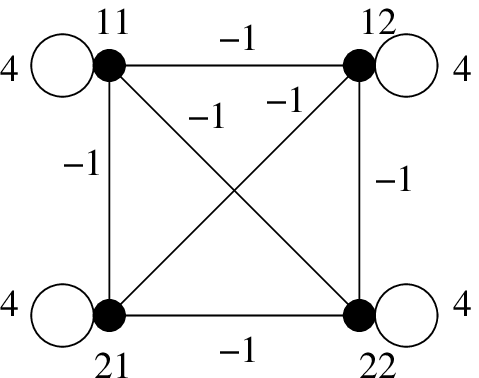}

Figure 2.9
\end{figure}
\benrr
\si(G_2, a_2) & = & \fr{1}{2} | 11 \ran \lan 11 | + \fr{1}{2} | \psi^+ \ran \lan \psi^+ | \\
& = & \fr{1}{4} \left[ \ba{cccc} 2 & 0 & 0 & 0 \\ 0 & 1 & 1 & 0 \\ 0 & 1 & 1 & 0 \\ 0 & 0 & 0 & 0 \ea \right].
\eenrr
whose graph is shown in Figure 2.10
\begin{figure}[!ht]
\includegraphics[width=4cm,height=3cm]{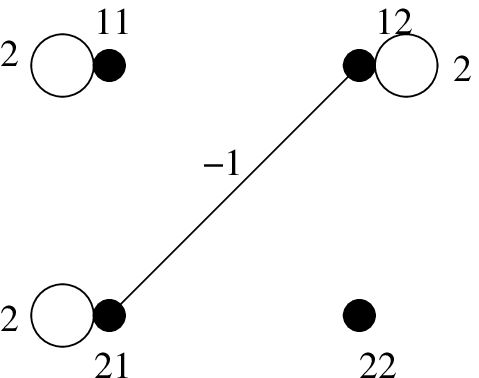}

Figure 2.10
\end{figure}

The graph corresponding to
\benrr
\si(G, a) & = & \fr{1}{3}  \si(G_1, a_1) + \fr{2}{3} \si(G_2, a_2) \\
& = & \fr{1}{12} \left[ \ba{cccc} 5 & 1 & 1 & 1 \\ 1 & 3 & 3 & 1 \\ 1 & 3 & 3 & 1 \\ 1 & 1 & 1 &1 \ea \right].
\eenrr
is given in Figure 2.11
\begin{figure}[!ht]
\includegraphics[width=4cm,height=3cm]{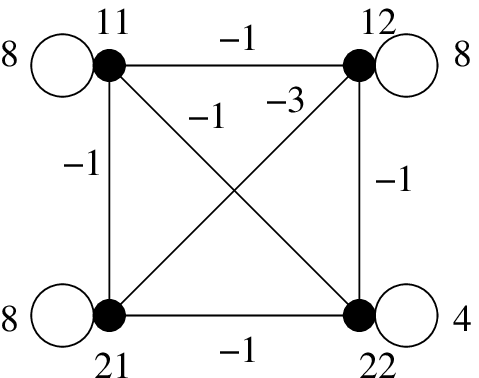}

Figure 2.11
\end{figure}

\ed

\noi {\bf Lemma 2.2.10 :} Let $(G_1, a_1)$, $(G_2, a_2)$ and  $(G, a)$ satisfy

$$(G, a) = (G_1, a_1) \sqcup (G_2, a_2) $$  or,
$$(G,a)= (G_1, a_1) \dotplus(G_2, a_2).$$

Then
$$ Q(G, a) = Q(G_1, a_1) + Q(G_2, a_2) $$
and
$$ \si(G, a) = \fr{\mathfrak{d}_{(G_1,a_1)}}{\mathfrak{d}_{(G, a)}}   \si(G_1, a_1) + \fr{\mathfrak{d}_{(G_2,a_2)}}{\mathfrak{d}_{(G, a)}}   \si(G_2, a_2).$$

\noi {\bf Proof :} For two factors of $(G, a), (G_1, a_1)$ and $(G_2, a_2)$ we have
\benrr
M(G, a) & = & M(G_1, a_1) + M(G_2, a_2) \\
\D(G, a) & = & \D(G_1, a_1) +  \D(G_2, a_2)\\
\D_0(G, a) & = & \D_0(G_1, a_1) + \D_0(G_2, a_2) \\
L(G, a) & = & \D(G, a) - M(G, a)\\
Q(G, a) & = & L(G, a) + \D_0(G, a)
\eenrr
Substitute  $M(G, a), \D(G, a), \D_0(G, a)$ and $L(G, a)$ in $Q(G, a)$ as above to get
$$Q(G, a) = Q(G_1, a_1) + Q(G_2, a_2)$$
and also
$$ \si(G, a) = \fr{\mathfrak{d}_{(G_1, a_1)}}{\mathfrak{d}_{(G, a)}} \si(G_1, a_1) + \fr{\mathfrak{d}_{(G_2, a_2)}}{\mathfrak{d}_{(G, a)}} \si(G_2, a_2).$$ \hspace{\stretch{1}}$ \blacksquare$

\noi {\bf Remark 2.2.11 :} Obviously, the operation $\sqcup$ is associative. We can apply  lemma 2.2.10 for more than two graphs,
 $$(G, a) = \sqcup_i (G_i, a_i) \Ra Q(G, a) = \sum_i Q(G_i, a_i)$$
and
$$ \si(G, a) = \fr{1}{\mathfrak{d}_{(G, a)}} \sum_i d(G_i, a_i) \si(G_i, a_i). $$

\noi {\bf Theorem 2.2.12 :} Every graph $(G, a)$ having a density matrix $\si(G, a)$ can be decomposed as $(G, a) = \sqcup_i (G_i, a_i)$ where $\si(G_i, a_i)$ is a pure state.

\noi {\bf Proof :} Every density matrix can be written as the convex combination of pure states $\si(G, a) = \sum\limits^k_{i=1} p_i | \psi_i \ran \lan \psi_i|$.

By applying algorithm 2.2.9, lemma 2.2.10 and remark 2.2.11, we get the result.\hspace{\stretch{1}}$ \blacksquare$

\subsection{Tracing out a part}

Consider a bipartite system with dimension $pq$. Let $\si(G,a)$ be a state of the system with graph $(G,a)$ having $pq$ vertices labeled by $(ij), i=1,\cdots, p $ and $j=1, \cdots, q$. If we trace out the second part with dimension $q$, we get the state of the first part which is $p \times p$ reduced density matrix of $\si(G,a)$. The corresponding graph $(G',a')$ has $p$ vertices indexed by $(i)$ and its weight function $a'$ is given by $$ a'_{ij}=\sum_{k=1}^q a_{ik,jk} , i\neq j$$ and $$a'_{ii}=\sum_{k=1}^q \mathfrak{d}_{ik}-\sum_{l\in V(G',a')} a'_{il},\;\; l\neq i,$$ where $\mathfrak{d}_{ik}$ is the degree of vertex $(ik)$ in original graph.

\noi{\it Example (2.2.4)} : Consider a graph $(G,a)$ as shown in Figure 2.12a in $\mathbb{R}^2 \otimes \mathbb{R}^2$. The corresponding density matrix is $$\si^{AB}(G, a) =  \fr{1}{16} \left[ \ba{cccc} 9 & -1 & -1 & 1\\ -1 & 3 & -1 & -1 \\ -1 & -1 & 3 & -1 \\ 1 & -1 & -1 & 1 \ea \right].$$\\
After tracing out the second particle the graph $(G',a')$ on two vertices becomes as in Figure 2.12b with corresponding density matrix $$\si^A(G', a') = \fr{1}{16} \left[ \ba{cc} 12 & -2\\ -2 & 4 \ea \right]=\fr{1}{8} \left[\ba{cc} 6 & -1\\ -1 & 2\ea \right]$$\\
which is the same as the reduced density matrix $\si^A$ of $\si^{AB}$.

\begin{figure}[!ht]
\includegraphics[width=6cm,height=3cm]{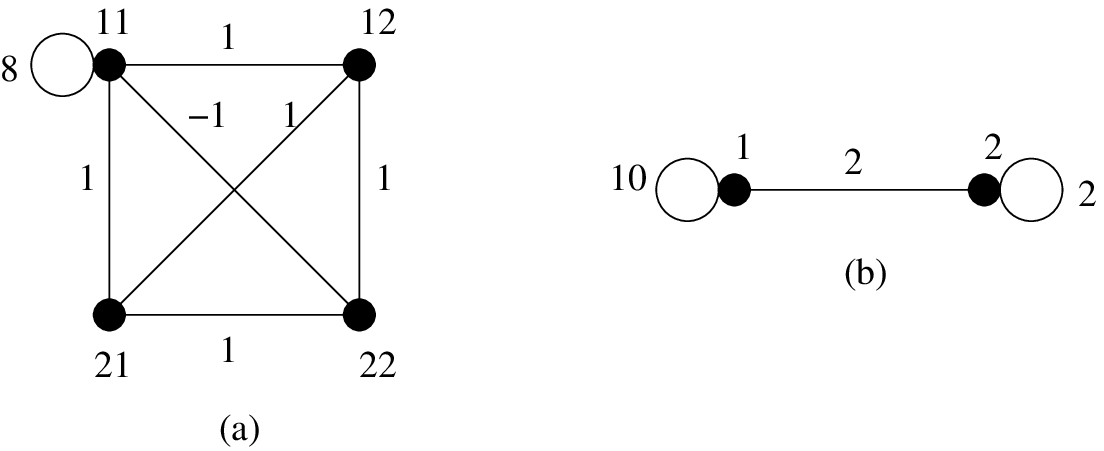}

Figure 2.12
\end{figure}

\section{Von Neumann entropy}

The Von Neumann entropy of the $n \times n$ density matrix $\si$ is
$$ S(\si) = - \sum^n_{i=1} \la_i(\si) \log_2 \la_i(\si) $$
It is conventional to define $0 \log 0 = 0$.  The Von Neumann entropy is a measure of mixedness of the density matrix.  For a pure state $\si, S(\si)= 0$.

\subsection{Maximum and minimum}

  Let
$$ (G, a) = \uplus^n_{i=1} (K^i_1, a_i) \eqno{(2.17)}$$
where $(K^i_1, a_i)$ is the graph on $i$-th vertex with a loop having weight $a_i > 0$.

\noi {\bf Lemma 2.3.1 :} Let $(G, a)$ be given by Eq.(2.17) with the additional constraint that $a_i = c = \fr{1}{n}, i = 1, 2, \cdots, n$.  The density matrix of the graph $(G, a)$ is the random mixture of pure states with $\si(G, a) =  \fr{1}{n} I_n$.

\noi {\bf Proof :} For the graph $(G, a)$, the first term in Eq.(2.14) vanishes. Then
$$ \si(G, a) = \fr{1}{\mathfrak{d}_{(G,a)}} \sum^n_{t=1} \D_0(H_{i_ti_t}, a)$$
where $\D_0(H_{i_ti_t}, a)$ is the $n \times n$ matrix with all elements zero except  the $(i_t, i_t)th$ element which is equal to $a$. This means
$$ \si(G, a) = \fr{a}{\mathfrak{d}_{(G,a)}} \left[ \ba{cccc} 1 & & & 0 \\ & 1 & & \\ & & \ddots \\ 0 & & & 1 \ea \right] = \fr{1}{n} I_n,$$
because $\mathfrak{d}_{(G,a)}=na$.\hspace{\stretch{1}}$ \blacksquare$

\noi {\bf Theorem 2.3.2 :} Let $(G, a)$ be a graph on $n$ vertices. Then
\bd
\i(i) $\max_{(G, a)} S(\si(G, a)) = \log_2 n$

\i(ii) $\min_{(G,a)} S(\si(G, a)) = 0$, and this value is attained if $\si(G, a)$ is pure.
\ed

\noi {\bf Proof :}

 (i) By lemma 2.3.1 $\si(G, a)$ defined in the lemma has eigenvalue $1/n$ with multiplicity $n$.  The corresponding Von Neumann entropy is $\log_2 n$.  Since $(G, a)$ is on $n$ vertices, the support of $\si(G, a)$ has dimension $\le n$.  Any matrix having dimension of support $\le n$ cannot have Von Neumann entropy $> \log_2 n$.

(ii) For a pure state $S(\si(G,a))=0$ and $S(\si(G,a))\nless 0 .$ \hspace{\stretch{1}}$ \blacksquare$

\section{Separability}

In this section we primarily deal with the graphs representing a bipartite quantum system with Hilbert space $\mathbb{R}^p \otimes \mathbb{R}^q$ of dimension $pq$. Obviously, the corresponding graph has $n = pq$ vertices.  We label the vertices using standard (product) basis  $\{| v_i \ran = | u_{s+1} \ran \otimes |w_t \ran \},\; 0 \le s \le p-1,\; 1 \le t \le q, i = sq + t$.

\subsection{Tensor product of weighted graphs}

The tensor product of two graphs $(G, a)$ and $(H, b)$ denoted $(G, a) \otimes (H, b)$ is defined as follows.

The vertex set of $(G, a) \otimes (H, b)$ is $V(G, a) \times V(H, b)$. Two vertices $(u_1, v_1)$ and $(u_2, v_2)$  are adjacent if $\{u_1, u_2\} \in E(G, a)$ and $\{v_1, v_2\} \in E(H, b)$.  The weight of the edge $\{ (u_1, v_1), (u_2, v_2)\}$ given by $a_{\{u_1, u_2\}} b_{\{v_1,v_2\}}$ and is denoted by $c(\{ (u_1, v_1), (u_2, v_2)\})$.  Note that either $u_1$ and $u_2$ or $v_1$ and $v_2$ or both can be identical, to include loops.

The adjacency, degree and the loops matrices of $(G, a) \otimes (H, b)$ are given by
$$M((G, a) \otimes (H, b)) = M(G, a) \otimes M(H, b) \eqno{(2.18a)}$$
$$ \D((G, a) \otimes (H, b)) = \D(G, a) \otimes \D(H, b) \eqno{(2.18b)}$$
$$ \D_0((G, a) \otimes (H, b)) = \D_0(G, a) \otimes \D_0(H, b) \eqno{(2.18c)}$$
Note that
$$ L((G, a) \otimes (H, b)) \ne L(G, a) \otimes L(H, b) $$
$$Q((G, a) \otimes (H, b)) \ne Q(G, a) \otimes Q(H, b).$$
In fact, in general, the tensor product of two graphs having a density matrix may not have a density matrix.

For two simple graphs $G$ and $H$ we know that \cite{ikb00,bgs06}
$$\mathfrak{d}_{G \otimes H} =  \mathfrak{d}_G \cdot  \mathfrak{d}_H.$$
This result is also satisfied by the tensor product of the weighted graphs.
$$ \mathfrak{d}_{(G, a) \otimes (H, b)} = \mathfrak{d}_{(G, a)} \cdot \mathfrak{d}_{(H, b)} . \eqno{(2.19)}$$

\subsection{Modified tensor product}

We modify the tensor product of graphs in order to preserve positivity of the generalized Laplacian of the resulting graph.

Given a graph $(G, a)$ we define $(G^\phi, a)$ by
$$ V(G^\phi, a) = V(G, a)$$
$$E[(G^\phi, a)] = E(G, a)\setminus \{ \{v_i,v_i\} :  \{v_i, v_i\} \in E(G, a)\}$$
That is, $(G^\phi, a)$ is obtained from $(G, a)$ by removing all loops.

Given a graph $(G, a)$ we define $(\widetilde{G}, a)$ by
$$ V(\widetilde{G}, a) = V(G, a)$$
$$E(\widetilde{ G}, a) = E(G, a) \setminus \{ \{v_i, v_j\} : i \ne j, \{v_i, v_j\} \in E(G, a)\}.$$
That is, $(\widetilde{ G}, a)$ is obtained by removing all edges connecting neighbors and keeping loops.

Note that in both $(G^\phi, a)$ and $(\widetilde{ G}, a)$, the weight function $a$ remains the same, only its support is restricted.

Given a graph $(G, a)$ we define $(-G, a) = (G, - a)$.  Given a graph $(G, a)$ we define $(G^\#, a')$
$$V(G^\#,a')= V(G,a)$$
$$(G^\#,a')=\uplus_i^n(K_i,a'_i),$$
where $K_i$ is the graph consisting of $i$th vertex with a loop and $a'_i$ is the weight of the loop on the $i$th vertex. If $a'_i=0$ then there is no loop on the $i$th vertex. $a'_i, i=1,2,\cdots n$ are given by $$a'_i=\sum_{v_k \in V(G,a)}a(\{v_i,v_k\}).\eqno{(2.20a)}$$
Note that the term $v_k=v_i$ is also included in the sum.

We now define the graph operators on the set of graphs
$$ \left. \ba{ll}  \mbox{(i)} & \eta : (G, a) \ra (- G, a) = (G ,-a) \\ \mbox{(ii)} & \cL : (G, a) \ra (G^\phi, a) \\ \mbox{(iii)} & \cN : (G, a) \ra (G^\#, a') \\ \mbox{(iv)} & \Om : (G, a) \ra (\widetilde{ G}, a) \ea \right\}. \eqno{(2.20b)}$$

Some properties of the graph operators defined in (2.20b) are,
$$ \ba{ll} \mbox{(i)} & M(\eta(G, a)) = - M(G, a) \\ & \D(\eta(G, a)) = - \D(G, a) \\ & \D_0(\eta(G, a)) = - \D_0(G, a) \ea \eqno{(2.21)}$$
$$ \mathfrak{d}_{\eta(G, a)} = - \mathfrak{d}_{(G, a)} $$
$$ \ba{ll} \mbox{(ii)} & M(\cL(G, a)) = M(G, a) - \D_0(G, a)  \\ & \D(\cL(G, a)) = \D(G, a) - \D_0(G, a)  \ea \eqno{(2.22)}$$
$$ \D_0(\cL(G, a)) = [0] $$
$$ \mathfrak{d}_{\cL(G, a)} = Tr(\D(G, a)) - Tr(\D_0(G, a)) = \mathfrak{d}_{(G^\phi, a)}$$
$$ \ba{ll} \mbox{(iii)} & M(\cN(G, a)) = \D(G, a) \\ & \D(\cN(G, a)) = \D(G, a) \\ & \D_0(\cN(G, a)) = \D(G, a)  \\ & \mathfrak{d}_{\cN(G, a)} = Tr(\D(G, a)) = \mathfrak{d}_{(G, a)} \ea \eqno{(2.23)}$$
$$ \ba{ll} \mbox{(iv)} &  M(\Om(G, a)) = \D_0(G, a) \\& \D(\Om(G, a)) =\D_0(G, a) \\& \D_0(\Om(G, a)) = \D_0(G, a) \\ &  \mathfrak{d}_{\Om(G, a)} = Tr(\D_0(G, a)) \ea \eqno{(2.24)}$$

\noi{\it Example (2.5)} :
Given a graph $(G,a)$ as shown in figure 2.13a, if we act by $\eta$,
 $\cL$ ,$\cN$ and $\Om$ on $(G,a)$, we get the graphs $\eta(G,a)$, $\cL(G,a)$, $\cN(G,a)$
 and $\Om(G,a),$  as shown in figures 2.13(b), (c), (d) and (e), respectively.

\begin{figure}
\includegraphics[width=12cm,height=5cm]{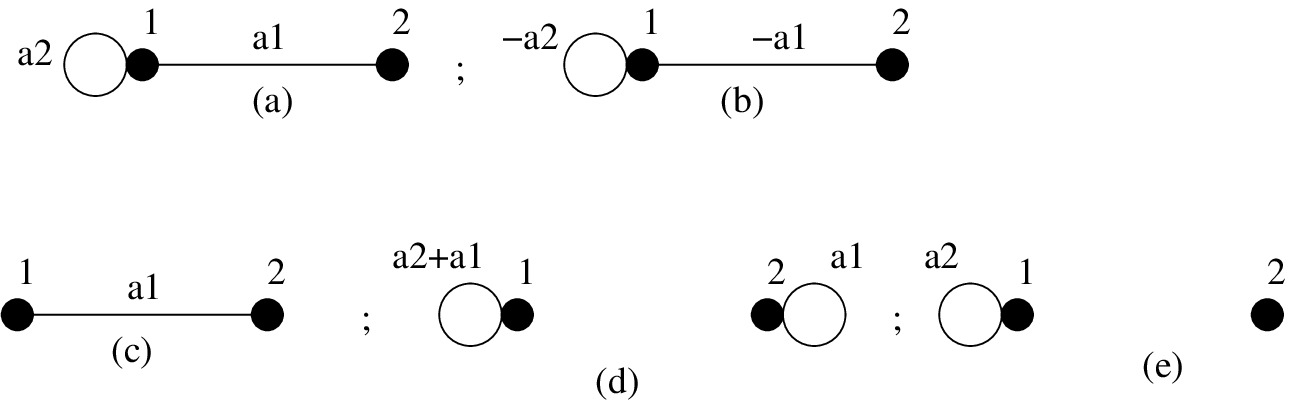}

Figure 2.13
\end{figure}

\noi {\bf Definition 2.4.1 :} Let $(G, a)$ and $(H, b)$ be two graphs with $p$ and $q\; (> p)$ vertices, respectively.  Then their modified tensor product is defined by
\benrr
(G, a) \boxdot (H, b) & = &\{ \cL(G, a) \otimes \cL \eta (H, b)\} \dotplus \{\cL(G, a) \otimes \cN(H, b)\}\\
& & \dotplus \{ \cN(G, a) \otimes \cL(H, b)\} \dotplus \{ \Om(G, a) \otimes \Om(H, b)\},~~~~~~~~~\mbox{(2.25)}
\eenrr
and
$$ V\{ (G, a) \boxdot (H, b)\} = V(G, a) \times V(H, b) $$
whose cardinality is $pq$.

$E\{ (G, a) \boxdot (H, b) \} $ = disjoint union of the edge set of each term in (2.25).

\noi {\bf Lemma 2.4.2 :}

(i)  $\D((G, a) \boxdot (H, b)) = \D(G, a) \otimes \D(H, b)$.

(ii) $\D_0((G, a) \boxdot (H, b)) = \D_0(G, a) \otimes \D_0(H, b)$.

\noi {\bf Proof :} Consider the degree matrix of the modified tensor product. We have
\benrr
 \D((G, a) \boxdot (H, b))& =& \D(\cL(G, a) \otimes \cL\eta(H, b)) + \D(\cL(G,a) \otimes \cN(H, b))\\
& & + \D(\cN(G, a) \otimes \cL(H, b)) + \D(\Om(G, a) \otimes \Om(H, b)). \\
\eenrr
This follows from lemma 2.2.10. Using  equation (2.18b) and equations (2.21) to (2.24)  to the terms on the RHS of the above equation  we get

$$\D((G, a) \boxdot (H,b)) = \D(G, a) \otimes \D(H, b).$$
Equation (ii) can be proved similarly.\hfill $\blacksquare$

\noi {\bf Corollary 2.4.3 :} $\mathfrak{d}_{(G,a) \boxdot (H,b)}(v_1,v_2)\; =\;\mathfrak{d}_{(G,a)}(v_1) \cdot \mathfrak{d}_{(H,b)}(v_2)$

\noi {\bf Proof :} This follows directly from equation (i) in lemma 2.4.2.\hfill $\blacksquare$

\noi {\bf Remark 2.4.4 :} From corollary  we get $\mathfrak{d}_{(G,a)\boxdot(H,b)}\; =\;\mathfrak{d}_{(G,a)} \cdot \mathfrak{d}_{(H,b)}$

\noi {\bf Theorem 2.4.5 :} Consider a bipartite system in $\mathbb{R}^p \otimes \mathbb{R}^q$ in the state $\si$. Then $\si = \si_1 \otimes \si_2$ if and only if $\si$ is the density matrix of the graph $(G, a) \boxdot (H, b)$, where $(G, a)$ and $(H, b)$ are the graphs having density matrices $\si_1$ and $\si_2,$ respectively.

\noi {\bf Proof :} \noi {\it If part :} Given $(G, a), (H, b)$ we want to prove
$$\si((G, a) \boxdot (H, b)) = \si_1(G, a) \otimes \si_2(H, b).$$

 From the definition of the modified tensor product we can write
$$ \si((G,a) \boxdot (H, b)) = \fr{1}{\mathfrak{d}_{(G, a) \boxdot (H, b)}} \{Q[\cL(G, a) \otimes \cL \eta(H, b)$$
$$ \dotplus \cL(G, a) \otimes \cN(H, b)\dotplus  \cN(G, a) \otimes \cL(H, b) \dotplus \Om(G, a) \otimes \Om(H, b)]\}$$
Using lemma 2.2.10, remark 2.2.11  and remark 2.4.4 we get
\benrr
\si((G, a) \boxdot (H, b)) & = & \fr{1}{\mathfrak{d}_{(G, a)} \cdot \mathfrak{d}_{(H, b)}} [ Q(\cL(G, a) \otimes \cL \eta(H, b)) \\
& & + Q(\cL(G, a) \otimes \cN(H, b)) + Q(\cN(G, a)\\
& &  \otimes \cL(H, b)) + Q(\Om(G, a) \otimes \Om(H, b))]. ~~~~~~~~~~~~~ \mbox{(2.26)}
\eenrr
We can calculate every term in (2.26) using (2.21) - (2.24) and substitute in (2.26) to get
$$ \si((G, a) \boxdot (H, b)) = \si(G, a) \otimes \si(H, b) .$$

\noi {\it Only if part :} Given $\si = \si_1 \otimes \si_2$, consider the graphs $(G, a)$ and $(H, b)$ for $\si_1$ and $\si_2$ respectively.  Then the graph of $\si$ has the generalized Laplacian

 $$[L(G, a) + \D_0(G, a)] \otimes [L(H, b) + \D_0(H, b)] $$
 $$ = L(G, a) \otimes L(H, b) + L(G, a) \otimes \D_0(G, a)$$
 $$ + \D_0(G, a) \otimes L(H, b) + \D_0(G, a) \otimes \D_0(H, b).$$

 Now it is straightforward to check that the graphs corresponding to each term are given by the corresponding terms in the definition of $(G, a) \boxdot (H, b)$. \hspace{\stretch{1}}$ \blacksquare$

 \noi {\bf Remark 2.4.6 :} Note that the proof of Theorem 2.4.5 does not depend in any way on the positivity or the Hermiticity of the associated generalized Laplacians. Therefore we have $$Q((G,a) \boxdot(H,b))\;=\;Q(G,a)\otimes Q(H,b)$$ for any two graphs $(G,a)$ and $(H,b)$

\noi {\bf Corollary 2.4.7 :} The modified tensor product is associative and distributive with respect to the disjoint edge union $\dotplus$.

\noi {\bf Proof :} Let $(G_1,a_1), (G_2,a_2)$ and $(G_3,a_3)$ be any graphs. Using theorem 2.4.5 and remark 2.4.6, we can write
\benrr
Q(((G_1,a_1) \boxdot(G_2,a_2))\boxdot(G_3,a_3))\\
& &=Q((G_1,a_1) \boxdot(G_2,a_2)) \otimes Q(G_3,a_3) \\
& &=\;(Q(G_1,a_1) \otimes Q(G_2,a_2))\otimes Q(G_3,a_3) \\
& &=\;Q(G_1,a_1) \otimes (Q(G_2,a_2)\otimes Q(G_3,a_3))\\
& &=\;Q(G_1,a_1) \otimes Q((G_2,a_2)\boxdot (G_3,a_3) )\\
& &=\;Q((G_1,a_1)\boxdot ((G_2,a_2)\boxdot (G_3,a_3) )
\eenrr
 Therefore,
$$((G_1,a_1) \boxdot(G_2,a_2))\boxdot(G_3,a_3)\;=\;(G_1,a_1)\boxdot ((G_2,a_2)\boxdot(G_3,a_3) ).$$

Similarly, using lemma 2.2.10 and distributive property of the matrix tensor product we get

$$Q((G_1,a_1) \boxdot((G_2,a_2))\dotplus (G_3,a_3)))\;$$
$$=\;Q((G_1,a_1) \boxdot(G_2,a_2)) \dotplus ((G_1,a_1)  \boxdot (G_3,a_3)), $$
 which gives
$$(G_1,a_1) \boxdot ((G_2,a_2)) \dotplus (G_3,a_3))\;=\; (G_1,a_1) \boxdot (G_2,a_2) \dotplus (G_1,a_1)  \boxdot (G_3,a_3). $$ \hspace{\stretch{1}}$ \blacksquare$

\noi {\bf Definition 2.4.8 :} The Cartesian product of two weighted graphs $(G,a)$ and $(H,b)$  is denoted $(G,a) \square (H,b)$ with weight function $c$ defined as follows:
$$V(G, a) \times V(H, b).$$
$E((G,a ) \square (H,b))=\{\{(u,v),(x,y)\} |\; u=x$ and $\{v,y\} \in E(H,b),\; v \ne y,\; c(\{(u,v),(u,y)\}) =  \mathfrak{d}_u \cdot b(\{v,y\}) $ or  $v = y$ and $\{u,x\} \in E(G,a), \; u \ne x,\; c(\{(u,v),(x,v)\})=  \mathfrak{d}_v \cdot a(\{u,x\}),$

where $ \mathfrak{d}_u$ and $ \mathfrak{d}_v$ are the degrees of the vertices $u \in V(G,a)$ and $v \in V(H,b),$ respectively. It is straightforward to check that

$$(G,a) \square (H,b)\;=\; \cL(G, a) \otimes \cN(H, b) \dotplus \cN(G, a) \otimes \cL(H, b),$$

which can be taken to be the definition of the Cartesian product of graphs in terms of the operators $\cL$ and $ \cN$ . We also note that

$$
(G, a) \boxdot (H, b)  = \{ \cL(G, a) \otimes \cL \eta (H, b)\} \dotplus \{(G,a) \square (H,b)\} \dotplus \{ \Om(G, a) \otimes \Om(H, b)\}. $$

Note that the isolated vertices in $(G,a)$ or $(H,b)$ do not contribute to $(G,a) \square (H,b)$ as their degree is zero.

\noi{\it Example(2.6)} : Consider $(G,a), (H,b)$ where $V(G,a)=\{1,2\}, E(G,a)=\{ \{1,2\}\} $ and $V(H,b)=\{1,2,3,4\}, E(H,b)=\{\{1,2\}, \{2,3\}, \{3,4\}\}$ with weight functions $a=b=1$,  as shown in figure 2.14a, b. The modified tensor product of these graphs is given by figures 2.15a, b, c, d, for each term in (2.25), and the resulting graph is as shown in figure 2.15e. The corresponding density matrix of the graph $(G,a)\boxdot(H,b)$ is

$$\si((G, a)\boxdot (H,b)) \; = \; \fr{1}{12} \left[ \ba{cccccccc} 1 & -1 & 0 & 0 & -1 & 1 &   0 & 0 \\-1 & 2 & -1 & 0 & 1 & -2 & 1 & 0 \\ 0 & -1 & 2 & -1 & 0 & 1 & -2 & 1 \\ 0 & 0 & -1 & 1 & 0 & 0 & 1 & -1 \\ -1 & 1 & 0 & 0 & 1 & -1 & 0 & 0 \\ 1 & -2 & 1 & 0 & -1 & 2 & -1 & 0 \\ 0 & 1 & -2 & 1 & 0 & -1 & 2 & -1 \\ 0 & 0 & 1 & -1 & 0 & 0 & -1 & 1 \ea \right],$$

 which is the same as  $\si(G,a)\otimes \si (H,b).$

\begin{figure}[!ht]
\includegraphics[width=6cm,height=3cm]{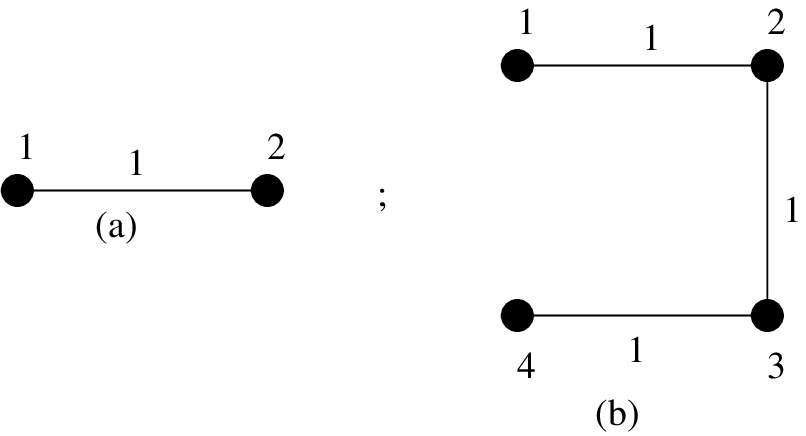}

Figure 2.14
\end{figure}

\begin{figure}[!ht]
\includegraphics[width=8cm,height=6cm]{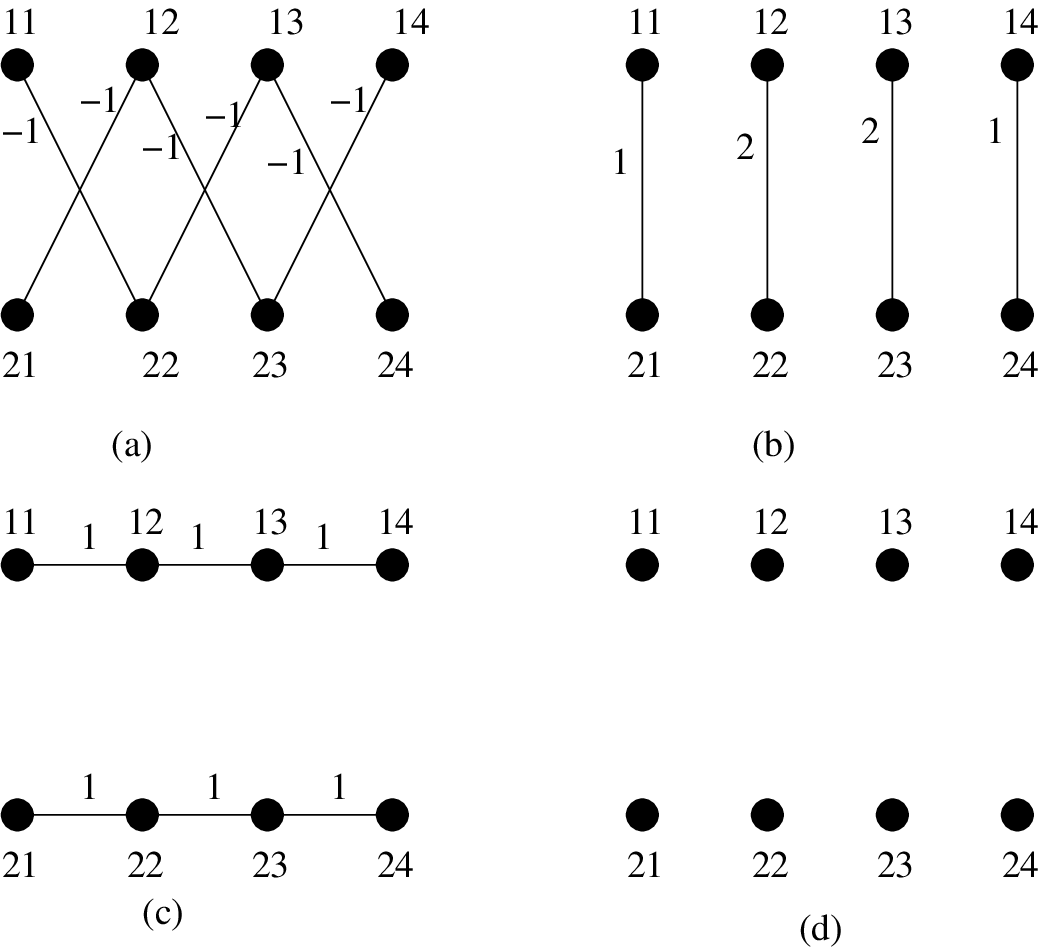}

\includegraphics[width=4cm,height=3cm]{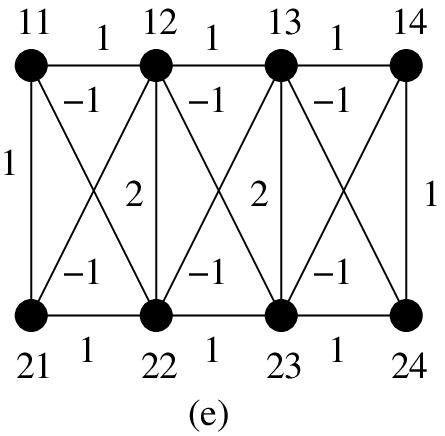}

Figure 2.15

\end{figure}

\noi {\bf Corollary 2.4.9 :} The density matrix of the modified tensor product of two graphs is separable.

\noi {\bf Proof :} From theorem 2.4.5, we see that $\si((G, a) \boxdot (H, b))$ is actually a product state.\hspace{\stretch{1}}$ \blacksquare$

\noi {\bf Corollary 2.4.10 :} $\si = \si_1 \otimes \si_2 \otimes \cdots \otimes \si_k$ for a $k$-partite system if and only if the graph of $\si$ is the modified tensor product of the graphs of $\si_1, \cdots, \si_k$.

\noi {\bf Proof :} Apply theorem 2.4.2 successively to $(\si_1 \otimes \si_2), ((\si_1 \otimes \si_2) \otimes \si_3)\cdots $ and then use the associativity of the modified tensor product corollary 2.4.7. \hspace{\stretch{1}}$ \blacksquare$

\noi {\bf Corollary 2.4.11 :} A state $\si$ of a $k$-partite system is separable if and only if the graph (G,a)  for  $\si$ has the form
$$ (G,a)\;=\; \sqcup_i \boxdot^k_{j=1} (G^j_i, a^j_i). $$

\noi {\bf Proof :} Let $\si$ be separable, i.e.,
$$ \si = \sum_i w_i \si_i^{(1)} \otimes \si_i^{(2)} \otimes \cdots \otimes \si_i^{(k)}, ~~ \sum_i w_i = 1.$$

By algorithm 2.2.9 and corollary 2.4.10 the graph of $\si$ has the form
$$(G, a) = \sqcup_i \boxdot^k_{j=1} (G^j_i, a^j_i).$$
Now let the graph of a $k$-partite state be
$$(G, a) = \sqcup_i \boxdot^k_{j=1} (G^j_i, a^j_i).$$
Then by lemma 2.2.10, remark 2.2.11 and the above corollary to theorem 2.4.5
$$ \si(G, a) = \sum_i w_i \si_1^{(1)} \otimes \si_1^{(2)} \otimes \cdots \otimes \si_i^{(k)}.$$ \hspace{\stretch{1}}$ \blacksquare$

  Corolary 2.4.11 says that Werner's definition \cite{wer89} of a separable state in the   $\mathbb{R}^{q_1} \otimes \mathbb{R}^{q_2} \otimes \mathbb{R}^{q_3} \otimes \cdots \otimes \mathbb{R}^{q_k}$ system can be expressed using corresponding graphs.

\noi {\bf Lemma 2.4.12 :} For any $n = pq$ the density matrix $\si(K_n, a)$ is separable in $\mathbb{R}^p \otimes \mathbb{R}^q$ if the weight function is constant $> 0$.

\noi {\bf Proof :} The proof is same as that given for corollary 2.4.3 in \cite{bgs06}, for simple graph.\hspace{\stretch{1}}$ \blacksquare$

\noi{\it Example (2.7)} : Consider the graph $(K_4, a)$.  The vertices of $(K_4, a)$ are denoted by 1, 2, 3, 4, where weight function is constant, say , $a = 3 > 0$ and has loops in vertices 1, 2.  We associate to these vertices the orthonormal basis $\{ |1\ran = |1\ran |1\ran, |2\ran = |1\ran |2\ran, |3\ran = |2\ran |1\ran, |4\ran = |2\ran |2\ran\}$.  In terms of this basis $\si(K_4, a)$ can be written as
$$ \si(K_4, a) = \fr{1}{42} \left[ \ba{rrrr} 12 & -3 & -3 & -3 \\ -3 & 12 & -3 & -3 \\ -3 & -3 & 9 & -3 \\ -3 & -3 & -3 & 9 \ea \right] = \fr{1}{14} \left[ \ba{rrrr} 4 & -1 & -1 & -1 \\ -1 & 4 & -1 & -1 \\ -1 & -1 & 3 & -1 \\ -1 & -1 & -1 & 3 \ea \right], $$
and from equation (2.15) we can write $\si(K_4, a)$ as
$$
\si(K_4, a) = \fr{1}{42} \{6P[|1\ran \fr{1}{\sq 2} (|1\ran - |2\ran)] + 6P[ \fr{1}{\sq 2} (|1\ran - |2\ran)|1\ran]$$
$$ + 6P[ \fr{1}{\sq 2} (|1\ran - |2\ran) |2\ran] + 6P[|2\ran \fr{1}{\sq 2} (|1\ran - |2\ran)] + 6P[\fr{1}{\sq 2} (|11\ran - |22\ran)] $$
$$ +6P[ \fr{1}{\sq 2} (|12\ran - |21\ran)] + 3P[|11\ran] + 3P[|12\ran]\}$$

\benrr
\si(K_4, a) & = & \fr{1}{7} P[|1\ran \fr{1}{\sq 2} (|1\ran - |2\ran)] + \fr{1}{7} P[ \fr{1}{\sq 2}(|1\ran - |2\ran) |1\ran] \\
& & + \fr{1}{7} P[ \fr{1}{\sq 2} ( |1\ran - |2\ran)|2\ran] + \fr{1}{7} P[|2\ran \fr{1}{\sq 2} (|1\ran - |2\ran)] \\
& & + \fr{2}{7} \{ \fr{1}{2} P[ \fr{1}{\sq 2} (|11\ran - |22\ran)] + \fr{1}{2} P[ \fr{1}{\sq 2} (|12\ran - |21\ran)]\} \\
& &+ \fr{1}{14} P[|11\ran] + \fr{1}{14} P[|12\ran].
\eenrr
Each of the first four terms in the above expression is a projector on a product state, and also the last two terms are  projectors, while the fifth and sixth terms give rise to the separable density matrix $\fr{1}{2} p[\mid - \ran \mid + \ran] + \fr{1}{2} p[| + \ran | - \ran]$, where $| \pm\ran \stackrel{def}{=} \fr{1}{\sq 2} (|1\ran \pm |2\ran)$ \cite{bgs06}.  Thus $\si(K_4, a)$, $a$ constant, is separable in $\mathbb{R}^2 \otimes \mathbb{R}^2$.

Note that there exists a graph which is complete with a  real weight function, which is entangled as the  following graph shows in figure 2.16

\begin{figure}[!ht]
\includegraphics[width=2cm,height=2cm]{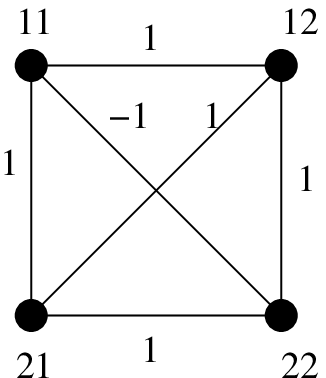}

Figure 2.16
\end{figure}

\noi {\bf Remark 2.4.13 :} The separability of $\si(K_n, a)$ with constant weight function $> 0$ does not depend upon the labeling of $V(K_n, a)$ provided every vertex has  a loop or there are no loops.  Given a graph, an isomorphism from $(G, a) \longmapsto (G, a)$ is called automorphism.  Under composition of maps, the set of automorphisms of $(G, a)$ form a group denoted $Aut(G, a)$.  If $\si(K_n, a)$ is separable, and if the $Aut(K_n, a) = S_n ,\; (G, a) \cong (K_n, a)$ is also separable. Note that $Aut(K_n, a) = S_n$ provided all weights are equal and either every vertex has a loop or there are no loops.

\noi {\bf Lemma 2.4.14 :} The complete graph $(K_n, a)$ on $n \geq 2$ vertices corresponding to a separable state with weight function $ constant > 0$ is not a modified tensor product of two graphs.

\noi {\bf Proof :} It is clear that if $n$ is prime then $(K_n, a)$ is not a  modified tensor product of graphs.  We then assume that $n$ is not a prime.  Suppose that there exist graphs $(G, b)$ and $(H, c)$ respectively on $p$ and $s$ vertices such that $(K_{ps}, a) = (G, b) \boxdot (H, c)$, where $ b$ and $c$ are constants.  From the definition of the modified tensor product
$$ a(\{ (u_1,v_1), (u_2, v_2)\}) = b(\{ u_1, u_2\}) \cdot c(\{ v_1, v_2\}),$$
the degree sum is
$$ \mathfrak{d}_{(G, b)} = \sum_{u \in V(G, b) }  \mathfrak{d}_u = \sum_{u \in V(G, b)} \sum_{w \in V(G, b)} b_{wv} = 2b|E(G, b)|.$$
We know that
$ \mathfrak{d}_{(G, b)} \le b(p(p-1))$ and also $\mathfrak{d}_{(H, c)} = 2c|E(H, b)| \le cs(s-1)$ and
$$ \mathfrak{d}_{(G, b)}\cdot \mathfrak{d}_{(H, c)} \le bcps(p-1) (s-1) = bcps(ps - p - s + 1) .\eqno{(2.27)}$$
Now observe that $V((G, b) \boxdot (H, c)) = ps$ and,
$$ \mathfrak{d}_{(G, b) \boxdot (H, c)} = aps(ps -1) \eqno{(2.28)}$$
because $(G, b) \boxdot (H, c) = (K_{ps}, a)$.

We know that
$$ \mathfrak{d}_{(G, b) \boxdot (H, c)} = \mathfrak{d}_{(G, b)}\cdot \mathfrak{d}_{(H, c)}. \eqno{(2.29)}$$
 Substituting from Eq.(2.27) and Eq.(2.28) we see that Eq.(2.29) is satisfied only when $ p = 1 = s $ , i.e. $n = ps = 1$.\hspace{\stretch{1}}$ \blacksquare$

Lemma 2.4.12, lemma 2.4.14 and theorem 2.4.5 together imply that a complete graph $(K_n,a)$ on $n \ge2$  vertices with $a=$ constant $>0$ is a separable state but  not a product state.

\noi{\bf Definition 2.4.15 :} Consider a graph $(G,a)$, without loops, pertaining to a bipartite system of dimension $pq$ . The partial transpose of $(G,a)$, denoted $(G^{\Gamma_B},a')$, is a graph defined as $V(G^{\Gamma_B},a')=V(G,a)$, $\{il,kj\}\in E(G^{\Gamma_B},a') \Longleftrightarrow \{ij,kl\}\in E(G,a)$ and $a'(\{il,kj\})=a(\{ij,kl\}).$

\noi{\bf Lemma 2.4.16 :} Consider a bipartite separable state $\si(G,a)$ with the associated graph $(G,a)$ without loops. Then $\Delta(G,a)=\Delta(G^{\Gamma_B},a'),$  where $(G^{\Gamma_B},a')$ is the partial transpose of $(G,a)$.

\noi{\bf Proof :} Let $Q(G,a)$ be the Laplacian of a graph $(G,a)$ with  real  weights without  loops, on $n$ vertices. Let $D$ be any $n\times n$ real diagonal matrix in the standard orthonormal basis $\{|v_i\ran\};i=1,2,\dots,n$, such that $D\neq 0$ and $Tr(D)=0$. This means that there is at least one negative entry in the diagonal of $D$. Denote this element by $D_{ii}=b_i$. Let $|\psi_0 \ran= \sum_j|v_j\ran$ and $|\phi\ran=\sum_j \chi_j|v_j \ran$  where
\begin{displaymath}
\chi_j =
\left\{ \begin{array}{ll}
 0 & \textrm{if $j\neq i$}\\
k \in R & \textrm{if $j=i$}
\end{array} \right.
\end{displaymath}
Let $|\chi \ran=|\psi\ran+|\phi\ran=\sum_{i=1}^n(1+\chi_j)|v_j\ran$. Then
\benrr
\lan\chi|Q(G,a)+D|\chi\ran& = & \lan\psi_0|Q(G,a)|\psi_0\ran+
\lan\psi_0|Q(G,a)|\phi\ran+\lan\phi|Q(G,a)|\psi_0\ran+\\
& &              \lan\phi|Q(G,a)|\phi\ran+\lan\psi_0|D|\psi_0\ran+\lan\psi_0|D|\phi\ran+\lan\phi|D|\psi_0\ran+\\
& &\lan\phi|D|\phi\ran
\eenrr
 Since $|\psi_0\ran$ is (unnormalized) vector having all components equal unity, from equation (2.9) it follows that $\lan\psi_0|Q(G,a)|\psi_0\ran=0$. Also $\lan\psi_0|D|\psi_0\ran=Tr(D)=0$.
  We have
 $$\lan\phi|Q(G,a)|\phi\ran=k^2(Q(G,a))_{ii}=k^2d_i$$
 $$\lan\psi_0|Q(G,a)|\phi\ran=\lan\phi|Q(G,a)|\psi_0\ran = 0.$$

 Finally, the remaining terms in the above equation are given by
 $$\lan\phi|D|\phi\ran=b_ik^2$$
 $$\lan\psi_0|D|\phi\ran = b_ik=\lan\phi|D|\psi_0\ran.$$
 Thus
 $$\lan\chi|Q(G,a)+D|\chi\ran = k^2(b_i+d_i)+ 2kb_i.$$
  So we can then always choose a positive $k$ , such that
 $$\lan\chi|Q(G,a)+D|\chi\ran < 0.$$
 It then follows $Q(G,a)+D \ngeq 0.$

  This expression is identical with that obtained in \cite{bgmsw}.
  For any graph $G$ on $n=pq$ vertices $$v_1=u_1w_1, v_2=u_1w_2, \dots, v_{pq}=u_pw_q,$$ consider  the degree condition $\Delta(G) = \Delta(G^{\Gamma_B}).$ Now $$(L(G))^{\Gamma_B}=(\Delta(G) - \Delta(G^{\Gamma_B}))+ L(G^{\Gamma_B}).$$
  Let $$D=\Delta(G) - \Delta(G^{\Gamma_B}).$$
  Then $D$ is an $n\times n$ real diagonal matrix with respect to the orthonormal basis $$|v_i\ran = |u_1\ran\otimes |w_1\ran, \dots, |v_{pq}\ran = |u_p\ran\otimes |w_q\ran.$$
  Also $$Tr(D)=Tr(\Delta(G))-Tr(\Delta(G^{\Gamma_B}))= 0.$$
  We have two possible cases : $D\neq 0$ or $D = 0$. If $D\neq 0$, that is the degree condition is not satisfied $(i.e. \Delta(G)\neq \Delta(G^{\Gamma_B})),$ we have seen that $L(G)+ D\ngeq 0$. As a consequence, $L(G^{\Gamma_B})+D\ngeq 0$ and then $(L(G))^{\Gamma_B}\ngeq 0.$ Hence $\si(G) $ is entangled.\hspace{\stretch{1}}$ \blacksquare$

\noi{\bf Lemma 2.4.17 :} A graph $(G,a)$  for a bipartite state corresponds to a separable state if $\{ij,kl\}$ $(i\neq k,j\neq l) \in E(G,a)\Longrightarrow \{il,kj\}\in E(G,a)$ and $a_{ij,kl}=a_{il,kj}$.

 \noi{\bf Proof :} Suppose $a_{ij,kl}=a_{il,kj}=a,i\neq k,j\neq l.$ The contribution of the corresponding two edges is $$a \{ P[\fr{1}{\sq2}(|ij\ran-|kl\ran)]+P[\fr{1}{\sq2}(|il\ran-|kj\ran)]\},$$ which is a separable state. Thus all such pairs contribute separable states.  Any other edge $\{ij,kl\}$ with $i=k$ or $ j=l$ has the contribution $a_{ij,kl}P[|i\ran \otimes (\fr{1}{\sq2}(|j\ran-|l\ran))]$ which is separable.  Loops contribute the product states $P[|ii\ran]$. \hspace{\stretch{1}}$ \blacksquare$

The reverse implication is not true in general. The counter-example is the graph (figure 2.12a) in example (2.4) which is separable.

\section{Graph Operators}

A graph operation is a map that takes a graph to another graph \cite{prisner95}. We deal with four cases, namely deleting and adding an edge and deleting and adding a vertex.

Deleting an edge $(\{ v_i, v_j\}, a_{v_iv_j})$ from a graph $(G, a)$ results in a graph\\ $(G, a) -  ( \{ v_i, v_j\}, a_{v_iv_j}) \stackrel{def}{=} (V(G, a), E(G, a) \setminus \{ v_i, v_j\})$ with $a_{v_iv_j} = 0$.  Note the possibility $v_i = v_j$ corresponding to the edge being a loop.  Addition of an edge $(\{ u_i, v_j\}, a_{ij})$ maps $(G, a)$ to the graph $(G, a) + (\{ v_i, v_j\}, a_{ij}) \stackrel{def}{=} [V(G, a), E(G, a) \cup \{ v_i, v_j\}]$ with $a_{v_iv_j} = a_{ij}$.  Deletion of a vertex $v_i$ maps $(G,a)$ to $(G, a) - \{ v_i\} \stackrel{def}{=} [V(G, a) \setminus \{ v_i\}, E(G, a) \setminus E_i]$ where $E_i$ is the set of all edges incident to $v_i$ (including the loop on $v_i)$ with the weight function zero for the edges in $E_i$.  Adding a vertex $v_i$ to $(G, a)$ maps $(G, a)$ to $(G, a) + \{ v_i\} \stackrel{def}{=} (V(G, a) \cup \{ v_i\}, E(G, a))$.

A very important point is that, in general, the set of graphs having a density matrix is not closed under these operations.  Addition of an edge with positive weight and deletion of an edge with negative weight preserves the positivity of the generalized Laplacian resulting in the graph having a density matrix.  However, addition (deletion) of an edge with negative (positive) weight may lead to a graph which does not have a density matrix.  In the next section, we give a method for addition and deletion of an edge which preserves the positivity of the generalized Laplacian.  Deletion and addition of vertices always preserves the positivity of the generalized Laplacian.

Let $\cB(\cH^n)$ be the space of all bounded linear operators on $\cH^n$.  A linear map $\Lambda : \cB(\cH^n) \ra \cB(\cH^m)$ is said to be hermiticity preserving if for every Hermitian operator $O \in \cB(\cH^n), \Lambda(O)$ is an Hermitian operator in $\cB(\cH^m)$.  A hermiticity preserving map $\Lambda : \cB(\cH^n) \ra \cB(\cH^m)$ is said to be positive if for any positive operator $O \in \cB(\cH^n), \Lambda(O)$ is a positive operator in $\cB(\cH^m)$.  A positive map $\Lambda : \cB(\cH^n) \ra \cB(H^m)$ is said to be completely positive if for each positive integer $k, (\Lambda \otimes I_{k^2}) : \cB(\cH^n \otimes \cH^k) \ra \cB(H^m \otimes \cH^k)$ is again a positive map.  A completely positive map $\Lambda : \cB(\cH^n) \ra \cB(\cH^m)$ is said to be trace preserving if $Tr(\Lambda(O)) = Tr(O)$, for all $O \in \cB(\cH^n)$.  A quantum operation is a trace preserving completely positive map (for short, TPCP) \cite{kraus83,ncb00}.  In standard quantum mechanics, any physical transformation of a quantum mechanical system is described by a quantum operation \cite{perb93}.  We are going to use the following result:

\noi {\bf (Kraus representation  Theorem)} \cite{preskill} : Given a quantum operation $\Lambda : \cB(\cH^n) \ra \cB(\cH^m)$, there exist $m \times n$ matrices $A_i$, such that $\Lambda(\rho) = \sum\limits_i A_i \rho A_i^\dagger$, where $\rho$ is any density matrix acting on $\cH^n$ and $\sum\limits_i A^\dagger_i A_i = I_m$ (the converse is true).  The matrices $A_i$'s are called Kraus operators.

A projective measurement $\cM = \{ P_i; i = 1, 2, \cdots, n\}$, on a quantum mechanical system $S$ whose state is $\rho$, consists of pairwise orthogonal projectors $P_i : \cH_s \ra \cH_s$, such that $\sum\limits^n_{i=1} P_i = I_{dim(\cH_s)}$.  The $i$-th outcome of the measurement occurs with probability $Tr(P_i \rho)$ and the post-measurement state of $S$ is $\fr{P_i\rho P_i}{tr(P_i \rho)}.$ Whenever the $i$-th outcome of the measurement occurs, we say that $P_i$ clicks. Last two paragraphs apply to complex Hilbert space and so also to real Hilbert space.

\subsection{Deletion and addition of an edge for a weighted graph with all weights $> 0$ }

Here we describe how to delete or add an edge by means of TPCP.  Our method of deleting an edge from a weighted graph with all positive weights is a simple generalization of the method in \cite{bgs06}.  Let $(G, a)$ be a graph on $n$ vertices $v_1, \cdots, v_n$ and $m$ edges $\{v_{i_1} v_{j_1}\} \cdots \{v_{i_m} v_{j_m}\}, i_k \ne j_k, k = 1, \cdots, m$ and $s$ loops $\{ v_{i_1} v_{i_1} \} \cdots \{ v_{i_s}v_{i_s}\}$.  Our purpose is to delete the edge $\{ v_{i_k} v_{j_k}\}, i_k \ne j_k$.  Then we have
\benrr
 \si(G,a) = \fr{1}{\mathfrak{d}_{(G,a)}} \big\{ \sum^m_{\ell=1} 2a_{i_\ell j_\ell} P[\fr{1}{\sq 2} (| v_{i_\ell} \ran - | v_{j_\ell} \ran)]  + \sum^s_{t=1} a_{i_ti_t} P[|v_{i_t} \ran ]\big\}
\eenrr
and
$$
\si((G, a) - \{ v_{i_k} v_{j_k}\})  = $$
$$ \fr{1}{\mathfrak{d}_{(G,a)} - 2a_{i_kj_k}}  \left\{ \sum^m_{\ba{c} \ell =1 \\ \ell \ne k \ea} 2a_{i_\ell j_{\ell}} P [ \fr{1}{\sq 2} (|v_{i_\ell} \ran - | v_{j_\ell} \ran)] + \sum^s_{t=1} a_{i_ti_t} P[|v_{i_t}\ran ]\right\}.
$$

A measurement in the basis $\cM = \{ \fr{1}{\sq 2} (| v_{i_k} \ran \pm | v_{j_k} \ran), |v_i\ran : i \ne i_k, j_k$ and $i = 1, 2, \cdots, n\}$ is performed on the system prepared in the state $\si(G, a)$.
The probability that $P_+ = P[\fr{1}{\sq 2} (|v_{i_k} \ran + | v_{j_k} \ran )]$ clicks is
$$
 Tr[P_+ \si(G, a)] = \sum^n_{i=1} \lan v_i | P_+ \si(G, a) | v_i \ran $$
 $$= \fr{1}{2 \mathfrak{d}_{(G, a)}} \{ \sum^m_{\ba{c} \ell = 1 \\ \ell \ne k \ea} a_{i_\ell j_\ell}[\del_{i_k i_\ell} - \del_{i_k j_\ell} + \del_{j_ki_\ell} - \del_{j_kj_\ell}]^2  + \sum^s_{t=1} a_{i_t i_t}(\del_{i_ti_k} +\del_{i_tj_k})^2\} \eqno{(2.30)}
$$

The state after the measurement is $P [ \fr{1}{\sq 2} (|v_{i_k} \ran + | v_{j_k} \ran]$.  Let $U^+_{k \ell} $ and $U^+_{kt}$ be $n \times n$ unitary matrices such that $U^+_{k\ell} [ \fr{1}{\sq 2} (| v_{i_k} \ran + | v_{j_k} \ran)] = \fr{1}{\sq 2} (|v_{i_\ell} \ran - | v_{j_\ell} \ran)$ for $\ell = 1, \cdots, k -1, k+1, \cdots, m$ and $U^+_{kt} [ \fr{1}{\sq 2} (| v_{i_k} \ran + | v_{j_k} \ran )] = | v_{i_t} \ran, t = 1, \cdots, s$. Now, with probability $2a_{i_\ell j_\ell}/(\mathfrak{d}_{(G, a)} - 2a_{i_kj_k})$
we apply $U^+_{k \ell}$ on $P [ \fr{1}{\sq 2} (|v_{i_k} \ran + | v_{j_k} \ran]$ for each $\ell = 1, \cdots, k -1, k+1, \cdots, m$, and with probability $a_{i_ti_t}/(\mathfrak{d}_{(G, a)} - 2a_{i_kj_k})$ we apply $U^+_{k t} $ on $P [ \fr{1}{\sq 2} (|v_{i_k} \ran + | v_{j_k} \ran]$ for each $t = 1, \cdots, s$. Finally we obtain $\si(( G, a) - \{ v_{i_k} v_{j_k} \})$ with probability given by Eq.(2.30). The probability that $ P[  \fr{1}{\sq 2} (|v_{i_k} \ran - | v_{j_k} \ran )]$ clicks is
$$
  \fr{1}{2 \mathfrak{d}_{(G, a)}} \{ \sum^m_{\ba{c} \ell = 1 \\ \ell \ne k \ea} a_{i_\ell j_\ell}[\del_{i_k i_\ell} - \del_{i_k j_\ell} - \del_{j_ki_\ell} + \del_{j_kj_\ell}]^2  + \sum^s_{t=1} a_{i_t i_t}(\del_{i_ti_k} - \del_{i_tj_k})^2\}  \eqno{(2.31)}
$$
the state after measurement is $ P[  \fr{1}{\sq 2} (|v_{i_k} \ran - | v_{j_k} \ran )]$. Let $U^-_{k \ell}$ and $U^-_{k t}$ be $n \times n$ unitary matrices such that
$$ U^-_{k \ell} \fr{1}{\sq 2} (|v_{i_k} \ran - | v_{j_k}\ran) = \fr{1}{\sq 2} (|v_{i_\ell} \ran - | v_{j_\ell} \ran)$$
for $\ell = 1, \cdots, k-1, k+1, \cdots, m$ and
$$ U^-_{k t} \fr{1}{\sq 2} (|v_{i_k} \ran - | v_{j_k}\ran) = |v_{i_t} \ran$$ for $t=1,.....\cdot,s$. With probability $2a_{i_\ell j_\ell}/(\mathfrak{d}_{(G, a)} - 2a_{i_kj_k})$
 we apply $U^-_{k\ell}$ on $P[\fr{1}{\sq 2} (| v_{i_k} \ran - | v_{j_k} \ran )]$ for each $\ell = 1, \cdots, k-1, k+1, \cdots, m$ and with probability $a_{i_ti_t}/(\mathfrak{d}_{(G, a)} - 2a_{i_kj_k})$ we apply $U^-_{kt}$ on $P[ \fr{1}{\sq 2} (| v_{i_k} \ran - | v_{j_k} \ran)]$ for each $t = 1, 2, \cdots, s$.  Finally we obtain $\si(( G, a) - \{ v_{i_k} v_{j_k} \})$ with probability given by Eq.(2.31).

The probability that $P[| v_i\ran]$ where $i \ne i_k,j_k$ and $i = 1, \cdots, n$ clicks is
$$ \fr{1}{\mathfrak{d}_{(G, a)}}\left\{ \sum^m_{\ba{c} \ell =1 \\ \ell \ne k\ea} a_{i_\ell j_\ell} (\del_{i i_\ell} - \del_{ij_\ell})^2 + \sum^s_{t=1} a_{i_ti_t}(\del_{ii_t})^2 \right\} \eqno{(2.32)}$$
and the state after measurement is $P[|v_i\ran]$. Let $U_{i \ell} $ and $U_{it}$ be $n \times n$ unitary matrices such that $U_{i \ell}[ | v_i\ran] = \fr{1}{\sq 2} (| v_{i_\ell} \ran - | v_{j_\ell}\ran]$ for $\ell = 1, \cdots k-1, k+1, \cdots, m$ and $U_{it} [| v_{i} \ran] = | v_{i_t} \ran$ for $t = 1, \cdots, s$.  With probability $2a_{i_\ell j_\ell} /(\mathfrak{d}_{(G, a)} - 2a_{i_k j_k})$ we apply $U_{i\ell}$ on $P[| v_i\ran]$ for each $\ell = 1, \cdots, k-1, k+1, \cdots, m$ and with probability $a_{i_t i_t}/(\mathfrak{d}_{(G, a)} - 2a_{i_k j_k})$ we apply $U_{it}$ on $P[|v_i\ran]$ for each $t = 1, \cdots, s$.

We obtain $\si((G, a) - \{ v_{i_k}, v_{j_k} \})$ with probability given by Eq.(2.32).  This completes the process.

The set of Kraus operators that realizes the TPCP for deleting the edge $\{ v_{i_k}, v_{j_k}\}$ is then
\benrr
& & \{ \sq{\fr{2a_{i_\ell j_\ell}}{\mathfrak{d}_{(G, a)} - 2a_{i_kj_k}}} U^+_{k\ell} P[ \fr{1}{\sq 2} (|v_{i_k} \ran + | v_{j_k}\ran)] ;  \ell = 1, \cdots, k-1, k+1, \cdots, m \}  \\
& &\cup \{\sq{\fr{a_{i_ti_t}}{\mathfrak{d}_{(G,a)} - 2a_{i_kj_k}}} U^+_{kt} P[\fr{1}{\sq 2} (|v_{i_k} \ran + | v_{j_k}\ran)]: t = 1, \cdots, s \} \\
& &\cup \{ \sq{\fr{2a_{i_\ell j_\ell}}{\mathfrak{d}_{(G, a)} - 2a_{i_kj_k}}} U^-_{k\ell} P[\fr{1}{\sq 2} (|v_{i_k} \ran - | v_{j_k}\ran)]: \ell  = 1, \cdots, k-1, k+1, \cdots m\}\\
& & \cup \{ \sq{\fr{a_{i_ti_t}}{\mathfrak{d}_{(G, a)} - 2a_{i_kj_k}}} U^-_{kt} P[\fr{1}{\sq 2} (|v_{i_k} \ran - | v_{j_k}\ran)] : t = 1, \cdots, s\} \\
& & \cup \{ \sq{\fr{2a_{i_\ell j_\ell}}{\mathfrak{d}_{(G, a)} - 2a_{i_k j_k}}} U_{i\ell} P[ | u_i\ran] : i = 1, \cdots, n, i \ne i_k, j_k ;\ell = 1, \cdots, k-1,\\
& &\; \; \;\;\;\;\;\;~~~~~~~~~~~~~~~~~~~~~~~~~k+1, \cdots, m\} \\
& & \cup \{ \sq{\fr{a_{i_t i_t}}{\mathfrak{d}_{(G, a)} - 2a_{i_kj_k}}} U_{it} P[|v_i\ran] : i = 1, \cdots, n, i \ne i_k, j_k; t = 1, \cdots, s \}
\eenrr
The set of Kraus operators that realizes TPCP for adding back edge $\{ v_{i_k}, v_{j_k} \}$ to $(G, a) - \{ v_{i_k} v_{j_k} \}$ is.
\benrr
& & \left\{ \sq{\fr{2a_{i_\ell j_\ell}}{\mathfrak{d}_{(G, a)} + 2a_{i_kj_k}}} V^+_{k\ell} P[\fr{1}{\sq 2} (|v_{i_k}\ran + |v_{j_k}\ran)] : \ell = 1, 2, \cdots, m\right\} \\
& &\cup \left\{ \sq{\fr{a_{i_ti_t}}{\mathfrak{d}_{(G, a)} + 2a_{i_kj_k}}} V^+_{kt} P[\fr{1}{\sq 2} (|v_{i_k}\ran + | v_{j_k}\ran)] : t = 1, \cdots, s\right\} \\
& & \cup \left\{ \sq{\fr{2a_{i_\ell j_\ell}}{\mathfrak{d}_{(G, a)} + 2a_{i_kj_k}}} V^-_{k\ell} P[\fr{1}{\sq 2} (|v_{i_k} \ran - |v_{j_k}\ran)] : \ell = 1, 2, \cdots, m\right\} \\
& & \cup \left\{ \sq{\fr{a_{i_ti_t}}{\mathfrak{d}_{(G, a)} + 2a_{i_kj_k}}} V^-_{kt} P[ \fr{1}{\sq 2} (|v_{i_k}\ran - |v_{j_k}\ran)] : t = 1,  \cdots, s\right\}\\
& & \cup \left\{ \sq{\fr{2a_{i_\ell j_\ell}}{\mathfrak{d}_{(G, a)} + 2a_{i_kj_k}}} V_{i\ell} P[|v_i\ran] : i = 1, 2, \cdots, n, i \ne i_k, j_k, \ell = 1, 2, \cdots, m\right\}\\
& & \cup \left\{ \sq{\fr{a_{i_ti_t}}{\mathfrak{d}_{(G, a)} + 2a_{i_kj_k}}} V_{it} P[|v_i\ran] : i = 1, 2, \cdots, n, i \ne i_k, j_k, t = 1, 2, \cdots, s\right\}
\eenrr
where $V^+_{k\ell}, V^-_{k\ell}, V^-_{kt}, V_{i\ell}, V_{it}$ are $n \times n$ unitary matrix defined as follows:
\benrr
& & V^+_{k\ell} \fr{1}{\sq 2}(|v_{i_k} \ran + | v_{j_k}\ran) = \fr{1}{\sq 2} (|v_{i_\ell}\ran - |v_{j_\ell}\ran), ~~ \mbox{for}~~ \ell = 1, 2, \cdots, m\\
& & V^+_{k t} \fr{1}{\sq 2} (|v_{i_k}\ran + |v_{j_k}\ran) = |v_{i_t}\ran,~~ \mbox{for}~~ t = 1, \cdots, s, \\
& & V^-_{k \ell} \fr{1}{\sq 2} (|v_{i_k} \ran - |v_{j_k}\ran) = \fr{1}{\sq 2} (|v_{i_\ell} \ran - |v_{j_\ell}\ran), ~ \mbox{for}~~ \ell = 1, 2, \cdots, m\\
& & V^-_{kt} \fr{1}{\sq 2} (|v_{i_k} \ran - |v_{j_k} \ran) = |v_{i_t}\ran) ,~~ \mbox{for}~~ t = 1, \cdots, s \\
& &V_{i\ell} |v_i\ran = \fr{1}{\sq 2} (|v_{i_\ell}\ran - |v_{j_\ell}\ran),~~ \mbox{for}~~ i = 1, 2, \cdots, n, i \ne i_k, j_k, \ell = 1, \cdots, m\\
& & V_{it} |v_i\ran = |v_{i_t}\ran,~~ \mbox{for}~~ i = 1, \cdots, n, i \ne i_k, j_k, t = 1, \cdots, s.
\eenrr

For deleting a loop $\{v_{i_{t'}}, v_{i_{t'}}\}$ a measurement in the basis $\{ | v_i \ran,  i = 1, \cdots,n \}$ is performed on the system prepared in the state $\si(G, a)$.  Then the probability that $P[| v_i \ran]$ clicks for $i = 1, \cdots n$ is
$$ \fr{1}{\mathfrak{d}_{(G, a)}} \left\{ \sum^m_{\ell =1} a_{i_\ell j_\ell}(\del_{ii_\ell} - \del_{ij_\ell} )^2 + \sum^s_{\ba{c} t = 1 \\ t \ne t' \ea} a_{i_ti_t} [\del_{ii_t}]^2 \right\}. \eqno{(2.33)}$$
The state after the measurement is $P[| v_i \ran]$.  Let $U_{i\ell}$ be $n \times n$ unitary matrices such that $U_{i\ell} [|v_i\ran] = \fr{1}{\sq 2} (| v_{i_\ell} \ran - | v_{j_\ell} \ran).$  For $ i = 1, \cdots, m$ and $U_{it}[| v_i \ran] = | v_{i_t} \ran$, for $ t = 1, \cdots, t' -1, t' + 1, \cdots, s$.  With probability $2a_{i_\ell j_\ell} /(\mathfrak{d}_{(G, a)} - a_{i_{t'}, i_{t'}})$ we apply $U_{i\ell}$ on $P[|v_i\ran]$ for each $\ell = 1, \cdots, m$ and with probability $a_{i_{t}i_{t}} / (\mathfrak{d}_{(G, a)} - a_{i_{t'}i_{t'}})$ we apply $U_{it}$ on $P[| v_i \ran]$ for each $t = 1, \cdots, t'-1, t'+1, \cdots s$.  We obtain $\si((G, a) - \{ v_{i_{t'}}, v_{i_{t'}}\})$ with probability given by Eq.(2.33).

The set of Kraus operators that realizes the TPCP for deleting the loop $\{ v_{i_{t'}}, v_{i_{t'}}\}$ is
$$ \left\{ \sq{\fr{2a_{i_\ell j_\ell}}{\mathfrak{d}_{(G, a)} - a_{i_{t'}, i_{t'}}}} U_{i\ell} P[ |v_i \ran ] ~~ i = 1, \cdots, m ,~~ \ell = 1, \cdots, m\right\}$$
$$ \cup \left\{ \sq{\fr{2a_{i_t i_t}}{\mathfrak{d}_{(G, a)} - a_{i_{t'}, i_{t'}}}} U_{it} P[ |v_i \ran ] ~~ i = 1, \cdots, m ,~~ t = 1, \cdots,  t' -1, t'+1, \cdots s\right\}.$$
The set of Kraus operators that realizes the TPCP for adding the loop $\{ v_{i_{t'}} v_{i_{t'}}\}$
$$ \left\{ \sq{\fr{2a_{i_\ell j_\ell}}{\mathfrak{d}_{(G, a)} + a_{i_{t'}i_{t'}}}} V_{i\ell} P[|v_i\ran] : i = 1, \cdots, n, \ell = 1, \cdots, m\right\}$$
$$ \cup\left\{ \sq{\fr{a_{i_ti_t}}{\mathfrak{d}_{(G, a)} + a_{i_{t'}i_{t'}}}} V_{it} P[|v_i\ran] : i = 1, \cdots, n, t = 1, \cdots, s \right\},$$
where $V_{i\ell}, V_{it}$ are $n \times n$ unitary matrices define as follows :
$$V_{i\ell} |v_i\ran = \fr{1}{\sq 2} (|v_{i_\ell}\ran - |v_{j_\ell}\ran),~~ \mbox{for}~~ \ell = 1, \cdots, m, i = 1, \cdots, n $$
$$V_{it} |v_i\ran = |v_{i_t}\ran,~~ \mbox{for}~~ t = 1, \cdots, s, i = 1, \cdots, n.$$

\subsection{Deletion and addition of an edge with real weight, which preserves the positivity of the generalized Laplacian}

Let $(G, a)$ be a graph with real weights on its edges not necessarily positive.   We are basically concerned here with the deletion of  $\{ v_i, v_j\}$ with $a_{v_iv_j} > 0$ and the addition of $\{ v_i,v_j\}$ with $a_{v_iv_j} < 0$, because in other cases the positivity of the Laplacian is preserved. We define the sets
$$ E^+ = \{ \{v_i, v_j\} \in E(G, a),a_{v_iv_j} > 0 \}, \eqno{(2.34)}$$
$$ E^- = \{ \{v_i, v_j\} \in E(G, a)  , a_{v_iv_j} < 0 \} \eqno{(2.35)}$$
and  $E = E^+ \cup E^-$.

We define a graph operator $\Xi$ as
$$ \Xi [E] = E \cup \{ \{v_i,v_i\}, \{v_j,v_j\} : a_{v_iv_i} = a_{v_jv_j} = 2|a_{v_iv_j}|~~\mbox{and}~~ \{v_i, v_j\} \in E^- \} \eqno{(2.36)}$$
Suppose we wish to delete a positive weighted edge $\{ v_{i_k}, v_{j_k}\} \in E^+,$ then we define the resulting graph as
$$ \Xi \cL ((G, a) - \{ v_{i_k}, v_{j_k} \} ) $$
where the graph operator $\cL$ is defined in (2.20b).

For adding a negative weighted edge between $v_i$ and $v_j, i \ne j$, we act on $E(G, a)$ by the appropriate element of the set of operators $\{ \in_{ij}\}, i, j = 1, \cdots, n, i \ne j$ defined as

$$\in_{ij} [E]  =  E \cup \{ \{ v_i, v_j\}, \{ v_i, v_i\}, \{ v_j, v_j\} : a_{v_iv_j} < 0, a_{v_iv_i} = 2|a_{v_iv_j}| = a_{v_jv_j}\} \eqno{(2.37)}$$

To obtain  the set of the corresponding TPCP operators we decompose the resulting graph, $(G', a')$ given by $\Xi \cL((G, a) - \{ v_{i_k}, v_{j_k}\}) (a_{v_{i_k }v_{j_k}} > 0)$ (Eq. (2.36) ) or by $\in_{ij}((G, a) + \{ v_i, v_j \}) (a_{v_iv_j} < 0)$ (Eq. (2.37)) or by $((G, a) - \{ u_{i_k}, v_{j_k} \}) (a_{v_{i_k} v_{j_k}} < 0)$ or by $(G, a) + \{ v_i, v_j \}) (a_{v_iv_j} > 0)$ into spanning subgraphs determined by the sets $E^+$ and $E^-$ and treat the spanning subgraph corresponding to $E^-$ replace the weights $a_{u_iv_j}$ of edges $\{v_i, v_j\} \in E^-$ by $- a_{v_iv_j}$, so that both the spanning subgraphs have only positive weights. For getting the Kraus operators we go through the following steps.

(a) First we determine the degree sums for the resulting graphs $(G', a')$ in four cases.

\bd
\i(i) Deletion of a positive weighted edge $\{ u_{i_k}, v_{j_k} \}$
$$ \mathfrak{d}_{(G',a')} = \mathfrak{d}_{(G, a)} - 2a_{v_{i_kj_k}} - \sum_i a_{ii} + 2 \sum_{\{u_i,v_j\} \in E^-} |a_{v_iv_j}|. \eqno{(2.38)}$$

\i(ii) Addition of a positive weighted edge $\{ v_i, v_j\}$.
$$ \mathfrak{d}_{(G', a')} = \mathfrak{d}_{(G, a)} + 2a_{v_iv_j}. \eqno{(2.39)} $$

\i(iii)  Deletion of a negative weighted edge $\{ v_{i_k}, v_{j_k}\}$
$$ \mathfrak{d}_{(G', a')} = \mathfrak{d}_{(G, a)} - 2a_{v_{i_k} v_{j_k}}. \eqno{(2.40)}$$

\i(iv) Addition of a negative weighted edge $\{ v_i, v_j\}$
$$ \mathfrak{d}_{(G', a')} = \mathfrak{d}_{(G, a)} + 2a_{v_i, v_j} + 4|a_{v_i,v_j}|. \eqno{(2.41)} $$
\ed

(b) We construct the Kraus operators separately for $G^+$ and $G^-$ for deleting the same edge $\{ v_i, v_j\}$ from $G^\pm \sqcup \{ v_i, v_j \}$ or adding the edge $\{ v_i, v_j\}$ to $G^\pm$, using the method given in section 2.5.1.  However, the probabilities of applying various unitary operator $U^\pm_{k\ell}$ and $U^\pm_{k\ell}, U_{i\ell}$ and $U_{it}$ are determined using $\mathfrak{d}_{(G', a')}$ as in step (a) above.

(c) Let $\{ A_i\}$ and $\{ B_i\}$ denote the sets of Kraus operators for the graph operations on $G^+$ and $G^-$ as described in (b).  Then
$$ \si(G', a') = \sum_i A_i \si(G, a) A^\dagger_i - \sum_j B_j \si(G, a) B^\dagger_j  \eqno{(2.42)}$$
and
$$ \sum_i A_i^\dagger A_i - \sum_j B_j^\dagger B_j = I  \eqno{(2.43)}$$
which can be justified by construction.

We comment here that it is possible to modify the graph,  after deleting a positive edge or adding a negative edge, which can preserve positivity in different ways, leading to different sets of Kraus operators.  The basic idea is to add new loops.  In our method we try to minimize the addition of loops.  Further, in our method we cannot reverse the graph operation for deleting a positive edge or adding a negative edge.  But this is not a problem since the quantum operations given by super operators are, in general, irreversible.

\subsection { Deleting Vertices}

 In order to delete a vertex $v_i$ from a graph (G,a),
\begin{enumerate}
\item[(i)] Delete edges, including loops, on $v_i$, one by one, by successively
applying the procedure in section 2.5.2.  The resulting graph $(G', a')$ has a density matrix with the $i$-th row and $i$-th column containing all zeros.

\item[(ii)] We now perform, on $\si(G', a')$, the projective measurement $M = \{ I_n -  P[|v_i\ran], P[|v_i\ran]\}$.  Since $P[|v_i\ran]$ is the matrix with all elements zero except the $i$-th diagonal element, while $\si(G', a')$ as all zeros in the $i$-th row and column, the probability that $P[|v_i\ran]$ clicks $ = Tr(\si P[| v_i\ran]) = 0$.  Thus when $M$ is performed on $\si(G', a')',  I_n - P[|v_i\ran]$ clicks with probability 1 and the state after measurement is $\si(G', a') - \{ v_i\})$ and is the same as $\si(G', a')$ without $i$-th row and $i$-th column.
\end{enumerate}

\subsection { Adding a Vertex } Let $(G, a)$ be a graph on $n$ vertices $v_1, \cdots, v_n$ and $m$ edges $\{ v_{i_k}, v_{j_k} \}, \; k = 1, \cdots, m, i_k \ne j_k$ and $s$ loops $\{ v_{i_t}, v_{i_t}\},\; t = 1, \cdots, s; \; 1 \le i_k, j_k, i_t \le n$.  Consider the following density operator
$$ \rho = \left( \fr{1}{2} \sum^2_{i=1} b_{ii} P[|u_i\ran]) \otimes (\si (G, a))\right),$$
where $\{ |u_1\ran, |u_2 \ran\}$ form an orthonormal basis of $\mathbb{C}^2$.  We associate vertices $u_i, i = 1, 2$ to the state $|u_i\ran$.  Consider the graph $$H = (\{ u_1, u_2\}, \{ \{ u_1,u_1\}, \{ u_2, u_2\}\})$$ with associated weights $b > 0$.  It is easy to check that $\si(H, b) = \fr{1}{2} \sum\limits^2_{i=1} b_{ii} P[|u_i\ran]$.  Also observe that
$$\rho  =  \si((H, b) \boxdot (G, a))  =  \si(H, b) \otimes \si(G, a).$$

Thus $(H, b) \boxdot (G, a)$ is the graph on $2n$ vertices labeled by $u_1 v_1, \cdots,u_1v_n, u_2v_1,$ $ \cdots, u_2v_n$ and with $2m$ edges and $2s$ loops (see section 2.4.2) $$\{ u_1v_{i_1}, u_1v_{j_1}\} \cdots \{ u_1v_{i_m}, u_1 v_{j_m}\}\\ \{ u_2v_{i_1}, u_2v_{j_1}\} \cdots \{ u_2v_{i_m}, u_2v_{j_m}\}$$ and loops $\{ u_1v_{i_t}, u_1 v_{i_t}\}, \{ u_2v_{i_t}, u_2v_{i_t}\}, t = 1, \cdots, s$.  So $(H, b) \boxdot (G, a) = (H_1,a_1) \uplus (H_2, a_2)$ where
$$ (H_1,a_1) = (\{ u_1v_1 \cdots u_1v_n\}, \{ \{ u_1v_{i_1} ,u_1v_{j_1}\} \cdots \{ u_1v_{i_m}, u_1v_j\} \})$$
$$ (H_2, a_2) = ( \{ u_2v_1 \cdots u_2v_n\}, \{ \{ u_2v_{i_1}, u_2v_{j_1}\} \cdots \{u_2v_{i_m}, u_2 v_{j_m}\} \}).$$
We first delete all edges and loops of $(H, b) \boxdot (G, a)$ which are incident to the vertex $u_2v_1 \in V(H_2, a_2)$ as in section 2.5.2.  Now we perform the following projective measurement on $\si((H, b) \boxdot (G, a)),$
$$ M = \{ I_{2n} - \sum^n_{i=2} P[|u_2 v_i\ran], \sum^n_{i=2} P[| u_2 v_i\ran] \}.$$
The probability that $I_{2n} - \sum\limits^n_{i=2} P[v_2 v_i\ran ]$ is 1 and the state after the measurement is $\si((H_1, a_1) + \{ u_2v_1\})$.

\noi{\it Example (2.8)} : Consider the graph as given in the figure 2.16 , we want  to delete the edge \{1,2\} with positive weight by means of TPCP.
Calculate the Kraus operators for $G^+$ and $G^-$ as in section 2.5.2 , where $A_i$ for $i=1,...,\cdots 24$ and $B_i$ for $G^-$ ,$i=1,.....,4$
and substitute in the following equation

$$ \si(G', a') = \sum_i A_i \si(G, a) A^\dagger_i - \sum_j B_j \si(G, a) B^\dagger_j$$
where
$$ \si(G, a) = \fr{1}{8} \left[ \ba{cccc} 1 & -1 & -1 & 1 \\ -1 & 3 & -1 & -1 \\ -1 & -1 &3 & -1 \\ 1 & -1 & -1 & 1 \ea \right] .$$
we get
$$ \si(G', a') = \fr{1}{10} \left[ \ba{cccc} 2 & 0 & -1 & 1 \\ 0 & 2 & -1 & -1 \\ -1 & -1 &3 & -1 \\ 1 & -1 & -1 & 3 \ea \right] .$$
and
$$ \sum_{i=1}^{24} A_i^\dagger A_i - \sum_{j=1}^4 B_j^\dagger B_j = I$$

\begin{figure}[!ht]
\includegraphics[width=2cm,height=2cm]{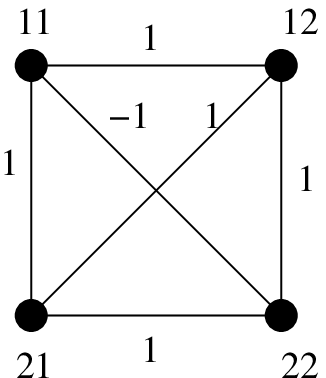}

Figure 2.16
\end{figure}

\section{Representation of a general Hermitian operator by a graph}

 In this section, we generalize sections 2.2 - 2.4, to quantum states in a complex Hilbert space, that is, to the density matrices with complex off-diagonal elements. We have also given rules to associate a graph to a general Hermitian operator. We believe that any further advance in the theory reported in this chapter will prominently involve graph operators and graphs associated with operators.

\subsection{Representation of a general density matrix with complex off diagonal elements}

Consider an $n \times n$ density matrix with complex off-diagonal elements.  We associate with this density matrix an oriented  graph $(G, a)$ on $n$ vertices, $m$ edges  and $s$ loops with weight function
$$ a : V(G) \times V(G) \ra \mathbb{C}.$$

The weight function $a$ has the following properties:

(i) $a(\{ u, v\}) \ne 0$ if $\{ u, v\} \in E(G, a)$ and $0$ otherwise.

(ii) $a(\{ u, v \}) = a^*(\{ v, u \})$

we write $a(\{u,v\})\; =\; |a(\{u,v\})| \; e^{i\phi_{uv}}, \phi_{vv} = 0$.

Note that when $\phi_{ij}=l \pi , l\;=\; 0,1, \cdots$, i.e. $a(\{u,v\})$ is real, positive when $l$ is even and real negative when $l$ is odd.

The degree $ \mathfrak{d}_v$ of vertex $v$ is given by
$$\mathfrak{d}_{(G,a)}(v)\;=\;   \mathfrak{d}_v = \sum_{u \in V(G, a) , \\ u \ne v} |a(\{u,v\})| \;+\; a(\{v,v\}) \eqno{(2.44)}$$

$$ \mathfrak{d}_{(G,a)}\;=\; \sum_{v \in V(G, a) }  \mathfrak{d}_v$$

The  adjacency matrix  $M(G, a)$ of a complex  weighted graph with $n$ vertices  is an $n \times n$ matrix whose rows and columns are indexed by vertices in $V(G,a)$:
$$M_{uv}\;=\; a(\{u,v\})\;=\;a^*(\{v,u\})\;=\;(M_{vu})^*.$$

The  degree matrix   $\D(G, a)$  of the complex weighted graph is an $n \times n$ real diagonal matrix, whose rows and columns are labeled by vertices in $V(G, a)$ and whose diagonal elements are the degrees of the corresponding vertices.
$$\D(G, a) = diag [ \mathfrak{d}_v; v \in V(G, a)]$$
where $ \mathfrak{d}_v$ is given by equation (2.44).

The loop matrix $\D_0(G, a)$ of a graph $(G,a)$  is an $n \times n$ real diagonal matrix with diagonal elements equal to the weights of the loops on the corresponding vertices
$$ [\D_0(G, a)]_{vv} = a_{vv} . $$

The   generalized Laplacian  of a graph $(G, a)$, which includes loops, is
$$ Q(G, a) = \D(G, a)\;+\; M(G, a) \;-\; \D_0(G, a) \eqno{(2.45)}$$

Note that $Q(G, a)$  is a Hermitian matrix . If  the generalized Laplacian $Q(G,a)$ is positive semidefinite, we can define the density matrix of the corresponding graph $(G, a)$ as
$$ \si(G, a) = \fr{1}{\mathfrak{d}_{(G, a)}} \; Q(G, a)  \eqno{(2.46)}$$
where $Tr(\si(G, a)) = 1$.

For any $n \times n$ density matrix $\si$ with complex off diagonal elements we can obtain the corresponding graph as follows:

\noi {\bf Algorithm 2.6.1 :}

\bd
\i(i) Label the $n$ vertices of the graph by the kets from the standard orthonormal basis.

\i(ii) For every nonzero $ij$th element with $j > i$ given by $ a(\{i,j\}) $ draw an edge between vertices labeled $|v_i\ran$ and $|v_j\ran$, with weight $a(\{i,j\})$.

\i(iii) Ensure that $\mathfrak{d}_{v_i} = \si_{ii}$ by adding loop of appropriate weight to $v_i$ if necessary.
\ed
\noi{\it Example (2.9)} :
(1)
\begin{figure}[!ht]
\includegraphics[width=10cm,height=4cm]{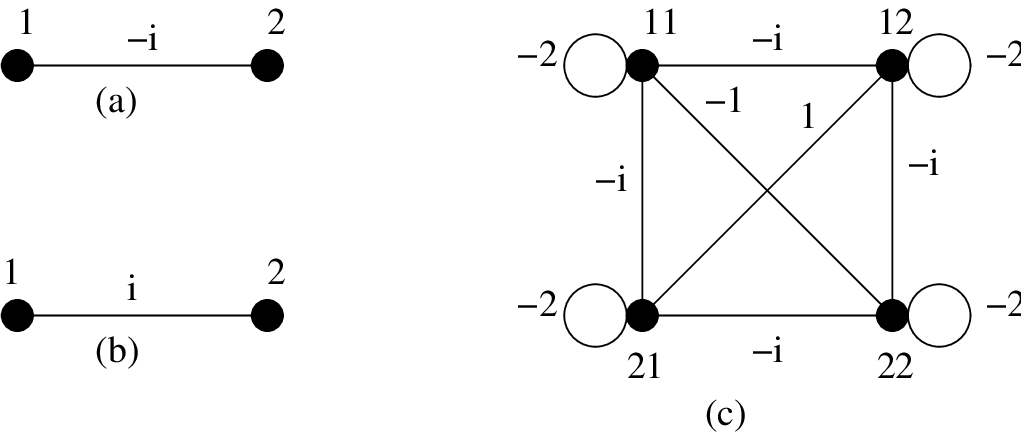}

Figure 2.17
\end{figure}
$$ P[|y, +\ran] = \fr{1}{2} \left[ \ba{cc} 1 & -i \\ i & 1 \ea \right] = \fr{1}{2} \left[ \ba{cc} 1 & e^{-i\pi/2}\\ e^{-i\pi/2} & 1 \ea \right] $$
where $|y, +\ran = \fr{1}{\sq 2} (|1\ran + i|2\ran)$ and the corresponding graph is as shown in figure 2.17a

(2)
$$ P[|y, - \ran] = \fr{1}{2} \left[ \ba{cc} 1 & i \\ -i & 1 \ea \right] = \fr{1}{2} \left[ \ba{cc} 1 & e^{i\pi/2} \\ e^{-i\pi/2} & 1 \ea \right] $$
where $|y, -\ran = \fr{1}{\sq 2} (|1\ran - i|2\ran)$ and the corresponding graph is as shown in figure 2.17b

(3)
$$ P[|y, + \ran|y,+\ran] = \fr{1}{4} \left[ \ba{cccc} 1 & -i & -i & -1 \\ i & 1 & 1 & -i \\i & 1 & 1 & -i\\-1 & i & i & 1\ea \right]  $$
The corresponding graph is as shown in figure 2.17c.

Note that remark 2.2.1 is valid also for complex weighted graphs.

\noi\textbf{Remark 2.6.2 :} Theorem 2.2.3 applies to complex weighted graphs with equation (2.11) changed to

$$ \sum^n_{i=1}  \mathfrak{d}^2_i + 2 \sum^m_{k=1} |a_{i_kj_k}|^2 =  \mathfrak{d}^2_{(G, a)} \eqno{(2.47)}$$
  also, lemma 2.2.4 applies to complex weighted graphs.

\noi {\bf Definition 2.6.3 :} A graph $(H, b)$ is said to be a factor of graph $(G, a)$ if $V(H, b) = V(G, a)$ and there exists a graph $(H', b')$ such that $V(H', b') = V(G, a)$ and $M(G, a) = M(H, b) + M(H', b')$.  Thus a factor is only a spanning subgraph.  Note that
$$ a_{v_iv_j} = \left\{ \ba{lll} b_{v_iv_j} & \mbox{if} & \{v_i,v_j\} \in E(H, b) \\ b'_{v_iv_j} & \mbox{if} & \{v_i,v_j\} \in E(H', b') \ea \right. $$

Now let  $(G, a)$ be a graph on $n$ vertices $v_1, \cdots, v_n$ having $m$ edges\\ $\{v_{i_1}, v_{j_1}\}, \cdots, \{v_{i_m}, v_{j_m}\}$ and $s$ loops $\{v_{i_1}, v_{i_1}\} \cdots \{v_{i_s}, v_{i_s}\}$ where $1 \le i_1j_1, $ $\cdots, i_m j_m \le n, 1 \le i_1 i_2 \cdots i_s \le n$.

Let $(H_{i_kj_k}, a_{i_kj_k})$ be the factor of $(G, a)$ such that
$$ [M(H_{i_kj_k}, a_{i_kj_k})]_{u,w} = \left\{ \ba{l} a_{i_kj_k} ~ \mbox{if}~ u = i_k~ \mbox{and}~ w = j_k ~\mbox{or}~  a^*_{i_kj_k} \mbox{if}~~u = j_k, w = i_k \\ 0  ~~ \mbox{otherwise} \ea \right. \eqno{(2.48)}$$
$$ [\D(H_{i_kj_k}, a_{i_kj_k})]_{u,w} = \left\{ \ba{l} |a_{i_kj_k}| ~~ \mbox{if}~~ u = i_k= w ~~\mbox{or} ~~u = j_k= w  \\ 0  ~~ \mbox{otherwise} \ea \right. \eqno{(2.49)}$$

Let $(H_{i_t,i_t}, a_{i_t i_t})$ be a factor of $(G, a)$ such that
$$ [M(H_{i_ti_t}, a_{i_t i_t})]_{u,w} =  [\D(H_{i_ti_t}, a_{i_ti_t})]_{u,w} = \left\{ \ba{l} a_{i_t i_t}~~ \mbox{when}~~ u = i_t = w \\ 0 ~~ \mbox{otherwise} \ea \right. \eqno{(2.50)}$$

\noi {\bf Theorem 2.6.4 :} The density matrix of a graph $(G, a)$ as defined above with factors given by equations (2.48), (2.49) and (2.50) can be decomposed as
$$ \si(G, a) = \fr{1}{\mathfrak{d}_{(G, a)}} \sum^m_{k=1} 2 |a(\{i_k,j_k\})| \si(H_{i_kj_k}, a_{i_kj_k}) + \fr{1}{\mathfrak{d}_{(G, a)}} \sum^s_{t=1} a_{i_ti_t} \si(H_{i_ti_t}, a_{i_ti_t}) $$
or
$$ \si(G, a) = \fr{1}{\mathfrak{d}_{(G, a)}} \sum^m_{k=1} 2|a(\{i_k,j_k\})| P[\fr{1}{\sq 2}(|v_{i_k}\ran -e^{i\phi_{i_kj_k}} |v_{j_k}\ran)] $$
$$+ \fr{1}{\mathfrak{d}_{(G, a)}}  \sum^s_{t=1} a_{i_ti_t} P[|v_{i_t}\ran]$$
Where $\phi_{i_kj_k} = \pi$ for any edge  $\{i_k,j_k\}$ with real positive weight and $\phi_{i_kj_k} = 0$  for any real negative weight.

\noi {\bf Proof :} From equations (2.48), (2.49), (2.50) and remark 2.6.2, the density matrix
$$\si (H_{i_kj_k}, a_{i_kj_k}) = \fr{1}{2|a_{i_kj_k}|} [ \D(H_{i_kj_k}, a_{i_kj_k}) + M(H_{i_kj_k},a_{i_kj_k})]$$
is a pure state.  Also,
$$ \si (H_{i_ti_t}, a_{i_ti_t}) = \fr{1}{a_{i_ti_t}} [ \D_0 (H_{i_t, i_t}, a_{i_ti_t})]$$
is a pure state.  Now
$$ \D(G, a) = \sum^m_{k=1} \D(H_{i_kj_k}, a_{i_kj_k}) + \sum^s_{t=1} \D_0(H_{i_ti_t}, a_{i_ti_t})$$
$$M(G, a) = \sum^m_{k=1} M(H_{i_kj_k}, a_{i_kj_k}) + \sum^s_{t=1} \D_0(H_{i_ti_t}, a_{i_ti_t}).$$
Therefore, from Eq. (2.46)

$$\si(G, a)  =  \fr{1}{\mathfrak{d}_{(G, a)}} \left[ \sum^m_{k=1} \D(H_{i_kj_k}, a_{i_kj_k}) + \sum^m_{k=1} M(H_{i_kj_k}, a_{i_kj_k})\right]$$
$$ + \fr{1}{\mathfrak{d}_{(G, a)}} \left[ \sum^s_{t=1} \D_0 (H_{i_ti_t}, a_{i_ti_t})\right]$$
$$=  \fr{1}{\mathfrak{d}_{(G, a)}} \sum^m_{k=1} [\D(H_{i_kj_k}, a_{i_kj_k}) + M(H_{i_kj_k}, a_{i_kj_k})] \\
 + \fr{1}{\mathfrak{d}_{(G, a)}} \sum^s_{t=1} \D_0(H_{i_ti_t}, a_{i_ti_t})$$
$$ =  \fr{1}{\mathfrak{d}_{(G, a)}} \sum_k 2|a(\{i_k,j_k\})| \si(H_{i_kj_k}, a_{i_kj_k}) \\
 + \fr{1}{\mathfrak{d}_{(G, a)}} \sum_t a_{i_ti_t} \si(H_{i_ti_t}, a_{i_ti_t}) \eqno{(2.51)}$$

 In terms of the standard basis, the $uw$-th element of matrices $\si(H_{i_kj_k}, a_{i_kj_k})$ and $\si(H_{i_ti_t}, a_{i_ti_t})$ are given by $\lan v_u | \si(H_{i_kj_k} , ,a_{i_kj_k}) | v_w \ran$ and $\lan v_u | \si (H_{i_ti_t} a_{i_ti_t} | v_w\ran,$ respectively.  In this basis
$$ \si(H_{i_kj_k}, a_{i_kj_k}) = P[ \fr{1}{\sq 2} ( | v_{i_k} \ran -e^{i\phi_{i_kj_k}} | v_{j_k} \ran )]$$
$$ \si(H_{i_ti_t}, a_{i_ti_t}) = P[| v_{i_t} \ran ] .$$

Therefore equation (2.51) becomes

$$\si(G, a)  =  \fr{1}{\mathfrak{d}_{(G, a)}} \sum^m_{k=1} 2|a(\{i_k,j_k\})| P[\fr{1}{\sq 2} (| v_{i_k}\ran -e^{i\phi_{i_kj_k}} | v_{j_k} \ran) $$
$$+ \fr{1}{\mathfrak{d}_{(G, a)}} \sum ^s_{t=1} a_{i_ti_t} P[ | v_{i_t} \ran]\eqno{(2.52)}$$
Where $\phi_{i_kj_k} = \pi$ for any edge  $\{i_k,j_k\}$ with real positive weight and $\phi_{i_kj_k} = 0$  for any real negative weight.$\hspace{\stretch{1}} \blacksquare$\\
\noi{\it Example (2.10)} :

(i) For a graph given in figure 2.17b, the density matrix is
\benrr
\si(G, a) & = & \fr{1}{2}\{2 P[\fr{1}{\sq 2} (|1\ran - e^{i\pi/2} |2\ran)]\}=P[\fr{1}{\sq 2} (|1\ran - i|2\ran)]
\eenrr

 (ii) For a graph given in figure 2.17c,  the density matrix is
\benrr
\si(G, a) & = & \fr{1}{4} \{ 2P[\fr{1}{\sq 2} (|11\ran - e^{-i\pi/2} |12\ran)] + 2P[\fr{1}{\sq 2}(|11\ran - e^{-i\pi/2} |21\ran)]\\
& & + 2 P[\fr{1}{\sq 2} (|11\ran - |22\ran)] + 2 P[\fr{1}{\sq 2} (|12\ran + |21\ran)] \\
& & + 2P[\fr{1}{\sq 2} (|12\ran - e^{-i\pi/2} |22\ran)] + 2P[\fr{1}{\sq 2} (|21\ran - e^{-i\pi/2} |22\ran)] \\
& & - 2P[|11\ran] - 2P[|22\ran] -2P[|12\ran]-2P[|21\ran]\}
\eenrr
$$ \si(G, a) = \fr{1}{4} \left[ \ba{cccc} 1 & -i & -i & -1 \\ i & 1 & 1 & -i \\ i & 1 &1 & -i \\ -1 & i & i & 1 \ea \right] .$$

\noi{\it Example (2.11)} : Consider the state $$ \si = \fr{1}{3} P[|y, +\ran |y, + \ran] + \fr{2}{3} P[|y, + \ran | \psi \ran]$$
where $ |y, +\ran  = \fr{1}{\sq 2} (|1\ran + i|2\ran)$ and $|\psi\ran = \fr{1}{\sq 3} (|1\ran + i\sq 2|2\ran)$
\benrr
\si  =  \fr{1}{36} \left[ \ba{cccc} 7 & -(3 + 4\sq 2)i & -7i & -(3 + 4\sq 2)\\ (3 + 4\sq 2)i & 11 & 3 + 4\sq 2 & -11 i \\ 7i & 3 + 4\sq 2 & 7 & -(3 + 4\sq 2)i \\ -(3 + 4\sq 2) & 11i & (3 + 4\sq 2)i & 11 \ea \right] \\
 =  \fr{1}{36} \left[ \ba{cccc} 7 & (3 + 4\sq 2)e^{-i\pi/2} & 7e^{-i\pi/2}  & -(3 + 4\sq 2)\\ (3 + 4\sq 2)e^{i\pi/2}  & 11 & 7 & 11e^{-i\pi/2} \\ 7e^{i\pi/2} & 3 + 4\sq 2 & 7 & (3 + 4\sq 2)e^{-\pi/2} \\ -(3 + 4\sq 2) & 11e^{i\pi/2} & (3 + 4\sq 2)e^{i\pi/2} & 11 \ea \right]
\eenrr
The corresponding graph is as shown in figure 2.18,
\begin{figure}[!ht]
\includegraphics[width=8cm,height=4cm]{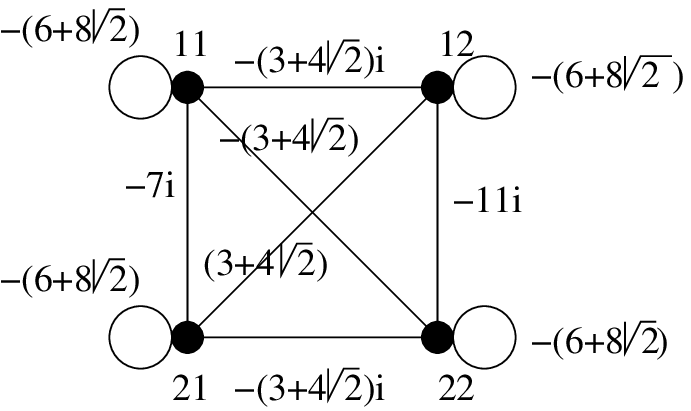}

Figure 2.18
\end{figure}
and using the equation (2.45) to get the matrix from graph in figure 2.18,
\benrr
\si(G, a) & = & \fr{1}{36} \{ 2(3 + 4 \sq 2)P[\fr{1}{\sq 2} (|11\ran - e^{-i\pi/2} |12\ran)] \\
& & + 2 \times 7 P[\fr{1}{\sq 2} (|11\ran - e^{-i\pi/2} |21\ran)] + (3 + 4\sq 2) P[\fr{1}{\sq 2}(|11\ran - |22\ran)] \\
& & + 2(3 + 4\sq 2) P[ \fr{1}{\sq 2} (|12\ran + |21 \ran)] + 2 \times 11 P[\fr{1}{\sq 2} (|12\ran - e^{-i\pi/2} |22\ran)] \\
& & + 2(3 + 4\sq 2) P[\fr{1}{\sq 2} (|21\ran - e^{-i\pi/2} |22\ran)]-(6+8 \sq{2})P[|11\ran] \\
& &- (6 + 8 \sq 2) P[|22\ran]- (6 + 8 \sq 2)P[|12\ran] - (6 + 8 \sq 2) P[|21\ran]\}.
\eenrr
We can check that
$$ \si(G, a) = \fr{1}{36} \left[\ba{cccc} 7 & -(3 + 4\sq 2)i & - 7i & -(3 + 4\sq 2) \\ (3 + 4\sq 2)i & 11 & 3 + 4 \sq 2 & -11i \\ 7i & 3 + 4 \sq 2 & 7 & -(3 + 4\sq 2)i \\ -(3 + 4\sq 2) & 11i & (3 + 4\sq 2)i & 11 \ea \right].$$
We can also check that this state is not pure by applying  remark 2.6.2 on the graph.\\
\noi\textbf{Remark 2.6.5:} Lemma 2.2.10 and remark 2.2.11 are valid for the complex weighted graphs with disjoint edge union $\dotplus$

\subsection{ Separability}

\noi\textbf{Remark 2.6.6:} The definition of the tensor product $(G,a) \otimes (H,b)$ of two complex weighted graphs $(G,a)$ and $(H,b)$ is the same as given before. However, note that $\{v_1,v_2\} \in E(G,a), \{w_1,w_2\} \in E(H,b)$ implies

$$c(\{(v_1,w_1),(v_2,w_2)\})\;=\; a(\{v_1,v_2\}) b(\{w_1,w_2\})$$ and $$c(\{v_1,w_2),(v_2,w_1)\})\;=\; a(\{v_1,v_2\}) b(\{w_2,w_1\})=a(\{v_1,v_2\}) b^*(\{w_1,w_2\}).$$

\noi\textbf{Remark 2.6.7 :} Equations (2.18a) and (2.18c) are valid for the tensor product of complex weighted graphs. Also, $Q((G,a) \otimes (H,b)) \ne Q(G,a) \otimes Q(H,b)$. Equation (2.18b) holds good only for graphs without loops, for graphs with only loops or when one factor has no loops and other factor has only loops.
For such graphs  equation (2.18b) immediately gives $$ \mathfrak{d}_{(G,a) \otimes (H,b)} (v,w) = \mathfrak{d}_{(G,a)} (v) \cdot \mathfrak{d}_{(H,b)} (w).$$

\subsection{ Modified tensor product}

The modified tensor product of two complex weighted graphs requires the operator $\cN$ to be redefined in the following way. We replace the equation (2.20a) by
 $$a'_i=\sum_{\begin{subarray}{I}
{v_k \in V(G,a)}\\
\hskip  .4cm {v_k \ne v_i}
\end{subarray}}
 |a(\{v_i,v_k\})| + a(\{ v_i,v_i\}) \eqno{(2.53)}$$
The definitions of the operators $\eta, \cL$ and $\Om$ remain the same. Equations (2.21) to (2.24) are satisfied by these operators on the complex weighted graphs. We further have
$$ \ba{ll} \mbox{(i)} & M(\cN \cL(G, a)) = \D(G, a)  - \D_0(G, a) \\ & \D(\cN \cL(G, a)) =  \D(G, a)  - \D_0(G, a) \\ & \D_0(\cN \cL(G, a)) = \D(G, a)  - \D_0(G, a)\\ & Q(\cN \cL(G,a))= \D(G, a)  - \D_0(G, a) \ea \eqno{(2.54)}$$
 The modified tensor product of two complex weighted graphs $(G,a)$ and $(H,b)$ with $p$ and $q\;(> p)$ vertices respectively is

$$(G,c)=(G, a) \boxdot (H, b)  =$$
$$ \cL(G, a) \otimes \cL (H, b) \dotplus \cL(G, a) \otimes \cN(H, b)
\dotplus  \cN(G, a) \otimes \cL(H, b)\dotplus$$
$$\{ \Om(G, a) \otimes \Om(H, b) \sqcup 2 \cN \cL(G,a) \otimes \cN \cL \eta (H,b)\} \eqno{(2.55)} $$

The weight function $c$ of $(G, a) \boxdot (H, b)$ is obtained via the definition of tensor product and the disjoint edge union.\\
\noi\textbf{Lemma 2.6.8 :} $\D((G, a) \boxdot (H, b)) = \D(G, a) \otimes \D(H, b)$.

\noi\textbf{Proof :} Since lemma 2.2.10 applies to disjoint edge union of complex weighted graphs,
$$ \D((G, a) \boxdot (H, b) ) = \D( \cL(G, a) \otimes \cL (H, b)) + \D(\cL(G, a) \otimes \cN(H, b))+$$
$$ \D( \cN(G, a) \otimes \cL(H, b))+ \D(  \Om(G, a) \otimes \Om(H, b)) + \D( 2 \cN \cL(G,a) \otimes \cN \cL \eta (H,b))$$
The last two terms are justified because the graphs involved are real weighted graphs. Using remark 2.6.7 we get

$$ \D((G, a) \boxdot (H, b) ) = \D( \cL(G, a)) \otimes \D( \cL (H, b)) + \D(\cL(G, a)) \otimes\D( \cN(H, b)) +$$
$$\D( \cN(G, a)) \otimes \D(\cL(H, b))+ \D(  \Om(G, a) ) \otimes \D(\Om(H, b)) +$$
$$ 2 \D( \cN \cL(G,a)) \otimes \D(\cN \cL \eta (H,b))$$

Using equations (2.21) to (2.24) and (2.54), we get, after some simplification,

 $$\D((G, a) \boxdot (H, b) ) = \D((G, a)) \otimes \D( (H, b))$$
\\
\noi\textbf{Corollary 2.6.9 :} $\mathfrak{d}_{(G,a) \boxdot (H,b)}(v,w)\; =\;\mathfrak{d}_{(G,a)}(v) \cdot \mathfrak{d}_{(H,b)}(w)$

and$$ \mathfrak{d}_{(G,a) \boxdot (H,b)}\; =\; \mathfrak{d}_{(G,a)} \cdot \mathfrak{d}_{(H,b)}$$

\noi {\bf Proof :}
The first result  follows directly from lemma 2.6.8. For the second  note that $$ Tr(\D((G, a) \boxdot (H, b) )) = Tr (\D(G, a) \otimes \D (H, b))= Tr( \D(G, a)) \cdot Tr(\D( H, b))$$ where  $Tr$ denotes the trace.  \hfill $\blacksquare$

\noi\textbf{Theorem 2.6.10 :} Consider  a bipartite syatem in $\mathbb{C}^p \otimes \mathbb{C}^q$ in the state $ \si $. Then $ \si = \si_1 \otimes \si_2$ if and only if $\si $ is the density matrix of the graph $(G,a) \boxdot (H,b)$ where $(G,a)$ and $(H,b)$ are the graphs having density matrices $\si_1$ and $\si_2,$ respectively.

\noi\textbf{Proof :} \noi {\it If part :} Given $(G, a), (H, b)$ we want to prove
$$\si((G, a) \boxdot (H, b)) = \si_1(G, a) \otimes \si_2(H, b).$$

 From the definition of the modified tensor product we can write
$$ \si((G,a) \boxdot (H, b)) = \fr{1}{\mathfrak{d}_{(G, a) \boxdot (H, b)}} \{Q[\cL(G, a) \otimes \cL (H, b)\dotplus$$
$$  \cL(G, a) \otimes \cN(H, b)\dotplus  \cN(G, a) \otimes \cL(H, b) \dotplus \{\Om(G, a) \otimes \Om(H, b)) \sqcup $$
$$2 \cN \cL(G,a) \otimes \cN \cL \eta (H,b) ]\}$$
Using remark 2.6.5 and corollary 2.6.9   we get \\

$\si((G, a) \boxdot (H, b))  =  \fr{1}{\mathfrak{d}_{(G, a)} \cdot \mathfrak{d}_{(H, b)}} [ Q(\cL(G, a) \otimes \cL (H, b))+ Q(\cL(G, a) \otimes \cN(H, b))+$
$$ Q(\cN(G, a)  \otimes \cL(H, b)) + Q(\Om(G, a) \otimes \Om(H, b) \sqcup 2 \cN \cL(G,a) \otimes \cN \cL \eta (H,b))]\eqno{(2.56)}$$

We can calculate every term in Eq.(2.56) using Eqs.(2.21) to (2.24) and Eq.(2.54) and substitute in Eq.(2.56) to get

$$ \si((G, a) \boxdot (H, b)) = \si(G, a) \otimes \si(H, b) .$$

\noi{\it Only if part :} Given $ \si = \si_1 \otimes \si_2$ consider the graphs $(G,a)$ and $(H,b)$ for $\si_1$ and $\si_2$ respectively. Then the graph of $\si$ has the generalized Laplacian
\benrr
Q(G,a) \otimes Q(H,b)& = &( \D(G,a)  + M(G,a)- \D_0(G,a)) \otimes (\D(H,b)+\\
& &M(H,b)- \D_0(H,b))\\
& = & \D(G,a) \otimes \D(H,b)+ \D(G,a) \otimes (M(H,b)- \\
& &\D_0(H,b)) +( M(G,a)-\D_0(G,a)) \otimes \D(H,b)+ \\
& &( M(G,a)- \D_0(G,a)) \otimes  (M(H,b)- \D_0(H,b)) ~~(2.57)
\eenrr
Using equations (2.21) to (2.24) and (2.54) we see that RHS of equation (2.57) is the generalized Laplacian for $(G,a) \boxdot (H,b)$
 \hfill $\blacksquare$

 \noi\textbf{Remark 2.6.11 :} The proof that the modified tensor product is associative and distributive with respect to the disjoint edge union is the same as that for the case of real weighted graphs (corollary 2.4.7).
\\
\noi\textbf{Remark 2.6.12 :} The definition of the Cartesian product of graphs is the same as given in definition 2.4.8.

\noi\textbf{Remark 2.6.13 :} Corollaries 2.4.9 and 2.4.10 apply to complex weighted graphs without any change.

\subsection{Convex combination of density matrices}

Consider two graphs $(G_1, a_1) $ and $(G_2, a_2)$ each on the same $n$ vertices, having $\si(G_1, a_1)$ and $\si(G_2, a_2)$ as their density matrices respectively, where $a_1$ and $a_2$ are complex weight functions. Let $(G,a)$ be the graph of the density matrix $\si(G, a)$ which is a convex combination of $\si(G_1, a_1) $ and $\si(G_2, a_2)$,

$$ \si(G, a) = \la \si(G_1, a_1) + (1 - \la) \si(G_2, a_2),\; 0 \le \la \le 1.$$

It is straightforward, using the definitions of the operators $\cN ,\cL$ and $\eta$, to verify that

$$ (G,a)=[\la \cN(G_1,a_1)\sqcup (1-\la)\cN(G_2,a_2)]\sqcup [\la \cL(G_1,a_1)\sqcup (1-\la)\cL(G_2,a_2)]$$
$$\sqcup \eta \cL[\la \cL(G_1,a_1)\sqcup (1-\la)\cL(G_1,a_2)]. \eqno{(2.58)}$$
We can apply this equation to any convex combination of density matrices. Let $$ \si(G, a) = \sum_i p_i \si(G_i, a_i) ,\;\; \sum_i p_i =1$$
Then,
$$ (G,a)=[\sqcup_i p_i\cN(G_i,a_i)]\sqcup [\sqcup_i p_i \cL(G_i,a_i)] $$
$$\sqcup \eta \cL[\sqcup_i p_i \cL(G_i,a_i)], \eqno{(2.59)}$$
where $a$ and $\{a_i\}$ are complex weight functions, $a{(\{v_l,v_k\})}=\sum_i a'_i{(\{v_l,v_k\})}$ and $a{(\{v_l,v_l\})}=\sum_i a'_i{(\{v_l,v_l\})}$ with $a'_i=p_i a_i.$\\
\noi\textbf{Lemma 2.6.14 :} Let $(G_1,a_1)$, $(G_2,a_2)$ and $(G,a)$ satisfy Eq.(2.58). Then
$$ \si(G, a) = \fr{\mathfrak{d}_{(G_1,a_1)}}{\mathfrak{d}_{(G, a)}}   \si(G_1, a_1) + \fr{\mathfrak{d}_{(G_2,a_2)}}{\mathfrak{d}_{(G, a)}}   \si(G_2, a_2).$$
\noi\textbf{Proof :} Similar to that of lemma 2.2.10.\hfill $\blacksquare$

In general, if  $(G,a)$ satisfies Eq.(2.59) for some set of graphs $\{(G_i,a_i)\}$, we have
$$ \si(G, a) = \fr{1}{\mathfrak{d}_{(G, a)}} \sum_i \mathfrak{d}_{(G_i,a_i)}  \si(G_i, a_i).\eqno{(2.60)}$$\\
\noi\textbf{Theorem 2.6.15 :} Every graph $(G,a)$ having a density matrix $\si(G,a)$ can be decomposed as in Eq.(2.59), where $\si(G_i,a_i)$ is a pure state.

\noi\textbf{Proof :} The same as that of theorem 2.2.12.
\hfill $\blacksquare$

\noi\textbf{Corollary 2.6.16 :} A state of a $k$-partite system is separable if and only if the graph $(G,a)$ for $\si$ has the form $$ (G,a)=[\sqcup_i \cN \boxdot^k_{j=1} (G^j_i, a^j_i)]\sqcup [\sqcup_i  \cL\boxdot^k_{j=1} (G^j_i, a^j_i)] $$
$$\sqcup \eta \cL[\sqcup_i \cL \boxdot^k_{j=1} (G^j_i, a^j_i)]. \eqno{(2.61)}$$\\
\noi\textbf{Proof :} The same as of corollary 2.4.11, where we refer to theorem 2.6.4 instead of theorem 2.4.5 and lemma 2.6.14 instead of lemma 2.2.10 and Eq.(2.60) instead of remark 2.2.11.\hfill $\blacksquare$

Corollary 2.6.16 says that Werner's definition \cite{wer89} of a separable state in  $\mathbb{C}^{q_1} \otimes \mathbb{C}^{q_2} \otimes \mathbb{C}^{q_3} \otimes \cdots \otimes \mathbb{C}^{q_k}$ system can be expressed using corresponding graphs.

\subsection{Representation of a Hermitian operator (observable) by a graph}
In order to represent a general Hermitian matrix corresponding to a quantum observable $A$ we lift the requirement that the Laplacian be positive semidefinite and $Tr[A]=1$. In other words we take the generalized Laplacian as the matrix for the graph.

Given a Hermitian matrix $A$, the algorithm 2.6.1 can be implemented to get its graph $(G, a)$.  The corresponding observable $\h A$ of a graph $(G, a)$ is
$$ \h A = \sum^m_{k=1} 2a_{i_kj_k}  P[\fr{1}{\sq 2} (|v_{i_k}\ran - e^{i\phi_{i_kj_k}}|v_{j_k}\ran)] + \sum^s_{t=1} a_{i_ti_t} P[|v_{i_t}\ran] \eqno{(2.62)}$$ \\
\noi{\it Example (2.12)} :  Give the graph of $\si_x$ and $\si_y$.

(1)~~~~ $\si_x = \left[ \ba{cc} 0 & 1 \\ 1 & 0 \ea \right] $.

The corresponding graph of $\si_x$ is shown in figure 2.19a

(2)~~~~$ \si_y = \left[ \ba{cc} 0 & -i \\ i & 0 \ea \right] = \left[ \ba{cc} 0 & e^{-i\pi/2} \\ e^{i\pi/2} & 0 \ea \right].$

The corresponding graph of $\si_y$ is shown in figure 2.19b

\begin{figure}[!ht]
\includegraphics[width=8cm,height=1cm]{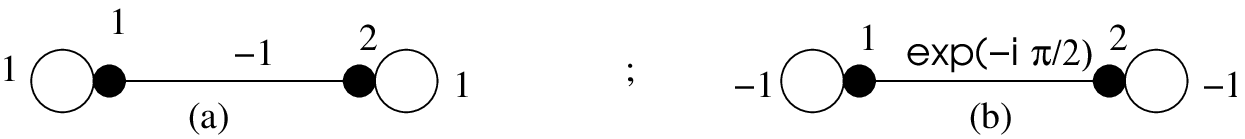}

Figure 2.19
\end{figure}

Using Eq. (2.62) to get the operators from graphs,

$$ \si_x = -2P[\fr{1}{\sq 2} (|1\ran - |2\ran)] + P[|1 \ran] + P[|2\ran] = |1\ran \lan 2| + |2\ran \lan 1| = \left[ \ba{cc} 0 & 1 \\ 1 & 0 \ea \right] $$
$$ \si_y = 2P[\fr{1}{\sq 2} (|1\ran - e^{-i\pi/2}  |2\ran)] - P[|1 \ran] - P[|2\ran] = -i|1\ran \lan 2| + i|2\ran \lan 1| = \left[ \ba{cc} 0 & -i\\ i& 0 \ea \right] $$

\section{Some graphical criteria for the positive semidefiniteness of the associated Laplacian}

 In this section we address the following question. Given a graph, can the positive semidefiniteness of the associated Laplacian be determined using the topological properties of the graph? A general answer to this question seems to be difficult because the theory of weighted graphs, with negative and complex weights is almost unavailable. Many results obtained for simple graphs do not apply to the weighted graphs with real or complex weights.  Nevertheless, we give here the above mentioned criteria in some special cases.\\
\noi{\bf Lemma 2.7.1 :} Let $(G,a)$ be a graph with real or complex weights, having one or more nonisolated vertices with degree zero. Then the Laplacian of $(G,a)$ is not positive semidefinite.\\
\noi {\bf Proof :} Such a graph $(G,a)$ has one or more zeros along the diagonal of its Laplacian with nonzero entries in the corresponding rows. However, a Hermitian matrix with one or more zeros in its diagonal has at least one negative eigenvalue unless all the elements in the corresponding rows and columns are zero \cite{satake75}.\hspace{\stretch{1}}$ \blacksquare$\\
\noi{\bf Lemma 2.7.2 :} Let $(G,a)$ be a $n$ vertex graph with real weights, having at least one loop and  let the weights on all the loops be negative. Then the Laplacian of $(G,a)$ is not positive semidefinite.\\
\noi {\bf Proof :} For the given $(G,a)$ and some $x$ in $\mathbb{R}^n$ we have
$$x^T[Q(G,a)]x=\sum_k a_{i_kj_k}(x_{i_k}-x_{j_k})^2-\sum_t|a_{i_ti_t}|x^2_{i_t}$$
where the first sum is over edges and the second sum is over loops. It is easy to check that
$x^T[Q(G,a)]x<0$ for $x=(1~1~1~\dots  ~1)^T$.\hspace{\stretch{1}}$ \blacksquare$\\
\noi{\bf Lemma 2.7.3 :} Let $(G,a)$ be a graph without loops satisfying $a(u,v)=a_{uv}e^{i\phi_{uv}}, (\phi_{uv}\neq 2\pi n)$. Then the associated Laplacian is positive semidefinite.\\
\noi {\bf Proof :} This follows directly from theorem 2.6.4.\hspace{\stretch{1}}$ \blacksquare$\\
\noi{\bf Observation 2.7.4 :} Let $(G,a)$ be a graph satisfying $a(u,v)=a_{uv}e^{i\phi_{uv}}$ and $a(\{v,v\})\geq 0$ for all vertices in $V(G,a)$. Then the associated Laplacian is positive semidefinite.\\
\noi {\bf Proof :} The Laplacian is a Hermitian matrix which is diagonally  dominant. Therefore, by Gersgoin circle criterion \cite{ltb85,hjb90,mmb92} it is positive semidefinite.\hspace{\stretch{1}}$ \blacksquare$

On a $n$ vertex graph $(G,a)$, we define a new graph operator $\Theta(u_i)$ which deletes the vertex $u_i$ and rolls the edges incident on $u_i$ into loops with same weights on the edges connecting neighbors of $u_i$ as shown in figure 2.20. We call the resulting subgraph {\it principal subgraph}.  The Laplacian of the principal subgraph obtained by operating $\Theta(u_i)$ on $(G,a)$ is the principal submatrix of the Laplacian of $(G,a)$ obtained by deleting $i$th row and $i$th column.
\begin{figure}[!ht]
\begin{center}
\includegraphics[width=8cm,height=3cm]{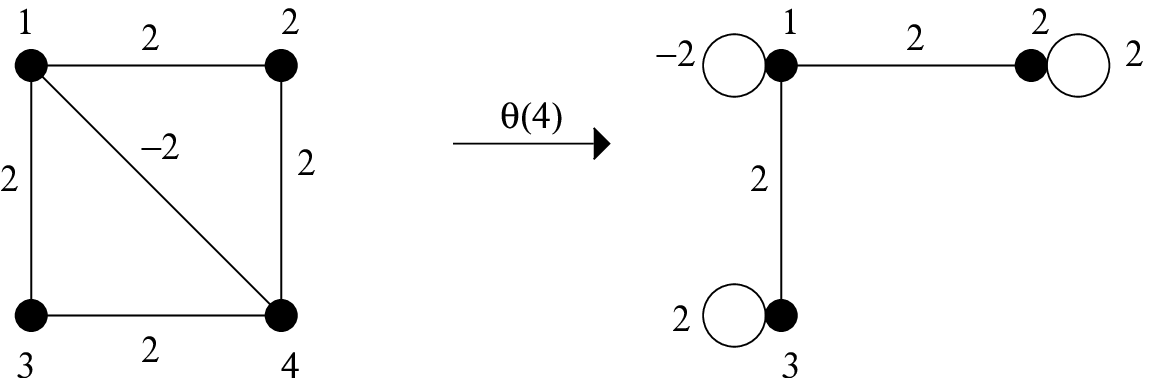}

Figure 2.20
\end{center}
\end{figure}\\
\noi{\bf Lemma 2.7.5 :} If one or more principal subgraphs of $(G,a)$ are not positive semidefinite, then $(G,a)$ is not positive semidefinite.\\
\noi{\bf Proof :} This follows from the result that all the principal submatrices of a positive semidefinite matrix are positive semidefinite \cite{hjb90}.\hspace{\stretch{1}}$ \blacksquare$

\noi{\bf Lemma 2.7.6 :} Let $(G,a)$ be either a $n$ vertex tree $(n\geq 2)$ or a $n$ vertex cycle $(n\geq 4)$. We assume that there are no loops in $(G,a)$ and that $a(u,v)$ is real for all $\{u,v\} \in E(G,a)$. Then $(G,a)$ has a positive semidefinite Laplacian if and only if $a(\{u,v\})>0$ for all $(u,v)\in E(G,a)$.

\noi{\bf Proof :} \noi{\it Only if part :} We prove that $a(\{u,v\})<0$ for any one $\{u,v\} \in E(G,a) \Longrightarrow Q(G,a)\ngeq 0$. Let $(T,a)$ be a tree with $v_1,\dots,v_n$ vertices, and let $\{v_i,v_{i+1}\}$ be an edge in $(T,a)$ with negative weight $a(\{v_i,v_{i+1}\})<0$. We operate on $(T,a)$ by $\Theta(v_{i+1})$. There are two possibilities. If $v_{i+1}$ is a leaf, we get only one component with a negative weighted loop on $v_i$. By lemma 2.7.2, the Laplacian of this principal subgraph is not positive semidefinite and by lemma 2.7.5 the Laplacian of $(T,a)$ is also not positive semidefinite. If $v_{i+1}$ is not a leaf then $\Theta(v_{i+1})$ will result in two or more principal subgraphs. The principal subgraph containing vertex $v_i$ is a graph having one loop with negative weight. By lemma 2.7.2 the Laplacian of this principal subgraph is not positive semidefinite and from lemma 2.7.5 $Q(T,a)\ngeq 0.$
Let $C_n$ be an $n$-cycle and let $a(\{v_i,v_{i+1}\})=a<0$. We operate by $\Theta(v_{i+1})$ which results in a $n-1$ vertex path, say $P_{n-1}$ with $v_i$ having a negative loop and $v_{i+2}$ having a positive or negative loop. If both the loops are negative we can use lemma 2.7.2 and lemma 2.7.5 in succession to show that $Q(C_n,a)\ngeq 0.$
Suppose the loop on $v_{i+2}$ is positive. Then for some $x\in R^{n-1}$ we have
$$x^TQ(P_{n-1},a)x=\sum_{k=1}^{n-2}a_{i_kj_k}(x_{i_k}-x_{j_k})^2+a(v_{i+1},v_{i+2})x_{v_{i+2}}^2-|a|x_{v_i}^2.$$
It is straightforward to check that $x^TQ(P_{n-1},a)x<0$ for $x^T=(1~\dots 0 ~ 1~\dots 1)$, that is a vector $x$ with all components $1$ except $v_{i+2}th$ component which is zero. Thus $Q(P_{n-1},a')\ngeq 0$. By lemma 2.7.5 $Q(G,a)\ngeq 0$.

\noi{\it If part :} Assume $Q(G,a)\ngeq 0$. This implies that there exists at least one $x\in R^n$ satisfying
$$x^TQ(G,a)x=\sum_k a(i_k,j_k)(x_{i_k}-x_{j_k})^2<0.$$

Since $(x_{i_k}-x_{j_k})^2\geq 0$ for all $k$, the above inequality is satisfied only when $a(i_k,j_k)<0$ for some $k$. This proves the if part.\hspace{\stretch{1}}$ \blacksquare$

We observe that the proof of if part applies to all graphs as it should.

\noi{\bf Lemma 2.7.7 :} Let all loops on a graph $(G,a)$ have real positive weights. Let every edge $\{u,v\}\in E(G,a)$ having $a(u,v)<0$ be common to pair of $C_3$. Let all such pairs of $C_3$, each containing a negative edge be disjoint. Let all the edges in each pair of $C_3$, other than the contained negative edge have positive weights satisfying $a(u,v)$ greater than the absolute value of the weight on the negative edge. Then the Laplacian of $(G,a)$ is positive semidefinite.

\noi{\bf Proof :} Consider a negative edge common to two $C_3$'s as shown in the figure 2.21.

\begin{figure}[!ht]
\begin{center}
\includegraphics[width=3cm,height=3cm]{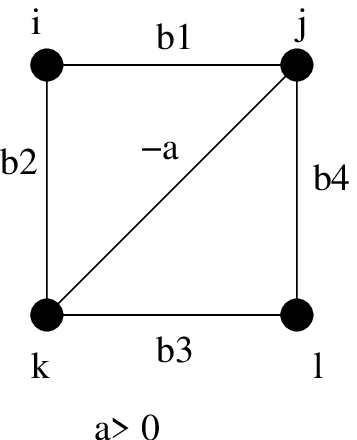}

Figure 2.21
\end{center}
\end{figure}

By hypothesis $b_i>a,\;i=1,2,3,4$. We can write $b_i=a+c_i,c_i>0,\;i=1,2,3,4$. We can decompose this graph as the edge union as shown in figure 2.22.
\begin{figure}[!ht]
\begin{center}
\includegraphics[width=12cm,height=7cm]{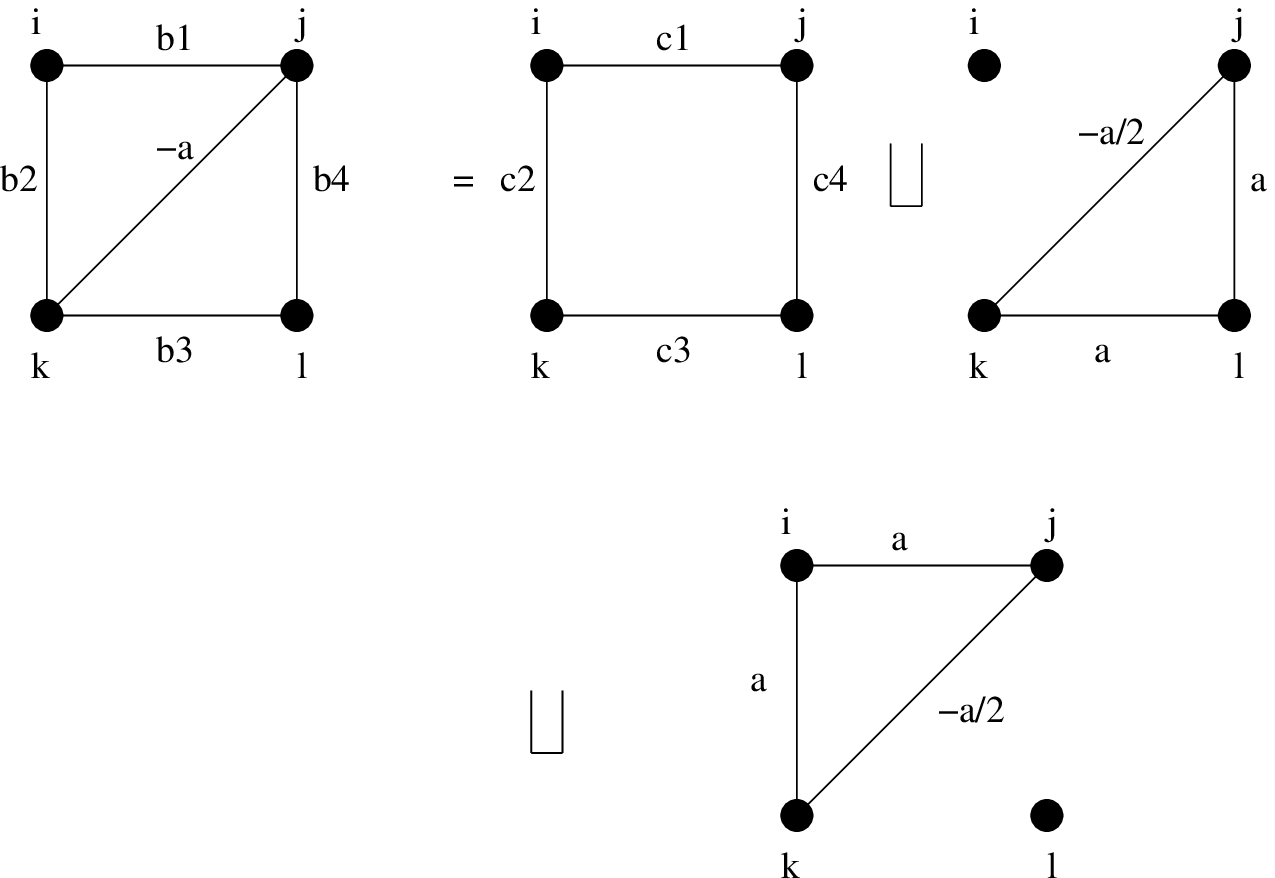}

Figure 2.22
\end{center}
\end{figure}

The first graph on RHS has all positive weights and hence has a positive semidefinite Laplacian. It is straighforword to check that the second and the third graphs on RHS correspond to the projectors $P[\fr{1}{\sqrt2}(|j\ran-2|l\ran+|k\ran)]$ and $P[\fr{1}{\sqrt2}(|j\ran-2|i\ran+|k\ran)]$ respectively. Hence they have positive semidefinite Laplacians. The Lapalcian of the graph on LHS is the sum of the Laplacian of the graphs on RHS (lemma 2.2.10), each of which is positive semidefinite. But we Know that the sum of positive semidefinite matrices is a positive semidefinite matrix \cite{hjb90}. Now the graph $(G,a)$ can be written as edge union of the factors (spanning subgraph) as figure 2.21 (possibly more than once) and the remaining factor which has all positive weights. The Laplacian of each factor is  positive semidefinite and the Laplacian of the given graph, being the sum of positive semidefinite matrices, is positive semidefinite.\hspace{\stretch{1}}$ \blacksquare$

\noi{\bf Lemma 2.7.8 :} If all the negative  edges of a real weighted  graph $(G,a)$ occur  as in  the  following   subgraph as shown in figure 2.23, where  $c_i>b\; ;\; i= 1, 2, \dots, 8$ and $b>a>0;$ then the associated Laplacian is positive semidefinite.

\begin{figure}[!ht]
\begin{center}
\includegraphics[width=8cm,height=6cm]{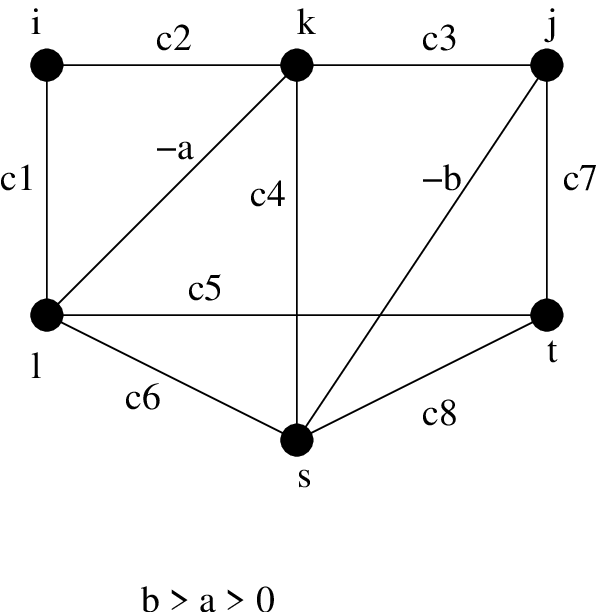}

Figure 2.23
\end{center}
\end{figure}

\noi{\bf Proof :} We can decompose the above graph into factors as shown in figure 2.24

\begin{figure}[!ht]
\begin{center}
\includegraphics[width=14cm,height=8cm]{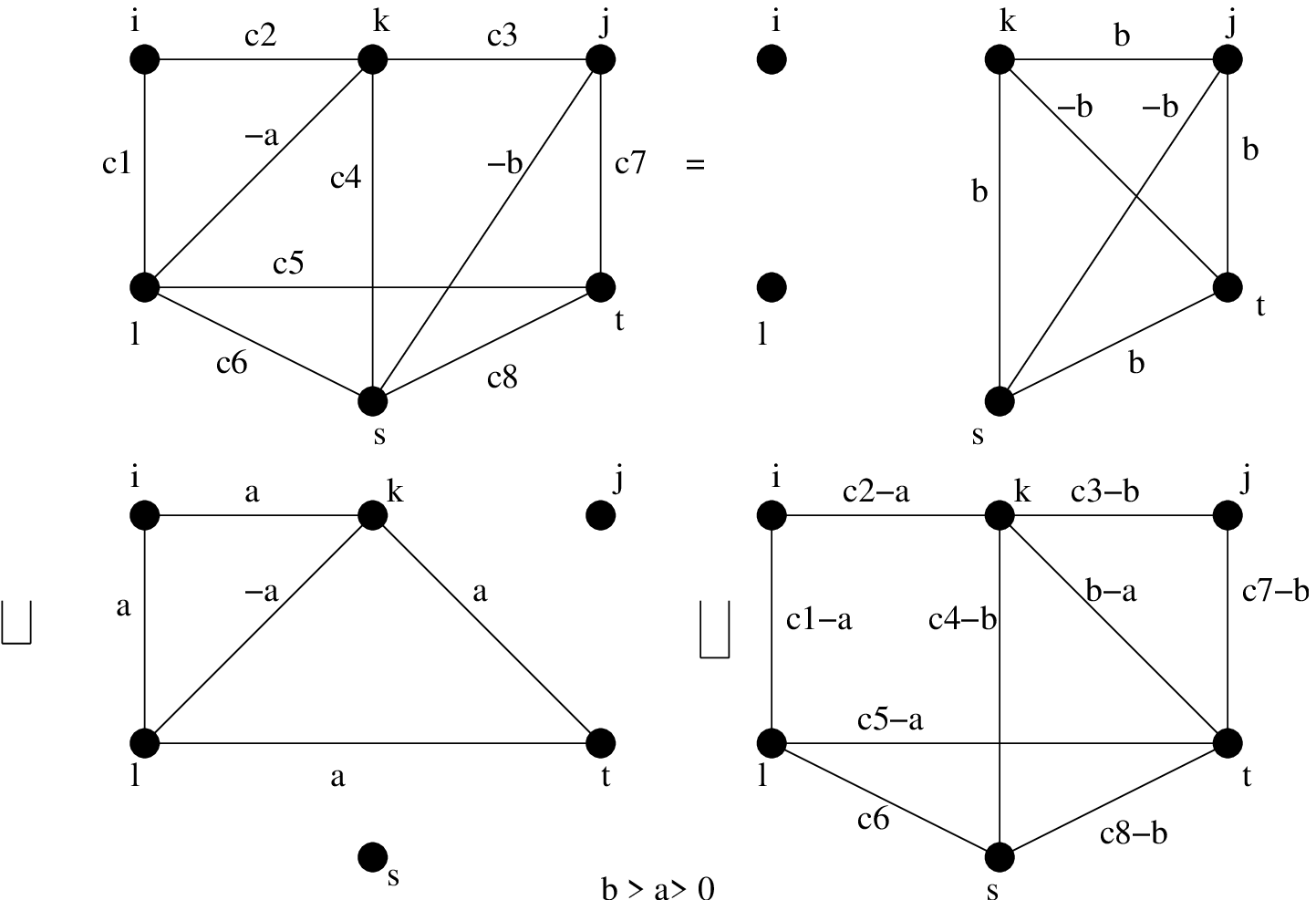}

Figure 2.24
\end{center}
\end{figure}

From the graphical equation in figure 2.24 we see that the first factor on RHS corresponds to $P[|-\ran|-\ran]$, the second factor has a positive semidefinite Laplacian from lemma 2.7.7 and the third factor has a positive semidefinite Laplacian as it has all positive weights.  Since this graph occurs (once or more) as disjoint subgraphs of $(G,a)$ it can be written as edge union of one or more of these subgraphs and the remaining graph containing only positive or complex edges. Since each of these has a positive semidefinite Laplacian, $(G,a)$ also has a positive semidefinite Laplacian.\hspace{\stretch{1}}$ \blacksquare$

\noi{\bf Lemma 2.7.9 :} Let $(G^{2^n},a)$ be a complete signed graph with weight function $a_{ij}\in \{-1,1\}$
without loops on $2^n$ vertices $n\geq 1$. Let $E_i$ denote the set of edges incident on $i$th vertex $(|E_i|=2^n-1)$ and let $E_i^+, E_i^-$ denote the sets of edges incident on the $i$th vertex with weight $+1$ and $-1$ respectively, $(E_i=E_i^+ +E_i^-)$. Let $(G^{2^n},a)$ satisfy the following condition
(i), $|E_i^-|=2^{n-1}-1, i=1, 2, \dots, 2^n$, so that the degree of every vertex$=1$.
Then $(G^{2^n},a)$ corresponds to a pure state in $2^n$ dimensional Hilbert space.

\noi{\bf Proof :} We need to prove that condition (i) in the statement of the lemma can be realized for all $n$ and that the resulting signed graph corresponds to a pure state for all $n$. We use induction on $n$. It is clear that the assertion is true for $n=1$ with the corresponding pure state given by $P[\fr{1}{\sqrt2}(|1\ran-|2\ran)]$.
Now assume that assertion (that is condition (i) and purity of the corresponding state) is true for $n=k$. For the graph corresponding to $n=k+1$ with $|V(G,a)|=2^{k+1},$ consider the modified tensor product
\benrr
(G^2,a)\boxdot (G^{2^k},a)& = & (G^{2^{k+1}},a)\\
& = &  \{ \cL(G^2, a) \otimes \cL \eta (G^{2^k}, a)\} \dotplus \{\cL(G^2, a) \otimes \cN(G^{2^k}, a)\}\dotplus\\
& & \{ \cN(G^2, a) \otimes \cL(G^{2^k}, a)\} \dotplus
 \{ \Om(G^2, a) \otimes \Om(G^{2^k}, a)\}\mbox{(2.63)}
\eenrr
where $\cL, \eta, \cN$ and $\Om$ are graph operators defined in equation (2.20b) and $G^2,  G^{2^k}$ are graphs with number of vertices  $2$ and $2^k$ respectively.
Note that the last term corresponds to an empty graph as $G^{2^k}$ does not have any loops. Since the modified tensor product of two complete graphs is also a complete graph,  $(G^{2^{k+1}},a)$ is a complete graph. Therefore $|E_i(G^{2^{k+1}},a)|=2^{k+1}-1,  i=1, 2, \dots, 2^{k+1}$.
To show that condition (i) is realized for $(G^{2^{k+1}},a)$ given the induction hypothesis,  we note that the first term in equation (2.63) contributes $|E^+_i(G^{2^k},a)|$ negative edges to $(1,i)$th vertex in $(G^{2^{k+1}})$ and the third term contributes $|E^-_i(G^{2^k},a)|$ negative edges, while the other two terms have no contribution. Therefore
$$|E^-_{1i}(G^{2^{k+1}},a)|=|E^+_i(G^{2^k},a)|+|E^-_i(G^{2^k},a)|=2^k-1$$
Similarly, the first three terms contribute $|E^-_i(G^{2^k},a)|, 1$ and $|E^+_i(G^{2^k},a)|$ positive edges to $(1,i)$th vertex. Therefore,
$$|E^+_{1i}(G^{2^{k+1}},a)|=|E^-_i(G^{2^k},a)|+|E^+_i(G^{2^k},a)|+1=2^{k-1}-1+2^{k-1}+1=2^k$$
and similarly for $E_{2i}.$
That $(G^{2^{k+1}},a)$ corresponds to pure state follows from the fact that the state corresponding to the modified tensor product of two graphs is the tensor product of the states corresponding to the factors. Since the state for $(G^{2^k},a)$ is pure by induction hypothesis and  $(G^2,a)$ is pure, the preceding statement means that $(G^{2^{k+1}},a)$ is a pure product state.

\hspace{\stretch{1}}$ \blacksquare$

\section{Summary and Comments}

Following is a brief summary of the main features of the chapter

\begin{verse}

(i) We have given rules to associate a graph with a quantum state and a quantum state to a graph, with a positive semidefinite generalized Laplacian, for states in real as well as complex Hilbert space sections (2.2 and 2.6).

(ii) We have shown that projectors involving states in the standard basis are associated with the edges of the graph (theorems 2.2.7 and 2.6.4)

(iii) We have given graphical criteria for a state being pure. In particular, we have shown that a pure state must have a graph which is a clique plus isolated vertices  (theorem 2.2.3, 2.2.4, remark 2.6.2)

(iv) For states in a real Hilbert space, we have given an algorithm to construct graph corresponding to a convex combination of density matrices, in terms of the graphs of these matrices algorithm 2.2.9.

(v) We have defined a modified tensor product of two graphs in terms of the graph operators $\cL, \eta, \cN, \Om$ and obtained the properties of these operators sections (2.4.2, 2.6.2). We have shown that this product is associative and distributive with respect to the disjoint edge union  of graphs (corollary 2.4.7, remark 2.6.11).

(vi) We have proved that the density matrix of the modified tensor product of two graphs is the tensor product matrices of the factors. (theorem 2.4.5,  2.6.10 ). For density matrices,  we show that a convex combination of the products of density matrices has a graph which is the edge union of the modified tensor products of the graphs for these matrices (corollary 2.4.11, 2.6.6). Thus we can code werner's definition of separability in terms of graphs.

(vii)  We have generalized the separability criterion given by  S. L. Braunstein, S. Ghosh,T. Mansour, S. Severini, R.C. Wilson \cite{bgmsw} to the real density matrices having graphs without loops (lemma 2.4.16).

(viii) We have found the quantum superoperators corresponding to the basic operations on graphs, namely addition and deletion of edges and vertices. it is straightforward to see that all quantum operations on states result in the addition / deletion of edges and/ or vertices, or redistribution of weights. However, addition / deletion of edges / vertices correspond to quantum operations which are irreversible, in general. Hence it seems to be difficult to encode a unitary operator, which has to be reversible, in terms of the operations on graphs. Further, graphs do not offer much advantage for quantum operations which only redistribute the weights, without changing the topology of the graph, as in this case the graph is nothing more than a clumsy way of writing the density matrix.

(ix) In section 2.6, we generalize the results obtained in sections 2.2 - 2.4, to quantum states in a complex Hilbert space, that is, to the density matrices with complex off-diagonal elements. In fact, all the results previous to section 2.5 go over to the complex case, except lemma 2.4.16. Many of these results have been explicitly dealt with (e.g. theorem 2.6.4, remark 2.6.2, section 2.6.2 etc). We have also given rules to associate a graph to a general Hermitian operator. We believe that any further advance in the theory reported in this chapter will prominently involve graph operators and graphs associated with operators.

(x)  Finally, we have given several graphical criteria for the positive semidefiniteness of the generalized Laplacian associated with a graph. Note that by lemma 2.7.3 and observation 2.7.4 all graphs with complex weights, either without loops or  with positive weighted loops, have positive semidefinite generalized Laplacians. This characterizes a large class of graphs coding quantum states.

 \end{verse}

 This chapter is essentially a generalization of the work by Braunstein, Ghosh and Severini \cite{bgs06} in which the idea of coding quantum mechanics of multipartite quantum systems in terms of graphs was implemented. The motivation in both Braunstein, Ghosh, Severini and this chapter is to explore the possibility of facilitating the understanding of mulipartite and mixed state bipartite entanglement using graphs and various operations on them. Whether such a goal can be reached is too early to say. In order to code arbitrary quantum states and observables in terms of graphs, we have to deal with weighted graphs with real or complex weights. Unfortunately, a mathematical theory of such graphs is lacking. Many results pertaining to simple graphs are not available for such weighted graphs. We hope that, through the need of understanding entanglement and related issues the mathematical structure of weighted graphs gets richer and in turn gives a feedback to our understanding of entanglement.

\chapter{On the degree conjecture for separability of multipartite quantum states}
\begin{center}
\scriptsize\textsc{Models are to be used, not believed.\\ H. Theil ( {\it Principles of Econometrics})}
\end{center}

\section{Introduction}

In chapter 2, we have given a formulation to use graphs to address the separability and related problems pertaining to multipartite quantum states. In this chapter, we use this formulation to settle a conjecture due to Braunstein {\it et al.} \cite{bgmsw,bgs06} regarding the separability of quantum states.

Braunstein {\it et al.} \cite{bgmsw} have made a conjecture, called degree conjecture, for the separability of multipartite quantum states. The conjecture states that a multipartite quantum state is separable if and only if the degree matrix of the graph associated with the state is equal to the degree matrix of the partial transpose (with respect to a subsystem, see below) of this graph. We call this to be the strong form of the conjecture. In its weak version, it requires only the necessity, that is, if the state is separable the corresponding degree matrices match. We prove the strong version for a $m$-partite {\it pure} state (section 3.2) and give a polynomial-time algorithm for factorization of a $m$-partite pure state (section 3.3). We show that the conjecture fails, in general, for mixed states (section 3.4). Finally, we prove the weak version of the conjecture for a class of multipartite mixed states (section 3.5).

\section{The Separability Criterion and its Proof}

 For the definition of a weighted graph with real or complex weights, denoted $(G,a),$ we refer the reader to chapter 2. We denote by $V(G,a)$ and $E(G,a)$ the vertex set and the edge set respectively of a weighted graph. The degree of a vertex $v \in V(G,a)$ is denoted by $\mathfrak{d}_v$ and $\mathfrak{d}_{(G,a)}=\sum_v \mathfrak{d}_v$ is the degree sum of the graph. The degree matrix of a weighted graph is denoted by $\Delta(G, a)$ (see chapter 2). The combinatorial Laplacian of the weighted graph is denoted $L(G,a)$ and the generalized Laplacian of $(G,a)$ is denoted $Q(G,a)$ (see chapter 2). If the generalized Laplacian of a graph $(G,a)$ is positive semidefinite, we can define the density matrix of the graph $(G,a)$ as $\si(G, a) = \fr{1}{\mathfrak{d}_{(G, a)}} Q(G, a).$  Conversely, given a density matrix, we can assign a graph to it (see chapter 2). A real weighted graph gets assigned to a density matrix with all real elements, while a complex weighted graph is assigned to a density matrix with complex off-diagonal elements. The vertices of the graph are labeled by the elements of the standard basis in the state space of the multipartite quantum system which is used to set up the density matrix. Also, in what follows, we use the definition and properties of the modified tensor product of graphs proved in chapter 2.

Let $(G,a)$ be the graph corresponding to a $m$-partite pure state in $d=d_1 d_2 \cdots d_m$ dimensional Hilbert space $\mathcal{H}=\mathcal{H}^{d_1}\otimes \mathcal{H}^{d_2} \otimes \cdots \otimes \mathcal{H}^{d_m}$, where $d_i$ is the dimension of  $\mathcal{H}^{d_i},\; 1\le i \le m$. Each vertex of  $(G,a)$ is labeled by an $m$ tuple $(v_1v_2\cdots v_m)$ where  $1 \le v_i \le d_i,\; i=1,2,\cdots,m$. In other words, we set up $(G,a)$ using the standard basis in $\mathcal{H}$.  Since $(G,a)$ is the graph of a pure state, it must be a clique on some subset of $V(G,a)$, all vertices not belonging to this subset being isolated (see chapter 2).
We divide the $m$ parts of the system in two nonempty disjoint subsets (partitions) whose union makes up  the whole system. We call them $s$ and $t$, where $t$ is the complement of $s$ in the set of all parts of the system. That is, $s$ and $t$ are the nonempty subsets of $\{1,2,\cdots,m\}$, $s \cup t =\{1,2,\cdots,m\},$ and $s \cap t = \phi$. This corresponds to $\mathcal{H}=\mathcal{H}^{(s)} \otimes \mathcal{H}^{(t)},$  where $\mathcal{H}^{(s)}=\mathcal{H}^{d_{i_1}} \otimes \mathcal{H}^{d_{i_2}} \otimes \mathcal{H}^{d_{i_3}}\otimes \cdots \otimes \mathcal{H}^{d_{i_s}},$ and $\mathcal{H}^{(t)}=\mathcal{H}^{d_{j_1}} \otimes \mathcal{H}^{d_{j_2}} \otimes \mathcal{H}^{d_{j_3}}\otimes \cdots \otimes \mathcal{H}^{d_{j_t}},$ with $\{ i_1 ,i_2,\cdots ,i_s \} = s,$ and $\{j_1,j_2, \cdots ,j_t\}  = t.$
As  a result of this division, we can  divide the $m$ tuple $v=(v_1 v_2\cdots v_m)$ into  the  corresponding partitions (strings) which we  call $v_s,v_t.$ Thus, $v_s= v_{i_1} v_{i_2} \cdots v_{i_s} $ and $v_t= v_{j_1} v_{j_2} \cdots v_{j_t}.$ We can label each vertex equivalently by $(v_s,v_t)$. We call $v_s$ The $s$ part and $v_t$ the $t$ part of the vertex label. For example,  consider a four partite system whose parts are labeled $1,2,3,4$. Let $s=\{1,4\} ,\; t=\{2,3\}$. Then, a vertex label $(1122)$ can be written as $(12,12)$. A vertex label $(v_1v_2v_3v_4)$ becomes $(v_1 v_4,v_2v_3).$

\subsection{Partial transpose with partition  $s$.}

This is a graph operator denoted $T_s$ operating on $E(G,a)$ which we define separately for the graphs with real and complex weights.

\noi\textbf{Definition 3.2.1} : Let $(G,a)$ be a graph with real weights. The operator $T_{s}$ is defined as follows:
$$T_s : (G,a) \longmapsto (G^{T_s},a^{\prime}),\; (G,a)\;\ni\{(v_s,v_t),(w_s,w_t)\}$$  $$\longmapsto\;  \{(w_s,v_t),(v_s,w_t)\}\;\in\;(G^{T_s},a^{\prime}),$$
with $$a^{\prime}(\{(w_s,v_t),(v_s,w_t)\})=a(\{(v_s,v_t),(w_s,w_t)\}),$$ that is, $ a^{\prime}(T_s e)= a(e)$. Note that, in general, $E(G,a)$ is not closed under  $T_s$.  The operator $T_t$  for the partition $t$ giving the complement of $s$ in the $m$-partite system is defined in the same way. Note that $T_s=T_t$.

\noi\textbf{Definition 3.2.2} : Let $(G,a)$ be a graph with complex weights. The operator $T_{s}$ is defined as follows.
$$T_s : (G,a) \longmapsto (G^{T_s},a^{\prime}),\; (G,a)\;\ni\{(v_s,v_t),(w_s,w_t)\}$$
$$ \longmapsto\;  \{(w_s,v_t),(v_s,w_t)\}\;\in\;(G^{T_s},a^{\prime}),$$
with $$|a^{\prime}(\{(w_s,v_t),(v_s,w_t)\})| = |a(\{(v_s,v_t),(w_s,w_t)\})|,$$ that is, $ |a^{\prime}(T_s e)|=|a(e)|.$  Note that, in general, $E(G,a)$ is not closed under  $T_s$.  The operator $T_t$  for the partition $t$ giving the complement of $s$ in the $m$-partite system is defined in the same way. Note that $T_s = T_t$.

\noi\textbf{Definition 3.2.3} : Partial transpose of $(G,a)$ with real or complex weights  with respect to partition $s$, denoted $(G^{T_s},a^{\prime}),$ is the graph obtained by acting $T_s$ on $E(G,a)$. Note that $V(G,a)\;=\;V(G^{T_s},a^{\prime})$,  but in general $E(G,a) \ne E(G^{T_s},a^{\prime})$.

 An edge joining the vertices, whose labels have either the same $s$ part or the same $t$ part  or both are fixed points of $T_s$. We have $$ T_s \{(v_s,v_t),(v_s,w_t)\}\;=\; \{(v_s,v_t),(v_s,w_t)\},$$
with the condition on weights  automatically satisfied.
Similarly
$$T_s(e) \;=\; T_s \{(v_s,v_t),(w_s,v_t)\} = \{(w_s,v_t),(v_s,v_t)\}\;=\; \{(v_s,v_t),(w_s,v_t)\},$$  with the condition on weights automatically satisfied. Note that  in the case of complex weighted graphs the action of $T_s$ on   $\{(v_s,v_t),(w_s,v_t)\}$ changes its orientation and hence  $a^{\prime}(T_s(e))\;=\;a^*(e).$ However, by the definition of $T_s$, this edge is still invariant under $T_s$. Further,  the definition of the operator $T_s$ leaves the phase of  $a^{\prime}(T_s(e))$, $e \in E(G,a),$ as a free parameter. We shall use this freedom later in fixing the phases of the graphs corresponding to the factors in the tensor product decomposition of a density matrix.

If both $s$ and $t$ parts of two vertices of $e \in E(G,a)$ are the same, then both vertices are identical and we have a loop, which is obviously preserved under $T_s$. Thus, $T_s$ divides $E(G,a)$ into two partitions, one containing all fixed points of $T_s$, that is, edges with same $s$ or $t$ part and all loops, which we call $\mathcal{F}$ set and the other containing the remaining edges which we call $\mathcal{C}$ set. In other words, $(G,a)$ is  the disjoint edge union of the two spanning subgraphs corresponding to  the $\mathcal{F}$ set and the $\mathcal{C}$ set.

Note that  $T_s^2$ is the identity operator $(T_s^2\;=\;1)$. Thus, $T_s$ is its own inverse and is one to one and onto.

\noi\textbf{Lemma 3.2.4} :\textit{ Let $(G,a)$ be a graph of a pure state in the Hilbert space $\mathcal{H}=\mathcal{H}^{d_1}\otimes \mathcal{H}^{d_2} \otimes \cdots \otimes \mathcal{H}^{d_m}$. Let $(G^{T_s},a^{\prime})$  be  the partial transpose of $(G,a)$ with respect to a partition $s$ as defined above. Then, $E(G,a)$ is  closed under $T_s$, $E(G,a)=E(G^{T_s} ,a^{\prime})$, if and only if $\Delta(G, a) = \Delta(G^{T_s}, a^{\prime})$.}

We  emphasize that the closure of $E(G,a)$ under $T_s$ means, for every $e \in E(G,a),\; T_s(e) \;= \; e' \in E(G,a),$ and $ a(e')\; =\; a(e)$ or  $ |a(e')|= |a(e)| $ as appropriate.

\noi\textbf{Proof :}
\noi{ \it Only if part} :
We are given that $E(G,a)$ is closed under $T_s$. We divide $E(G,a)$ into two partitions $\mathcal{C}$  and  $\mathcal{F}$   as above. Note that if $E(G,a)$ is closed under $T_s$ then the sets $\mathcal{C}$  and  $\mathcal{F}$  are separately closed under $T_s$. Consider now the set of edges incident on a vertex in $V(G,a)$. The edges in this set which belong to $\mathcal{F}$ set  are not shifted by $T_s$.  Since $E(G,a)$ is closed under $T_s$  and $T_s^2 =1$,  every incident edge belonging to $\mathcal{C}$  is the image of an edge in $\mathcal{C}$ with same weight  ( or same absolute value for the weight) under the action of  $T_s$  on  $\mathcal{C}$. Thus, the  degree of each vertex is preserved under the action of $T_s$ on $E(G,a)$, for both, real and complex weighted $(G,a)$ (see chapter 2),  so that $ \Delta (G,a) = \Delta(G^{T_s},a^{\prime}).$

\noi{\it If  part }:  We are given $ \Delta (G,a) = \Delta (G^{T_s},a^{\prime})$. The edges in  $ E(G,a)$ belonging to $\mathcal{F}$ remain in $E(G,a)$ under the action of $T_s$. Now suppose that the set $\mathcal{C}$  is not closed under $T_s$.  Since $T_s$  is its own inverse, the edge  $e$ for which $T_s (e) \notin E(G,a)$ cannot be the image of any other edge in $E(G,a)$  under $T_s$. Therefore, the degree of the end vertices of $e$ is changed under the action of $T_s$.  This contradicts the assumption  $ \Delta (G,a) = \Delta (G^{T_s},a^{\prime}).$\hfill $\blacksquare$\\

\noi\textbf{Lemma 3.2.5} :\textit{ Let $(G,a)$ be the graph of a pure state in the Hilbert space   $\mathcal{H}=\mathcal{H}^{d_1}\otimes \mathcal{H}^{d_2} \otimes \cdots \otimes \mathcal{H}^{d_m}$ of a $m$-partite quantum system. Then,  $(G,a)=(G_s,b) \boxdot (G_t,c) $, where $(G_s,b)$ is the graph of a pure state in the Hilbert space made up of $s$ factors $\mathcal{H}^{(s)}=\mathcal{H}^{d_{i_1}} \otimes \mathcal{H}^{d_{i_2}} \otimes \mathcal{H}^{d_{i_3}}\otimes \cdots \otimes \mathcal{H}^{d_{i_s}}$ and $(G_t,c)$ is the graph of a pure state in the Hilbert space made up of $t=m-s$ factors  $\mathcal{H}^{(t)}=\mathcal{H}^{d_{j_1}} \otimes \mathcal{H}^{d_{j_2}} \otimes \mathcal{H}^{d_{j_3}}\otimes \cdots \otimes \mathcal{H}^{d_{j_t}}$, if and only if the edge set $E(G,a)$ is closed under $T_s,\; s=\{i_1, i_2, i_3, \cdots, i_s\}.$ }

\noi\textbf{Proof :} \noi{\it Only if part} : {\it Case I} : Graphs with real  weights.
 We are given \\

$(G,a)=(G_s,b) \boxdot (G_t,c) $
$$=  \cL(G_s,b) \otimes \cL \eta (G_t, c) \dotplus \cL(G_s,b) \otimes \cN(G_t,c) \dotplus  \cN(G_s,b) \otimes \cL(G_t,c)$$
$$\dotplus  \Om(G_s, b) \otimes \Om(G_t,c)  \eqno{(3.1)}$$

Using the definition of the operators $ \cL, \eta, \cN, \Om$ (see chapter 2) and that of the tensor product of graphs, we can make the following observations. The second term is  a spanning subgraph of $(G,a)$ each of whose edges has a common $t$ part and hence is fixed point of $T_s$. The  third term is a spanning subgraph of $(G,a)$ each of whose edges has a common $s$ part, so that each edge is a fixed point of $T_s$. The fourth term is a spanning subgraph of $(G,a)$ which contains only loops all of which are fixed points of $T_s$. Again, from the definition of the tensor product, we see that in the first term, any edge $\{v_s,w_s\}$ in $\cL(G_s,b)$ with weight $b(\{v_s,w_s\})$ and any $\{v_t,w_t \}$ in  $\cL \eta(G_t,c)$ with weight  $c(\{v_t,w_t\})$  gives us, under the tensor product, two edges $\{(v_s,v_t),(w_s,w_t)\}$ and $\{(v_s,w_t),(w_s,v_t)\}$  with the same weight $b(\{v_s,w_s\}) c(\{v_t,w_t\})$, which are the images of each other under $T_s$. This proves that $E(G,a)$ is closed under $T_s$.

{\it Case II} : Graphs with complex weights .
 We are given \\

$(G,a)=(G_s,b) \boxdot (G_t,c) $
$$=  \cL(G_s,b) \otimes \cL (G_t, c) \dotplus \cL(G_s,b) \otimes \cN(G_t,c) \dotplus \cN(G_s,b) \otimes \cL(G_t,c)$$
  $$ \dotplus \{ \Om(G_s, b) \otimes \Om(G_t,c) \sqcup 2 \cN \cL (G_s,b) \otimes \cN \cL \eta (G_t,c)\}  \eqno{(3.2)}$$
Note  that the first three terms are similar to those in Eq. (3.1) and the arguments corresponding to these terms in the paragraph following Eq.(3.1) apply, except that we require $|a(T_s e)| = |a(e)| $. Again, fourth term correspond to graph with loops (and no edges) which are fixed points of $T_s$. This proves that $E(G,a)$ is closed under $T_s$.

\noi{\it If part} : We begin by noting that the graph $(G,a)$ has the structure of a clique and isolated vertices, with $|V(G,a)| = d = d_1 d_2 \cdots d_m$,  where $d_i$ is the dimension of $\mathcal{H}^{d_i},\; 1 \le i \le m$, because it is the graph of a pure state in  $\mathcal{H}=\mathcal{H}^{d_1}\otimes \mathcal{H}^{d_2} \otimes \cdots \otimes \mathcal{H}^{d_m}$. Let $(K_n,a)$ denote the clique in $(G,a)$ and $V_k(G,a) $ be the set of vertices on the clique. Let $|V_k(G,a)|= n$. We are given that $E(G,a)$ is closed under $T_s$. Note that all loops are on the clique and no loops are on the isolated vertices. Consider a vertex $(v_s,v_t)$ on $(K_n,a)$. Let $q$  denote the number of vertices in $(K_n,a)$  having the same $s$ part as $(v_s,v_t)$ and  $p$ denote the number of vertices on $(K_n,a)$ with the same $t$ part as $(v_s,v_t)$. We note that $p$ and $q$  are the same for all vertices on $(K_n,a)$, otherwise the set $\mathcal{C}$ is not closed under $T_s$. We draw $(K_n,a)$ as a lattice of $p$ rows and $q$ columns, such that all vertices in one row have common $s$ part and all vertices in one column have common $t$ part. Since $(K_n,a)$ is a complete graph, from figure 3.1, we see that any vertex $(v_s,v_t)$ has $(p-1)+(q-1)$  neighbors giving edges in the $\mathcal{F}$ set and $(p-1)(q-1)$ neighbors giving edges in the $\mathcal{C}$ set. Since $(K_n,a)$ is complete $(v_s,v_t)$ has $n-1$ neighbors giving $n=pq$.

\begin{figure}[!ht]
\begin{center}
\includegraphics[width=10cm,height=10cm]{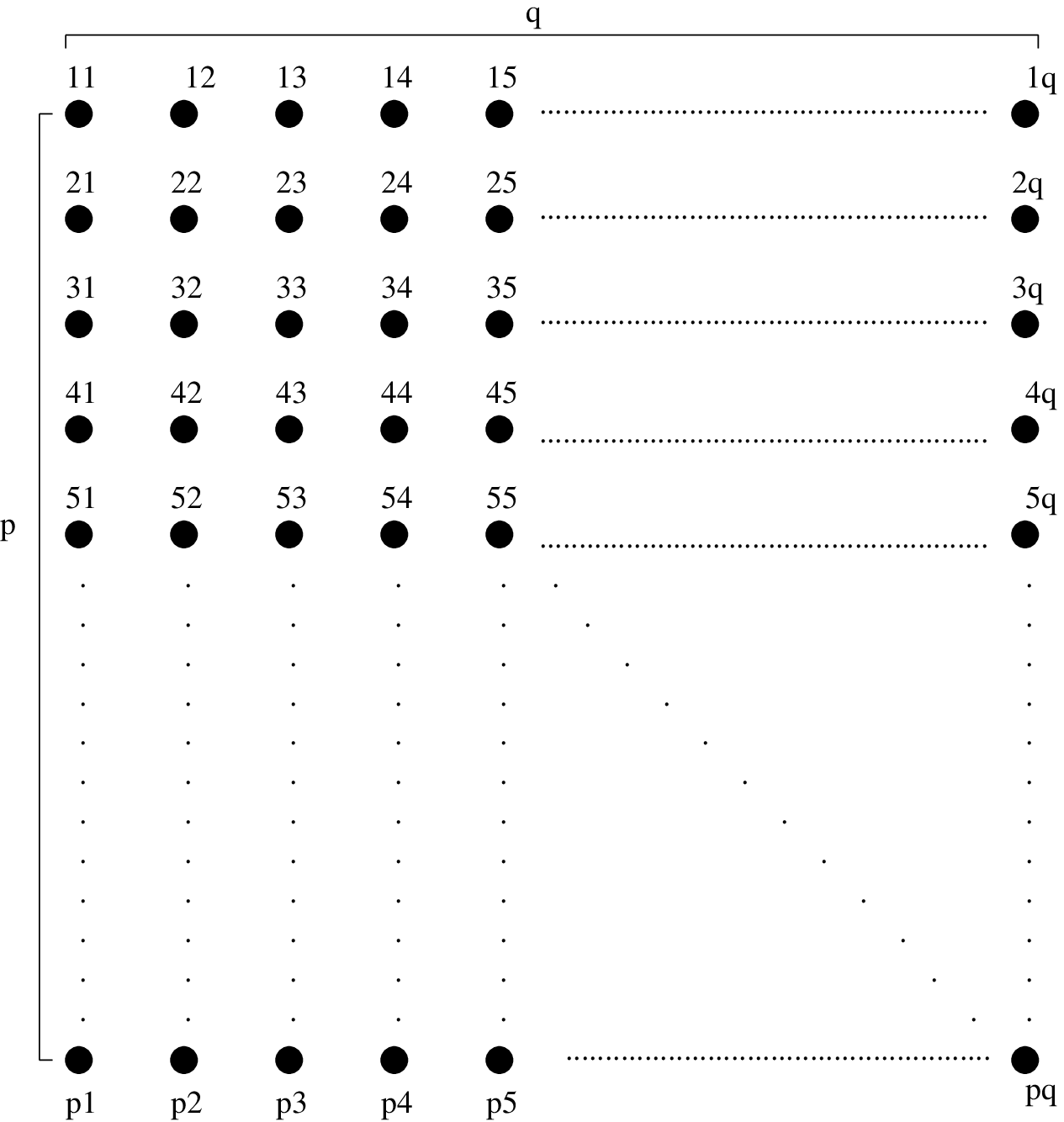}

Figure 3.1: Every row contains vertices with same $ s$ part, and every column contains vertices with same $t$ part. In a vertex label, first number stands for the $s$ part and the second for the $t$ part. For example, the edges between the vertex 34 and all the vertices in the 3rd row and 4th column are in the set  $\mathcal{F}$, while the edges between vertex 34 and all the vertices in the four blocks obtained by deleting third row and fourth column are in the set  $\mathcal{C}$.
\end{center}
\end{figure}

Now consider $\mathcal{C}$ set on $(K_n,a)$. From the definition of the tensor product of weighted graphs (see chapter 2), we can factorize each pair $\{(v_s,v_t)(w_s,w_t)\}$,
$\{(w_s,v_t),(v_s,w_t)\}$  in the set $\mathcal{C}$ as the tensor product of two edges $\{v_s,w_s\}$ and $\{v_t,w_t\}$ with weights $b'$ and $c'$ satisfying  $$a(\{(v_s,v_t),(w_s,w_t)\})=  b'(\{v_s,w_s\}) \cdot c'(\{v_t,w_t\}) =a(\{(w_s,v_t),(v_s,w_t)\})$$ or $$|a(\{(v_s,v_t),(w_s,w_t)\})| =  |b'(\{v_s,w_s\})| \cdot |c'(\{v_t,w_t\})| = |a(\{(w_s,v_t),(v_s,w_t)\})|.$$  Writing each pair $\{e,T_s e\}$ in $\mathcal{C}$   set in this way and taking the disjoint edge union (see chapter 2) in all these tensor products, we get $(G_{\mathcal{C}},a)= (G'_1,b') \otimes (G'_2,c')$  where $(G_{\mathcal{C}},a)$ is the spanning subgraph on $(K_n,a)$ corresponding to set $\mathcal{C}$. $(G'_1,b')$ and $(G'_2,c')$ are graphs on $p$ and $q$ vertices, respectively. Again, from the definition of the tensor product of weighted graphs, we know that isolated vertices in the factors produce isolated vertices in the product. Therefore, we can add $(d_s-p)$ isolated vertices to $(G'_1,b')$ and $(d_t-q)$ isolated vertices to $(G'_2,c')$, where $d_s=d_{i_1} d_{i_2} \cdots d_{i_s}$ and $d_t=d_{j_1} d_{j_2} \cdots d_{j_t}$. We call this graphs $\cL(G'_s,b')$ and $\cL(G'_t,c')$  (the operator $\cL$ defined in chapter 2 removes loops from a graph). The tensor product $\cL(G'_s,b') \otimes \cL(G'_t,c')$ gives the spanning subgraph of $(G,a)$ corresponding to set $\mathcal{C}$. Now consider a row in figure 3.1 containing vertices with common  $s$ part  say $v_s$. This row generates $q(q-1)/2$ edges of the form $\{(v_s,v_t),(v_s,w_t)\}$
 all in the set $\mathcal{F}$. By the definition of the Cartesian product of the weighted graphs (see chapter 2), each of these edges is the Cartesian product of the vertex $v_s$ in, say, $(G_1,b)$ with  the edge $\{v_t,w_t\}$, in, say, $(G_2,c)$, where $a(\{(v_s,v_t),(v_s,w_t)\}) =\mathfrak{d}_{v_s} c(\{v_t,w_t\}).$  Thus, the graph $(G_2,c)$ is a graph of $q$ vertices obtained by projecting each of the $q(q-1)/2$ edges $\{(v_s,v_t),(v_s,w_t)\}$ to $\{v_t,w_t\}$, with corresponding weight assignments, and thus is a complete graph on $q$ vertices. As we vary $v_s$ through its $p$ possible values, one row corresponding to each value,  the definition of the Cartesian product of weighted graphs generates the same $(G_2,c) $ possibly with different weights on edges.
In exactly the same way, $q$ columns in figure 3.1 generate the complete graph on $p$ vertices $(G_1,b)$ by employing the Cartesian product of weighted graphs. Noting that Cartesian product of an isolated vertex in $(G_1,b)$ with an edge in $(G_2,c)$ or vice versa gives an isolated vertex in the product graph,  we have shown that the spanning subgraph of $(G,a)$ corresponding to the $\mathcal{F}$ set edges gets generated by $(G_s,b) \square (G_t,c)$ containing $d_s$ and $d_t$ vertices, respectively.  Finally, note that there are no loops in $(G_s,b) \square (G_t,c)$ because loops contribute to the Cartesian product of weighted graphs only via the degrees $\mathfrak{d}_{v_s}$ and $\mathfrak{d}_{v_t}$.
In  the above analysis, question can be raised regarding the weight functions of the factors. Thus, more than one weight functions can generate the spanning subgraph corresponding to the $\mathcal{C}$ set, while the $\mathcal{F}$ set edges on different rows in figure 3.1 may be generated by different weight functions on the factors in the Cartesian product. These points will be addressed later in this proof.
We  denote by $G_1$ and $G'_1$  to be the graphs underlying  $(G_1,b)$ and $(G'_1,b')$ and similarly define $G_2$,  $G'_2$, $G$, $G'_1 \otimes G'_2$ and $G_1 \square G_2$. We first note that $|V(G_1)| =p=|V(G'_1)|$ and $|V(G_2)|=q=|V(G'_2)|$. In fact, we can identify the set of vertices for $ G_1$ and $G'_1$ and for $G_2$ and $G'_2$, respectively, that is, $V(G_1)= V(G'_1)$ and $V(G_2)= V(G'_2)$. We now show that the identity map on $V(G_1)$ is an automorphism taking $G_1$ to $G'_1$. In other words, $G_1$ and $G'_1$ are identical.  Consider  a vertex $(v_s,v_t)$ in $G$ and let $N_G(v_s,v_t)$ denote its neighborhood in $G$. We denote by $N_{G_1 \square G_2}(v_s,v_t)$ and $N_{G'_1 \otimes G'_2}(v_s,v_t)$  the neighborhoods of $(v_s,v_t)$ in $G_1 \square G_2$ and $G'_1 \otimes G'_2$, respectively.
 In other words, $N_{G'_1 \otimes G'_2}(v_s,v_t)$ contains neighbors of $(v_s,v_t)$ with edges in set $\mathcal{C}$  and  $N_{G_1 \square G_2}(v_s,v_t) $  contains the neighbors of $ (v_s,v_t)$ with edges in set   $\mathcal{F}$. Clearly, $N_{G'_1 \otimes G'_2}(v_s,v_t)$ and   $N_{G_1 \square G_2}(v_s,v_t) $  partition  $N_G(v_s,v_t)$.
 Let $V_K(G_1)$ and $V_K(G_2)$ be the set of vertices on the clique in $G_1$ and $G_2$, respectively.  Consider \cite{ikb00}

  $N_{G'_1 \otimes G'_2}(v_s,v_t)= V_K(G) - \{(v_s,v_t)\} - N_{G_1 \square G_2}(v_s,v_t) $

$ =V_K(G) - \{v_s\} \times \{v_t\} -\{ \{v_s\} \times  N_{G_2}(v_t) \cup N_{G_1} (v_s) \times \{v_t\}\} $

 $ =V_K(G_1) \times V_K(G_2) - \{v_s\} \times \{v_t\} -\{ \{v_s\} \times  N_{G_2}(v_t) \cup N_{G_1} (v_s) \times \{v_t\}\} $

 $ =\{N_{G_1} (v_s) \cup \{v_s\} \} \times \{N_{G_2}(v_t) \cup \{v_t\}\} - \{v_s\} \times \{v_t\} -\{ \{v_s\} \times  N_{G_2}(v_t) \cup N_{G_1} (v_s) \times \{v_t\}\} $

$ =N_{G_1} (v_s) \times N_{G_2}(v_t)  \cup N_{G_1} (v_s) \times  \{v_t\}  \cup \{v_s\}  \times N_{G_2}(v_t) \cup\{v_s \} \times \{v_t\} - \{v_s\} \times \{v_t\} $
    $  -\{ \{v_s\} \times  N_{G_2}(v_t) \cup N_{G_1} (v_s) \times \{v_t\}\} $

$ =N_{G_1} (v_s) \times N_{G_2}(v_t) =N_{G_1 \otimes G_2}(v_s,v_t). $

Therefore \cite{ikb00},  $ N_{G'_1} (v_s) \times N_{G'_2}(v_t) = N_{G_1} (v_s) \times N_{G_2}(v_t)$

or  $N_{G'_1} (v_s) =N_{G_1} (v_s) $  and $N_{G'_2}(v_t) = N_{G_2}(v_t).$\\

Therefore,  $G_1$ and $G'_1$ and $G_2$ and $G'_2$  are identical.  In particular, $G_1$ and $G_2$ are cliques. $G_s=G'_s$ and $G_t=G'_t$ are cliques plus isolated vertices.  Thus, we get
$$ (G,a) =\cL(G_s,b') \otimes \cL(G_t,c') \dotplus(G_s,b)\square (G_t,c) \dotplus (G'',a), \eqno{(3.3)}$$
where $(G'',a)$ is the graph obtained from $(G,a)$ by removing all edges and keeping loops.
\\

The only remaining gap is to show that a consistent assignment of weights to the factors $G_s$ and $G_t$ is possible so as to express $(G,a)$ as modified tensor product. To get the required weight assignments, we use the requirement that both the factors in the modified tensor product must correspond to pure states.  Indeed, we know that both $G_s$ and $G_t$ have the form of clique plus isolated vertices as required for them to represent pure states. We know that the graph $(G,a)$ corresponds to pure state. Therefore, its weight function $a$ must satisfy (see chapter 2), {\it assuming $(G,a)$ to be a real weighted graph},

$$\sum_{(v_s,v_t) \in V(G,a)} \mathfrak{d}^2_{(v_s,v_t)} + 2\sum_{e \in E(G,a))} a^2(e) = \mathfrak{d}^2_{(G,a)}.  \eqno{(3.4)}$$

Splitting $E(G,a)$ into   $\mathcal{C}$ and $\mathcal{F}$ sets and using the definitions of the tensor and Cartesian products of weighted graphs, we get [note that a paired label is for a vertex in $V(G,a)$ and single labels with suffix $s$ and $t$ are vertices in $V(G_s,b)$ and $V(G_t,c)$, respectively]
$$
\sum_{(v_s,v_t) } \mathfrak{d}^2_{(v_s,v_t)} + 2 \sum_{(v_s,v_t) }
 \sum_{\begin{subarray}{I}
     \hskip  .5cm       (w_s,w_t)\\
     \hskip .2 cm { w_s \ne v_s , w_t \ne v_t}
   \end{subarray}} 							
 (a^2(\{(v_s,v_t),(w_s,w_t)\}) +a^2(\{(v_s,w_t),(w_s,v_t)\}))	
$$
  $$
+ 2 \sum_{(v_s,v_t)}
\sum_{\begin{subarray}{|}
  \hskip .2 cm  {(v_s,w_t)} \\
   \hskip .2 cm {  w_t \ne v_t}
\end{subarray}}
a^2(\{(v_s,v_t),(v_s,w_t)\})
+2 \sum_{(v_s,v_t) }
 \sum_{\begin{subarray}{|}
     \hskip .2 cm (w_s,v_t)\\
    \hskip .2 cm { w_s \ne v_s }
\end{subarray}}
a^2(\{(v_s,v_t),(w_s,v_t)\}) =$$
$$ \mathfrak{d}^2_{(G,a)}.
$$

 Since $E(G,a)$ is closed under $T_s$,  we can write

$$
\sum_{(v_s,v_t) } \mathfrak{d}^2_{(v_s,v_t)} + 4 \sum_{(v_s,v_t) }
\sum_{\begin{subarray}{I}
  \hskip  .5cm   (w_s,w_t) \\
   \hskip  .2cm   { w_s \ne v_s , w_t \ne v_t}
\end{subarray}}
 a^2(\{(v_s,v_t),(w_s,w_t)\})
$$
 $$
+ 2 \sum_{(v_s,v_t) }
\sum_{\begin{subarray}{I}
 \hskip  .1cm   (v_s,w_t) \\
 \hskip  .1cm   {  w_t \ne v_t}
\end{subarray}}
a^2(\{(v_s,v_t),(v_s,w_t)\})
+2 \sum_{(v_s,v_t) }
\sum_{\begin{subarray}{I}
 \hskip  .1cm   (w_s,v_t)\\
 \hskip  .1cm   { w_s \ne v_s }
\end{subarray}}
 a^2(\{(v_s,v_t),(w_s,v_t)\}) = $$
 $$\mathfrak{d}^2_{(G,a)}. \eqno{(3.5)}$$

 Using splitting of the weight functions in $\mathcal{C}$ set and $\mathcal{F}$ set, we get

$$
\sum_{(v_s,v_t) } \mathfrak{d}^2_{(v_s,v_t)} + 4 \sum_{v_s }
\sum_{\begin{subarray}{I}
\hskip  .4 cm {w_s }\\
\hskip  .1cm { w_s \ne v_s }
\end{subarray}}
b'^2(\{v_s,w_s\}) \sum_{v_t }
\sum_{\begin{subarray}{I}
\hskip  .4cm {w_t }\\
\hskip  .1cm { w_t \ne v_t }
\end{subarray}}
c'^2(\{v_t,w_t\})
$$

 $$
 + 2 \sum_{v_s }  \mathfrak{d}^2_{v_s} \sum_{v_t }
\sum_{\begin{subarray}{I}
\hskip  .4cm{w_t }\\
\hskip  .1cm{w_t \ne v_t}
\end{subarray}}
 c^2(\{v_t,w_t\}
+2 \sum_{v_t } \mathfrak{d}^2_{v_t}  \sum_{v_s }
 \sum_{\begin{subarray}{I}
\hskip  .4cm {w_s }\\
\hskip  .1cm {w_s \ne v_s}
\end{subarray}}
 b^2(\{v_s,w_s\}) =\mathfrak{d}^2_{(G,a)}. \eqno{(3.6)}$$

Since the graphs $(G_s,b),(G_t,c) ,(G_s,b')$, and $(G_t,c')$  correspond to pure states,  $b' , c', b$ and $c$ must satisfy

$$
\sum_{v_s} \mathfrak{d}^2_{v_s} + 2 \sum_{v_s }
\sum_{\begin{subarray}{I}
\hskip  .4cm {w_s}\\
\hskip  .1cm { w_s \ne v_s }
\end{subarray}}
b^2(\{v_s,w_s\})= \mathfrak{d}^2_{(G_s,b)}\eqno{(3.7)}$$

$$
\sum_{v_s } \mathfrak{d}^2_{v_s} + 2 \sum_{v_s }
\sum_{\begin{subarray}{I}
\hskip  .4cm {w_s }\\
\hskip  .1cm { w_s \ne v_s }
\end{subarray}}
 b'^2(\{v_s,w_s\})= \mathfrak{d}^2_{(G_s,b')}\eqno{(3.8)}$$

$$
\sum_{v_t } \mathfrak{d}^2_{v_t} + 2 \sum_{v_t }
 \sum_{\begin{subarray}{I}
\hskip  .4cm {w_t }\\
\hskip  .1cm{ w_t \ne v_t }
\end{subarray}}
c^2(\{v_t,w_t\})= \mathfrak{d}^2_{(G_t,c)}\eqno{(3.9)}$$

$$
\sum_{v_t } \mathfrak{d}^2_{v_t} + 2 \sum_{v_t}
 \sum_{\begin{subarray}{I}
\hskip  .4cm {w_t}\\
\hskip  .1cm { w_t \ne v_t }
\end{subarray}}
c'^2(\{v_t,w_t\})= \mathfrak{d}^2_{(G_t,c')}.\eqno{(3.10)}$$

We see that Eqs. (3.7)-(3.10) are consistent with Eq. (3.6) provided

\begin{verse}
(i)~~~$ \mathfrak{d}_{(v_s,v_t)}  = \mathfrak{d}_{v_s} \mathfrak{d}_{v_t}$ for all $(v_s,v_t) \in V(G,a)$ and consequently $\mathfrak{d}_{(G,a)} = \mathfrak{d}_{(G_s,b)} \mathfrak{d}_{(G_t,c)}$
and

(ii)~~~$ b^2(\{v_s,w_s\}) = b'^2(\{v_s,w_s\})$ and $c^2(\{v_t,w_t\}) = c'^2(\{v_t,w_t\})$.
\end{verse}

We first fix a vertex $(v_s,v_t) \in V(G,a)$ and obtain its degree $ \mathfrak{d}_{(v_s,v_t)}$. Summing the edges with the same $s$ part, we get

$$
\sum_{\begin{subarray}{I}
\hskip  .1cm (v_s,w_t)\\
\hskip  .1cm{ w_t \ne v_t}
\end{subarray}}
 a(\{(v_s,v_t),(v_s,w_t)\}) = \mathfrak{d}_{v_s}
\sum_{\begin{subarray}{I}
\hskip  .4cm {w_t} \\
\hskip  .1cm { w_t \ne v_t }
\end{subarray}}
c(\{v_t,w_t\}). \eqno{(3.11)}
$$

Adding edges with the same $t$ part, we have

$$
\sum_{\begin{subarray}{I}
\hskip  .1cm{(v_s,w_t)}\\
\hskip  .1cm{  w_s \ne v_s}
\end{subarray}}
 a(\{(v_s,v_t),(w_s,v_t)\}) = \mathfrak{d}_{v_t}
\sum_{\begin{subarray}{I}
\hskip  .4cm{w_s }\\
\hskip  .1cm{ w_s \ne v_s }
\end{subarray}}
b(\{v_s,w_s\}). \eqno{(3.12)}$$

Adding over edges in the  $\mathcal{C}$ set, we get

$$
\sum_{\begin{subarray}{I}
\hskip  .1cm (v_s,w_t) \\
\hskip  .1cm {  w_t \ne v_t}
\end{subarray}}
 a(\{(v_s,v_t),(w_s,w_t)\}) =
\sum_{\begin{subarray}{I}
\hskip  .4cm {w_s}\\
\hskip  .1cm{ w_s \ne v_s }
\end{subarray}}
b'(\{v_s,w_s\})
\sum_{\begin{subarray}{I}
\hskip  .4cm {w_t}\\
\hskip  .1cm { w_t \ne v_t }
\end{subarray}}
c'(\{v_t,w_t\}). \eqno{(3.13)}$$

Adding these three terms and the weight of the loop on $(v_s,v_t)$, we get $\mathfrak{d}_{(v_s,v_t)}$. The requirement that $\mathfrak{d}_{(v_s,v_t)} = \mathfrak{d}_{v_s} \mathfrak{d}_{v_t}$ is  satisfied provided

\begin{verse}

(iii)~~~$ b'(\{v_s,w_s\}) = b(\{v_s,w_s\})$ and $c'(\{v_t,w_t\}) = -c(\{v_t,w_t\})$  which leads to

(iv)~~~$\mathfrak{d}_{(v_s,v_t)} - a(\{(v_s,v_t),(v_s,v_t)\}) = \mathfrak{d}_{v_s} \mathfrak{d}_{v_t} - b(\{v_s,v_s\}) c(\{v_t,v_t\}).$
\end{verse}

 This is satisfied provided $ a(\{(v_s,v_t),(v_s,v_t)\}) = b(\{v_s,v_s\}) \cdot c(\{v_t,v_t\})$. This requirement is consistent with $ (G'',a) = \Om (G_s,b) \otimes \Om(G_t,c).$

We can write, therefore, $$ (G, a)   =  \cL(G_s, b) \otimes \cL  (G_t,- c) \dotplus (G_s, b) \square (G_t,c)\\
\dotplus  \Om(G_s, b) \otimes \Om(G_t,c)$$  $$ = (G_s,b) \boxdot (G_t,c).$$

Now, let us deal with the case where $(G,a)$ \textit{ is a graph with complex weights} (see chapter 2). In this case Eq. (3.4) is replaced by

$$\sum_{(v_s,v_t) \in V(G,a)} \mathfrak{d}^2_{(v_s,v_t)} + 2\sum_{e \in E(G,a))} |a(e)|^2 = \mathfrak{d}^2_{(G,a)}  \eqno{(3.4')}$$

and Eqs. (3.5)-(3.10)  become

$$
\sum_{(v_s,v_t) } \mathfrak{d}^2_{(v_s,v_t)} + 4 \sum_{(v_s,v_t) }
\sum_{\begin{subarray}{I}
  \hskip  .5cm   (w_s,w_t) \\
   \hskip  .2cm   { w_s \ne v_s , w_t \ne v_t}
\end{subarray}}
| a(\{(v_s,v_t),(w_s,w_t)\})|^2
$$
 $$
+ 2 \sum_{(v_s,v_t) }
\sum_{\begin{subarray}{I}
 \hskip  .1cm   (v_s,w_t) \\
 \hskip  .1cm   {  w_t \ne v_t}
\end{subarray}}
|a(\{(v_s,v_t),(v_s,w_t)\})|^2
+2 \sum_{(v_s,v_t) }
\sum_{\begin{subarray}{I}
 \hskip  .1cm   (w_s,v_t)\\
 \hskip  .1cm   { w_s \ne v_s }
\end{subarray}}
| a(\{(v_s,v_t),(w_s,v_t)\})|^2 =$$
$$ \mathfrak{d}^2_{(G,a)}, \eqno{(3.5')}$$
\\

$$
\sum_{(v_s,v_t) } \mathfrak{d}^2_{(v_s,v_t)} + 4 \sum_{v_s }
\sum_{\begin{subarray}{I}
\hskip  .4 cm {w_s }\\
\hskip  .1cm { w_s \ne v_s }
\end{subarray}}
|b'(\{v_s,w_s\})|^2 \sum_{v_t }
\sum_{\begin{subarray}{I}
\hskip  .4cm {w_t }\\
\hskip  .1cm { w_t \ne v_t }
\end{subarray}}
|c'(\{v_t,w_t\})|^2
$$
 $$
 + 2 \sum_{v_s }  \mathfrak{d}^2_{v_s} \sum_{v_t }
\sum_{\begin{subarray}{I}
\hskip  .4cm{w_t }\\
\hskip  .1cm{w_t \ne v_t}
\end{subarray}}
 |c(\{v_t,w_t\} )|^2
+2 \sum_{v_t } \mathfrak{d}^2_{v_t}  \sum_{v_s }
 \sum_{\begin{subarray}{I}
\hskip  .4cm {w_s }\\
\hskip  .1cm {w_s \ne v_s}
\end{subarray}}
 |b(\{v_s,w_s\})|^2 =\mathfrak{d}^2_{(G,a)}, \eqno{(3.6')}$$
\\

$$
\sum_{v_s} \mathfrak{d}^2_{v_s} + 2 \sum_{v_s }
\sum_{\begin{subarray}{I}
\hskip  .4cm {w_s}\\
\hskip  .1cm { w_s \ne v_s }
\end{subarray}}
|b(\{v_s,w_s\})|^2 = \mathfrak{d}^2_{(G_s,b)},\eqno{(3.7')}$$
\\

$$
\sum_{v_s } \mathfrak{d}^2_{v_s} + 2 \sum_{v_s }
\sum_{\begin{subarray}{I}
\hskip  .4cm {w_s }\\
\hskip  .1cm { w_s \ne v_s }
\end{subarray}}
| b'(\{v_s,w_s\})|^2 = \mathfrak{d}^2_{(G_s,b')},\eqno{(3.8')}$$
\\

$$
\sum_{v_t } \mathfrak{d}^2_{v_t} + 2 \sum_{v_t }
 \sum_{\begin{subarray}{I}
\hskip  .4cm {w_t }\\
\hskip  .1cm{ w_t \ne v_t }
\end{subarray}}
|c(\{v_t,w_t\})|^2 = \mathfrak{d}^2_{(G_t,c)},\eqno{(3.9')}$$
\\

$$
\sum_{v_t } \mathfrak{d}^2_{v_t} + 2 \sum_{v_t}
 \sum_{\begin{subarray}{I}
\hskip  .4cm {w_t}\\
\hskip  .1cm { w_t \ne v_t }
\end{subarray}}
|c'(\{v_t,w_t\})|^2= \mathfrak{d}^2_{(G_t,c')}.\eqno{(3.10')}$$

We see that Eqs. $(3.7'), (3.8'), (3.9')$, and $(3.10')$ are consistent with Eq. $(3.6')$ provided

\begin{verse}

(v)~~~$\mathfrak{d}_{(v_s,v_t)}  = \mathfrak{d}_{v_s} \mathfrak{d}_{v_t}$ for all $(v_s,v_t) \in V(G,a)$ and consequently $\mathfrak{d}_{(G,a)} = \mathfrak{d}_{(G_s,b)} \mathfrak{d}_{(G_t,c)}$ and

(vi)~~~$|b(\{v_s,w_s\})|^2\; =\; |b'(\{v_s,w_s\})|^2$ and $ |c(\{v_t,w_t\})|^2 \; =\; |c'(\{v_t,w_t\})|^2$.
\end{verse}

 Eqs. (3.11)-(3.13)  become

$$\sum_{\begin{subarray}{I}
(v_s,w_t) \\
{ w_t \ne v_t}
\end{subarray}}
|a(\{(v_s,v_t),(v_s,w_t)\})| \; =\; \mathfrak{d}_{v_s}
\sum_{\begin{subarray}{I}
\hskip  .4cm {w_t} \\
{ w_t \ne v_t }
\end{subarray}}
 |c(\{v_t,w_t\})|,  \eqno{(3.11')}$$
\\

$$\sum_{\begin{subarray}{I}
(v_s,w_t) \\
{  w_s \ne v_s}
\end{subarray}}
 |a(\{(v_s,v_t),(w_s,v_t)\})| \; =\; \mathfrak{d}_{v_t}
 \sum_{\begin{subarray}{I}
\hskip  .4cm {w_s }\\
{ w_s \ne v_s }
\end{subarray}}
 |b(\{v_s,w_s\})|,  \eqno{(3.12')}$$
\\

$$\sum_{\begin{subarray}{I}
(v_s,w_t)\\
{ w_t \ne v_t}
\end{subarray}}
|a(\{(v_s,v_t),(w_s,w_t)\})|
=\sum_{\begin{subarray}{I}
\hskip  .4cm {w_s }\\
{ w_s \ne v_s }
\end{subarray}}
|b'(\{v_s,w_s\})|
  \sum_{\begin{subarray}{I}
\hskip  .4cm {w_t }\\
{ w_t \ne v_t }
\end{subarray}}
 |c'(\{v_t,w_t\})|.  \eqno{(3.13')}$$

Adding these three terms and the weight of the loop on $(v_s,v_t)$, we get $\mathfrak{d}_{(v_s,v_t)}$. The requirement that $\mathfrak{d}_{(v_s,v_t)} = \mathfrak{d}_{v_s} \mathfrak{d}_{v_t}$ is  satisfied provided

\begin{verse}

(vii)~~~$|b'(\{v_s,w_s\})| \;=\; |b(\{v_s,w_s\})| $ and  $ |c'(\{v_t,w_t\})| \;=\; |c(\{v_t,w_t\})|$ and

(viii)~~~$a(\{(v_s,v_t),(v_s,v_t)\})=$
$$b(\{v_s,v_s\}) c(\{v_t,v_t\})-
2 \sum_{\begin{subarray}{I}
\hskip  .4cm {w_s}\\
{ w_s \ne v_s }
\end{subarray}}
 |b(\{v_s,w_s\})|
\sum_{\begin{subarray}{I}
\hskip  .4cm {w_t}\\
{ w_t \ne v_t }
\end{subarray}}
 |c(\{v_t,w_t\})|.$$

 \end{verse}

Requirement (viii) is consistent with
$$(G^{''},a)\;=\;  \Om(G_s, b) \otimes \Om(G_t,c) \sqcup 2 \cN \cL (G_s,b) \otimes \cN \cL \eta (G_t,c). \eqno{(3.14)}$$

We now choose phases of the weight functions $b$ and  $c$.  Consider the edges $e=\{(v_s,v_t),(w_s,w_t)\}$ and $T_s e \;=\;\{(w_s,v_t),(v_s,w_t)\}$  in $E(G,a)$. We know that $|a(e)| \;=\; |a(T_s e)|$ . Let $ e^{i \th_1} $ and $ e^{i \th_2}$ be the phases of $a_1=a(e)$ and $a_2= a(T_s e)$, respectively. If we require that these two edges in $E(G,a)$ be produced by the tensor product of the edge $\{v_s,w_s\} $ in $(G_s,b)$ with the edge $\{v_t,w_t\} $ in $(G_t,c)$  (see figure 3.2) then the phases of weights $b$ and $c$ on the corresponding edges must be $\phi_1 \;=\; (\th_1 + \th_2 )/2$ and $ \phi_2 \;=\; (\th_1 - \th_2 )/2$, respectively.

\begin{figure}[!ht]
\begin{center}
\includegraphics[width=16cm,height=4cm]{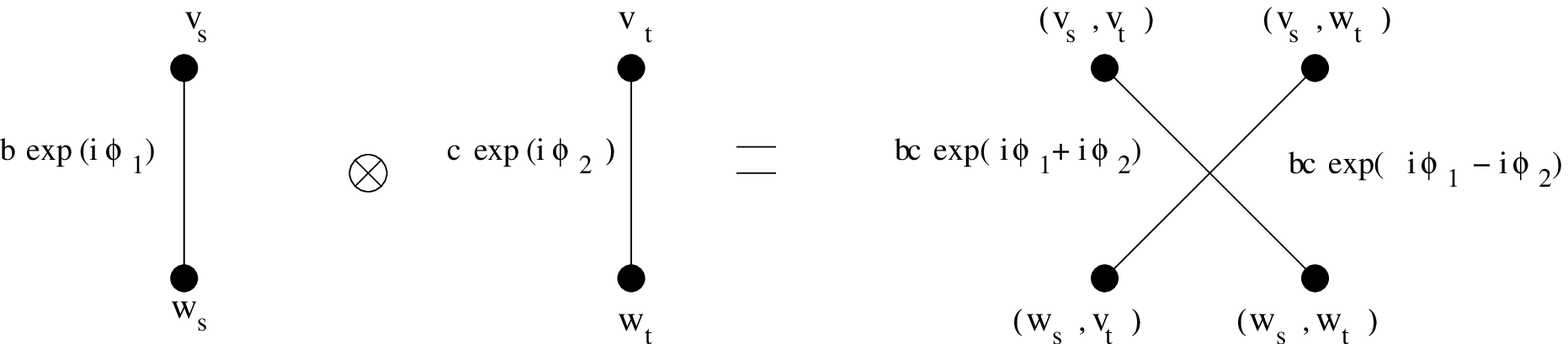}

Figure 3.2
\end{center}
\end{figure}

This completely fixes the weight functions $b$ and $c$ on $(G_s,b)$ and $(G_t,c)$, respectively. We now have, for every edge $e$ in $E(G,a)$, the corresponding edges $e_1,e_2$ in $(G_s,b)$ and $(G_t,c)$, respectively, such that $ e=e_1 \otimes e_2$ and $a(e)=b(e_1) c(e_2)$. Thus, we have, finally, from Eq. (3.3), (vii), and Eq.(3.14),

$$ (G,a)\; =\; \cL(G_s,b) \otimes \cL (G_t, c) \dotplus \cL(G_s,b) \otimes \cN(G_t,c) \dotplus \cN(G_s,b) \otimes \cL(G_t,c)$$
  $$ \dotplus \{ \Om(G_s, b) \otimes \Om(G_t,c) \sqcup 2 \cN \cL (G_s,b) \otimes \cN \cL \eta (G_t,c)\}  $$
$$\;=\;(G_s,b) \boxdot (G_t,c).\eqno{(3.2')}$$
\hfill $\blacksquare$

\noi\textbf{ Lemma 3.2.6} :\textit{ Let $ \si , \si_s $ and $ \si_t$ be density matrices for pure states. Then $ \si = \si_s \otimes \si_t$ if and only if $(G,a) = (G_s,b) \boxdot (G_t,c)$ where $(G,a),\; (G_s,b) ,$ and  $(G_t,c)$ are the graphs for  $ \si,\; \si_s $ and $ \si_t$, respectively.}

\noi\textbf{Proof :}  This lemma is identical with theorems  2.4.5, 2.6.10 in chapter 2.

\hfill $\blacksquare$

\noi\textbf{Theorem 3.2.7} : \textit{Let $(G,a)$ be the graph of a $m$-partite pure state $\si$ in the Hilbert space $\mathcal{H}^{d_1}\otimes \mathcal{H}^{d_2} \otimes \cdots \otimes \mathcal{H}^{d_m}$.  Let $\mathcal{H}^{(s)}=\mathcal{H}^{d_{i_1}} \otimes \mathcal{H}^{d_{i_2}} \otimes \mathcal{H}^{d_{i_3}}\otimes \cdots \otimes \mathcal{H}^{d_{i_s}},\; \{ i_1 ,i_2,\cdots ,i_s\} = s$  correspond to an  $s$ set,  $\mathcal{H}^{(t)}=\mathcal{H}^{d_{j_1}} \otimes \mathcal{H}^{d_{j_2}} \otimes \mathcal{H}^{d_{j_3}}\otimes \cdots \otimes \mathcal{H}^{d_{j_t}},\;  \{j_1,j_2, \cdots ,j_t  \} =t$, correspond to the $t$ set which is the complement of $s$ set in $\{1,2, \cdots,m\}$. Then, $\si = \si_s \otimes \si_t$,  where
 $\si_s $ and $ \si_t$ are pure states in $\mathcal{H}^{(s)}$ and $\mathcal{H}^{(t)}$  with graphs $(G_s,b)$ and $(G_t,c)$,  respectively, if and only if $\Delta (G,a) =\Delta (G^{T_s},a')$.}

\noi\textbf{Proof :} Using lemmas 3.2.4, 3.2.5 and 3.2.6, we have

$\si = \si_s \otimes \si_t  \Longleftrightarrow  (G,a) = (G_s,b) \boxdot (G_t,c)  \Longleftrightarrow  E(G,a)  $ is closed under $ T_s \Longleftrightarrow  \Delta (G,a) =\Delta (G^{T_s},a')$. A state $\si$ is entangled if  $ \Delta (G,a) \ne \Delta (G^{T_s},a')$  in every partition $s$ and $t$ of $\{1,2, \cdots,m\}$. \hfill $\blacksquare$

\section{ Algorithm}

While proving the if part of lemma 3.2.5, we have shown that the number of vertices in the cliques of the factors $G_1$ and $G_2$ defined there ($p$ and $q$, respectively) are the factors of the number of vertices on the clique in $(G,a)$, that is, $n=pq$. This means that the $m$-partite pure state $|\psi \ran$ corresponding to $(G,a)$ has two factors   $|\psi_1 \ran$  and  $|\psi_2 \ran$, corresponding to $G_1$ and $G_2$, respectively, such that  $|\psi_1 \ran$ lives in a $p$-dimensional subspace of $\mathcal{H}^{d_{i_1}} \otimes \mathcal{H}^{d_{i_2}} \otimes \mathcal{H}^{d_{i_3}}\otimes \cdots \otimes \mathcal{H}^{d_{i_s}}$ and $|\psi_2 \ran$ lives in a $q$-dimensional subspace of $\mathcal{H}^{d_{j_1}} \otimes \mathcal{H}^{d_{j_2}} \otimes \mathcal{H}^{d_{j_3}}\otimes \cdots \otimes \mathcal{H}^{d_{j_t}}$. If  the weighted versions of  $G_1$ and $G_2$, namely, $(G_s,b)$ and $(G_t,c)$, can be further factorized, the dimensions of the corresponding subspaces will be the factors of $p$ and $q$, respectively. This procedure can be iterated at most until the dimensions of the subspaces for the factors of $|\psi \ran$  are the prime factors of $n$. Therefore, the dimension of the subspaces containing the factors of $|\psi \ran$ are the prime factors of $n$ or the products of such factors. This fact can be used to get a polynomial algorithm to find the full separability of $|\psi \ran$ in the following way. By full separability, we mean expressing $|\psi \ran$ as a product state whose further factorization is impossible. Denote by $p_1 \geq p_2 \geq \cdots \geq p_k$ the prime factors of $n$. Let $d_1 \ge d_2 \ge d_3 \ge \cdots \ge d_m $ be the dimensions of the Hilbert spaces of $m$ parts arranged in a nonincreasing order. Let $s_1 > 0$ be the least integer satisfying $d_1 d_2 \cdots d_{s_1} \ge p_1$. We implement our algorithm (theorem 3.2.7) on partitions $(s,t)$ with $s_1\leq s \leq m-1$. The total number of times the algorithm has to run, in the worst case, is
\begin{displaymath}
\binom{m}{s_1}\;+\;\binom{m}{s_1+1}\;+\; \cdots+\;\binom{m}{m-1}
\end{displaymath}

 which is a polynomial of degree $s_1$ in $m$. Thus, we have a polynomial algorithm to check separability of a m-partite system. Suppose we get the separability as $\mathcal{H}^{(s)}\otimes \mathcal{H}^{(t)}$. Then, the factor in $\mathcal{H}^{(s)}$ cannot be further factorized as it corresponds to the largest prime factor of $n$ and $\mathcal{H}^{(t)}$ contains factors corresponding to $ p_2 \geq p_3 \cdots \geq p_k$. We repeat the above algorithm on $\mathcal{H}^{(t)}$ with $p_2$ as the largest prime factor. Its worst case complexity is given by a polynomial of degree $(m-s)^{s_2}$, where $s_2$ is defined like $s_1$ above. We carry out these iterations until full separability is obtained. Thus, if we do not get any factorization in the first iteration, corresponding to the largest prime factor $p_1$, then the state is fully entangled, such as GHZ or W state. Unless the factorization carries up to $m$ factors, the factors of the state contain one or more entangled states involving less than $m$ parts. The total algorithm is polynomial in $m$. Note that, if $n$ is prime, then all that is necessary is to look for some $v_i$ common to the $m$ tuples for {\it all} vertices on the clique. If, say, $v_i$ is common, then $$| \psi \ran\;=\;|\phi\ran\otimes |v_i\ran,$$ with $|\phi\ran \in \mathcal{H}^{d_1}\otimes \cdots \otimes  \mathcal{H}^{d_{i-1}}\otimes \mathcal{H}^{d_{i+1}}\otimes \cdots \otimes \mathcal{H}^{d_m}$ and $|v_i \ran \in \mathcal{H}^{d_i}$. Otherwise, the given state $| \psi \ran$ is entangled.

\section{ A CounterExample }

We  note that theorem 3.2.7 may not apply to mixed states  as the following example shows. Consider the bipartite separable state $$\si\;=\; 1/2 |y,- \ran |y,- \ran \lan y,-| \lan y,-| + 1/2 |x,+ \ran |x,+ \ran \lan x,+| \lan x,+|,$$ where $|y,- \ran = \fr{1}{ \sqrt2} (|0 \ran - i |1 \ran)$ and $   |x,+ \ran = \fr{1}{ \sqrt2} (|0 \ran + |1 \ran)$. The corresponding density matrix in standard  basis is $$\si = \fr{1}{4} \left[ \ba{rrrr} 2 & 1+i & 1+i & 0 \\1-i & 2 & 2 & 1+i \\1-i & 2 & 2 & 1+i \\ 0 &1-i &1-i & 2 \ea \right], $$ and  the corresponding graph is shown in figure 3.3. We see that the $\mathcal{C}$ set contains only one edge  for all possible partitions and hence cannot be closed under any $T_s$. We show, in section 3.5, that the degree conjecture applies to states with real weighted graphs without loops. Therefore, the above example shows that the degree conjecture does not apply to all mixed states.\\

\begin{figure}[!ht]
\begin{center}
\includegraphics[width=6cm,height=5cm]{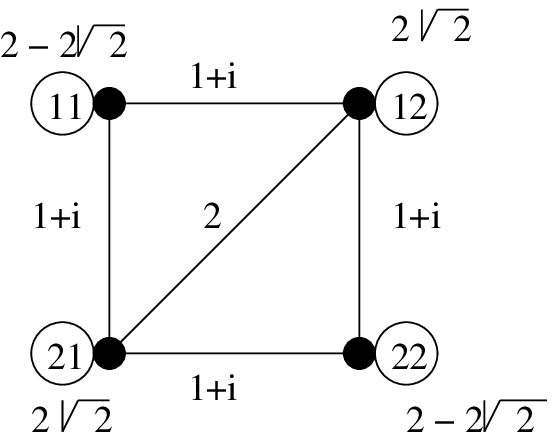}

Figure 3.3  \\
A counterexample.
\end{center}
\end{figure}

\newpage
\section{ Proof of Degree Criterion for A Class of Multipartite Mixed States}

In this section, we prove the degree criterion for the class of states whose graphs have no loops and have real weights.

\noi\textbf{Theorem 3.5.1 :} Let $\si(G,a)$ be a density matrix acting on $\mathcal{H}^{d_1} \otimes \mathcal{H}^{d_2} \otimes \cdots \otimes \mathcal{H}^{d_m}$, with a real weighted graph $(G,a)$, on $d=d_1 d_2 \dots d_m$ vertices,  having no loops. If $\si(G,a)$ is separable in $s,t$ cut where $s,t$ is a partition of $\{1,2,\dots,m\}$, so that $\si(G,a)=\sum_i p_i \si_i^{(s)}\otimes \si_i^{(t)}$, where  $\sigma_i^{(s)}$ and $ \sigma_i^{(t)}$ are density matrices acting on $\mathcal{H}^{(s)}$ and $\mathcal{H}^{(t)}$ with graphs $(G_i^{(s)},b)$ and $(G_i^{(t)},c)$ respectively, then $\Delta(G,a)=\Delta(G^{T_s},a')$.

\noi{\bf Proof :} Let $Q(G,a)$ be the Laplacian of a graph $(G,a)$ with  real  weights without  loops on $d$ vertices. For a graph without loops, $Q(G,a)=L(G,a)$ (see chapter 2). Let $D$ be any $d\times d$ real diagonal matrix in the standard orthonormal basis $\{|v_i\ran\}, \; i=1,2,\dots,d$, such that $D\neq 0$ and $Tr(D)=0$, where $Tr$ is the trace of $D$. This means that there is at least one negative entry in the diagonal of $D$. Denote this element by $D_{ii}=b_i < 0$. Let $|\psi_0 \ran= \sum_j|v_j\ran$ and $|\phi\ran=\sum_j \chi_j|v_j \ran$,  where
\begin{displaymath}
\chi_j =
\left\{ \begin{array}{ll}
 0 & \textrm{if $j\neq i$}\\
k  & \textrm{if $j=i$},\; \mbox{$k$ real.}
\end{array} \right.
\end{displaymath}
Let $|\chi \ran=|\psi_0\ran+|\phi\ran=\sum_{j=1}^d(1+\chi_j)|v_j\ran$. Then,
\benrr
\lan\chi|L(G,a)+D|\chi\ran& = & \lan\psi_0|L(G,a)|\psi_0\ran+
\lan\psi_0|L(G,a)|\phi\ran+\lan\phi|L(G,a)|\psi_0\ran+\\
& &              \lan\phi|L(G,a)|\phi\ran+\lan\psi_0|D|\psi_0\ran+\lan\psi_0|D|\phi\ran+\lan\phi|D|\psi_0\ran+\\
& &
\lan\phi|D|\phi\ran. ~~~~~~~~~~~~~~~~~~~~~~~~~~~~~~~~~~~~~~~~~~~~~~~~(3.15)
\eenrr

 Since $L(G, a)$ is positive semidefinite, we must have, for the quadratic form associated with $L(G,a)$ of a graph without loops (see chapter 2) and \cite{mohar91},
$$ x^T L(G, a)x = \sum_{\{j,l\}\in E(G,a)} a_{jl} (x_j - x_l)^2 \ge 0,  \eqno{(3.16)}$$
for every $x \in I\!\!R^d.$
Since $|\psi_0\ran$ is (unnormalized) vector having all components equal to unity, from Eq.(3.16) it follows that $\lan\psi_0|L(G,a)|\psi_0\ran=0$. Also $\lan\psi_0|D|\psi_0\ran=Tr(D)=0$.
  We have
 $$\lan\phi|L(G,a)|\phi\ran=k^2(L(G,a))_{ii}=k^2\mathfrak{d}_i,$$
 where $\mathfrak{d}_i$ denotes the degree of $ith$ vertex. For a real weighted graph without loops, the sum of the elements in any row of its Laplacian is zero. This leads to
 $$\lan\psi_0|L(G,a)|\phi\ran=\lan\phi|L(G,a)^T|\psi_0\ran = \lan\phi|L(G,a)|\psi_0\ran = 0.$$

 Finally, the remaining terms in Eq. (3.15) are given by
 $$\lan\phi|D|\phi\ran=b_ik^2,$$
 $$\lan\psi_0|D|\phi\ran = b_ik=\lan\phi|D|\psi_0\ran.$$
 Thus,
 $$\lan\chi|L(G,a)+D|\chi\ran = k^2(b_i+\mathfrak{d}_i)+ 2kb_i.$$
  So we can then always choose a positive $k$, such that
 $$\lan\chi|L(G,a)+D|\chi\ran < 0.$$
 Then, it follows that $$L(G,a)+D \ngeq 0.\eqno{(3.17)}$$

  This expression is identical with that obtained in chapter 2 and \cite{bgmsw}.
  For any graph $(G,a)$ on $d=d_sd_t$ vertices, $$v_1=(u_1,w_1), v_2=(u_1,w_2), \dots, v_{d}=(u_{d_s},w_{d_t}),$$ consider  the degree condition $\Delta(G,a) = \Delta(G^{T_s},a^{\prime}).$  Since $M^{T_s}(G,a)=M(G^{T_s},a^{\prime})$, where $M$ is the adjacency matrix, we get,
  $$(L(G,a))^{T_s}=(\Delta(G,a) - \Delta(G^{T_s},a^{\prime}))+ L(G^{T_s},a^{\prime}).\eqno{(3.18)}$$
  Let $$D=\Delta(G,a) - \Delta(G^{T_s},a^{\prime}).$$

  Then, $D$ is an $d\times d$ real diagonal matrix with respect to the orthonormal basis, $$|v_i\ran = |u_1\ran\otimes |w_1\ran, \dots, |v_{d}\ran = |u_{d_s}\ran\otimes |w_{d_t}\ran.$$

  As the degree sum of a graph is invariant under partial transpose,
  $$Tr(D)=Tr(\Delta(G,a))-Tr(\Delta(G^{T_s},a^{\prime}))= 0.$$
  We have two possible cases : $D\neq 0$ or $D = 0$. If $D\neq 0$, that is, the degree condition is not satisfied $[i.e.  \Delta(G,a)\neq \Delta(G^{T_s},a^{\prime})]$, we can write, via Eq.(3.17),
   $L(G^{T_s},a^{\prime})+D\ngeq 0$ because $(G^{T_s},a^{\prime})$ is a real weighted graph without loops and $D\ne 0$ with $Tr(D) = 0.$ Using Eq.(3.18), this means  $(L(G,a))^{T_s}\ngeq 0.$ As  $\si(G,a)=\fr{1}{\mathfrak{d}_{(G,a)}}L(G,a)$ (see chapter 2), $\si(G,a)$ is entangled \cite{per96b}.

\hfill $\blacksquare$\\

 In order to test the separability of the $m$-partite state $\si(G,a)$, we have to apply the degree criterion to all the bipartite cuts $(s,t)$ of $m$-partite system. This procedure may not detect the full separability of  $m$-partite state \cite{hhhh07}. However, all the known separability criteria for multipartite states test separability only in bipartite cuts, and hence are not enough to guarantee full separability.

In this chapter, we have settled the degree conjecture for the separability of multipartite quantum states. Recently, Hildebrand {\it et al.} have proved that the degree criterion is equivalent to the PPT criterion for separability \cite{hms06}. However, the importance of degree conjecture ensues from the opportunity it offers to test the strengths and limitations of a nascent approach to the separability problem. We see that this approach has contributed to test the separability of a class of multipartite mixed states (theorem 3.5.1) as well as to the efficient factorization of multipartite pure states.

\chapter{Separability criterion for Multipartite Quantum States Based on The Bloch Representation of Density Matrices}
\begin{center}
\scriptsize\textsc{We must all hang together, or assuredly we shall all hang separately. \\{\it B. Franklin}}
\end{center}

\section{Introduction}

In chapters 2 and 3, we presented the approach based on graphs to the problem of separability of quantum states. In this and the next chapter, we give another approach to the separability problem, based on the Bloch representation of a density operator.

A $N$-partite state acting on  $\mathcal{H}=\mathcal{H}^{d_1}\otimes \mathcal{H}^{d_2} \otimes \cdots \otimes \mathcal{H}^{d_N}$ is separable \cite{wer89} ( or fully separable) if it can be written as a convex sum of tensor products of subsystem states
 $$\rho =\sum_w p_w \rho_w^{(1)}\otimes \rho_w^{(2)}\dots \otimes \rho_w^{(N)}=\sum_w p_w \bigotimes_{j=1}^N \rho_w^{(j)} , \; p_w > 0; \; \sum_w p_w=1 .\eqno{(4.1)}$$

 A state is called $k$ separable if we can write
  $$\rho =\sum_w p_w \rho_w^{(a_1)}\otimes \rho_w^{(a_2)}\dots \otimes \rho_w^{(a_k)} \eqno{(4.2)}$$
   where  $a_i ; \; i=1,2,\dots ,k $ are the disjoint subsets of $\{1,2,\dots,N \}$ and $\rho^{(a_i)}$  acts on the tensor product space made up by the factors of $\mathcal{H}$ labeled by the members of $a_i$.
    The understanding of multipartite entanglement has progressed by dealing with some special classes of states like the density operators supported on the symmetric subspace of $\mathcal{H}$ \cite{dpr07}. A lower bound on concurrence on the multipartite mixed states is obtained \cite{mkbdbkm}. K. Chen and L. Wu have given a generalized partial transposition and realignment criterion to detect entanglement of a multipartite quantum state \cite{cw02}.

 There are two definitions commonly used for the entanglement of multipartite quantum states, the one from Ref. \cite{abls01} (ABLS) and the one introduced in \cite{dct99} (DCT). In DCT, all possible partitions of N parties are considered and it is tested for each partition if the state is fully separable there or not. A state is called N partite entangled if it is not separable for any partition. If a state is separable for a bipartite partition, it is called biseparable.
In ABLS, a state is called biseparable if it is a convex combination of biseparable states, possibly concerning different partitions. A $N$-partite entangled state is one which is not biseparable.

 In this chapter we derive a necessary  condition for the  separability of multipartite quantum states for arbitrary finite  dimensions of the subsystem Hilbert spaces and without any further restriction on them. The criterion is based on the Bloch representation of a multipartite quantum state, which has been used in previous works to characterize the separability of bipartite density matrix, in particular, our work is a generalization of de Vicente's work on bipartite systems \cite{vic07a}. We make use of the algebra of higher order tensors, in particular the matrization of a tensor \cite{lmv00,lmv00b,kold06,kold01,krus77,zg01,bk04,bk06}.

 The chapter is organized  as follows. In section 4.2 we present the Bloch representation of a $N$-partite quantum state. In section 4.3 we obtain the main results on separability of a $N$-partite quantum state. In section 4.4 we give a sufficient condition for the separability of a 3-partite quantum state generalizable to the case $N > 3$.
 In section 4.5 we investigate our separability criterion for mixed states, in particular, bound entangled states.
 Finally we summarize in section 4.6.

\section{ Bloch Representation of a $N$-Partite Quantum State}

Bloch representation \cite{bloc46,kk05,kim03,bk03,mwb95} of a density operator acting on the Hilbert space of a $d$-level quantum system $\mathbb{C}^d$ is given by \cite{vic07a} $$ \rho = \fr{1}{d} (I_d + \sum_i s_i \lambda_i) \eqno{(4.3)}$$
 Eq.(4.3) is the expansion of $\rho$ in the Hilbert-Schmidt basis $\{I_d,\lambda_i; i=1,2,\dots,d^2-1\}$ where $\lambda_i$ are the traceless hermitian generators of $SU(d)$ satisfying $Tr(\lambda_i \lambda_j)=2\delta_{ij}$
  and are characterized by the structure constants of the corresponding Lie algebra, $f_{ijk}, g_{ijk}$ which are,
     respectively, completely antisymmetric and completely symmetric.
 $$\lambda_i \lambda_j =\fr{2}{d} \delta_{ij} I_d + i f_{ijk}\lambda_k +g_{ijk}\lambda_k    \eqno{(4.4)}$$

 $\textbf{s} = (s_1,s_2,\dots,s_{d^2-1})$ in Eq.(4.3) are the vectors in $\mathbb{R}^{d^2-1}$, constrained by the positive semidefiniteness of $\rho$,  called Bloch vectors \cite{bk03}. The set of all Bloch vectors that constitute a density operator is known as the Bloch vector space $B(\mathbb{R}^{d^2-1})$. The problem of determining $B(\mathbb{R}^{d^2-1})$ where  $d\ge 3$ is still open \cite{kk05,kim03}. However, for pure states $(\rho=\rho^2)$ the following relations hold.
 $$||\textbf{s}||_2 = \sqrt\fr{d(d-1)}{2};\;\;\; s_i s_j g_{ijk}=(d-2)s_k     \eqno{(4.5)}$$
  where $||.||_2$ is the Euclidean norm in $\mathbb{R}^{d^2-1}$.

  It is known \cite{har78,kos03} that $B(\mathbb{R}^{d^2-1})$ is a subset of the ball $D_R(\mathbb{R}^{d^2-1})$ of radius $R=\sq{\fr{d(d-1)}{2}}$ , which is the minimum ball containing it, and that the ball $D_r(\mathbb{R}^{d^2-1})$ of radius $r=\sq{\fr{d}{2(d-1)}}$ is included in $B(\mathbb{R}^{d^2-1})$. That is,

  $$D_r(\mathbb{R}^{d^2-1})\subseteq B(\mathbb{R}^{d^2-1}) \subseteq D_R(\mathbb{R}^{d^2-1}) \eqno{(4.6)}$$

  In order to give the Bloch representation of a density operator acting on the Hilbert
  space $\mathbb{C}^{d_1} \otimes \mathbb{C}^{d_2} \otimes \cdots \otimes \mathbb{C}^{d_N}$
  of a $N$-partite quantum system, we introduce following notation. We use $k$, \; $k_i \; (i=1,2,\cdots)$ to denote a subsystem chosen from $N$ subsystems, so that $k$,\; $k_i \; (i=1,2,\cdots)$ take values in the set  $\mathcal{N}=\{1,2,\cdots,N\}$. The variables $\alpha_k \;\mbox{or} \; \alpha_{k_i}$ for a given $k$ or $k_i$ span the set of generators of $SU(d_k)\; \mbox{or}\;  SU(d_{k_i})$ group (Eqs.(4.3) and (4.4)) for the $k$th or $k_i$th subsystem, namely the set $\{\la_{1_{k_i}},\la_{2_{k_i}},\cdots,\la_{d_{k_i}^2-1}\}$ for the $k_i$th subsystem. For two subsystems $k_1$ and $k_2$ we define

   $$\lambda^{(k_1)}_{\alpha_{k_1}}=(I_{d_1}\otimes I_{d_2}\otimes \dots \otimes \lambda_{\alpha_{k_1}}\otimes I_{d_{k_1+1}}\otimes \dots \otimes I_{d_N})   $$
   $$\lambda^{(k_2)}_{\alpha_{k_2}}=(I_{d_1}\otimes I_{d_2}\otimes \dots \otimes \lambda_{\alpha_{k_2}}\otimes I_{d_{k_2+1}}\otimes \dots \otimes I_{d_N})  $$
   $$\lambda^{(k_1)}_{\alpha_{k_1}} \lambda^{(k_2)}_{\alpha_{k_2}}=(I_{d_1}\otimes I_{d_2}\otimes \dots \otimes \lambda_{\alpha_{k_1}}\otimes I_{d_{k_1+1}}\otimes \dots \otimes \lambda_{\alpha_{k_2}}\otimes I_{d_{k_2+1}}\otimes I_{d_N})   \eqno{(4.7)}$$

  where  $\lambda_{\alpha_{k_1}}$ and $\lambda_{\alpha_{k_2}}$ occur at the $k_1$th and $k_2$th places (corresponding to $k_1$th and $k_2$th subsystems respectively) in the tensor product and are the $\alpha_{k_1}$th and  $\alpha_{k_2}$th generators of $SU(d_{k_1}),\; SU(d_{k_2}) ,\alpha_{k_1}=1,2,\dots,d_{k_1}^2-1\; \mbox{and} \; \alpha_{k_2}=1,2,\dots,d_{k_2}^2-1$ respectively. Then we can write

$$\rho=\fr{1}{\Pi_k^N d_k} \{\otimes_k^N I_{d_k}+ \sum_{k \in \mathcal{N}}\sum_{\alpha_{k}}s_{\alpha_{k}}\lambda^{(k)}_{\alpha_{k}} +\sum_{\{k_1,k_2\}}\sum_{\alpha_{k_1}\alpha_{k_2}}t_{\alpha_{k_1}\alpha_{k_2}}\lambda^{(k_1)}_{\alpha_{k_1}} \lambda^{(k_2)}_{\alpha_{k_2}}+\cdots +$$
$$\sum_{\{k_1,k_2,\cdots,k_M\}}\sum_{\alpha_{k_1}\alpha_{k_2}\cdots \alpha_{k_M}}t_{\alpha_{k_1}\alpha_{k_2}\cdots \alpha_{k_M}}\lambda^{(k_1)}_{\alpha_{k_1}} \lambda^{(k_2)}_{\alpha_{k_2}}\cdots \lambda^{(k_M)}_{\alpha_{k_M}}+ \cdots+$$
$$\sum_{\alpha_{1}\alpha_{2}\cdots \alpha_{N}}t_{\alpha_{1}\alpha_{2}\cdots \alpha_{N}}\lambda^{(1)}_{\alpha_{1}} \lambda^{(2)}_{\alpha_{2}}\cdots \lambda^{(N)}_{\alpha_{N}}\}.\eqno{(4.8)}$$

 where $\textbf{s}^{(k)}$ is a Bloch vector corresponding to $k$th subsystem,  $\textbf{s}^{(k)} =[s_{\alpha_{k}}]_{\alpha_{k}=1}^{d_k^2-1} $ which is a tensor of order one defined by
 $$s_{\alpha_{k}}=\fr{d_k}{2} Tr[\rho \lambda^{(k)}_{\alpha_{k}}]= \fr{d_k}{2} Tr[\rho_k \lambda_{\alpha_{k}}],\eqno{(4.9a)}$$ where $\rho_k$ is the reduced density matrix for the $k$th subsystem. Here $\{k_1,k_2,\cdots,k_M\},\; 2 \le M \le N,$ is a subset of $\mathcal{N}$ and can be chosen in $\binom{N}{M}$  ways, contributing $\binom{N}{M}$ terms in the sum $\sum_{\{k_1,k_2,\cdots,k_M\}}$ in Eq.(4.8), each containing a tensor of order $M$. The total number of terms in the Bloch representation of $\rho$ is $2^N$. We denote the tensors occurring in the sum $\sum_{\{k_1,k_2,\cdots,k_M\}},\; (2 \le M \le N)$ by $\mathcal{T}^{\{k_1,k_2,\cdots,k_M\}}=[t_{\alpha_{k_1}\alpha_{k_2}\cdots \alpha_{k_M}}]$ which  are defined by

 $$t_{\alpha_{k_1}\alpha_{k_2}\dots\alpha_{k_M}}=\fr{d_{k_1}d_{k_2}\dots d_{k_M}}{2^M} Tr[\rho \lambda^{(k_1)}_{\alpha_{k_1}} \lambda^{(k_2)}_{\alpha_{k_2}}\cdots \lambda^{(k_M)}_{\alpha_{k_M}}]$$

$$ =\fr{d_{k_1}d_{k_2}\dots d_{k_M}}{2^M} Tr[\rho_{k_1k_2\dots k_M} (\lambda_{\alpha_{k_1}}\otimes\lambda_{\alpha_{k_2}}\otimes\dots \otimes\lambda_{\alpha_{k_M}})]   \eqno{(4.9b)}$$
where $\rho_{k_1k_2\dots k_M}$ is the reduced density matrix for the subsystem $\{k_1 k_2\dots k_M\}$. We call  The tensor in last term in Eq. (4.8) $\mathcal{T}^{(N)}$.

\section{ Separability Conditions}

 Before we obtain the main results we need following definition. Throughout the chapter, we use the bold letter for vector and normal letter for components of a vector, matrix and tensor elements.\\

A rank-1 tensor is a tensor that consists of the outer product of a number of vectors. For $M$th order tensor $\mathcal{T}^{(M)}$ and $M$ vectors \; $\mathbf{u}^{(1)},\mathbf{u}^{(2)},\dots,\mathbf{u}^{(M)}$ this means that
$t_{i_1i_2\dots i_M}=u_{i_1}^{(1)}u_{i_2}^{(2)}\dots u_{i_M}^{(M)}$
for all values  of the indices. This  is concisely written  as
$\mathcal{T}^{(M)}=\mathbf{u}^{(1)}\circ \mathbf{u}^{(2)} \circ \dots \circ \mathbf{u}^{(M)}$ \cite{bk06,lmv00b}.

Also, given two tensors $\mathcal{T}^{(M)}$  and $\mathcal{S}^{(N)}$ of  order $M$ and $N$ respectively, with
 dimensions $I_1 \times I_2 \times \dots \times I_M$ and $J_1\times J_2 \times \dots\times J_N$ respectively, their outer product is defined as \cite{bk04,lmv00}

 $$(\mathcal{T}^{(M)}\circ \mathcal{S}^{(N)})_{i_1 i_2 \dots i_M j_1 j_2 \dots j_N}= t_{i_1 i_2\dots i_M}s_{j_1 j_2 \dots j_N} \eqno{(4.10)}$$

\noi\textbf{ Proposition 4.3.1 }: A  pure $N$-partite quantum state with Bloch representation Eq.(4.8) is fully separable (product state) if and only if      $$\mathcal{T}^{\{k_1,k_2,\cdots,k_M\}}= \mathbf{s}^{(k_1)}\circ \mathbf{s}^{(k_2)} \circ \dots \circ \mathbf{s}^{(k_M)} \eqno{(4.11)}$$
 for $2\le M\le N. $
     In particular $\mathcal{T}^{(N)}=\mathbf{s}^{(1)}\circ \mathbf{s}^{(2)} \circ \dots \circ \mathbf{s}^{(N)}$
    holds. Here $\{k_1,k_2,\dots,k_M\}\subset \{1,2,\dots,N\}$, and $ \mathbf{s}^{(k)}$ is the Bloch vector of $k$th subsystem reduced density matrix.
 \\

\noi\textbf{ Proof} : Notice that Eq.(4.8) can be rewritten as

  $$\rho=\rho^{(1)} \otimes\rho^{(2)} \otimes \dots \otimes \rho^{(N)}+\fr{1}{d_1 d_2 \cdots d_N}\{\sum_{\{k_1,k_2\}}\sum_{\alpha_{k_1}\alpha_{k_2}}[t_{\alpha_{k_1}\alpha_{k_2}}-s_{\alpha_{k_1}} s_{\alpha_{k_2}}]\lambda^{(k_1)}_{\alpha_{k_1}} \lambda^{(k_2)}_{\alpha_{k_2}}+$$
  $$\cdots +\cdots+\sum_{\alpha_{1}\alpha_{2}\cdots \alpha_{N}}[t_{\alpha_{1}\alpha_{2}\cdots \alpha_{N}}-s_{\alpha_{1}}s_{\alpha_{2}}\dots s_{\alpha_{N}}]\lambda^{(1)}_{\alpha_{1}} \lambda^{(2)}_{\alpha_{2}}\cdots \lambda^{(N)}_{\alpha_{N}}\}. \eqno{(4.12)} $$

   For full separability, the sum of all the  terms apart from the first term must vanish.
   Note that for every subsystem $k=1,2,\dots,N$ the set $\{I_d,\lambda_i; i=1,2,\dots,d_{k}^2-1\}$ forms an orthonormal Hilbert-Schmidt basis for the $k$th subsystem. Hence $ \lambda^{(k)}_{\alpha_{k}} ; \lambda^{(k_1)}_{\alpha_{k_1}} \lambda^{(k_2)}_{\alpha_{k_2}} \dots;
    \lambda^{(k_1)}_{\alpha_{k_1}} \lambda^{(k_2)}_{\alpha_{k_2}}\cdots \lambda^{(k_M)}_{\alpha_{k_M}};\dots ; \lambda^{(1)}_{\alpha_{1}} \lambda^{(2)}_{\alpha_{2}}$

    $\cdots\lambda^{(N)}_{\alpha_{N}} $ are the vectors belonging to the orthonormal product basis of the Hilbert-Schmidt space of the whole $N$-partite system. By orthonormality of the tensor product of $\lambda$'s occurring in different terms, the required sum will vanish if and only if coefficients of each term vanish separately,  that is if and only if

   $ t_{\alpha_{k_1}\alpha_{k_2}\cdots \alpha_{k_M}}=s_{\alpha_{k_1}}s_{\alpha_{k_2}}\dots s_{\alpha_{k_M}}; \; 2 \le M \le N,$\\

that is,\\

$\mathcal{T}^{\{k_1,k_2,\cdots,k_M\}}=\mathbf{s}^{(k_1)}\circ \mathbf{s}^{(k_2)} \circ \dots \circ \mathbf{s}^{(k_M)}   \; ; 2 \le M \le N.$ $\hspace{\stretch{1}} \blacksquare$\\

In fact, the condition (4.11) for all $N$ parts is enough to decide the separability of pure $N$-partite quantum states, as the following proposition shows.

\noi\textbf{ Proposition 4.3.1a }: A  pure $N$-partite quantum state with Bloch representation (4.8) is fully separable (product state) if and only if   $$\mathcal{T}^{(N)}=\mathbf{s}^{(1)}\circ \mathbf{s}^{(2)} \circ \dots \circ \mathbf{s}^{(N)},$$
     where $ \mathbf{s}^{(k)}$ is the Bloch vector of $k$th subsystem reduced density matrix.\\

\noi\textbf{ Proof} : Suppose $\rho$ is a product state $\rho= \rho_1 \otimes \rho_2 \otimes \cdots \otimes \rho_N.$ Then,
$$t_{\alpha_{1}\alpha_{2}\dots\alpha_{N}}=\fr{d_{1}d_{2}\dots d_{N}}{2^N} Tr[(\rho_1 \otimes \rho_2 \otimes \cdots\otimes \rho_N )(\lambda_{\alpha_{1}}\otimes \lambda_{\alpha_{2}}\otimes\cdots \otimes\lambda_{\alpha_{N}})]$$
$$=\fr{d_{1}d_{2}\dots d_{N}}{2^N} Tr[(\rho_1\lambda_{\alpha_{1}}) \otimes (\rho_2 \lambda_{\alpha_{2}})\otimes \cdots\otimes (\rho_N \lambda_{\alpha_{N}})]$$

$$=\fr{d_{1}d_{2}\dots d_{N}}{2^N} [Tr(\rho_1\lambda_{\alpha_{1}}) Tr(\rho_2 \lambda_{\alpha_{2}}) \cdots Tr( \rho_N \lambda_{\alpha_{N}})]$$
$$=[s_{\alpha_{1}}s_{\alpha_{2}}\cdots s_{\alpha_{N}}].$$

Suppose the condition holds, that is, $[s^{(1)}\circ s^{(2)}\circ \cdots \circ s^{(N)}]_{\alpha_{1}\alpha_{2}\dots\alpha_{N}}= t_{\alpha_{1}\alpha_{2}\dots\alpha_{N}}$. Then,

$$[s^{(1)}\circ s^{(2)}\circ \cdots \circ s^{(N)}]_{\alpha_{1}\alpha_{2}\dots\alpha_{N}}=\fr{d_{1}d_{2}\dots d_{N}}{2^N}[Tr(\rho_1\lambda_{\alpha_{1}}) Tr(\rho_2 \lambda_{\alpha_{2}}) \cdots Tr( \rho_N \lambda_{\alpha_{N}})]$$
$$=\fr{d_{1}d_{2}\dots d_{N}}{2^N} Tr[(\rho_1\lambda_{\alpha_{1}}) \otimes (\rho_2 \lambda_{\alpha_{2}})\otimes \cdots\otimes (\rho_N \lambda_{\alpha_{N}})]$$
$$=\fr{d_{1}d_{2}\dots d_{N}}{2^N} Tr[(\rho_1 \otimes \rho_2 \otimes \cdots\otimes \rho_N )(\lambda_{\alpha_{1}}\otimes \lambda_{\alpha_{2}}\otimes\cdots\otimes \lambda_{\alpha_{N}})]$$

$$=t_{\alpha_{1}\alpha_{2}\dots\alpha_{N}}= \fr{d_{1}d_{2}\dots d_{N}}{2^N}Tr[\rho(\lambda_{\alpha_{1}} \otimes \lambda_{\alpha_{2}} \otimes \cdots \otimes \lambda_{\alpha_{N}})].$$
The equality
$$Tr[(\rho_1 \otimes \rho_2 \otimes \cdots\otimes \rho_N )(\lambda_{\alpha_{1}}\otimes \lambda_{\alpha_{2}}\otimes \cdots \otimes \lambda_{\alpha_{N}})]= Tr[\rho(\lambda_{\alpha_{1}}\otimes \lambda_{\alpha_{2}} \otimes \cdots  \otimes \lambda_{\alpha_{N}})]$$
is satisfied for all elements in the orthonormal basis $\{\otimes_{k=1}^N \lambda_{\alpha_k}\},\; 0 \le \alpha_k \le d_k^2-1,\; (\alpha_k=0$ for $I_{d_k})$ where $\{\lambda_{\alpha_k}\}$ are the $d^2_k-1$ generators of $SU(d_k)$. This means that the joint probabilities obtained from the ensemble of measurements of $(\lambda_{\alpha_1} \cdots \lambda_{\alpha_N}) $ for states $\rho$ and $\rho=\rho_1 \otimes \rho_2 \otimes \cdots\otimes \rho_N$ are equal.
This implies $$\rho=\rho_1 \otimes \rho_2 \otimes \cdots\otimes \rho_N.$$ $\hspace{\stretch{1}} \blacksquare$\\

Note that this criterion is easily amenable with experiments. In order to check it for an element of $\mathcal{T}^{(N)}$ we have to measure the corresponding generators on each subsystem and then check whether the product of the averages equals the average of the products.

Thus in order to check whether a given pure state is a product state we have to check whether $\mathcal{T}^{(N)}=\mathbf{s}^{(1)}\circ \mathbf{s}^{(2)} \circ \dots \circ \mathbf{s}^{(N)}$, where the Bloch vectors $\mathbf{s}^{(1)},  \mathbf{s}^{(2)}, \dots , \mathbf{s}^{(N)}$ can be constructed from the reduced density matrices $\rho_{1}, \rho_{2}, \cdots, \rho_{N}$ for subsystems $1,2,\cdots, N \; (s_{\alpha_{k}}=\fr{d_k}{2} Tr(\rho_{k} \la_{\alpha_{k}}),\; k \in \{1,2,\cdots,N\}$, see Eq.(4.9a)).

In the case of  mixed states we can characterize separability from the Bloch representation point of view as follows.

  {\it A $N$-partite quantum state with Bloch representation Eq.(4.8) is fully separable if and only if there exist vectors $u_w^{(k)} \in \mathbb{R}^{d_{k}^2-1}$ satisfying Eq.(4.5), and weights $p_w$  satisfying $0 \le p_w \le 1$ and $\sum_w p_w =1$ such that  $$ \mathcal{T}^{(N)}=\sum_w^R p_w \bigcirc_{k=1}^N u_w^{(k)}, \; \; \;  \mathbf{s}^{(k)}=\sum_w p_w u_w^{(k)} \eqno{(4.13a)}$$   and

 $$\mathcal{T}^{\{k_1,k_2,\cdots,k_M\}} = \sum_w^R p_w \bigcirc_{i=1}^M u_w^{(k_i)}\eqno{(4.13b)}$$ for $2\le M\le N $ ; for all subsets $\{k_1,k_2,\dots,k_M\}\subset \{1,2,\dots,N\},$} \\

 where $\mathbf{s}^{(k)}$  is the Bloch vector of the mixed state density matrix  for $k$th subsystem  and $u_w^{(k)}$ represent the Bloch vector of the pure state of the $k$th subsystem contributing to the $w$th term in Eq. (4.1).

 This follows from proposition 4.3.1 and Eq. (4.1). However, in view of proposition 4.3.1a, the necessary and sufficient condition is given by Eq.(4.13a), so that Eq.(4.13b) can be dropped. The above result can not be used directly, as it amounts to rewriting Werner's definition of separability in a different way.
  However, it allows us to derive a necessary condition for separability for $N$-partite quantum states.

  We need some concepts in multilinear algebra. Consider a tensor $\mathcal{T}^{(N)}\in \mathbb{R}^{I_1 \times I_2 \times \dots \times I_N}$, where $I_k=d_k^2-1$. The $n$th matrix unfolding of $\mathcal{T}^{(N)}$ $(n=1,2,\cdots,N)$ \cite{lmv00} is a matrix
 $T^{(N)}_{(n)}\in \mathbb{R}^{I_n \times (I_{n+1} I_{n+2}  \dots  I_N I_1 I_2 \dots I_{n-1})}$. $T^{(N)}_{(n)}$ contains the element $t_{i_1 i_2 \dots i_N} $ at the position with row index $i_n$  $(i_n=1,2,\cdots,I_n)$ and column index

 $$(i_{n+1} - 1)I_{n+2}I_{n+3} \dots I_N I_1 I_2 \dots I_{n-1} + (i_{n+2} - 1)I_{n+3}I_{n+4}\dots I_N I_1 I_2 \dots I_{n-1} $$
 $$+ \dots +(i_N - 1)I_1 I_2 \dots I_{n-1} + (i_1 - 1)I_2I_3 \dots I_{n-1} + (i_2 - 1)I_3I_4 \dots I_{n-1} +\dots + i_{n-1}.$$

  For $n=1$, we take the last term $i_{n-1}=i_{0}=i_N$. This ordering is called backward cyclic \cite{bk04}.  To facilitate understanding, put $N$ points on a circle and label them successively by  $i_1,i_2,\cdots,i_N.$ The consecutive terms in the expression for the column index in $T^{(N)}_{(n)}$ corresponding to $t_{i_1,i_2,\cdots,i_N}$ become quite apparent using this circle, for more detials see section 1.5.

  For  $\mathcal{T}^{(3)} \in \mathbb{R}^{I_1 \times I_2 \times I_3}$ the matrix unfolding $T^{(3)}_{(1)}$ contains the elements $t_{i_1i_2i_3} \; (i_k=1,2,\cdots,I_k ; \; k=1,2,3)$ at the position with row number $i_1$ and column number equal to $(i_2-1)I_3+i_3$, $T^{(3)}_{(2)}$ contains $t_{i_1i_2i_3}$ at the position with row number $i_2$ and column number equal to $(i_3-1)I_1+i_1$ and $T^{(3)}_{(3)}$ contains $t_{i_1i_2i_3}$ at the position with row number $i_3$ and column number equal to $(i_1-1)I_2+i_2.$ Example see section 1.5

 Note that there are $N$ possible matrix unfoldings of $\mathcal{T}^{(N)}$. The matrix unfolding is called the matrization of the tensor \cite{lmv00,bk06}. We can now define the Ky Fan norm of the tensor $\mathcal{T}^{(N)}$ (of order $N$)   over $N$ matrix unfoldings of a tensor, as $$||\mathcal{T}^{(N)}||_{KF} = max\{||T^{(N)}_{(n)}||_{KF}\}, \; n=1,\dots,N ; \eqno{(4.14)}$$
 where $||T^{(N)}_{(n)}||_{KF}$ is the  Ky Fan norm of matrix $T^{(N)}_{(n)}$ defined as the sum of singular values of $T^{(N)}_{(n)}$ \cite{hjb91}. It is straightforward to check that $||\mathcal{T}^{(N)}||_{KF}$ defined in Eq.(4.14) satisfies all the conditions of a norm and is also unitarily invariant \cite{vic07a,hjb91}.\\

  The tensors in Eq.(4.13a) are called  Kruskal tensors with the restriction $0 \le p_w\le 1,\;  \sum_w p_w=1$ \cite{krus77,bk04}. We are interested in finding the matrix unfoldings and Ky Fan norms of $\mathcal{T}^{(N)}$ occurring in Eq.(4.13a).    The  $k$th matrix unfolding for Kruskal tensor is \cite{bk06}

 $$T_{(k)}^{(N)} = U^{(k)} \Sigma (U^{(N)}\odot U^{(N-1)}\odot \dots \odot U^{(k+1)}\odot U^{(k-1)} \odot \dots \odot U^{(1)})^T . \eqno{(4.15)}$$

  Here  $U^{(k)}=[\mathbf{u}_1^{(k)}  \mathbf{u}_2^{(k)} \dots \mathbf{u}_R^{(k)}] \in  \mathbb{R}^{I_{k} \times R}; k =1,2,\dots N$ and  $R$ is the rank of Kruskal tensor \cite{krus77,kold06,bk06}, i.e. the number of terms in Eq.(4.13a). $\mathbf{u}_i^{(k)}$ is a vector in  $\mathbb{R}^{I_{k}}$ and is the $i$th column vector in the matrix $U^{(k)}.$ $\Sigma$ is the $R \times R$ diagonal matrix,  $\Sigma =$diag$[p_1 \dots p_R]$. The symbol $\odot$ denotes the Khatri-Rao product of matrices \cite{bk06} $U \in  \mathbb{R}^{I\times R}$   and $V \in \mathbb{R}^{J\times R}$ defined as $U\odot V =[ \mathbf{u}_1\otimes \mathbf{v}_1 \; \mathbf{u}_2 \otimes \mathbf{v}_2 \; \dots \; \mathbf{u}_R \otimes \mathbf{v}_R] \in \mathbb{R}^{IJ\times R}$ where $\mathbf{u}_i$ and $\mathbf{v}_i ,\; i=1,2,\dots R $ are column vectors of matrices $U$ and $V$ respectively. Eq.(4.15) can be rewritten as

 $$T_{(k)}^{(N)} = U^{(k)} \Sigma [\mathbf{v}_1^{(\bar{k})} \mathbf{v}_2^{(\bar{k})}\dots \mathbf{v}_R^{(\bar{k})}]^T = U^{(k)} \Sigma V^{(\bar{k})^T}  \eqno{(4.16)}$$
 where $\mathbf{v}_i^{(\bar{k})} ; i=1,2,\dots , R$ are the column vectors of the matrix \\
 $V^{(\bar{k})^T} \in \mathbb{R}^{I_{N} I_{N-1} I_{N-2}  \dots  I_{k+1}I_{k-1}\dots I_{1} \times R}$   and  $ \mathbf{v}_i^{\bar{(k)}} = \mathbf{u}_i^{(N)} \otimes \mathbf{u}_i^{(N-1)}\otimes \mathbf{u}_i^{(N-2)}\otimes \dots \otimes \mathbf{u}_i^{(k+1)} \otimes \mathbf{u}_i^{(k-1)} \otimes \dots \otimes \mathbf{u}_i^{(1)}.$
  Using Eq.(4.16) we can write $T_{(k)}^{(N)}$ as  $$T_{(k)}^{(N)}=\sum_{w=1}^R p_w \mathbf{u}_w^{(k)} \mathbf{v}_w^{(\bar{k})^T} ;\; \;k=1,2,\dots ,N.  \eqno{(4.17)}$$

\noi\textbf{ Theorem 4.3.2 }: If a $N$-partite quantum state of $d_1 d_2 \dots d_N$ dimension with Bloch representation Eq.(4.8) is fully  separable, then $$||\mathcal{T}^{(N)}||_{KF} \le \sqrt{\fr{1}{2^N} \Pi_{k=1}^N d_{k}(d_{k}-1)}. \eqno{(4.18)}$$

\noi\textbf{ Proof} : If the state $\rho$ is separable then $\mathcal{T}^{(N)}$ has to admit a decomposition of the form Eq.(4.13) with $||\mathbf{u}_w^{(k)}||_2= \sqrt{\fr{d_{k}(d_{k}-1)}{2}}, k = 1,2,\dots,N.$
   From definition of KF norm of tensors, Eq.(4.14),

    $$||\mathcal{T}^{(N)}||_{KF} =max \{||T_{(k)}^{(N)}||_{KF}\}\; ;\; k=1,\dots, N.$$
   From Eq.(4.17),

    $$||\mathcal{T}^{(N)}||_{KF} =max \{||\sum_w p_w \mathbf{u}_w^{(k)} \mathbf{v}_w^{(\bar{k})^T} ||_{KF}\} \; ;\; k=1,\dots, N$$
    $$\le max \{\sum_w p_w ||\mathbf{u}_w^{(k)} \mathbf{v}_w^{(\bar{k})^T} ||_{KF}\}
     =max\{ \sum_w p_w \sqrt{{\fr{1}{2^N}} \Pi_k^N d_{k}(d_{k}-1)}|| \mathbf{\tilde{u}}_w^{(k)} \mathbf{\tilde{v}}_w^{(\bar{k})^T} ||_{KF}\}$$

       where $\mathbf{\tilde{u}}_w^{(k)}, \mathbf{\tilde{v}}_w^{(\bar{k})}$ are unit vectors in
   $ \mathbb{R}^{d_{k}^2-1}$ and $\mathbb{R}^{d_{N}^2-1}\otimes \mathbb{R}^{d_{N-1}^2-1}\otimes \dots \otimes \mathbb{R}^{d_{k+1}^2-1}\otimes  \mathbb{R}^{d_{k-1}^2-1}\otimes \dots \otimes \mathbb{R}^{d_1^2-1}$ respectively, so that
 $|| \mathbf{\tilde{u}}_w^{(k)} \mathbf{\tilde{v}}_w^{(\bar{k})^T} ||_{KF}=1$ for all $k=1,2,\dots,N$. Using $\sum_w p_w=1$ we get $||\mathcal{T}^{(N)}||_{KF} \le \sqrt{\fr{1}{2^N} \Pi_{k=1}^N d_{k}(d_{k}-1)}$.

 $\hspace{\stretch{1}} \blacksquare$\\
 For a subsystem we get,

\noi\textbf{ Corollary 4.3.3 }: If the reduced density matrix of a subsystem consisting of $M$ out of $N$ parts is separable then
 $||\mathcal{T}^{\{k_1,k_2,\cdots,k_M \}}||_{KF} \le \sqrt{\fr{1}{2^M} \Pi_{k=1}^M d_{k}(d_{k}-1)}$.

  The negation of the above condition, that is, $$||\mathcal{T}^{(N)}||_{KF} > \sqrt{\fr{1}{2^N} \Pi_{k=1}^N d_{k}(d_{k}-1)},$$  is a sufficient condition of entanglement of $N$-partite quantum state. This leads to a hierarchy of inseparability conditions  which test entanglement in all the subsystems.

For $N=2$ the condition $||\mathcal{T}^{(N)}||_{KF} \le \sqrt{\fr{1}{2^2} d_1(d_1-1)d_2(d_2-1)} $ has been shown in Ref. \cite{vic07a}, to be a sufficient condition for entanglement associated with any bipartite density matrix. Note that for $N$-qubits, ${d_i=2,\; i=1,2,\dots,N}$, the above criterion  becomes, for a separable state, $||\mathcal{T}^{(N)}||_{KF} \le 1$.

  Consider a $N$ qudit system $\mathcal{H}_s=\otimes_{k=1}^N\mathcal{H}^{d}_k$ in a state $\rho$, supported in the symmetric subspace of $\mathcal{H}_s$. It is straightforward to see that all the tensors in the Bloch representation of $\rho$ are supersymmetric, that is (see Eqs.(4.8) and (4.9)),
$ t_{\alpha_{k_1}\alpha_{k_2} \cdots \alpha_{k_M}}= t_{P(\alpha_{k_1})P(\alpha_{k_2}) \cdots P(\alpha_{k_M})}, \; 2 \le M \le N,$  where  $P$ is any permutation over indices  $\{\alpha_{k_1},\alpha_{k_2},\cdots ,\alpha_{k_M}\}$. We have, neglecting the constant multipliers,

\benrr
t_{\alpha_{k_1}\alpha_{k_2} \cdots \alpha_{k_M}}&=& Tr[\rho_{k_1k_2\cdots k_M}  \lambda_{\alpha_{k_1}}\otimes\lambda_{\alpha_{k_2}}\otimes\dots \otimes\lambda_{\alpha_{k_M}}]\\
&=& Tr[\rho_{k_1k_2\cdots k_M}P P^T \lambda_{\alpha_{k_1}}\otimes\lambda_{\alpha_{k_2}}\otimes\dots \otimes\lambda_{\alpha_{k_M}}PP^T]\\
 &=& Tr[(P^T\rho_{k_1k_2\cdots k_M}P) (P^T \lambda_{\alpha_{k_1}}\otimes\lambda_{\alpha_{k_2}}\otimes\dots \otimes\lambda_{\alpha_{k_M}}P)]\\
&= & Tr[\rho_{k_1k_2\cdots k_M} ( \lambda_{P(\alpha_{k_1})}\otimes\lambda_{P(\alpha_{k_2})}\otimes\dots \otimes\lambda_{P(\alpha_{k_M})})]\\
 &=& t _{P(\alpha_{k_1})P(\alpha_{k_2}) \cdots P(\alpha_{k_M})}
\eenrr

  where $P$ is the appropriate permutation matrix permuting the $\la$ matrices within the tensor product \cite{hjb91}, $P^T$ being the transpose of $P$ satisfying $P^T=P^{-1}$. In particular $\mathcal{T}^{(N)}$ is supersymmetric. All matrix unfoldings of a supersymmetric tensor have the same set of singular values \cite{lmv00} and hence the same KF norm. Thus, for a $N$-qudit system in a state supported in the symmetric subspace, it is enough to calculate the KF norm for any one of the $N$ matrix unfoldings to get  $\max \{||T_{(k)}^{(N)}||_{KF}\}.$

 \section{ A Sufficient Condition for Separability of a 3-Partite Quantum State}

 Consider the Bloch representation of a tripartite state $\rho$ acting on $\mathcal{H}^{d_1}\otimes \mathcal{H}^{d_2} \otimes \mathcal{H}^{d_3}$,\; $d_1 \le d_2 \le d_3$.

 $$\rho = \fr{1}{d_1 d_2 d_3}(\otimes_{k=1}^3 I_{d_k}+\sum_{\alpha_1} r_{\alpha_1} \lambda^{(1)}_{\alpha_1}+\sum_{\alpha_2} s_{\alpha_2} \lambda^{(2)}_{\alpha_2}+\sum_{\alpha_3} q_{\alpha_3} \lambda^{(3)}_{\alpha_3}+\sum_{\alpha_1\alpha_2}t_{\alpha_1\alpha_2}  \lambda^{(1)}_{\alpha_1} \lambda^{(2)}_{\alpha_2}$$

 $$+\sum_{\alpha_1\alpha_3}t_{\alpha_1\alpha_3}  \lambda^{(1)}_{\alpha_1} \lambda^{(3)}_{\alpha_3}+\sum_{\alpha_2\alpha_3}t_{\alpha_2\alpha_3}  \lambda^{(2)}_{\alpha_2} \lambda^{(3)}_{\alpha_3} +\sum_{\alpha_1\alpha_2\alpha_3}t_{\alpha_1\alpha_2\alpha_3}  \lambda^{(1)}_{\alpha_1} \lambda^{(2)}_{\alpha_2}\lambda^{(3)}_{\alpha_3},\eqno{(4.19a)}$$

 where $\mathbf{r},\; \mathbf{s}$ and $\mathbf{q}$ are the Bloch vectors of three subsystems respectively , $T^{\{\mu,\nu\}}=[t_{\alpha_{\mu}\alpha_{\nu}}]$ the correlation matrix between the subsystems $\mu, \nu;  \{\mu,\nu\} \subset \{1,2,3\}$ and $\mathcal{T}^{(3)}=[ t_{\alpha_1\alpha_2\alpha_3}]$ the correlation tensor among three subsystems. Before stating proposition 4.4.1, we need the following definition and result.\\

Kruskal decomposition of a tensor $\mathcal{T}^{(N)}$

$$\mathcal{T}^{(N)}=\sum_{j=1}^R \xi_j \mathbf{u}_j^{(1)}\circ\mathbf{u}_j^{(2)}\circ\cdots \circ\mathbf{u}_j^{(N)}$$

is called completely orthogonal if $\lan u_k^{(i)},u_l^{(i)}\ran = \delta_{kl},\;i=1,2,\cdots,N;\;k,l=1,2,\cdots,R$ \cite{kold01}, where $\lan,\ran$ denotes the scalar product of two vectors.
If $\mathcal{T}^{(N)}$  has completely orthogonal Kruskal decomposition, then it is straightforward to show that $$||\mathcal{T}^{(N)}||_{KF}=\sum_{j=1}^R \xi_j,\eqno{(4.20)}$$
where $R$ is the rank of $\mathcal{T}^{(N)}$  and $\xi_j,\;j=1,2,\cdots,R$ are the coefficients occurring in the completely orthogonal Kruskal decomposition of $\mathcal{T}^{(N)}$. In the proof of proposition 4.4.1, we assume that completely orthogonal Kruskal decomposition of $\mathcal{T}^{(k)},\; k>2$ is available. A completely orthogonal Kruskal decomposition may not be available for an arbitrary tensor \cite{kold01}.  The general conditions under which the completely orthogonal Kruskal decomposition is possible is an open problem. We conjecture that completely orthogonal kruskal decomposition is available for all tensors in the Bloch representation of a quantum state, but we do not have a proof.
As it stands, this issue has to be settled case by case.

\noi\textbf{ Proposition 4.4.1 }: If a tripartite state $\rho$ acting on  $\mathcal{H}^{d_1}\otimes \mathcal{H}^{d_2} \otimes \mathcal{H}^{d_3}$,\; $d_1 \le d_2 \le d_3$, with Bloch representation Eq.(4.19a), where $\mathcal{T}^{(3)}$ has the completely orthogonal Kruskal decomposition, satisfies

$$\sq{\fr{2(d_1-1)}{d_1}} ||\mathbf{r}||_2+\sq{\fr{2(d_2-1)}{d_2}} ||\mathbf{s}||_2+\sq{\fr{2(d_3-1)}{d_3}} ||\mathbf{q}||_2+$$
$$\sum_{\{\mu,\nu\}}\sq{\fr{4(d_{\mu}-1)(d_{\nu}-1)}{d_{\mu}d_{\nu}}}||T^{\{\mu,\nu\}}||_{KF}+ \sq{\fr{8(d_1-1)(d_2-1)(d_3-1)}{d_1d_2d_3}} ||\mathcal{T}||_{KF} \le 1,  \eqno{(4.21a)}$$

 then $\rho$ is separable.

\noi\textbf{ Proof} : The idea of the proof is as follows.\\

(i)~~ We first decompose all the tensors in the Bloch representation of $\rho$ as the completely orthogonal Kruskal decomposition in terms of the outer products of the vectors in the Bloch spaces of the subsystems (coherence vectors).

(ii)~~We prove that we can decompose $\rho$ using the Kruskal decompositions described in (i) above, as the linear combination of separable density matrices, which is a convex combination if the coefficient of identity is positive. This condition is the same as the condition stated in the proposition.

Let $T^{\{\mu,\nu\}} ;\{\mu,\nu\} \subset \{1,2,3\}$ in Eq.(4.19a) have singular value decomposition $T^{\{\mu,\nu\}} = \sum_i \si_i \mathbf{a}^{(\mu)}_i ({\mathbf{a}^{(\nu)}_i})^T \; ; \; $ with $||\mathbf{a}_i^{(\mu)}||_2=||\mathbf{a}_i^{(\nu)}||_2=1$ , for $\{\mu, \nu \}\subset \{1,2,3\}$ and  let $\mathcal{T}$ in Eq. (4.19a) have the completely orthogonal Kruskal decomposition $\mathcal{T}= \sum_j \xi_j \mathbf{u}_j\circ \mathbf{v}_j \circ \mathbf{w}_j$ \cite{bk06,krus77,cw06} with $||\mathbf{u}_j||_2 = ||\mathbf{v}_j||_2=||\mathbf{w}_j||_2=1$. We define\\

 $\mathbf{\tilde{a}}^{(\mu)}_i=\sq{\fr{d_{\mu}}{2(d_{\mu}-1)}}\;\mathbf{a}^{(\mu)}_i$  \; , \;  $\mu  \in \{1,2,3\}$\\

 so that we can rewrite  $$T^{\{\mu,\nu\}} = \sq{\fr{4(d_{\mu}-1)(d_{\nu}-1)}{d_{\mu}d_{\nu}}} \sum_i \si_i \mathbf{\tilde{a}}^{\mu}_i (\mathbf{\tilde{a}}^{\nu}_i)^T.\eqno{(4.22a)}$$

  Similarly, we define \\

 $\mathbf{\tilde{u}}_j=\sq{\fr{d_1}{2(d_1-1)}} \; \mathbf{u}_j$\; ; \; $\mathbf{\tilde{v}}_j=\sq{\fr{d_2}{2(d_2-1)}} \; \mathbf{v}_j$\; ; \;  $\mathbf{\tilde{w}}_j=\sq{\fr{d_3}{2(d_3-1)}} \; \mathbf{w}_j$,   so that we can write $$\mathcal{T}=\sq{\fr{8(d_1-1)(d_2-1)(d_3-1)}{d_1d_2d_3}} \sum_j \xi_j \mathbf{\tilde{u}}_j\circ \mathbf{\tilde{v}}_j \circ \mathbf{\tilde{w}}_j \eqno{(4.22b)}$$

 If we substitute Eqs.(4.22a) and (4.22b) in $\rho$ Eq.(4.19a), we get

 $$\rho = \fr{1}{d_1 d_2 d_3}(\otimes_{k=1}^3 I_{d_k}+\sum_{\alpha_1} r_{\alpha_1} \lambda^{(1)}_{\alpha_1}+\sum_{\alpha_2} s_{\alpha_2} \lambda^{(2)}_{\alpha_2}+\sum_{\alpha_3} q_{\alpha_3} \lambda^{(3)}_{\alpha_3}$$
$$+\sum_{\{\mu,\nu\}}\sum_{\alpha_{\mu}\alpha_{\nu}}\sq{\fr{4(d_{\mu}-1)(d_{\nu}-1)}{d_{\mu}d_{\nu}}}\sum_i\si_i(\mathbf{\tilde{a}}^{(\mu)}_{i})_{\alpha_{\mu}}(\mathbf{\tilde{a}}^{(\nu)}_i)_{\alpha_{\nu}}\lambda^{(\mu)}_{\alpha_{\mu}} \lambda^{(\nu)}_{\alpha_{\nu}}$$
$$+\sq{\fr{8(d_1-1)(d_2-1)(d_3-1)}{d_1d_2d_3}}\sum_{\alpha_1\alpha_2\alpha_3}\sum_j\xi_j(\mathbf{\tilde{u}}_j)_{\alpha_1}(\mathbf{\tilde{v}}_j)_{\alpha_2}(\mathbf{\tilde{w}}_j)_{\alpha_3}  \lambda^{(1)}_{\alpha_1} \lambda^{(2)}_{\alpha_2}\lambda^{(3)}_{\alpha_3}) \eqno{(4.19b)}$$

 The coherence vectors $\mathbf{\tilde{a}}^{(\mu)}_i$ occur in $D_r(\mathbb{R}^{d_{\mu}^2-1})$, $\mathbf{\tilde{a}}^{(\nu)}_i$ occur in $D_r(\mathbb{R}^{d_{\nu}^2-1})$, $\mathbf{\tilde{u}}_j$ occur in $D_r(\mathbb{R}^{d_1^2-1})$, $\mathbf{\tilde{v}}_j$ occur in $D_r(\mathbb{R}^{d_2^2-1})$ and $\mathbf{\tilde{w}}_j$ occur in $D_r(\mathbb{R}^{d_3^2-1})$ (see Eq.(4.6)), so that they correspond to Bloch vectors.

 We can decompose $\rho$ Eq.(4.19b) as the following convex combination of the density matrices

  $ \rho_j \; , \; \rho'_j \; ,\; \rho''_j \;,\; \rho'''_j$ ; $\varrho_i \;,\; \varrho'_i\;,\; \tau_i\;,\;\tau'_i\;,\; \pi_i\;,\;\pi'_i$ ; $\rho_r \;,\; \rho_s \; ,\; \rho_q$ and $\fr{1}{d_1d_2d_3} I_{d_1d_2d_3}$;

  $$\rho=\sum_j\sq{\fr{8(d_1-1)(d_2-1)(d_3-1)}{d_1d_2d_3}} \fr{\xi_j}{4}(\rho_j +\rho'_j +\rho''_j + \rho'''_j)+$$
  $$\sum_i\sq{\fr{4(d_1-1)(d_2-1)}{d_1d_2}} \fr{\si_i}{2}(\varrho_i+ \varrho'_i)
  +\sum_i\sq{\fr{4(d_1-1)(d_3-1)}{d_1d_3}} \fr{\si'_i}{2}(\tau_i+ \tau'_i)+$$
  $$\sum_i\sq{\fr{4(d_2-1)(d_3-1)}{d_2d_3}} \fr{\si''_i}{2}(\pi_i+ \pi'_i)
  +\sq{\fr{2(d_1-1)}{d_1}}||\mathbf{r}||_2 \rho_r+$$
  $$\sq{\fr{2(d_2-1)}{d_2}}||\mathbf{s}||_2 \rho_s+
  \sq{\fr{2(d_3-1)}{d_3}}||\mathbf{q}||_2 \rho_q+(1-\sq{\fr{2(d_1-1)}{d_1}} ||\mathbf{r}||_2-$$
$$\sq{\fr{2(d_2-1)}{d_2}}||\mathbf{s}||_2-
\sq{\fr{2(d_3-1)}{d_3}}||\mathbf{q}||_2-
\sum_{\{\mu,\nu\}}\sq{\fr{4(d_{\mu}-1)(d_{\nu}-1)}{d_{\mu}d_{\nu}}} ||T^{\{\mu,\nu\}}||_{KF} $$
$$-\sq{\fr{8(d_1-1)(d_2-1)(d_3-1)}{d_1d_2d_3}} ||\mathcal{T}||_{KF})\fr{I_{d_1d_2d_3}}{d_1d_2d_3}. \eqno{(4.23)}$$

   where $ \rho_j$ in Bloch representation is

   $$\rho_j= \fr{1}{d_1 d_2 d_3}\Big(\otimes_{k=1}^3 I_{d_k}+\sum_{\alpha_1} (\mathbf{\tilde{u}}_j)_{\alpha_1} \lambda^{(1)}_{\alpha_1}+\sum_{\alpha_2} (\mathbf{\tilde{v}}_j)_{\alpha_2} \lambda^{(2)}_{\alpha_2}+\sum_{\alpha_3} (\mathbf{\tilde{w}}_j)_{\alpha_3} \lambda^{(3)}_{\alpha_3}$$
$$+\sum_{\alpha_1\alpha_2}(\mathbf{\tilde{u}}_j)_{\alpha_1}(\mathbf{\tilde{v}}_j)_{\alpha_2}\lambda^{(1)}_{\alpha_{1}} \lambda^{(2)}_{\alpha_{2}}+\sum_{\alpha_1\alpha_3}(\mathbf{\tilde{u}}_j)_{\alpha_1}(\mathbf{\tilde{w}}_j)_{\alpha_3}\lambda^{(1)}_{\alpha_{1}} \lambda^{(3)}_{\alpha_{3}}+\sum_{\alpha_2\alpha_3}(\mathbf{\tilde{v}}_j)_{\alpha_2}(\mathbf{\tilde{w}}_j)_{\alpha_3}\lambda^{(2)}_{\alpha_{2}} \lambda^{(3)}_{\alpha_{3}}$$
$$+\sum_{\alpha_1\alpha_2\alpha_3}(\mathbf{\tilde{u}}_j)_{\alpha_1}(\mathbf{\tilde{v}}_j)_{\alpha_2}(\mathbf{\tilde{w}}_j)_{\alpha_3}  \lambda^{(1)}_{\alpha_1} \lambda^{(2)}_{\alpha_2}\lambda^{(3)}_{\alpha_3} \Big)$$

 $$= \fr{1}{d_1 d_2 d_3}(I_{d_1}+\sum_{\alpha_1}(\mathbf{\tilde{u}}_j)_{\alpha_1}\lambda^{(1)}_{\alpha_1})\otimes (I_{d_2}+\sum_{\alpha_2}(\mathbf{\tilde{v}}_j)_{\alpha_2}\lambda^{(2)}_{\alpha_2})\otimes(I_{d_3}+\sum_{\alpha_3}(\mathbf{\tilde{w}}_j)_{\alpha_3}\lambda^{(3)}_{\alpha_3}). \eqno{(4.24)}$$

  Note that $||\mathcal{T}||_{KF}$ in Eq.(4.23) is defined via Eq.(4.20), which is based on completely orthogonal Kruskal decomposition of  $\mathcal{T}.$

  The Bloch vectors, correlation matrices and correlation tensors of the density matrices $ \rho_j \; , \; \rho'_j \; ,\; \rho''_j \;,\; \rho'''_j$ ; $\varrho_i \;,\; \varrho'_i\;,\; \tau_i\;,\;\tau'_i\;,\; \pi_i\;,\;\pi'_i$ ; $\rho_r \;,\; \rho_s \; ,\; \rho_q$ are:

For $\rho_j$,

  $\mathbf{r}_j= \mathbf{\tilde{u}}_j \; ,\; \mathbf{s}_j=\mathbf{\tilde{v}}_j \; ,\; \mathbf{q}_j=\mathbf{\tilde{w}}_j \; ,\;  T^{\{1,2\}}_j=\mathbf{\tilde{u}}_j {\mathbf{\tilde{v}}_j}^T \;,\; T^{\{1,3\}}_j=\mathbf{\tilde{u}}_j {\mathbf{\tilde{w}}_j}^T \; ,\; T^{\{2,3\}}_j=\mathbf{\tilde{v}}_j {\mathbf{\tilde{w}}_j}^T \mathcal{T}_j=\mathbf{\tilde{u}}_j\circ \mathbf{\tilde{v}}_j\circ \mathbf{\tilde{w}}_j$.\\

  For $\rho'_j$,

$\mathbf{r'}_j= \mathbf{\tilde{u}}_j \; ,\; \mathbf{s'}_j=-\mathbf{\tilde{v}}_j \; ,\;\mathbf{q'}_j=-\mathbf{\tilde{w}}_j \; ,\;  T'^{\{1,2\}}_j=-\mathbf{\tilde{u}}_j {\mathbf{\tilde{v}}_j}^T \;,\; T'^{\{1,3\}}_j=-\mathbf{\tilde{u}}_j {\mathbf{\tilde{w}}_j}^T $
$T'^{\{2,3\}}_j=\mathbf{\tilde{v}}_j {\mathbf{\tilde{w}}_j}^T\;,\; \mathcal{T'}_j=\mathbf{\tilde{u}}_j\circ \mathbf{\tilde{v}}_j\circ \mathbf{\tilde{w}}_j$.\\

 For $\rho''_j$,

 $\mathbf{r''}_j= -\mathbf{\tilde{u}}_j \; ,\; \mathbf{s''}_j=\mathbf{\tilde{v}}_j \; ,\; \mathbf{q''}_j=-\mathbf{\tilde{w}}_j \; ,\;  T''^{\{1,2\}}_j=-\mathbf{\tilde{u}}_j {\mathbf{\tilde{v}}_j}^T \;,\; T''^{\{1,3\}}_j=\mathbf{\tilde{u}}_j {\mathbf{\tilde{w}}_j}^T$

 $ T''^{\{2,3\}}_j=-\mathbf{\tilde{v}}_j {\mathbf{\tilde{w}}_j}^T\;,\; \mathcal{T''}_j=\mathbf{\tilde{u}}_j\circ \mathbf{\tilde{v}}_j\circ \mathbf{\tilde{w}}_j$.\\

For $\rho'''_j$,

$\mathbf{r'''}_j=- \mathbf{\tilde{u}}_j \; ,\; \mathbf{s'''}_j=-\mathbf{\tilde{v}}_j \; ,\;\mathbf{q'''}_j=\mathbf{\tilde{w}}_j \; ,\;  T'''^{\{1,2\}}_j=\mathbf{\tilde{u}}_j {\mathbf{\tilde{v}}_j}^T \;,\; T'''^{\{1,3\}}_j=-\mathbf{\tilde{u}}_j {\mathbf{\tilde{w}}_j}^T$
$ T'''^{\{2,3\}}_j=-\mathbf{\tilde{v}}_j {\mathbf{\tilde{w}}_j}^T\;,\; \mathcal{T'''}_j=\mathbf{\tilde{u}}_j\circ \mathbf{\tilde{v}}_j\circ \mathbf{\tilde{w}}_j$.\\

For $\varrho_i$,

$\mathbf{r}^{\varrho}_i=\mathbf{\tilde{a}}_i^{(1)} \;,\; \mathbf{s}^{\varrho}_i=\mathbf{\tilde{a}}_i^{(2)} \; ,\; \mathbf{q}^{\varrho}_i=0\; ,\; T^{\varrho\{1,2\}}_i=\mathbf{\tilde{a}}_i^{(1)} {\mathbf{\tilde{a}}_i^{(2)T}} \;,\; T^{\varrho\{1,3\}}_i=0$

$ T^{\varrho\{2,3\}}_i=0\;,\; \mathcal{T}^{\varrho}_i=0$.\\

For $\varrho'_i$,

$\mathbf{r}^{\varrho'}_i=-\mathbf{\tilde{a}}_i^{(1)} \;,\; \mathbf{s}^{\varrho'}_i=-\mathbf{\tilde{a}}_i^{(2)} \; ,\; \mathbf{q}^{\varrho'}_i=0\; ,\; T^{\varrho'\{1,2\}}_i=\mathbf{\tilde{a}}_i^{(1)} {\mathbf{\tilde{a}}_i^{(2)T}}$

$T^{\varrho'\{1,3\}}_i=0 \; ,\; T^{\varrho'\{2,3\}}_i=0\;,\; \mathcal{T}^{\varrho'}_i=0$.\\

For $\tau_i$,

$\mathbf{r}^{\tau}_i=\mathbf{\tilde{a}}_i^{(1)} \;,\; \mathbf{s}^{\tau}_i=0 \; ,\; \mathbf{q}^{\tau}_i=\mathbf{\tilde{a}}_i^{(3)}\; ,\; T^{\tau^\{1,2\}}_i=0\;,\; T^{\tau\{1,3\}}_i=\mathbf{\tilde{a}}_i^{(1)} {\mathbf{\tilde{a}}_i^{(3)T}}$

$ T^{\tau\{2,3\}}_i=0\;,\; \mathcal{T}^{\tau}_i=0$.\\

For $\tau'_i$,

$\mathbf{r}^{\tau'}_i=-\mathbf{\tilde{a}}_i^{(1)} \;,\;\mathbf{s}^{\tau'}_i=0 \; ,\; \mathbf{q}^{\tau'}_i=-\mathbf{\tilde{a}}_i^{(3)}\; ,\; T^{\tau'\{1,2\}}_i=0\;,\; T^{\tau'\{1,3\}}_i=\mathbf{\tilde{a}}_i^{(1)} {\mathbf{\tilde{a}}_i^{(3)T}}$

$ T^{\tau'\{2,3\}}_i=0\;,\; \mathcal{T}^{\tau'}_i=0$.\\

For $\pi$,

$\mathbf{r}^{\pi}_i=0 \;,\; \mathbf{s}^{\pi}_i=\mathbf{\tilde{a}}_i^{(2)} \; ,\; \mathbf{q}^{\pi}_i=\mathbf{\tilde{a}}_i^{(3)}\; ,\; T^{\pi}_i{\{1,2\}}=0\;,\; T^{\pi\{1,3\}}_i=0$

$ T^{\pi\{2,3\}}_i=\mathbf{\tilde{a}}_i^{(2)} {\mathbf{\tilde{a}}_i^{(3)T}}\;,\; \mathcal{T}^{\pi}_i=0$.\\

For $\pi'$,

$\mathbf{r}^{\pi'}_i=0 \;,\; \mathbf{s}^{\pi'}_i=-\mathbf{\tilde{a}}_i^{(2)} \; ,\;\mathbf{q}^{\pi'}_i=-\mathbf{\tilde{a}}_i^{(3)}\; ,\; T^{\pi'}_i\{1,2\}=0\;,\; T^{\pi'\{1,3\}}_i=0$

$T^{\pi'\{2,3\}}_i=\mathbf{\tilde{a}}_i^{(2)} {\mathbf{\tilde{a}}_i^{(3)T}}\;,\; \mathcal{T}^{\pi'}_i=0$.\\

For $\rho_r$,

$\mathbf{r}_r=\sq{\fr{d_1}{2(d_1-1)}} \fr{\mathbf{r}}{||\mathbf{r}||_2} \;,\; \mathbf{s}_r=0 \;, \; \mathbf{q}_r=0  \;,\; T^{\{\mu,\nu\}}_r=0\;;\; \forall \{\mu,\nu\} \subset \{1,2,3\} \;,\; \mathcal{T}_r=0$.\\

For $\rho_s$,

$\mathbf{r}_s=0 \;,\; \mathbf{s}_s= \sq{\fr{d_2}{2(d_2-1)}} \fr{\mathbf{s}}{||\mathbf{s}||_2} \;,\; \mathbf{q}_s=0 \;,\; T^{\{\mu,\nu\}}_s=0 \;;\; \forall \{\mu,\nu\} \subset \{1,2,3\} \;,\; \mathcal{T}_s=0$.\\

For $\rho_q$,

$\mathbf{r}_q=0 \;,\;\mathbf{s}_q=0 \;,\; \mathbf{q}_q=\sq{\fr{d_3}{2(d_3-1)}} \fr{\mathbf{q}}{||\mathbf{q}||_2} \;,\; T^{\{\mu,\nu\}}_q=0\;;\; \forall \{\mu,\nu\} \subset \{1,2,3\} \;,\; \mathcal{T}_q=0$.\\

If we write all matrices $\rho'_j, \; \; \rho''_j, \;\; \rho'''_j$; $\varrho_i, \;\; \varrho'_i,\;\; \tau_i,\;\;\tau'_i,\;\; \pi_i,\;\;\pi'_i$; $\rho_r ,\;\; \rho_s, \; \; \rho_q$ (as we have done for $\rho_j$ in Eq.(4.24)) in the Bloch representation and substitute them in Eq.(4.23) we get $\rho$ as in Eq.(4.19b).

To understand this let us see how the first term in Eq.(4.23) adds up to give the last term in Eq.(4.19b). The definition of $\rho_j,\;\rho'_j,\;\rho''_j,\;\rho'''_j$ (denoting the Bloch vectors by $s_1, s_2, s_3, s_4,....$) can be summarized in the tabular form\\

\bc
{\bf Table 4.1}
Correspondence between the first term in Eq.(4.23) and the last term in Eq. (4.19b).
\vspace{.2in}\\

\begin{tabular}{||c|c|c|c||c|c|c|c||}
\hline
&$s_1$ & $s_2$ & $s_3$ &$s_1s_2$&$s_1s_3$&$s_2s_3$&$s_1s_2s_3$\\
\hline
$\rho_j$& $\tilde{u}_j$&$\tilde{v}_j$ &$\tilde{w}_j$ & $\tilde{u}_j\tilde{v}_j$&$\tilde{u}_j\tilde{w}_j$&$\tilde{v}_j\tilde{w}_j$&$\tilde{u}_j\tilde{v}_j\tilde{w}_j$\\
\hline
$\rho'_j$& $\tilde{u}_j$&$-\tilde{v}_j$ &$-\tilde{w}_j$ & $-\tilde{u}_j\tilde{v}_j$&$-\tilde{u}_j\tilde{w}_j$&$\tilde{v}_j\tilde{w}_j$&$\tilde{u}_j\tilde{v}_j\tilde{w}_j$\\
\hline
$\rho''_j$& $-\tilde{u}_j$&$\tilde{v}_j$ &$-\tilde{w}_j$ & $-\tilde{u}_j\tilde{v}_j$&$\tilde{u}_j\tilde{w}_j$&$-\tilde{v}_j\tilde{w}_j$&$\tilde{u}_j\tilde{v}_j\tilde{w}_j$\\
\hline
$\rho'''_j$& $-\tilde{u}_j$&$-\tilde{v}_j$ &$\tilde{w}_j$ & $\tilde{u}_j\tilde{v}_j$&$-\tilde{u}_j\tilde{w}_j$&$-\tilde{v}_j\tilde{w}_j$&$\tilde{u}_j\tilde{v}_j\tilde{w}_j$\\
\hline
 \end{tabular}\\

\ec

\vspace{.2in}

The contribution of each column to  $\rho_j+\rho'_j+\rho''_j+\rho'''_j$ is zero except the last column which reproduces the last term in Eq.(4.19b). We can get the contributions of each term in $\rho_j,\;\rho'_j,\;\rho''_j,\;\rho'''_j$ to their sum by just keeping track of their signs. Thus we only need the following table (dropping $j$)

\newpage
\bc
{\bf Table 4.2}
Contributions of various terms in $\rho, \rho', \rho'',\rho'''$  to their sum.
\vspace{.2in}\\

\begin{tabular}{||c|c|c|c||c|c|c|c||}
\hline
&$s_1$ & $s_2$ & $s_3$ &$s_1s_2$&$s_1s_3$&$s_2s_3$&$s_1s_2s_3$\\
\hline
$\rho$& $+$&$+$ &$+$ & $+$&$+$&$+$&$+$\\
\hline
$\rho'$& $+$&$-$ &$-$ & $-$&$-$&$+$&$+$\\
\hline
$\rho''$& $-$&$+$ &$-$ &$-$&$+$& $-$&$+$\\
\hline
$\rho'''$& $-$&$-$ &$+$&$+$&$-$&$-$&$+$\\
\hline

\end{tabular}\\

\ec
\vspace{.2in}

In the same way, the contributions of the terms involving $\varrho,\;\tau,\;\pi$ are obtained by using the table corresponding to table 4.2  for the bipartite case \cite{vic07a}. $\varrho,\;\tau,\;\pi$ which contain tensors of order two correspond to three 2-partite subsystems 12,13 and 23 . The corresponding tables are

\vspace{.2in}
\bc
{\bf Table 4.3}
Contributions to $\varrho+ \varrho'$
\vspace{.2in}\\
\begin{tabular}{||c|c|c|c||c|c|c|c||}
\hline
&$s_1$ & $s_2$ & $s_3$ &$s_1s_2$&$s_1s_3$&$s_2s_3$&$s_1s_2s_3$\\
\hline
$\varrho$& $+$&$+$ &$0$ & $+$&$0$&$0$&$0$\\
\hline
$\varrho'$& $-$&$-$ &$0$ & $+$&$0$&$0$&$0$\\
\hline

\end{tabular}\\

\ec

\vspace{.2in}

\bc
{\bf Table 4.4}
Contributions to $\tau+\tau'$
\vspace{.1in}\\

\begin{tabular}{||c|c|c|c||c|c|c|c||}
\hline
&$s_1$ & $s_2$ & $s_3$ &$s_1s_2$&$s_1s_3$&$s_2s_3$&$s_1s_2s_3$\\
\hline
$\tau$& $+$&$0$ &$+$ & $0$&$+$&$0$&$0$\\
\hline
$\tau'$& $-$&$0$ &$-$ & $0$&$+$&$0$&$0$\\
\hline

\end{tabular}\\

\ec

\vspace{.1in}

\bc
{\bf Table 4.5}
Contributions to $\pi+ \pi'$
\vspace{.2in}\\
\begin{tabular}{||c|c|c|c||c|c|c|c||}
\hline
&$s_1$ & $s_2$ & $s_3$ &$s_1s_2$&$s_1s_3$&$s_2s_3$&$s_1s_2s_3$\\
\hline
$\pi$& $0$&$+$ &$+$ & $0$&$0$&$+$&$0$\\
\hline
$\pi'$& $0$&$-$ &$-$ & $0$&$0$&$+$&$0$\\
\hline

\end{tabular}\\

\ec

\vspace{.2in}

Tables 4.2, 4.3, 4.4, 4.5 encode the procedure to construct the possible separable state given in Eq.(4.23).

 We now note the following points
\begin{verse}

(i) If the condition (4.21a) holds, then the coefficient of the matrix $I_{d_1d_2d_3}$ in Eq.(4.23) is positive which ensures that the decomposition (4.23) of $\rho$ is positive semidefinite.\\

(ii) By virtue of Eq.(4.6), all the coherence vectors occurring in  $\rho'_j \; ,\; \rho''_j \;,\; \rho'''_j$ ; $\varrho_i \;,\; \varrho'_i\;,\; \tau_i\;,\;\tau'_i\;,\; \pi_i\;,\;\pi'_i$ ; $\rho_r \;,\; \rho_s \; ,\; \rho_q$ belong to the corresponding Bloch spaces.
\end{verse}

By (i) and (ii) we conclude that  $\rho'_j \; ,\; \rho''_j \;,\; \rho'''_j$ ; $\varrho_i \;,\; \varrho'_i\;,\; \tau_i\;,\;\tau'_i\;,\; \pi_i\;,\;\pi'_i$ ; $\rho_r \;,\; \rho_s \; ,\; \rho_q$ constitute density matrices. Further, all these matrices satisfy condition (4.11) so that, via proposition 4.3.1, all these matrices correspond to pure separable states, equal to the tensor products of their reductions. Therefore, they constitute density matrices and they are separable and so must be $\rho$. $\hspace{\stretch{1}} \blacksquare$\\

We can generalize proposition 4.4.1 to the $N$-partite case by constructing the tables successively for $N=4,5,6,\cdots$. First note that the number of $\rho$ s in the first term of Eq.(4.23) lifted to the $N$-partite case is $2^{N-1}$. For $N=4$ we have eight. The corresponding table is

\vspace{.6in}
\bc
{\bf Table 4.6}
Generalization of Table 4.1 to $N=4$.
\vspace{.2in}\\

\begin{tabular}{||c|c|c|c|c||c|c|c|c|c|c|c|c|}
\hline
&$s_1$ & $s_2$ & $s_3$&$s_4$ &$s_1s_2$&$s_1s_3$&$s_1s_4$&$s_2s_3$&$s_2s_4$&$s_3s_4$&$s_1s_2s_3$&$s_1s_2s_4$\\
\hline
$\rho^{(1)}$& $+$&$+$ &$+$ & $+$&$+$&$+$&$+$& $+$&$+$ &$+$ & $+$&$+$\\
\hline
$\rho^{(2)}$& $+$&$+$ &$-$ & $-$&$+$&$-$&$-$& $-$&$-$ &$+$ & $-$&$-$\\
\hline
$\rho^{(3)}$& $+$&$-$ &$+$ & $-$&$-$&$+$&$-$& $-$&$+$ &$-$ & $-$&$+$\\
\hline
$\rho^{(4)}$& $+$&$-$ &$-$ & $+$&$-$&$-$&$+$& $+$&$-$ &$-$ & $+$&$-$\\
\hline
$\rho^{(5)}$& $-$&$+$ &$+$ & $-$&$-$&$-$&$+$& $+$&$-$ &$-$ & $-$&$+$\\
\hline
$\rho^{(6)}$& $-$&$+$ &$-$ & $+$&$-$&$+$&$-$& $-$&$+$ &$-$ & $+$&$-$\\
\hline
$\rho^{(7)}$& $-$&$-$ &$+$ & $+$&$+$&$-$&$-$& $-$&$-$ &$+$ & $+$&$+$\\
\hline
$\rho^{(8)}$& $-$&$-$ &$-$ & $-$&$+$&$+$&$+$& $+$&$+$ &$+$ & $-$&$-$\\
\hline

\end{tabular}\\

\ec
\vspace{.2in}
\bc
{\it (Table 4.6. Continued)}\\

\vspace{.1in}
\begin{tabular}{|c|c|c||}
\hline
$s_1s_3s_4$&$s_2s_3s_4$&$s_1s_2s_3s_4$\\
\hline
$+$&$+$& $+$\\
\hline
$+$&$+$& $+$\\
\hline
$-$&$+$& $+$\\
\hline
$-$&$+$& $+$\\
\hline
$+$&$-$& $+$\\
\hline
$+$&$-$& $+$\\
\hline
$-$&$-$& $+$\\
\hline
$-$&$-$& $+$\\
\hline

\end{tabular}\\

\ec
\vspace{.2in}

We see that the contribution of each column to the sum $\sum_i \rho^{(i)}$ is zero except the last one corresponding to the Kruskal decomposition of $\mathcal{T}^{(N)}$ occurring in the Bloch representation of the given state $\rho$. For general case of $N$-partite state we construct the table for $\rho^{(i)},\; i=1,2,\cdots,2^{N-1}$ as follows. First column consists of $2^{N-2}$ plus signs followed by  $2^{N-2}$ minus signs. Second column comprises alternating $2^{N-3}$ plus and minus signs. Continuing in this way upto $2^{N-N}=1$ we get alternating plus and minus signs in the $(N-1)$th column. We set the $N$th column to ensure that there are zero or even number of minus signs in each row. Rest  of the columns can be constructed by appropriate multiplications. This procedure can be checked on table 4.6. We denote the  sequence of such tables for $N=2,3,4,\cdots$ as $T_i,\; i=2,3,4,\cdots$.

The tables corresponding to $(N-1), (N-2),...,2$ partite subsystems giving rise to the remaining terms in the Eq. (4.23), lifted to $N$-partite case, are obtained from $T_{N-1},T_{N-2},...,T_3,T_2,$ exactly as described in the proof of  proposition 4.4.1. In this way we can lift Eq.(4.23) to the $N$-partite case, with the total numbers of terms $\sum_{i=0}^{N-1}\binom{N}{i}2^{N-1-i} +1$. Once this is done, the rest of the proof for $N$-partite case follows as in proposition 4.4.1. Thus we have\\

\noi\textbf{Proposition 4.4.1a}: If a $N$-partite state $\rho$ acting on  $\mathcal{H}=\mathcal{H}^{d_1}\otimes \mathcal{H}^{d_2} \otimes \cdots \otimes \mathcal{H}^{d_N},\; d_1 \le d_2 \le \cdots \le d_N$ with  Bloch representation Eq.(4.8), where all $\mathcal{T}^{(k)},\; k>2$ have the completely orthogonal Kruskal decomposition, satisfy

$$\sum_k\sq{\fr{2(d_k-1)}{d_k}} ||\mathbf{s_k}||_2+\sum_{\{\mu,\nu\}}\sq{\fr{4(d_{\mu}-1)(d_{\nu}-1)}{d_{\mu}d_{\nu}}} ||T^{\{\mu,\nu\}}||_{KF}$$

$$+\sum_{\{\mu,\nu,\kappa\}}\sq{\fr{8(d_{\mu}-1)(d_{\nu}-1)(d_{\kappa}-1)}{d_{\mu}d_{\nu}d_{\kappa}}} ||\mathcal{T}^{\{\mu,\nu,\kappa\}}||_{KF}+\cdots+$$
$$\sum_{\{k_1,k_2,\cdots,k_M\}}\sq{\fr{2^M\Pi_{k_i}(d_{k_i}-1)}{\Pi_{k_i}d_{k_i}}}||\mathcal{T}^{\{k_1,k_2,\cdots,k_M\}}||_{KF} +\cdots+$$
$$\sq{\fr{2^N\Pi_{i}^N(d_{i}-1)}{\Pi_{i}^Nd_{i}}}||\mathcal{T}^{(N)}||_{KF} \le 1,  \eqno{(4.21b)}$$\\

 then $\rho$ is separable.$\hspace{\stretch{1}} \blacksquare$\\
For a $N$-qubit system  theorem 4.3.2 and proposition 4.4.1a  together imply \\
\noi\textbf{Corollary 4.4.2 :} Let a $N$-qubit state have a Bloch representation
 $$\rho= \fr{1}{2^N}(\otimes_{k=1}^N  I_2^{(k)}+\sum_{\alpha_1 \cdots \alpha_N} t_{\alpha_1 \cdots \alpha_N} \la_{\alpha_1}^{(1)}\la_{\alpha_2}^{(2)} \cdots \la_{\alpha_N}^{(N)}),$$
 and let the tensor in the second term have the completely orthogonal Kruskal decomposition.
 Then $\rho$ is separable if and only if $||\mathcal{T}^{(N)}||_{KF} \le 1.$$\hspace{\stretch{1}} \blacksquare$

\section{ Examples}

 We now investigate our separability criterion (4.18) for mixed states.
  We consider $N$-qubit state $$\rho^{(N)}_{noisy}\;=\;\fr{1-p}{2^N}I+p|\psi\ran\lan\psi|, \;\;\;0\le p\le 1  \eqno{(4.25)}$$

where $|\psi\ran$ is a $N$-qubit $W$ state or GHZ state. We test for $N=3,4,5$ and $6$ qubits. We get,

\vspace{.2in}
\bc
{\bf Table 4.7}
The values of $p$ above which the states are entangled.
\vspace{.1in}\\

\begin{tabular}{||c|c||c||}
\hline
$|GHZ\ran$ & $|W\ran$ & $N$\\
$p >$  & $p >$  & \\
\hline
 0.35355 & 0.3068 & 3\\
\hline
 0.2 &  0.3018 & 4\\
\hline
 0.17675 &  0.30225 & 5\\
\hline
 0.1112 &  0.3045  & 6\\
\hline
\end{tabular}\\
\vspace{.2in}

 \ec
\vspace{.2in}

  Entanglement in various partitions of $W$ noisy state Eq.(4.25) is obtained by using $(N-n)$ qubit reduced $W$ noisy state
  $$\rho_{noisy}^{(N-n)}(W)=\fr{1-p}{2^{N-n}} I_{N-n} +\fr{n}{N} p |0_{N-n}\ran \lan 0_{N-n}|+\fr{N-n}{N} p |W_{N-n}\ran \lan W_{N-n}|  \eqno{(4.26)}$$

    For $N=6$ and $n=2$ we found that the state is entangled for $0.491 < p \le 1.$

  For $N$ qutrits $(d=3)$ we test for $$\rho_{noisy}^{(N)}\;=\;\fr{1-p}{3^N}\;I\;+\; p|\psi\ran\lan\psi|\eqno{(4.27)}$$ where    $|\psi \ran = \fr{1}{\sqrt{d}}\sum_{k=1}^d |kkk \dots  \ran $ is the maximally entangled state for $N$ qutrits.

   For $N=3$ and $N=4$ (qutrits) the state $\rho_{noisy}^{(N)}$ in Eq. (4.27) is entangled for

   $$0.2285 < p \le 1,  \; \; \; \; N=3$$

    $$0.2162 < p \le 1, \; \; \; \; N=4  \eqno{(4.28)}$$

  The state $$\rho_{noisy}\;=\;\fr{1-p}{24}\;I\;+\; p|\psi\ran\lan\psi| \eqno{(4.29)}$$  where  $|\psi\ran=\fr{1}{2}(|112\ran+|123\ran+|214\ran+|234\ran) $ in the space $\mathbb{C}^2\otimes\mathbb{C}^3\otimes \mathbb{C}^4$ is found to be entangled for $0.24152 < p\le 1.$

   All of the above examples involve NPT states. Now we apply our criterion to PPT entangled states for which PPT criterion is not available.\\

  We apply our criterion to the three qutrit bound entangled state considered by L. Clarisse and P. Wocjan \cite{cw06}, given by $\rho_c \otimes|\psi\ran \lan \psi|$ where $\rho_c$ is the chess-board state given in \cite{cw06} and $|\psi\ran$ is an uncorrelated ancilla. Our criterion detects the entanglement of this state as $||\mathcal{T}^{(12)}|| =3.75 >3.$
  Further, the four qutrit state $\rho = (1-\beta)\rho_c \otimes \rho_c + \beta I/81$ considered by the same authors yields entanglement for $0 \le \beta \le 0.2,$ after tracing out either subsystems 1 and 2 or subsystems 3 and 4.

  Now we consider the important example of the Smolin state \cite{hhhh07,smo01}, which is a four qubit bound entangled state given by $$ \rho_{ABCD}^{unlock}=\frac{1}{4}\sum_{i=1}^4 |\psi_{AB}^i\ran \lan \psi_{AB}^i|\otimes |\psi_{CD}^i\ran \lan \psi_{CD}^i| \eqno{(4.30)}$$

  where $|\psi_{AB}^i\ran$ and $|\psi_{CD}^i\ran$ are the Bell states. $ \rho_{ABCD}^{unlock}$ has the Bloch representation $\rho_{ABCD}^{unlock}=\frac{1}{16}(I^{\otimes4}+\sum_{i=1}^3 \si_i^{\otimes4})$ so that Corollary 4.4.2 applies (note that the requirement of completely orthogonal Kruskal decomposition is trivially satisfied). We find for this state $||\mathcal{T}^{(4)}||_{KF} = 3 > 1$ confirming its entanglement.

  Our last example is the four qubit bound entangled state due to W. D\"ur \cite{dur01,acin02}
  $$\rho_4^{BE}= \frac{1}{5}(|\psi\ran \lan \psi|+\frac{1}{2}\sum_{i=1}^4(P_i+\overline P_i))$$
  where $|\psi\ran$ is a 4-party (GHZ) state , $P_i$ is the projector onto the state $|\phi_i\ran$, which is a product state equal to $|1\ran$ for party $i$ and $|0\ran$ for the rest , and $\overline P_i$ is obtained from  $P_i$ by replacing all zeros by ones and vice versa. We get $||\mathcal{T}^{(4)}||_{KF} = 1.4 > 1$ confirming the entanglement of this state.

\section{Summary}

  In conclusion we have presented a new criterion for separability of $N$ partite quantum states based on the Bloch representation of states. This criterion is quite general, as it  applies to all $N$-partite quantum states living in $\mathcal{H}=\mathcal{H}^{d_1}\otimes \mathcal{H}^{d_2} \otimes \cdots \otimes \mathcal{H}^{d_N},$ where, in general, the Hilbert space dimensions of various parts are not equal. Most of the previous such criteria had restricted domain of applicability like the states supported on symmetric subspace \cite{dpr07} or, are, in general, restricted to bipartite case.
   In proposition 4.4.1, we have given a sufficient condition for the separability of a tripartite state under the condition that the tensors occurring in the Bloch representation of the state have completely orthogonal Kruskal decomposition. This result can be generalized to the $N$-partite case. Via corollary 4.4.2, we give a necessary and sufficient condition to test the separability of a class of $N$-qubit states which includes $N$-qubit PPT states. Smolin state Eq.(4.30) is an important example in this class. The key idea in our work is the matrization of multidimensional tensors, in particular, Kruskal decomposation. We have defined a new tensor norm as the maximum of the KF norms of all the matrix unfoldings of a tensor, which is easily computed. We have also shown that this norm can be calculated even more efficiently  for a $N$-qudit state supported in the symmetric subspace.
   It will be interesting to seek a relation of this tensor norm with other entanglement measures. Again, the entanglement measures like concurrence known so far are successfully applied to pure states, bipartite or multipartite, while our tensor norm can be easily computed for arbitrary $N$-partite quantum state. Finally, our result on full separability (proposition 4.3.1) of $N$-partite pure states can be easily moulded for the $k$-separability of an $N$-partite pure state. In fact it is straightforward to construct an algorithm giving the complete factorization of the $N$-partite pure state (see the paragraph following the proof of proposition 4.3.1). It is also easy to see that theorem 4.3.2 can be applied to any partition of a $N$-partite system via the Bloch representation in terms of the generators of the appropriate $SU$ groups. Most important is the observation that all the tensors in the Bloch representation can be computed using the measured values of the basis operators $\{\lambda_{\alpha_k}\}$ so that our detectiblity criterion is experimentally implementable.

\chapter{Experimentally accessible geometric measure for entanglement in N-qubit pure states}
\begin{center}
\scriptsize \textsc{First we guess it. Then we compute the consequences of the guess to see what would be implied if the law we guess is right. Then we compare the result of the computation to nature, with experiment or experience, compare it directly with observation, to see if it works. If it disagrees with experiment it is wrong. In that simple statement is the key to science. it does not make any difference how beautiful your guess is. It does not make any difference how smart you are, who made the guess, or what your name is-if it disagrees with experiment it is wrong. That is all there is to it. {\it Richard Feynman}}
\end{center}

\section{Introduction}

In the previous chapter, we dealt with the problem of detecting separable (entangled) states based on the Bloch representation of states. In this chapter, we deal with the problem of quantification of entanglement of multipartite quantum states, that is, to find a measure of entanglement (see chapter 1) for these states.

Quantification of entanglement of multipartite quantum  states is fundamental to the whole field of quantum information and in general, to the physics of multicomponent quantum systems. The principal achievements regarding this problem are in the setting of bipartite systems. Among these, one highlights Wootter's formula for the entanglement of formation of two qubit mixed states \cite{woot98}, which still awaits a viable generalization to multiqubit case. Others include corresponding results for highly symmetric states \cite{vw01,tv00,dpr07}. The issue of entanglement in multipartite states is far more complex. Notable achievements in this area include applications of the relative entropy \cite{vp98}, negativity \cite{zhsl,vw02} Schimidt measure \cite{ebheb} and the global entanglement measure proposed by Meyer and Wallach \cite{mw02}.

A measure of entanglement is a function on the space of states of a multipartite system, which is invariant on individual parts. Thus a complete characterization of entanglement is the characterization of all such functions. Under the most general local operations assisted by classical communication (LOCC), entanglement is expected to decrease. A measure of entanglement that decreases under LOCC is called an entanglement monotone. On bipartite pure states the sums of the $k$ smallest eigenvalues of the reduced density matrix are entanglement monotones. However, the number of independent invariants (i.e., the entanglement measures) increase exponentially as the number of particles $N$ increases and complete characterization rapidly becomes impractical. A pragmatic approach would be to seek a measure which is defined for any number of particles (scalable), which is easily calculated and which provides physically relevant information or equivalently which passes the tests expected of a {\it good} entanglement measure \cite{pv07,zb06}.

In this chapter, we present a global entanglement measure for $N$-qubit pure states which is scalable, which passes most of the tests expected of a good measure and whose value for a given system can be determined experimentally, without having a detailed {\it prior} knowledge of the state of the system. The measure is based on the Bloch representation of multipartite quantum states (see chapter 4).

The chapter is organized as follows. In section 5.2 we give the Bloch representation of an $N$-qubit quantum state and define our measure $E_{\mathcal{T}}.$ In section 5.3 we compute $E_{\mathcal{T}}$ for different classes of $N$-qubit states, namely, the Greenberger-Horne-Zeilinger ($GHZ$) and $W$ states and their superpositions. In section 5.4 we prove various properties of $E_{\mathcal{T}},$ including its monotonicity, expected of a good entanglement measure. In section 5.5 we extend $E_{\mathcal{T}}$ to $N$-qubit mixed states via convex roof and establish its monotonicity. In section 5.6 we introduce a related measure which is additive and shares all other properties with $E_{\mathcal{T}}.$ Finally, we conclude in section 5.7.

\section{Bloch Representation of An $N$-qubit State and The Definition of The Measure}

Consider the generators $\{I,\si_x,\;\si_y,\;\si_z\}\equiv \{\si_0,\si_1,\si_2,\si_3\}$ of $SU(2)$ group (Pauli matrices). These Hermitian operators form a orthogonal basis (under the Hilbert-Schmidt scalar product) of the Hilbert space of operators acting on a single qubit state space. The $N$ times tensor product of this basis with itself generates a product basis of the Hilbert space of operators acting on the $N$-qubit state space. Any $N$-qubit density operator $\rho$ can be expanded in this basis. The corresponding expansion is called the Bloch representation of $\rho$ (see chapter 4).

  In order to give the Bloch representation of a density operator acting on the Hilbert
  space $\mathbb{C}^{2} \otimes \mathbb{C}^{2} \otimes \cdots \otimes \mathbb{C}^{2}$
  of an $N$-qubit quantum system, we introduce the following notation. We use $k$, $k_i \; (i=1,2,\cdots)$ to denote a qubit chosen from $N$ qubits, so that $k$,\; $k_i \; (i=1,2,\cdots)$ take values in the set  $\mathcal{N}=\{1,2,\cdots,N\}$. The variables $\alpha_k \;\mbox{or} \; \alpha_{k_i}$ for a given $k$ or $k_i$ span the set of generators of the $SU(2)$ group for the $k$th or $k_i$th qubit, namely, the set $\{I_{k_i},\si_{1_{k_i}},\si_{2_{k_i}},\si_{3_{k_i}}\}$ for the $k_i$th qubit. For two qubits $k_1$ and $k_2$ we define

   $$\si^{(k_1)}_{\alpha_{k_1}}=(I_{2}\otimes I_{2}\otimes \dots \otimes \si_{\alpha_{k_1}}\otimes I_{2}\otimes \dots \otimes I_{2})   $$
   $$\si^{(k_2)}_{\alpha_{k_2}}=(I_{2}\otimes I_{2}\otimes \dots \otimes \si_{\alpha_{k_2}}\otimes I_{2}\otimes \dots \otimes I_{2})  $$
   $$\si^{(k_1)}_{\alpha_{k_1}} \si^{(k_2)}_{\alpha_{k_2}}=(I_{2}\otimes I_{2}\otimes \dots \otimes \si_{\alpha_{k_1}}\otimes I_{2}\otimes \dots \otimes \si_{\alpha_{k_2}}\otimes I_{2}\otimes I_{2})   \eqno{(5.1)}$$

  where  $\si_{\alpha_{k_1}}$ and $\si_{\alpha_{k_2}}$ occur at the $k_1$th and $k_2$th places (corresponding to $k_1$th and $k_2$th qubits, respectively) in the tensor product and are the $\alpha_{k_1}$th and  $\alpha_{k_2}$th generators of $SU(2),\; (\alpha_{k_1}=1,2,3\; \mbox{and} \; \alpha_{k_2}=1,2,3)$, respectively. Then we can write

$$\rho=\fr{1}{2^N} \{\otimes_k^N I_{2}+ \sum_{\{k \}\subset \mathcal{N}}\sum_{\alpha_{k}}s_{\alpha_{k}}\si^{(k)}_{\alpha_{k}} +\sum_{\{k_1,k_2\}}\sum_{\alpha_{k_1}\alpha_{k_2}}t_{\alpha_{k_1}\alpha_{k_2}}\si^{(k_1)}_{\alpha_{k_1}} \si^{(k_2)}_{\alpha_{k_2}}+\cdots +$$
$$\sum_{\{k_1,k_2,\cdots,k_M\}}\sum_{\alpha_{k_1}\alpha_{k_2}\cdots \alpha_{k_M}}t_{\alpha_{k_1}\alpha_{k_2}\cdots \alpha_{k_M}}\si^{(k_1)}_{\alpha_{k_1}} \si^{(k_2)}_{\alpha_{k_2}}\cdots \si^{(k_M)}_{\alpha_{k_M}}+ \cdots+$$
$$\sum_{\alpha_{1}\alpha_{2}\cdots \alpha_{N}}t_{\alpha_{1}\alpha_{2}\cdots \alpha_{N}}\si^{(1)}_{\alpha_{1}} \si^{(2)}_{\alpha_{2}}\cdots \si^{(N)}_{\alpha_{N}}\}.\eqno{(5.2)}$$

 where $\textbf{s}^{(k)}$ is a Bloch vector (see below) corresponding to $k$th subsystem,  $\textbf{s}^{(k)} =[s_{\alpha_{k}}]_{\alpha_{k}=1}^{3},$ which is a tensor of order 1 defined by
 $$s_{\alpha_{k}}= Tr[\rho \si^{(k)}_{\alpha_{k}}]= Tr[\rho_k \si_{\alpha_{k}}],\eqno{(5.3)}$$
 where $\rho_k$ is the reduced density matrix for the $k$th qubit. Here $$\{k_1,k_2,\cdots,k_M\},\; 1 \le M \le N,$$ is a subset of $\mathcal{N}$ and can be chosen in $\binom{N}{M}$  ways, contributing $\binom{N}{M}$ terms in the sum $\sum_{\{k_1,k_2,\cdots,k_M\}}$ in Eq.(5.2), each containing a tensor of order $M$. The total number of terms in the Bloch representation of $\rho$ is $2^N$. We denote the tensors occurring in the sum $\sum_{\{k_1,k_2,\cdots,k_M\}},\; (1 \le M \le N)$ by $\mathcal{T}^{\{k_1,k_2,\cdots,k_M\}}=[t_{\alpha_{k_1}\alpha_{k_2}\cdots \alpha_{k_M}}]$, which  are defined by

 $$t_{\alpha_{k_1}\alpha_{k_2}\dots\alpha_{k_M}}= Tr[\rho \si^{(k_1)}_{\alpha_{k_1}} \si^{(k_2)}_{\alpha_{k_2}}\cdots \si^{(k_M)}_{\alpha_{k_M}}]$$

$$ = Tr[\rho_{k_1k_2\dots k_M} (\si_{\alpha_{k_1}}\otimes\si_{\alpha_{k_2}}\otimes\dots \otimes\si_{\alpha_{k_M}})]   \eqno{(5.4)}$$

where $\rho_{k_1k_2\dots k_M}$ is the reduced density matrix for the subsystem $\{k_1 k_2\dots k_M\}$. We call  The tensor in last term in Eq. (5.2) $\mathcal{T}^{(N)}$.

From Eq.(5.4) we see that all the correlations between $M$ out of $N$-qubits are contained in  $\mathcal{T}^{\{k_1,k_2,\cdots,k_M\}}$ and all the $N$-qubit correlations are contained in  $\mathcal{T}^{(N)}$.
If $\rho$ is a $N$-qubit pure state we have $$Tr(\rho^2)=\fr{1}{2^N}(1+\sum_{k=1}^N ||\mathbf{s^{(k)}}||^2+\sum_{\{k_1,k_2\}}||\mathcal{T}^{\{k_1,k_2\}}||^2+\cdots+\sum_{\{k_1,k_2,\cdots,k_M\}}||\mathcal{T}^{\{k_1,k_2,\cdots,k_M\}}||^2$$
$$+\cdots+||\mathcal{T}^{(N)}||^2)=1 \eqno{(5.5)}$$
Any state $\rho=|\psi\ran\lan\psi|$ existing  in a $d^2$-dimensional Hilbert space of operators acting on a $d$-dimensional Hilbert space of kets can be expanded in the basis comprising $d^2-1$ generators of SU(d) and the identity operator. The set of coefficients in this expansion, namely, $\{Tr(\rho\lambda_i)\},\;i=1,2,\cdots,d^2-1,$ is a vector in $\mathbb{R}^{d^2-1}$ and is the Bloch vector of $\rho.$ The set of Bloch vectors and the set of density operators are in one-to-one correspondence with each other. The set of Bloch vectors for a given system forms a subspace of $\mathbb{R}^{d^2-1}$ denoted  $B(\mathbb{R}^{d^2-1}).$ The specification of this subspace for $d\geq 3$ is an open problem \cite{kk05,kim03}. However, for pure states, following results are known \cite{bk03}: $$\Arrowvert s \Arrowvert_2=\sqrt{\fr{d(d-1)}{2}}\eqno(5.6)$$ $$D_{r}(\mathbb{R}^{d^2-1})\subseteq B(\mathbb{R}^{d^2-1})\subseteq D_{R}(\mathbb{R}^{d^2-1}),$$ where $D_r$ and $D_R$ are the balls of radii $r=\sqrt{\fr{d}{2(d-1)}}$ and $R=\sqrt{\fr{d(d-1)}{2}},$ respectively, in $\mathbb{R}^{d^2-1}.$

We propose the following measure for an $N$-qubit pure state entanglement $$E_{\mathcal{T}}(|\psi\ran)=(||\mathcal{T}^{(N)}||-1), \eqno{(5.7)}$$

where $\mathcal{T}^{(N)}$ is given by Eq.(5.4) for ($M=N$) in the Bloch representation of $\rho=|\psi\ran \lan\psi|$. The norm of the tensor $\mathcal{T}^{(N)}$ appearing in definition (5.7) is the Hilbert-Schmidt (Euclidean) norm $||\mathcal{T}^{(N)}||^2=(\mathcal{T}^{(N)},\mathcal{T}^{(N)})= \sum_{\alpha_{1}\alpha_{2}\cdots \alpha_{N}}t_{\alpha_{1}\alpha_{2}\cdots \alpha_{N}}^2$. Throughout this chapter, by norm, we mean the Hilbert-Schmidt (Euclidean) norm. We comment on the normalization of $E_{\mathcal{T}}(|\psi\ran)$ below.\\

\section{ $GHZ$ and $W$ States}

Before proving various properties of $E_{\mathcal{T}}(|\psi\ran)$, we evaluate it for states in the $N$-qubit $GHZ$ or $W$ class. A general $N$-qubit $GHZ$ state is given by $$|\psi\ran=\sq{p}|000\cdots0\ran+\sq{1-p}|111\cdots1\ran, \;\;N\ge 2 \eqno{(5.8)}$$

A general element of $\mathcal{T}^{(N)}$ is given by $t_{i_1i_2\cdots i_N}=\lan\psi|\si_{i_1}\otimes\si_{i_2}\otimes\cdots\otimes\si_{i_N}|\psi\ran,\;i_k=1,2,3,\;k=1,2,\cdots ,N.$
The nonzero elements of $\mathcal{T}^{(N)}$ are $t_{11\cdots1}=2\sq{p(1-p)}$, $t_{33\cdots3}=p+(-1)^N(1-p)$.
Other nonzero elements of $\mathcal{T}^{(N)}$ are those with $2k\si_2$'s and $(N-2k)\si_1$'s, $k=0,1,\cdots,\lfloor \fr{N}{2} \rfloor$, where $\lfloor x \rfloor$ is the greatest integer less than or equal to $x$ (e.g., for $N=3,\; t_{122}\; etc.)$. These are equal to $(-1)^k \;2 \sq{p(1-p)}.$ This gives $$||\mathcal{T}^{(N)}||^2=4p(1-p)+[p+(-1)^N(1-p)]^2+4p(1-p)\sum_{k=1}^{\lfloor \fr{N}{2} \rfloor}\binom{N}{2k}. \eqno{(5.9)}$$

Thus we get, for  $E_{\mathcal{T}}(|\psi\ran),$

$$E_{\mathcal{T}}(|\psi\ran)=||\mathcal{T}^{(N)}||-1$$ $$=\sq{4p(1-p)+[p+(-1)^N(1-p)]^2+4p(1-p)\sum_{k=1}^{\lfloor \fr{N}{2} \rfloor}\binom{N}{2k}}-1 \eqno{(5.10)}$$

Equation (5.8), with $N=2,$ represents a general two-qubit entangled state in its Schmidt decomposition, $$\arrowvert\psi\ran=\sq{p}\arrowvert00\ran+\sq{1-p}\arrowvert11\ran.$$
Thus Eq.(5.10) gives the entanglement in a two-qubit pure state. Using Eq.(5.10) it is straightforward to see that $E_{\mathcal{T}}(|\psi\ran)$ for an arbitrary two-qubit pure state is related to concurrence by
$$E_{\mathcal{T}}(|\psi\ran)=\sq{1+2C^2}-1,$$ where concurrence $C$ for such a state is $2\sq{p(1-p)}.$

Figure 5.1  plots $E_{\mathcal{T}}(\psi\ran)$ in Eq. (5.10) as a function of $p$ for $N=3$. For the $N$-qubit $GHZ$ (maximally entangled) state $p=1/2$, so that
$$R_N=E_{\mathcal{T}}(|GHZ\ran)=\sq{1+\fr{1}{4}[1+(-1)^N]^2+\sum_{k=1}^{\lfloor \fr{N}{2} \rfloor}\binom{N}{2k}}-1 \eqno{(5.11)}$$

\begin{figure}[!ht]
\begin{center}
\includegraphics[width=15cm,height=12cm]{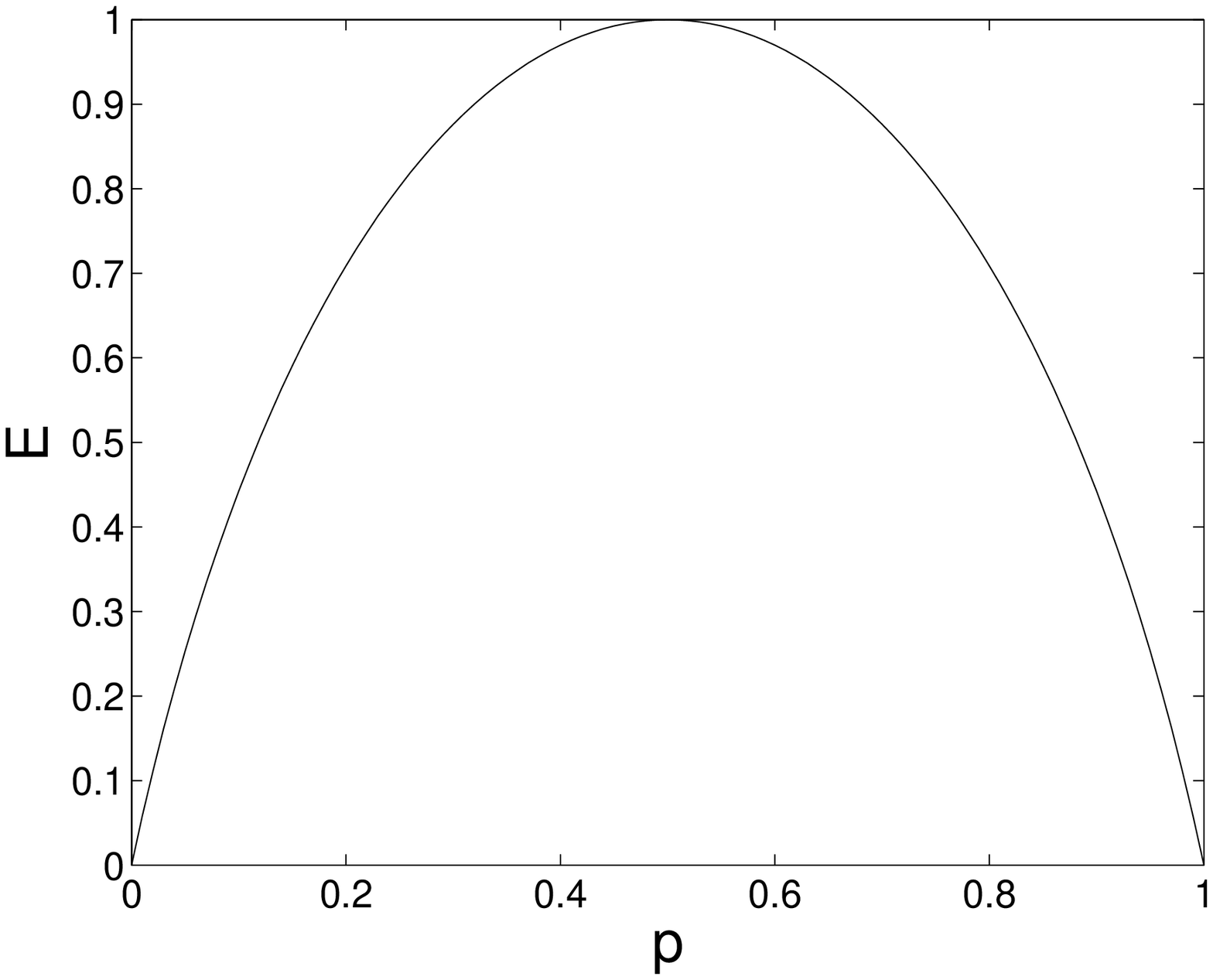}

Figure 5.1. Variation of $E_{\mathcal{T}}(|\psi\ran)$ [Eq. (5.10)] for $N=3,$ express in units of $R_3,$ with parameter $p$.
\end{center}
\end{figure}

We see that, as a function of $N$, $E_{\mathcal{T}}(|GHZ\ran)$ increases as a polynomial of degree $\lfloor \fr{N}{2} \rfloor$. Figure 5.2 plots $E_{\mathcal{T}}(|GHZ\ran)$ as a function of $N$. $E_{\mathcal{T}}(|GHZ\ran)$ increases sharply with $N$ as expected. Note that $E_{\mathcal{T}}(|GHZ\ran)\ge 0$ for $GHZ$ class of states. Whenever appropriate, we normalize the entanglement of an $N$-qubit state $|\psi\ran$, $E_{\mathcal{T}}(|\psi\ran)$, by dividing by $R_N=E_{\mathcal{T}}(|GHZ\ran).$

\begin{figure}[!ht]
\begin{center}
\includegraphics[width=15cm,height=12cm]{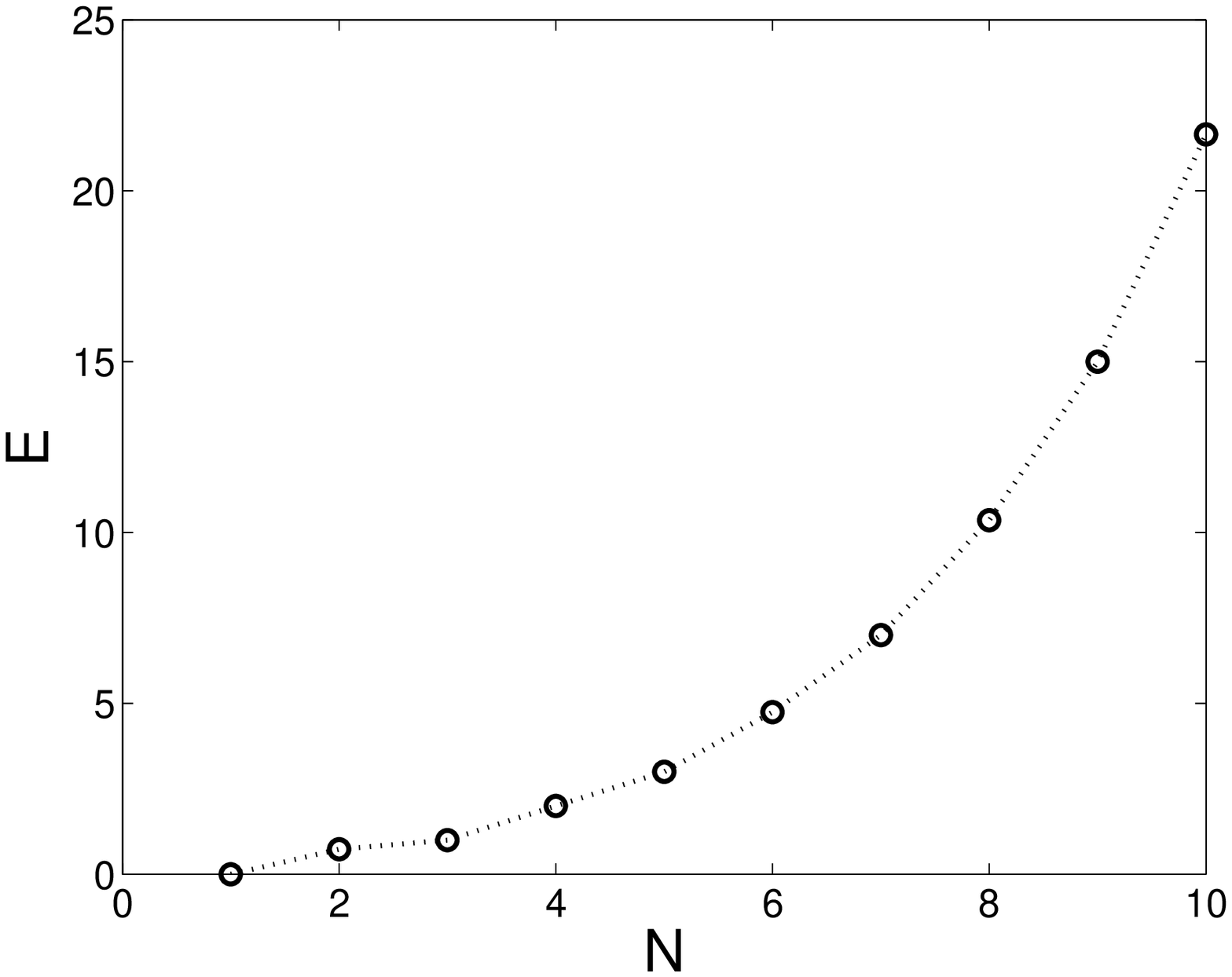}

Figure 5.2. Variation of $R_N=E_{\mathcal{T}}(|GHZ\ran)$ [Eq. (5.11)] with number of qubits $N$.
\end{center}
\end{figure}

\vspace{.2in}

The $N$-qubit W state is given by $$|W\ran=\fr{1}{\sq{N}}\sum_j|00\cdots 1_j 0 \cdots 00\ran,\; \;N \ge 3$$
where the $j$th term has a single 1 at the $j$th bit. The state $|\widetilde{W}\ran=\otimes_{k=1}^N \si_1^{(k)} |W\ran$ is  given by $|\widetilde{W}\ran=\fr{1}{\sq{N}}\sum_j |11\cdots 0_j 1 \cdots 11\ran,\; N\ge 3$, and has a single 0 at the $j$th bit. We note that $|\widetilde{W}\ran$ is locally unitarily connected to $|W\ran$ so that their entanglements must have the same value. The general element of $\mathcal{T}^{(N)}$ for the state $\rho=|W\ran \lan W|$ is

$$t_{i_1i_2\cdots i_N}=\fr{1}{N}\sum_{j=1}^{N}\lan00\cdots 1_j\cdots00|\si_{i_1}\otimes\si_{i_2}\otimes\cdots\otimes\si_{i_N}|00\cdots 1_j\cdots00\ran$$

$$+\fr{1}{N}\sum_{j,l=1;j\ne l}^{N}\lan00\cdots1_j\cdots00|\si_{i_1}\otimes\si_{i_2}\otimes\cdots\otimes\si_{i_N}|00\cdots 1_l\cdots00\ran$$
Only the first term contributes to  $t_{33\cdots33}=-1$. Other nonzero elements have the form $t_{3\cdots31_j3\cdots 31_l3\cdots3}=\fr{2}{N}=t_{3\cdots32_j3\cdots 32_l3\cdots3}.$

There are $\binom{N}{2}$ elements of each of these two types, so that $$||\mathcal{T}^{(N)}||^2=1+2\Big(\fr{2}{N}\Big)^2 \binom{N}{2}=1+4\fr{N-1}{N},\eqno{(5.12)}$$

$$E_{\mathcal{T}}(|W\ran)=||\mathcal{T}^{(N)}||-1 =\sq{1+4\fr{N-1}{N}}-1 \eqno{(5.13)}$$

It is straightforward to check that $E_{\mathcal{T}}(|W\ran)=E_{\mathcal{T}}(|\widetilde{W}\ran)$ as expected. Note that $E_{\mathcal{T}}(|W\ran)\ge 0.$

Next we consider a superposition of $|W\ran$ and $|\widetilde{W}\ran$ states,
$|\psi_{s,\phi}\ran=\sq{s}|W\ran+\sq{1-s}e^{i\phi}|\widetilde{W}\ran$.
It is clear that the entanglement of $|\psi_{s,\phi}\ran$ cannot depend on the relative phase $\phi$, as $|\psi_{s,\phi}\ran$ is invariant under the local unitary transformation $\{|0\ran,|1\ran\} \rightarrow \{|0\ran,e^{i\phi}|1\ran \}$ upto an overall phase factor. As we shall prove below, $E_{\mathcal{T}}$ is invariant under local unitary transformations. Figure 5.3 shows the entanglement of $|\psi_{s,\phi}\ran$ as a function of $s$, calculated using our measure.

\begin{figure}[!ht]
\begin{center}
\includegraphics[width=15cm,height=12cm]{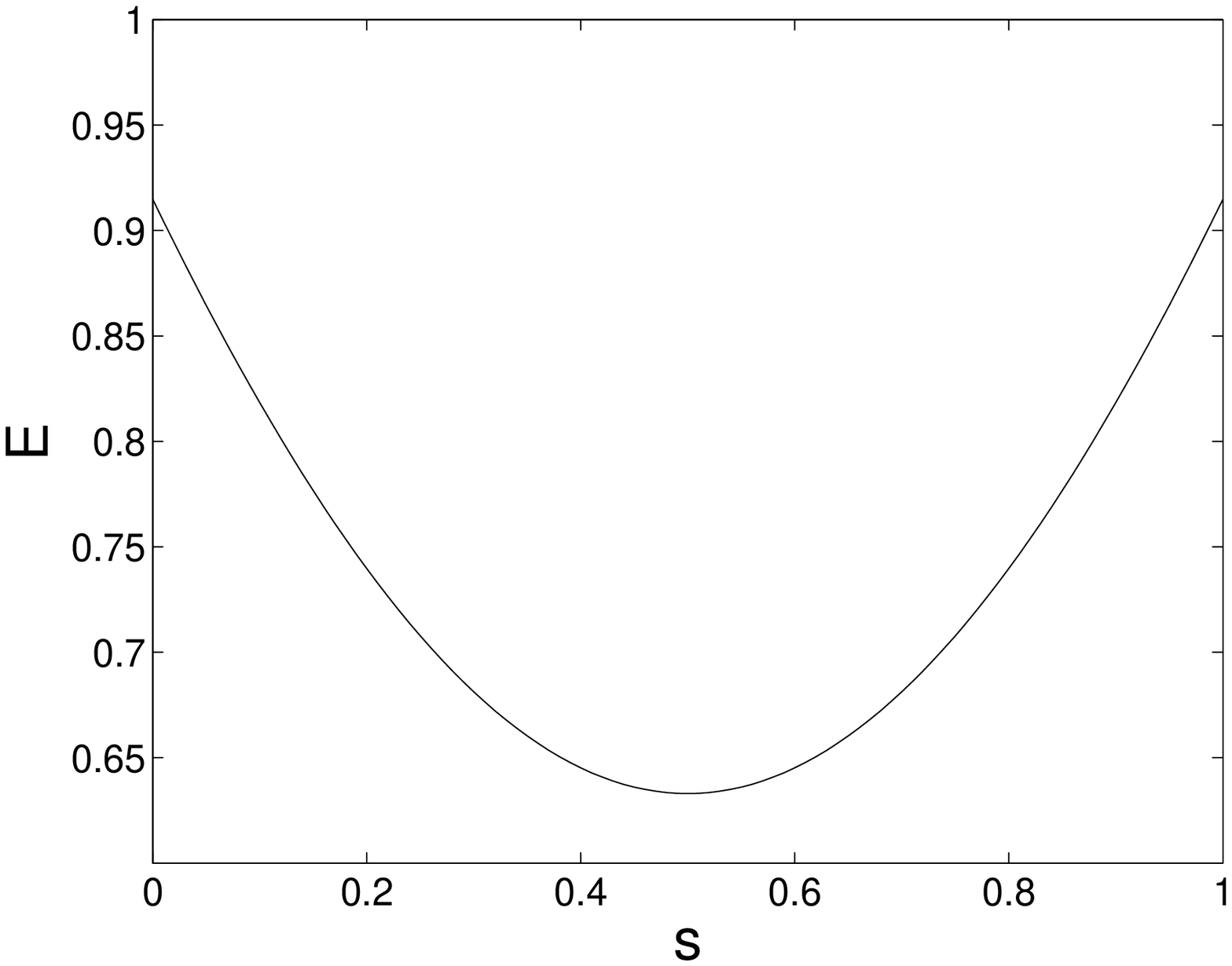}

Figure 5.3. Variation of  $E_{\mathcal{T}}(|\psi_{s,\phi}\ran),$ expressed in units of $R_N,$ with the superposition parameter $s$, for $N=3.$
\end{center}
\end{figure}

An important example of a $W$ state and its generalizations is the one-dimensional spin-$\fr{1}{2}$ Heisenberg antiferromagnet, on a lattice of size $N,$ with periodic boundary conditions, given by the Hamiltonian
$$ H_{N}= \sum^{N}_{j=1}X_{j}X_{j+1}+ Y_{j}Y_{j+1}+Z_{j}Z_{j+1}\eqno{(5.14)}$$ where the subscripts are $mod \;N $ and $X,Y,Z$ denote Pauli operators $\si_x,\si_y,\si_z$, respectively. $H_N$ commutes with $S_z=\sum Z_j$, so the eigenstate of $H_N$ is a superposition of basis vectors $|b_1\cdots b_n\ran$ where $s$ of $b_1\cdots b_N$ are ones and $N-s$ are zeros for some fixed $0 \le s \le N.$ When $s=1$, the translational invariance of $H_N$ implies that the eigenstates are $$|\psi^{(k)}_N\ran=\fr{1}{\sq{N}}\sum_{j=0}^{N-1} e^{ikj}|00\cdots 1_j0\cdots 0\ran \eqno{(5.15)}$$
where the $j$th summand has a single 1 at $j$th bit just like the $W$ state and the wave number $k=\fr{2 \pi m}{N}$ for some integer $0\le m \le N-1$. The state $|\psi^{(k)}_N\ran$ is locally unitarily transformed to the $W$ state so that it has the same value of $E_{\mathcal{T}}(|W\ran)$ or $E_{\mathcal{T}}(\widetilde{W}\ran).$

For $ s \ge 2$ the eigenstates of $H_N$ have the form $$|\psi_N(s)\ran=\fr{1}{\sq{\binom{N}{s}}}\sum_{\{j_1 \cdots j_s\}} |00\cdots 1_j0 \cdots 1_{j_s} 0\ran, \eqno{(5.16)}$$
where 1 occurs at $j_1\cdots j_s$, $\{j_1\cdots j_s\}\subseteq \mathcal{N}=\{1,2,\cdots,N\}$, and can be chosen in $\binom{N}{s}$ ways. We see that for $|\psi_N(s)\ran$, $t_{33\cdots3}=(-1)^s.$ For even $N$, $t_{1\cdots12\cdots23\cdots3}$ with $x$ $1$'s and $y$ $2$'s, corresponding to the average of $x$ $\si_x$'s, $y$ $\si_y$'s, and $[N-x-y]$ $\si_z$'s, we get, for even $x$ and even $y$,
$$t_{1\cdots12\cdots23\cdots3}=\Bigg[2\binom{x}{\fr{x}{2}}\binom{y}{\fr{y}{2}}-\binom{x+y}{\fr{(x+y)}{2}}\Bigg]\binom{N-x-y}{s-\fr{(x+y)}{2}}.$$
Since $|\psi_N(s)\ran$ is a symmetric state, any permutation of its indices does not change the value of an element of $\mathcal{T}^{(N)}$ (see chapter 4), so that
$${\Arrowvert\mathcal{T}^{(N)}_{|\psi_{N}(s)\ran}\Arrowvert}^2=$$
$$1+\fr{1}{{\binom{N}{s}}^2}\Bigg[
\sum_{\begin{subarray}{I}
  \hskip  .1cm  {x+y=2}\\
 \hskip  .1cm   {x,y\; even}
\end{subarray}} ^{2s} {\Bigg[2\binom{x}{\fr{x}{2}}\binom{y}{\fr{y}{2}}-\binom{x+y}{\fr{(x+y)}{2}}\Bigg]}^2 {\binom{N-x-y}{s-\fr{(x+y)}{2}}}^2 \binom{N}{x}
\binom{N-x}{y}\Bigg].$$
Figure 5.4 shows the variation of $E_{\mathcal{T}}(|\psi_{N}(s)\ran)=\Arrowvert{\mathcal{T}}_{|\psi_{N}(s)\ran}^{(N)}\Arrowvert
-1$ with $s$. We see that it is maximum at $s=\fr{N}{2}$, which is a characteristic of the ground state of $H_N$, as expected. Note that
$E_{\mathcal{T}}(|\psi_{N}(\fr{N}{2})\ran)$ for the ground state $(s=\fr{N}{2})$ rises far more rapidly than the entanglement of the $N$-qubit $GHZ$ state $R_{N}=E_{\mathcal{T}}(|GHZ\ran)$ [Eq.(5.11)] with the number of spins (qubits) $N.$ This can be understood by noting that $|\psi_{N}(\fr{N}{2})\ran$ for $s=\fr{N}{2}$ can be written as a superposition of $\fr{1}{2}\binom{N}{\fr{N}{2}}$ $N$-qubit $GHZ$ states. For example, $|\psi_{4}(2)\ran$ can be written as the superposition of three four-qubit $GHZ$ states,
$$|\psi_{4}(2)\ran=\fr{1}{\sq{3}}\Big[\fr{1}{\sq{2}}(\arrowvert 0011\ran+\arrowvert1100\ran)+ \fr{1}{\sq{2}}
(\arrowvert 0101\ran+\arrowvert1010\ran)+\fr{1}{\sq{2}}(\arrowvert 1001\ran+\arrowvert0110\ran)\Big].$$
As $N$ increases, initially $E_{\mathcal{T}}(|\psi_{N}(\fr{N}{2})\ran)$ is comparable to $R_{N},$ but after $N=16$ the ratio
 $\fr{E_{\mathcal{T}}(|\psi_{N}(\fr{N}{2})\ran)}{R_{N}}$ increases very rapidly, reaching $10^7$ for $100$ qubits. Also, as $N$ increases, $E_{\mathcal{T}}(|\psi_{N}(s)\ran)$ falls off more rapidly as $s$ deviates from $\fr{N}{2}$. We are presently trying to understand this behavior.

\begin{figure}[!ht]
\begin{center}
\includegraphics[width=10cm,height=8cm]{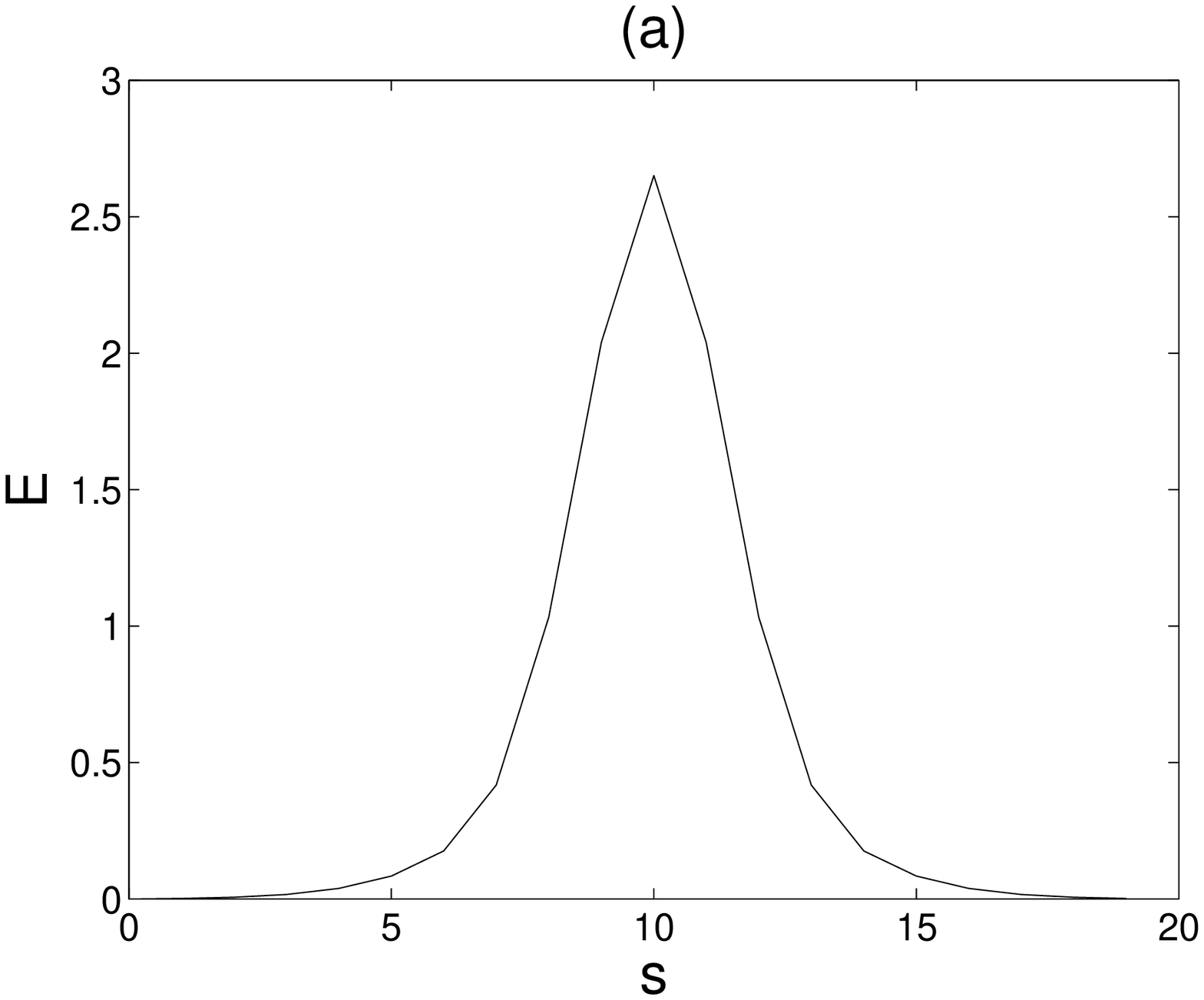}

\includegraphics[width=10cm,height=8cm]{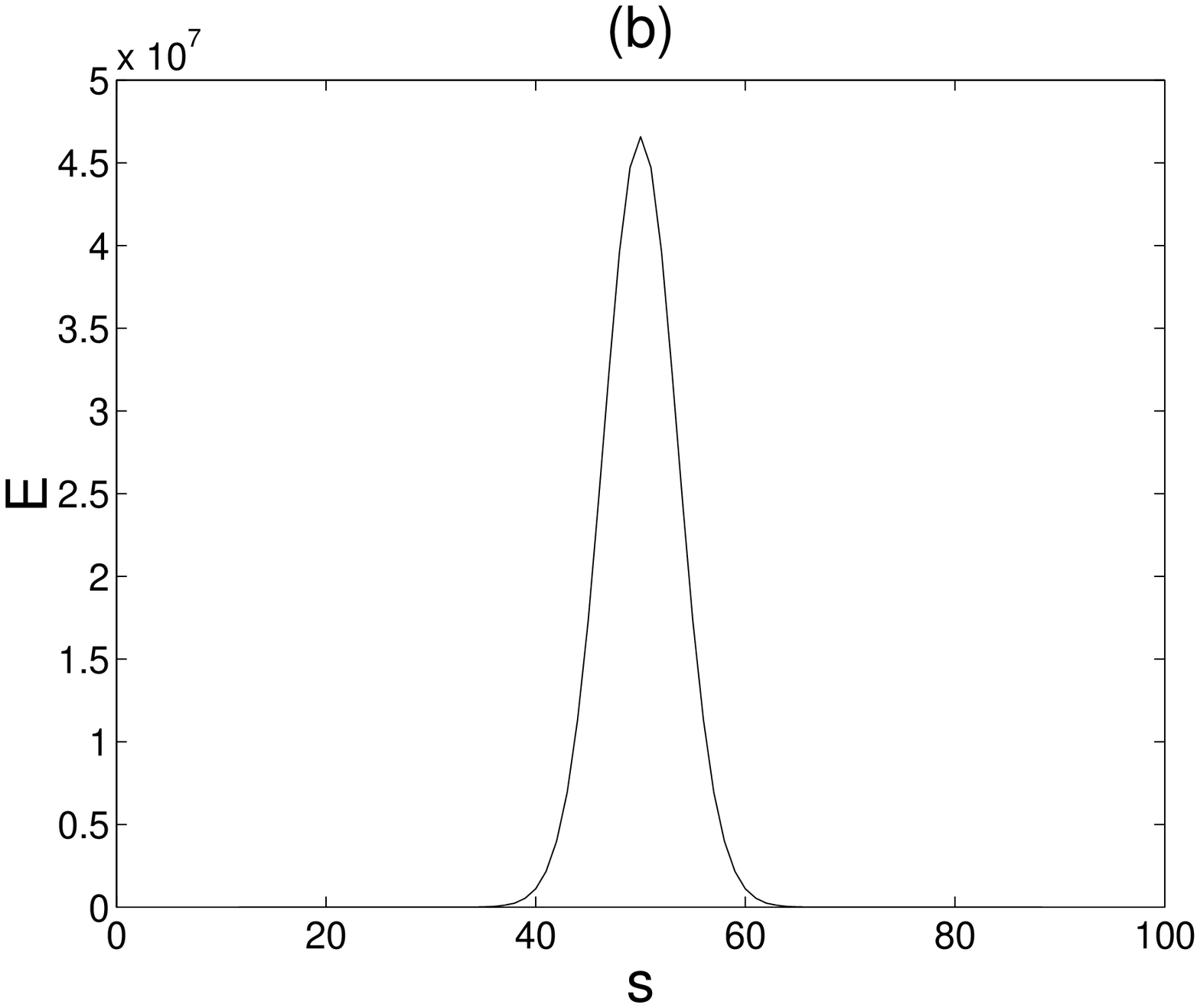}\\

Figure 5.4. Variation of $E_{\mathcal{T}}(|\psi_{N}(s)\ran),$ (in units of $R_N$), with $s$, for  $N= (a) 20$ and  $ (b)100$ (see text).

\end{center}
\end{figure}

Finally, in this section, we consider the superpositions of $W$ and $GHZ$ states,
$$|\psi_{W+GHZ}(s,\phi)\ran=\sq{s}|GHZ\ran+\sq{1-s}\; e^{i\phi}|W\ran, \eqno{(5.17)}$$ also considered in \cite{wg03}. For three qubits, $N=3$, a direct calculation gives, for this state, $${\Arrowvert\mathcal{T}^{(N)}\Arrowvert}^2\;=\;4s^2+6s(1-s)+\fr{11}{3}(s-1)^2 \;\;(0\leq s\leq 1), $$
$$E_{\mathcal{T}}(|\psi_{W+GHZ}(s,\phi)\ran)=\Arrowvert{\mathcal{T}}_{|\psi_{N}(s)\ran}^{N}\Arrowvert
-1, \eqno{(5.18)}$$
which coincides with the corresponding values of $W\;(s=0)$ and $GHZ\;(s=1)$ states. Note that $E_{\mathcal{T}}(|\psi_{W+GHZ}(s,\phi)\ran)$
is independent of the phase $\phi,$ in contrast to the entanglement measure used in \cite{wg03}. Figure 5.5 shows the dependence of
$E_{\mathcal{T}}(|\psi_{W+GHZ}(s,\phi)\ran)$ on $s$.\\

\begin{figure}[!ht]
\begin{center}
\includegraphics[width=15cm,height=12cm]{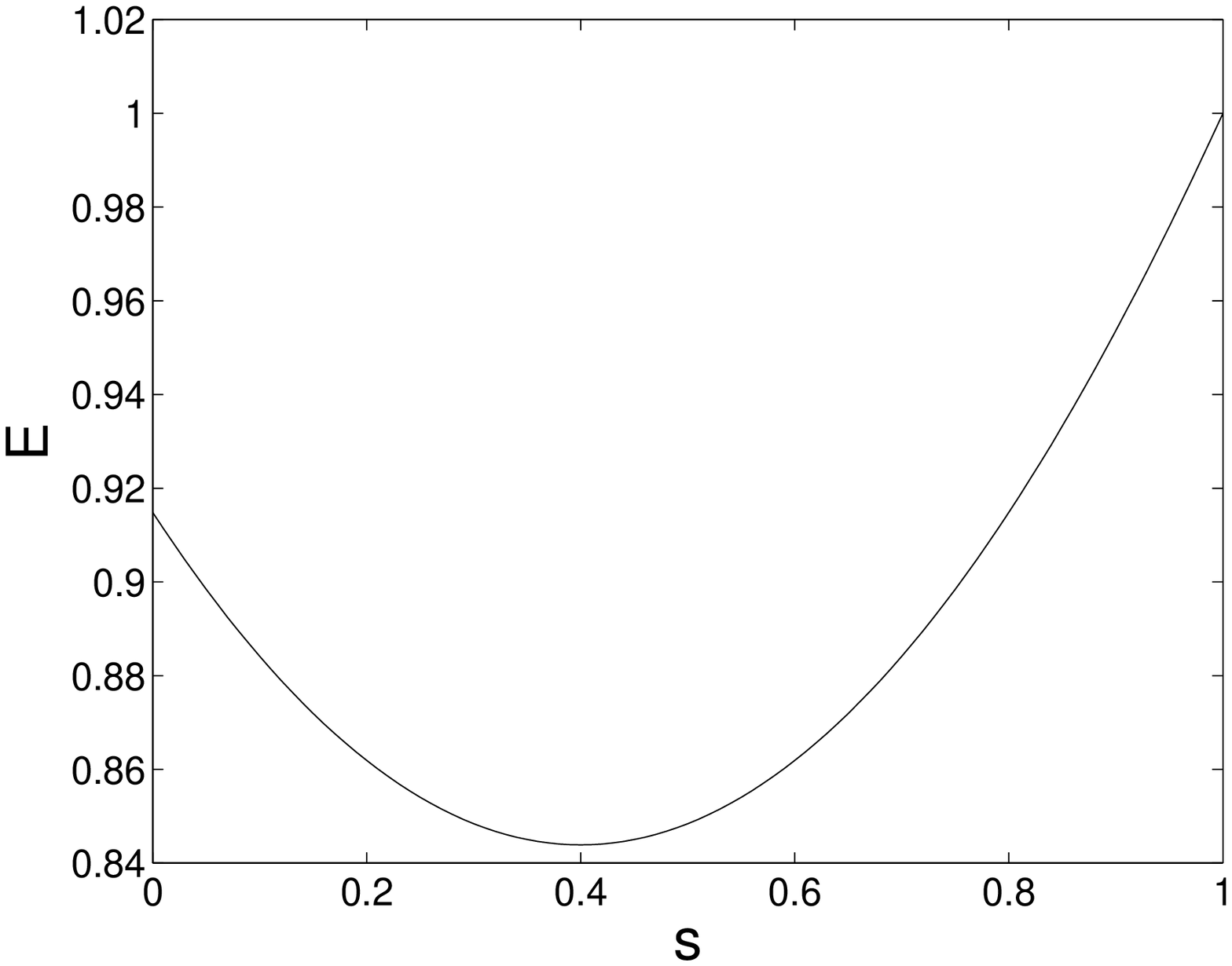}

Figure 5.5. Variation of $E_{\mathcal{T}}(|\psi_{W+GHZ}(s,\phi)\ran)$, expressed in units of $R_N$, with the superposition parameter $s$, for $N=3$.
\end{center}
\end{figure}

\section{Properties of $E_{\mathcal{T}}(|\psi\ran)$}

To be a valid entanglement measure, $E_{\mathcal{T}}(|\psi\ran)$ must have the following properties \cite{vida00,bzb06}.

(a) ~~~(i) {\it Positivity} : $E_{\mathcal{T}}(|\psi\ran)\ge 0$ for all $N$-qubit pure state $|\psi\ran$.

~~~~~~(ii) {\it Discriminance}: $E_{\mathcal{T}}(|\psi\ran)=0$ if and only if $|\psi\ran$ is separable (product) state.

(b) {\it $LU$ invariance} : $E_{\mathcal{T}}(|\psi\ran)$ must be  invariant under local unitary (LU) operations.

(c) {\it Monotonicity} : local operators and classical communication do not increase the expectation value of $E_{\mathcal{T}}(|\psi\ran)$.

 We prove the above properties  for $E_{\mathcal{T}}(|\psi\ran)$. We also prove the following additional properties for $E_{\mathcal{T}}(|\psi\ran)$:

(d) {\it Continuity} : $||(|\psi\ran\lan\psi|-|\phi\ran\lan\phi|)|| \rightarrow 0 \Rightarrow \Big{|}E(|\psi\ran)-E(|\phi\ran)\Big{|}\rightarrow 0 $.

(e) {\it Superadditivity} : $E_{\mathcal{T}}(|\psi\ran \otimes |\phi\ran )\ge E_{\mathcal{T}}(|\psi\ran)+ E_{\mathcal{T}}(|\phi\ran).$ \\
We need the following result, which we have proved in chapter 4.\\
\noi\textbf{ Proposition 5.4.0 }: A  pure $N$-partite quantum state is fully separable (product state) if and only if   $$\mathcal{T}^{(N)}=\mathbf{s}^{(1)}\circ \mathbf{s}^{(2)} \circ \dots \circ \mathbf{s}^{(N)},\eqno{(5.19)}$$
     where $ \mathbf{s}^{(k)}$ is the Bloch vector of $k$th subsystem reduced density matrix.  The symbol $\circ$ stands for  the outer product of vectors defined as follows.

Let $\mathbf{u}^{(1)},\mathbf{u}^{(2)},\dots,\mathbf{u}^{(M)}$ be vectors in $\mathbb{R}^{d_1^2-1},\mathbb{R}^{d_2^2-1},\cdots,\mathbb{R}^{d_M^2-1}.$
The outer product $\mathbf{u}^{(1)}\circ \mathbf{u}^{(2)} \circ \dots \circ \mathbf{u}^{(M)}$ is a tensor of order $M$, (M-way array), defined by

$ t_{i_1 i_2\cdots i_M}=\mathbf{u}^{(1)}_{i_1} \mathbf{u}^{(2)}_{i_2} \dots  \mathbf{u}^{(M)}_{i_M},\; 1\le i_k \le d_k^2-1,\; k=1,2,\cdots,M.$  \\
\noi\textbf{ Proposition 5.4.1 }: Let $|\psi\ran$ be an $N$-qubit pure state. Then, $||\mathcal{T}^{(N)}_{\psi}||=1$ if and only if $|\psi\ran$ is a separable (product) state.\\
\noi\textbf{Proof :} By proposition 5.4.0, $|\psi\ran$ is separable (product) if and only if $$\mathcal{T}^{(N)}=\mathbf{s}^{(1)}\circ \mathbf{s}^{(2)} \circ \dots \circ \mathbf{s}^{(N)},$$

As shown in \cite{kold06,kold01}, see also section 1.5, $$( \bigcirc_{k=1}^N s^{(k)},\bigcirc_{k=1}^N s^{(k)})=\Pi_{k=1}^N ( s^{(k)},s^{(k)}),\eqno{(5.20)}$$

where $(,)$ denotes the scaler product. This immediately gives, for qubits, $$||\mathcal{T}^{(N)}||^2=( \mathcal{T}^{(N)},\mathcal{T}^{(N)}) =\Pi_{k=1}^N ( s^{(k)},s^{(k)})= \Pi_k ||s^{(k)}||^2=1.$$
Proposition 5.4.1 immediately gives the following proposition.\\
\noi\textbf{ Proposition 5.4.2 } Let $|\psi\ran$ an $N$-qubit pure state. Then $E_{\mathcal{T}}(|\psi\ran)= 0$ if and only if $|\psi\ran$ is a product state.\\
\noi\textbf{Proposition 5.4.3 }: Let $|\psi\ran$ be an $N$-qubit pure state. Then $ ||\mathcal{T}^{(N)}||\ge 1.$
It is instructive to show this result by direct computation of $\mathcal{T}^{(N)}$ for the cases of two- and three-qubit states. First, consider a general two-qubit state $$|\psi\ran=a_1|00\ran+a_2|01\ran+a_3|10\ran+a_4|11\ran,\; \sum_k|a_k|^2=1,$$
by direct computation we get $$||\mathcal{T}^{(2)}||^2=1+8(|a_2a_3|-|a_1a_4|)^2 \ge 1. \eqno{(5.21)}$$
This means, via proposition 5.4.2, that $|\psi\ran$ is a product state if $|a_2a_3|=|a_1a_4|.$
Next, consider a three-qubit state in the general Schmidt form \cite{aacjlt} $$|\psi\ran=\la_0|000\ran+\la_1e^{i\phi}|100\ran+\la_2|101\ran+\la_3|110\ran+\la_4|111\ran \eqno{(5.22)}$$
where $\la_i \ge 0, \; i=0,1,2,3,4$ and $\sum_i \la_i^2=1.$
By direct calculation of $||\mathcal{T}^{(3)}||$ we get
$$||\mathcal{T}^{(3)}||^2 \ge1+12\la_0^2\la_4^2+8\la_0^2\la_2^2+8\la_0^2\la_3^2+8(\la_1\la_4-\la_2\la_3)^2 \ge 1. \eqno{(5.23)}$$
Here the conditions for product state become $\la_1\la_4=\la_2\la_3 $ and $ \la_0=0$.
We now prove proposition 5.4.3 for a general $N$-qubit state $|\psi\ran.$

If $|\psi\ran$ is not a product of $N$ single-qubit states (i.e., $|\psi\ran$ is not $N$-separable) then it is $(N-k)$-separable, $k=2,3,\cdots,N$. Viewing the $N$-qubit system as a system comprising $N-k$ qubits, each with Hilbert space of of dimension 2, and $k$ entangled qubits with the Hilbert space of dimension $2^k$, we can apply proposition 5.4.0 to this separable system of $N-k+1$ parts in the state $|\psi\ran.$ We get $ \mathcal{T}^{(N)}_{|\psi\ran}= \mathbf(s^{(1)})\circ\mathbf(s^{(2)})\circ \cdots\mathbf(s^{(N-k)})\circ\mathbf(s^{(N-k+1)}).$

This implies, as in proposition 5.4.1, via Eq. (5.20) and Eq. (5.6) that $$||\mathcal{T}^{(N)}_{|\psi\ran}||^2=\Pi_{i-1}^{N-k+1} ||\mathbf(s^{(i)})||^2=\fr{d(d-1)}{2} > 1 \; \;\;(d=2^k).\eqno{(5.24)}$$
If $k=N$ we attach an ancilla qubit in an arbitrary state $|\phi\ran$ and apply proposition 5.4.0 to $(N+1)$-qubit system in the state $|\psi\ran\otimes|\phi\ran$ where $|\psi\ran$ is the $N$-qubit entangled state. This result, combined with proposition 5.4.1,  completes the proof.
Proposition 5.4.3 immediately gives the following proposition. \\
\noi\textbf{Proposition 5.4.4} : $E_{\mathcal{T}}(|\psi\ran)\ge 0.$

We now prove that $E_{\mathcal{T}}(|\psi\ran)$ is nonincreasing under local operations and classical communication. Any such local action can be decomposed into four basic kinds of operations \cite{bdsw} (i) appending an ancillary system not entangled with the state of original system, (ii) performing a unitary transformation, (iii) performing measurements, and (iv) throwing away, i.e. tracing out, part of the system. It is clear that appending ancilla cannot change
$\Arrowvert\mathcal{T}^{(N)}\Arrowvert.$ We prove that  $E_{\mathcal{T}}(|\psi\ran)$ does not increase under the remaining three local operations.\\
\noi\textbf{Proposition 5.4.5} : Let $U_i \; (i=1,2,\cdots,N)$ be local unitary operator acting on the Hilbert space of subsystems  $\mathcal{H}^{(i)}\;\; (i=1,2,\cdots,N),$ respectively.
Let $$\rho=(\otimes_{i=1}^N U_i)\rho'(\otimes_{i=1}^N U_i^{\dag})  \eqno{(5.25)}$$
for  density operators $\rho$ and $\rho'$ acting on $\mathcal{H}=\otimes_{i=1}^{N}\mathcal{H}^{(i)}$ and let $\mathcal{T}^{(N)}$ and $\mathcal{T'}^{(N)}$ denote the $N$-partite correlation tensors for $\rho$ and $\rho'$, respectively. Then

 $||\mathcal{T'}^{(N)}||=||\mathcal{T}^{(N)}||,$ so that $E_{\mathcal{T}}(\rho)=E_{\mathcal{T}}(\rho').$\\
\noi\textbf{Proof :} Let $U$ denote a one-qubit unitary operator; then it is straightforward to show that $U \si_{\alpha}U^{\dag}=\sum_{\beta} O_{\alpha \beta}\si_{\beta}$,
 where $[O_{\alpha \beta}]$ is a real matrix satisfying $O O^T =I= O^T O$.
 It is an element of the rotation group $O(3)$. Now consider $$t'_{i_1i_2\cdots i_N}=Tr(\rho'\si_{i_1}\otimes\si_{i_2}\otimes\cdots \otimes\si_{i_N})$$
 $$=Tr \big[\rho(\otimes_{i=1}^N U_i) \si_{i_1}\otimes\si_{i_2}\otimes\cdots \otimes\si_{i_N}(\otimes_{i=1}^N U_i^{\dag})\big]$$
 $$=Tr(\rho U_1\si_{i_1}U_1^{\dag}\otimes U_2\si_{i_2}U_2^{\dag}\otimes\cdots \otimes U_N\si_{i_N}U_N^{\dag})$$
 $$=\sum_{\alpha_1\cdots\alpha_N}Tr(\rho\si_{\alpha_1}\otimes\si_{\alpha_2}\otimes\cdots \otimes\si_{\alpha_N})O^{(1)}_{i_1\alpha_1}O^{(2)}_{i_2\alpha_2}\cdots O^{(N)}_{i_N\alpha_N}$$
 $$=\sum_{\alpha_1\cdots\alpha_N}t_{\alpha_1\cdots\alpha_N}O^{(1)}_{i_1\alpha_1}O^{(2)}_{i_2\alpha_2}\cdots O^{(N)}_{i_N\alpha_N}$$
 $$=(\mathcal{T}^{(N)}\times_1 O^{(1)}\times_2 O^{(2)}\cdots \times_N O^{(N)})_{i_1i_2\cdots i_N},$$
where $\times_k $ is the $k$-mode product of a tensor $\mathcal{T}^{(N)}\in \mathbb{R}^{3\times 3 \times \cdots 3}$ by the orthogonal matrix $O^{(k)}\in \mathbb{R}^{3\times 3}$ \cite{kold06,kold01,lmv00}, see section 1.5. Therefore,

$$\mathcal{T'}^{(N)}=\mathcal{T}^{(N)}\times_1 O^{(1)}\times_2 O^{(2)}\cdots \times_N O^{(N)}$$
By proposition 1.5.3 in chapter 1, we get $$||\mathcal{T'}^{(N)}||= ||\mathcal{T}^{(N)}\times_1 O^{(1)}\times_2 O^{(2)}\cdots \times_N O^{(N)}||=||\mathcal{T}^{(N)} ||.$$ \hfill $\blacksquare$\\
\noi\textbf{Proposition 5.4.6} : If a multipartite pure state $|\psi\ran$ is subjected to a local measurement on the $k$th qubit giving outcomes $i_k$ with probabilities $p_{i_k}$ and leaving residual $N$-qubit pure state $|\phi_{i_k}\ran$, then the expected entanglement $\sum_{i_k} p_{i_k} E_{\mathcal{T}}(|\phi_{i_k}\ran)$ of residual state is not greater then $ E_{\mathcal{T}}(|\psi\ran)$,

$$\sum_{i_k} p_{i_k} E_{\mathcal{T}}(|\phi_{i_k}\ran) \le E_{\mathcal{T}}(|\psi\ran). \eqno{(5.26)}$$\\
\noi\textbf{Proof :} Local measurements can be expressed as the tensor product matrix $\bar{D}=\bar{D}^{(1)}\otimes\bar{D}^{(2)}\otimes\cdots \otimes \bar{D}^{(N)}$  on the expanded coherence vector $\mathcal{T}$ \cite{zlwwt}. The expanded coherence vector $\mathcal{T}$ is the extended correlation tensor $\mathcal{T}$ (defined below) viewed as a vector in the real space of appropriate dimension. The extended correlation tensor $\mathcal{T}$ is defined by the equation
$$\rho= \fr{1}{2^N}\sum_{i_1i_2\cdots i_N=0}^3 \mathcal{T}_{i_1i_2\cdots i_N} \si_{i_1}\otimes \si_{i_2}\otimes\cdots\otimes \si_{i_N},  \eqno{(5.27)}$$
where $\si_{i_k} \in \{I,\si_x,\si_y,\si_z\}$ is the $i_k$th local Pauli operator on the $k$th qubit $(\si_0 =I)$ and the real coefficients $\mathcal{T}_{i_1i_2\cdots i_N} $ are the components of the extended correlation tensor $\mathcal{T}.$  Equation (5.2) and Eq.(5.27) are equivalent with $\mathcal{T}_{000\cdots 0}=1 $ , $\mathcal{T}_{i_100\cdots 0}= s^{(1)}_{i_1},\; \cdots $, $\mathcal{T}_{i_1i_2\cdots i_M 00\cdots 0} = \mathcal{T}^{\{1,2,\cdots M\}}_{i_1i_2\cdots i_M}, \cdots $
and  $\mathcal{T}_{i_1i_2\cdots i_N} =\mathcal{T}^{(N)}_{i_1i_2\cdots i_N} ,\; i_1, i_2,\cdots ,i_N \ne 0 .$ $\bar{D}^{(k)};\; k=1,2,\cdots n$ are $4\times 4$ matrices. Without losing generality, we can assume the local measurements to be positive operator valued measures (POVMs), in which case $\bar{D}^{(k)}= diag(1,D^{(k)})$ and the $3  \times 3$ matrix $D^{(k)}$ is contractive, $D^{(k)T} D^{(k)} \le I$ \cite{zlwwt}. The local POVMs acting on an $N$-qubit state $\rho$ correspond to the map
$\rho\longmapsto\mathcal{M}(\rho)$ given by
$$\mathcal{M}(\rho)=\sum_{i_1i_2\cdots i_N}L^{(1)}_{i_1}\otimes L^{(2)}_{i_2}\otimes \cdots \otimes L^{(N)}_{i_N}\rho L^{(1)\dag}_{i_1}\otimes L^{(2)\dag}_{i_2}\otimes \cdots \otimes L^{(N)\dag}_{i_N},$$
where $L^{(k)}_{i_k}$ are the linear, positive, trace-preserving operators satisfying $\sum_{i_k}L^{(k)\dag}_{i_k}L^{(k)}_{i_k}=I$ and $[L^{(k)}_{i_k},L^{(k)\dag}_{i_k}]=0.$ The resulting correlation tensor of $\mathcal{M}(\rho)$ can be written as

$$\mathcal{T'}^{(N)}=\mathcal{T}^{(N)}\times_1 D^{(1)}\times_2 D^{(2)}\times \cdots \times_N D^{(N)}$$
where $D^{(k)}$ is $3\times3$ matrix and $D^{(k)T} D^{(k)} \le I.$

The action of POVM on $k$th qubit corresponds to the map $\mathcal{M}_{k}(|\psi\ran\lan\psi|)=\sum_{i_k} M_{i_k}\rho M_{i_k}^{\dag}$, where $M_{i_k}=I \otimes \cdots L^{(k)}_{i_k}\otimes \cdots I$, $\sum_{i_k}L^{(k)\dag}_{i_k}L^{(k)}_{i_k}=I$ and $[L^{(k)}_{i_k},L^{(k)\dag}_{i_k}]=0$,
with the resulting mixed state $\sum_{i_k}p_{i_k}|\phi_{i_k}\ran\lan\phi_{i_k}|,$ where $|\phi_{i_k}\ran$ is the $N$-qubit pure state which results after the the outcome $i_k$ with probability $p_{i_k}.$
The average entanglement of this state is $$\sum_{i_k} p_{i_k} E_{\mathcal{T}}(|\phi_{i_k}\ran\lan\phi_{i_k}|)=\sum_{i_k}p_{i_k} ||\mathcal{T}^{(N)}_{|\phi_{i_k}\ran}||-1$$
$$=\sum_{i_k} p_{i_k}||\mathcal{T}^{(N)}_{|\psi\ran}\times_k D^{(k)}||-1$$
$$=\sum_{i_k}p_{i_k}||D^{(k)}T_{(k)}(|\psi\ran)||-1$$
where, by proposition 1.5.2 in chapter 1, $ D^{(k)}T_{(k)}(|\psi\ran)$ is the $k$th matrix unfolding (see chapters 1 and 4) of $\mathcal{T}^{(N)}_{|\psi\ran}\times_k D^{(k)}.$
Therefore, from the definition of the Euclidean norm of a matrix, $||A||=\sq{Tr(A A^{\dag})}$,  \cite{hjb90} we get
$$\sum_{i_k} p_{i_k}  E_{\mathcal{T}}(|\phi_{i_k}\ran\lan\phi_{i_k}|)=\sum_{i_k} p_{i_k}\big{\{}Tr\big{[}D^{(k)}T_{(k)}(|\psi\ran)T_{(k)}^{\dag}(|\psi\ran)D^{(k)T}\big{]}\big{\}}^{\fr{1}{2}}-1$$

$$=\sum_{i_k} p_{i_k} \big{\{}Tr \big{[} D^{(k)T}D^{(k)}T_{(k)}(|\psi\ran)T_{(k)}^{\dag}(|\psi\ran)\big{]}\big{\}}^{\fr{1}{2}}-1$$

 $$\le \sum_{i_k}p_{i_k}\sq{Tr\big{[}T_{(k)}(|\psi\ran)T_{(k)}^{\dag}(|\psi\ran)\big{]}}-1$$
 $$=||\mathcal{T}^{(N)}_{|\psi\ran}||-1 = E_{\mathcal{T}}(|\psi\ran),$$
 because $D^{(k)T} D^{(k)} \le I$, and $\sum_{i_k}p_{i_k} = 1.$ We have also used the fact that Euclidean norm of a tensor equals that of any of its matrix unfoldings. \hfill $\blacksquare$

 As an example, we consider the four-qubit state \cite{byw07}
 $$|\psi\ran_{ABCD}=\fr{1}{\sq{6}}(|0000\ran +|0011\ran +|0101\ran +|0110\ran +|1010\ran +|1111\ran). \eqno{(5.28)}$$
 A POVM $\{A_1,A_2\}$ is performed on the subsystem $A$, which has the form $A_1=U_1 diag\{\alpha, \beta \}V$
  and $A_2= U_2 diag\{\sq{1-\alpha^2}, \sq{1-\beta^2} \}V.$  Due to LU invariance of    $\Arrowvert\mathcal{T}^{(N)}\Arrowvert$ we need only consider the diagonal matrices in which the parameters are chosen to be $\alpha=0.9$ and $\beta=0.2.$ After the POVM, two outcomes $|\phi_1\ran=A_1|\psi\ran/\sq{p_1}$ and $|\phi_2\ran=A_2|\psi\ran/\sq{p_2}$ are obtained, with the probabilities as $p_1=0.5533$ and $p_2=0.4467.$ We find
 $$E_{\mathcal{T}}(|\psi\ran)=0.7802,\;E_{\mathcal{T}}(|\phi_1\ran)=0.0725/p_1,\;E_{\mathcal{T}}(|\phi_2\ran)=0.0436/p_2.$$ This gives $$E_{\mathcal{T}}(|\psi\ran)-[p_1 E_{\mathcal{T}}(|\phi_1\ran)+p_2 E_{\mathcal{T}}(|\phi_2\ran)]=0.6641\;>\;0.$$ This is to be contrasted with the similar calculation in \cite{byw07}, with the same state $|\psi\ran$ in Eq.(5.28) and the same POVM given above. \\
\noi\textbf{Proposition 5.4.7 :} Let $|\psi\ran$ be an $N$-qubit pure state. Let $\rho$ denote the reduced density matrix after tracing out one qubit from the state $|\psi\ran$. Then $$|| \mathcal{T}^{(N-1)}_{\rho}|| \le ||\mathcal{T}^{(N)}_{|\psi\ran}||$$
with equality only when $|\psi\ran=|\phi\ran\otimes|\chi\ran$, where $|\chi\ran$ is the state of the qubit which is traced out.\\
\noi\textbf{Proof :} we prove this for a special case whose generalization is straightforward. Let $|\psi\ran=a|b_1\cdots b_N\ran+b|b'_1\cdots b'_N\ran,\;\; |a|^2+|b|^2=1.$
Here $|b_i\ran$ and $|b'_i\ran$ are the eigenstates of $\si_z^{(i)}$ operating on the $i$th qubit. Now consider sets of  $N$-fold tensor products of qubit operators $\{\si_{\alpha}\},\; \alpha=1,2,3$, namely,  $S=\{\si_{\alpha_1}\otimes\si_{\alpha_2}\otimes \cdots \otimes \si_{\alpha_N}\},\; \alpha_1\cdots \alpha_N =1,2,3.$
Choosing $\alpha_1,\cdots, \alpha_N =3$ we get $\si_{3}\otimes\si_{3}\otimes \cdots\otimes \si_{3}|b_1\cdots b_N\ran=\pm |b_1\cdots b_N\ran.$
We can choose an operator from $S$, denoted $B$, such that $B|b_1\cdots b_N\ran = \pm |b'_1\cdots b'_N\ran.$
If $B$ contains $q \le N \; \si_x$ operators we can replace $k \le q$ of them by $\si_y$ operators. We denote the resulting tensor product operator by $B_k \; (B_0=B)$. We have, $B_k|b_1\cdots b_N\ran = \pm (i)^k |b'_1\cdots b'_N\ran.$
Then, $$\lan b_1\cdots b_N\arrowvert\si_3\otimes\cdots\otimes\si_3\arrowvert b_1\cdots b_N\ran = \pm 1 = \lan b^{\prime}_1\cdots b^{\prime}_N\arrowvert\si_3\otimes\cdots\otimes\si_3\arrowvert b^{\prime}_1\cdots b^{\prime}_N\ran$$
$$\lan b_1\cdots b_N\arrowvert B \arrowvert b^{\prime}_1\cdots b^{\prime}_N\ran = \pm 1 = \lan b^{\prime}_1\cdots b^{\prime}_N\arrowvert B \arrowvert b_1\cdots b_N\ran$$
$$\lan b^{\prime}_1\cdots b^{\prime}_N\arrowvert B_k \arrowvert b_1\cdots b_N\ran = \pm (i)^{k}$$
$$\lan b_1\cdots b_N\arrowvert B_k \arrowvert b^{\prime}_1\cdots b^{\prime}_N\ran = \pm (-i)^{k}$$
Now,$$ t_{\alpha_1\cdots\alpha_N} = \lan \psi\arrowvert\si_{\alpha_1}\otimes\cdots\otimes\si_{\alpha_N}\arrowvert\psi\ran$$
$$ = \arrowvert a\arrowvert^2 \lan b_1\cdots b_N\arrowvert\si_{\alpha_1}\otimes\cdots\otimes\si_{\alpha_N}\arrowvert b_1\cdots b_N\ran+|b|^2\lan b^{\prime}_1 \cdots b^{\prime}_N|\si_{\alpha_1}\otimes\cdots\otimes\si_{\alpha_N}|b^{\prime}_1 \cdots b^{\prime}_N\ran$$
$$ +a^* b \lan b_1\cdots b_N\arrowvert\si_{\alpha_1}\otimes\cdots\otimes\si_{\alpha_N}|b^{\prime}_1 \cdots b^{\prime}_N\ran+a b^* \lan b^{\prime}_1 \cdots b^{\prime}_N|\si_{\alpha_1}\otimes\cdots\otimes\si_{\alpha_N}\arrowvert b_1\cdots b_N\ran$$
The nonzero elements of $t_{\alpha_1 \cdots \alpha_N} $ are $t_{33\cdots 3}=\pm |a|^2 \pm |b|^2$,
$ t_B=\pm a b^* \pm a^* b= \pm 2 |a| |b| cos (\phi_a-\phi_b)$,

 \begin{displaymath}
t_{B_k}=\pm (i)^k a b^* \pm (-i)^k a^* b
 =\left\{ \begin{array}{ll}
 \pm 2 |a|\; |b| cos (\phi_a-\phi_b) & \textrm{ if $k$ is even,} \\

\pm 2 |a|\; |b| sin (\phi_a-\phi_b) & \textrm{ if $k$ is odd.} \\
\end{array} \right.
\end{displaymath}
We get $\sum _{k=0}^q \binom{q}{2k}$ elements with $cos (\phi_a-\phi_b)$ and $\sum _{k=0}^q \binom{q}{2k+1}$ elements with  $sin(\phi_a-\phi_b)$. If $q$ is odd (for the given state$|\psi\ran$) the number of cosines and the number of sines are equal. When $q$ is even the number of cosines exceeds by 1. Finally  we get
$$|| \mathcal{T}^{(N)}_{|\psi\ran}||^2 = (\pm |a|^2 \pm |b|^2)^2 + 4 |a|^2 |b|^2 cos^2 (\phi_a-\phi_b) \sum_{k=0}^q \binom{q}{2k}+$$
$$4 |a|^2 |b|^2 sin^2 (\phi_a-\phi_b) \sum_{k=0}^q \binom{q}{2k+1}$$
Note that, using  $|a|^2+|b|^2 =1$, it is easy to see that $||\mathcal{T}^{(N)}_{|\psi\ran}|| \ge 1$, showing that $E_{\mathcal{T}} \ge 0.$
 Next we consider $$|\psi\ran \lan \psi|= |a|^2 \arrowvert b_1\cdots b_N  \ran  \lan b_1\cdots b_N \arrowvert + |b|^2 |b^{\prime}_1 \cdots b^{\prime}_N \ran \lan b^{\prime}_1 \cdots b^{\prime}_N | +$$
 $$ a b^* \arrowvert b_1\cdots b_N  \ran \lan b^{\prime}_1 \cdots b^{\prime}_N |
 + a^* b |b^{\prime}_1 \cdots b^{\prime}_N \ran  \lan b_1\cdots b_N \arrowvert $$
 and trace out the $N$th qubit to get the $(N-1)$-qubit reduced density matrix
 $$\rho = |a|^2 \arrowvert b_1\cdots b_{N-1}  \ran  \lan b_1\cdots b_{N-1} \arrowvert + |b|^2 |b^{\prime}_1 \cdots b^{\prime}_{N-1} \ran \lan b^{\prime}_1 \cdots b^{\prime}_{N-1} | $$
 $$+  a b^* \arrowvert b_1\cdots b_{N-1}  \ran \lan b^{\prime}_1 \cdots b^{\prime}_{N-1} | \lan b_N|b^{\prime}_N\ran + a^* b |b^{\prime}_1 \cdots b^{\prime}_{N-1} \ran  \lan b_1\cdots b_{N-1} \arrowvert \lan b^{\prime}_N|b_N\ran $$
 Now
 $$t_{\alpha_1\cdots \alpha_{N-1}}= Tr(\rho \si_{\alpha_1}\otimes \si_{\alpha_2}\otimes \cdots \si_{\alpha_{N-1}})$$
 $$= \arrowvert a\arrowvert^2 \lan b_1\cdots b_{N-1}\arrowvert\si_{\alpha_1}\otimes\cdots\otimes\si_{\alpha_{N-1}}\arrowvert b_1\cdots b_{N-1}\ran$$
$$ +|b|^2\lan b^{\prime}_1 \cdots b^{\prime}_{N-1}|\si_{\alpha_1}\otimes\cdots\otimes\si_{\alpha_{N-1}}|b^{\prime}_1 \cdots b^{\prime}_{N-1}\ran   $$
$$+a^* b \lan b_1\cdots b_{N-1}\arrowvert\si_{\alpha_1}\otimes\cdots\otimes\si_{\alpha_{N-1}}|b^{\prime}_1 \cdots b^{\prime}_{N-1}\ran  \lan b_N|b^{\prime}_N\ran  $$
 $$+a b^* \lan b^{\prime}_1 \cdots b^{\prime}_{N-1}|\si_{\alpha_1}\otimes\cdots\otimes\si_{\alpha_{N-1}}\arrowvert b_1\cdots b_{N-1}\ran  \lan b^{\prime}_N|b_N\ran.$$
 We have for $N-1$ tensor product operators $ \si_3\otimes \si_3\otimes \cdots \otimes \si_3 |b_1 \cdots b_{N-1}\ran=\pm |b_1\cdots b_{N-1}\ran.$
 We construct the operators $D$ and $D_k$ corresponding to $B$ and $B_k$ acting on $N-1$ qubits. We then get $D |b_1 \cdots b_{N-1}\ran= \pm |b^{\prime}_1 \cdots b^{\prime}_{N-1}\ran $ and
  $ D_k |b_1 \cdots b_{N-1}\ran= \pm (i)^k |b^{\prime}_1 \cdots b^{\prime}_{N-1}\ran.$
 Now, the nonzero elements of $\mathcal{T}^{(N-1)}_{\rho}$ are
 $t_{33\cdots 3}=\pm |a|^2 \pm |b|^2$,\\
 $ t_D= \pm a b^* \lan b_N|b^{\prime}_N\ran \pm a^* b \lan b^{\prime}_N | b_N\ran = 2 |a| |b| |\lan b^{\prime}_N | b_N\ran| cos (\phi_a-\phi_b-\alpha)$, \\
 $t_{D_k}= \pm (i)^k a b^*  \lan b_N|b^{\prime}_N\ran \pm  (-i)^k a^* b \lan b^{\prime}_N | b_N\ran$
  \begin{displaymath}
 =\left\{ \begin{array}{ll}
 \pm 2 |a| |b|\;|\lan b^{\prime}_N | b_N\ran| cos (\phi_a-\phi_b-\alpha) & \textrm{ if $k$ is even,} \\

\pm 2 |a| |b|\; |\lan b^{\prime}_N | b_N\ran| sin (\phi_a-\phi_b-\alpha) & \textrm{ if $k$ is odd.} \\
\end{array} \right.
\end{displaymath}
  Finally we get  $$|| \mathcal{T}^{(N-1)}_{\rho}||^2 = (\pm |a|^2 \pm |b|^2)^2 + 4 |a|^2 |b|^2 |\lan b^{\prime}_N | b_N\ran|^2 cos^2 (\phi_a-\phi_b-\alpha) \sum_{k=0}^{q'} \binom{q'}{2k}$$
  $$+4 |a|^2 |b|^2 |\lan b^{\prime}_N | b_N\ran|^2 sin^2 (\phi_a-\phi_b-\alpha) \sum_{k=0}^{q'} \binom{q'}{2k+1},$$
  where $q' \le q$ is the number of $\si_1$ operators in $D$. Since $|\lan b^{\prime}_N | b_N\ran|^2 \le 1$ we see that $$|| \mathcal{T}^{(N-1)}_{\rho}||^2 \le || \mathcal{T}^{(N)}_{|\psi\ran}||^2,$$
  equality occurring when $|b_N\ran =|b^{\prime}_N\ran$, in which case $|\psi\ran=|\phi\ran\otimes|b_N\ran.$
  It is straightforward, but tedious to elevate this proof for the general case
  $$|\psi\ran=\sum_{\alpha_1 \cdots \alpha_N} a_{\alpha_1 \cdots \alpha_N} |b_{\alpha_1}\cdots b_{\alpha_N}\ran, \; \alpha_i=0,1.$$
  Basically we have to keep track of $\binom{r}{2}$ $B$ type of operators, where $r$ is the number of terms in the expansion of $|\psi\ran$, in order to obtain all nonzero elements of $\mathcal{T}^{(N)}_{|\psi\ran}$. When $N$th particle is traced out, the corresponding elements of $\mathcal{T}^{(N-1)}_{\rho}$ get multiplied by the overlap amplitudes, which leads to the required result.\hfill $\blacksquare$

\noi\textit{Continuity of $E_{\mathcal{T}}$}: We show that for $N$-qubit pure states $||(|\psi\ran\lan\psi|-|\phi\ran\lan\phi|)||\rightarrow 0 \Rightarrow \Big{|}E_{\mathcal{T}}(|\psi\ran)-E_{\mathcal{T}}(|\phi\ran)\Big{|}\rightarrow 0.$\\
\noi\textbf{Proof :} $||(|\psi\ran\lan\psi|-|\phi\ran\lan\phi|)||\rightarrow 0$
$\Rightarrow ||\mathcal{T}^{(N)}_{|\psi\ran}-\mathcal{T}^{(N)}_{|\phi\ran}||\rightarrow 0.$

But  $||\mathcal{T}^{(N)}_{|\psi\ran}-\mathcal{T}^{(N)}_{|\phi\ran}|| \ge \Big{|}||\mathcal{T}^{(N)}_{|\psi\ran}||-||\mathcal{T}^{(N)}_{|\phi\ran}||\Big{|}.$

Therefore $||\mathcal{T}^{(N)}_{|\psi\ran}-\mathcal{T}^{(N)}_{|\phi\ran}||\rightarrow 0 \Rightarrow \big{|}||\mathcal{T}^{(N)}_{|\psi\ran}||-||\mathcal{T}^{(N)}_{|\phi\ran}||\big{|} \rightarrow 0$

   $\Rightarrow \Big{|}E_{\mathcal{T}}(|\psi\ran)-E_{\mathcal{T}}(|\phi\ran)\Big{|}\rightarrow 0. $ \hfill $\blacksquare$\\

\subsection{Entanglement of Multiple Copies of a Given State}

{\it LU invariance. } We show that  $E_{\mathcal{T}}$ for multiple copies of $N$-qubit pure state $|\psi\ran$ is $LU$ invariant.
Consider a system of $N\times k$ qubits in the state $|\chi\ran=|\psi\ran\otimes |\psi\ran \otimes \cdots\otimes |\psi\ran$ ($k$ copies). It is straightforward to check that (see chapter 4)

$$  \mathcal{T}^{(N)}_{|\chi\ran}= \mathcal{T}^{(N)}_{|\psi\ran}\circ \mathcal{T}^{(N)}_{|\psi\ran}\circ \cdots \circ \mathcal{T}^{(N)}_{|\psi\ran} \eqno{(5.29)}$$

 This implies, in a straightforward way, that $$|| \mathcal{T}^{(N)}_{|\chi\ran}||=||  \mathcal{T}^{(N)}_{|\psi\ran}||^k.$$
Since by proposition 5.4.5 $||  \mathcal{T}^{(N)}_{|\psi\ran}||$ is $LU$ invariant, so is $|| \mathcal{T}^{(N)}_{|\chi\ran}||$.

    Let $|\psi\ran$ be a $N$-qubit pure state and $|\chi\ran=|\psi\ran\otimes|\psi\ran$. Then $E_{\mathcal{T}}(|\chi\ran)$ is expected to satisfy $$E_{\mathcal{T}}(|\chi\ran)\ge E_{\mathcal{T}}(|\psi\ran).$$
    We again use the fact that $$  \mathcal{T}^{(N)}_{|\chi\ran}= \mathcal{T}^{(N)}_{|\psi\ran}\circ \mathcal{T}^{(N)}_{|\psi\ran},$$

    which gives $$|| \mathcal{T}^{(N)}_{|\chi\ran}||=||  \mathcal{T}^{(N)}_{|\psi\ran}||^2.$$

    Since $||  \mathcal{T}^{(N)}_{|\psi\ran}||\ge 1$ we get
   $ || \mathcal{T}^{(N)}_{|\chi\ran}||\ge ||  \mathcal{T}^{(N)}_{|\psi\ran}||$ or,  $$E_{\mathcal{T}}(|\chi\ran)\ge E_{\mathcal{T}}(|\psi\ran).$$

\noi\textit{Superadditivity }: We have to show, for $N$qubit states   $|\psi\ran$ and $|\phi\ran$ that $$E_{\mathcal{T}}(|\psi\ran \otimes |\phi\ran )\ge E_{\mathcal{T}}(|\psi\ran)+ E_{\mathcal{T}}(|\phi\ran). \eqno{(5.30)}$$
We already know that for $|\chi\ran=|\psi\ran\otimes|\phi\ran$
$$|| \mathcal{T}^{(N)}_{|\chi\ran}||= ||  \mathcal{T}^{(N)}_{|\psi\ran}||\; ||  \mathcal{T}^{(N)}_{|\phi\ran}||$$

Thus Eq. (5.30) gets transformed to $$||  \mathcal{T}^{(N)}_{|\psi\ran}||\; ||  \mathcal{T}^{(N)}_{|\phi\ran}||-1 \ge ||  \mathcal{T}^{(N)}_{|\psi\ran}||+ ||  \mathcal{T}^{(N)}_{|\phi\ran}||-2$$

which is true for $||  \mathcal{T}^{(N)}_{|\psi\ran}||\ge 1$ and $||\mathcal{T}^{(N)}_{|\phi\ran}||\ge 1$. \hfill $\blacksquare$\\

\subsection{ Computational Considerations}

Computation or experimental determination of $E_{\mathcal{T}}$ involves $3^{N}$ elements of $\mathcal{T}^{(N)}$ so that it increases exponentially with the number of qubits $N$. However, for many important classes of states, $E_{\mathcal{T}}$ can be easily computed and increases only polynomially with $N$. We have already computed $E_{\mathcal{T}}$ for the class of $N$-qubit $W$ states, $GHZ$ states, and their superpositions. We have also computed
$E_{\mathcal{T}}$ for an important physical system like one-dimensional Heisenberg antiferromagnet. For symmetric or antisymmetric states, $\mathcal{T}^{(N)}$ is supersymmetric, that is, the value of its elements are invariant under any permutation of its indices (see chapter 4). This reduces the problem to the computation of $\fr{1}{2}(N+1)(N+2)$ distinct elements of $\mathcal{T}^{(N)},$ which is quadratic in $N$.

\subsection{ Entanglement Dynamics : Grover Algorithm}

We show that $E_{\mathcal{T}}$ can quantify the evolution of entanglement. We consider Grover's algorithm. The goal of Grover's algorithm is to convert the initial state of $N$ qubits, say $|0\cdots 0\ran,$ to a state that has probability bounded above $\frac{1}{2}$ of being in the state $|a_1\cdots a_{N}\ran,$ using
$$U_a|b_1\cdots b_{N}\ran=(-1)^{\Pi\delta_{a_jb_j}}|b_1\cdots b_{N}\ran$$
 the fewest times possible. Grover showed that this can be done with $O(\sq{2^{N}})$ uses of $U_a$ by starting with the state $$\fr{1}{\sq{2^{N}}}\sum_{x=0}^{2^{N}-1}|x\ran \;=\;H^{\otimes N}|0\cdots0\ran,$$ where

\begin{displaymath}
H=\fr{1}{\sq{2}}
\left(\begin{array}{cc}
1& 1  \\
1 & -1\\
\end{array}\right),
\end{displaymath}
and then iterating the transformation $H^{\otimes N}U_a H^{\otimes N}U_a$ on this state \cite{mw02}. The initial state is a product state as is the target state, but intermediate states $\psi(k)$ are entangled for $k>0$ iterations. Figure 5.6 shows the development of $E_{\mathcal{T}}(|\psi(k)\ran)$ with number of iterations $k,$ for six qubits. The values of $k$ for which $E_{\mathcal{T}}$ vanishes are the iterations at which the probability of measuring $|a_1\cdots a_{N}\ran$ is close to $1.$ Thus $E_{\mathcal{T}}$ can be used to quantify the evolution of a $N$-qubit entangled state.\\

\begin{figure}[!ht]
\begin{center}
\includegraphics[width=15cm,height=12cm]{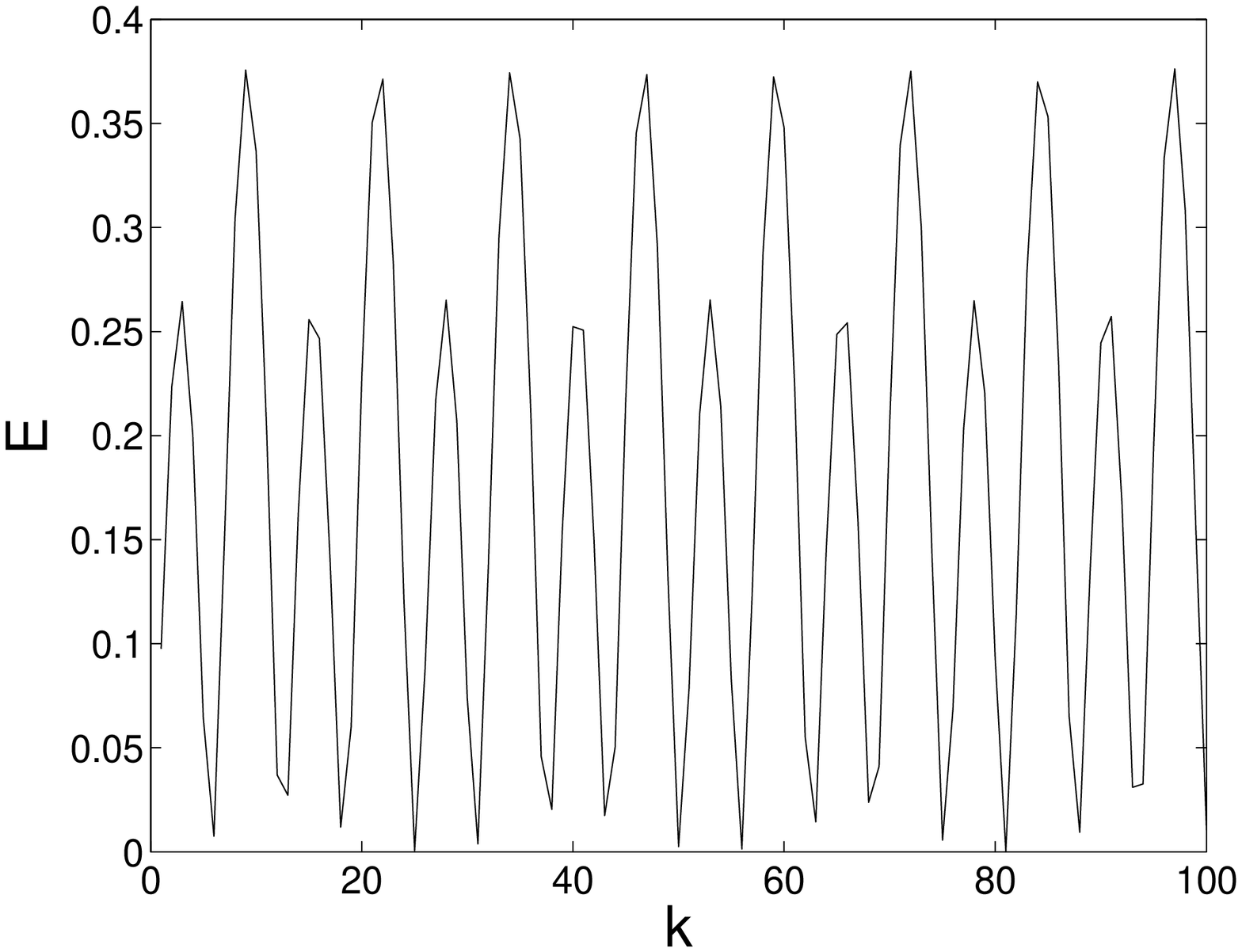}

Figure 5.6. Entanglement in Grover's algorithm for six qubits as a function of number of iterations.
\end{center}
\end{figure}

\section{ Extension to  Mixed States}

The extension of $E_{\mathcal{T}}$ to mixed states $\rho$ can be made via the use of the {\it convex roof} or {\it (hull)} construction as was done for the entanglement of formation \cite{woot98}. We define $E_{\mathcal{T}}(\rho)$ as a minimum over all decompositions $\rho=\sum_{i}p_i\arrowvert\psi_i\ran\lan\psi_i\arrowvert$ into pure states i.e.,
$$E_{\mathcal{T}}(\rho)=\min_{\begin{subarray}{I}
  \hskip  .1cm  {\{p_i,\psi_i\}}
 \end{subarray}}
\sum_{i}p_iE_{\mathcal{T}}(\arrowvert\psi_i\ran).\eqno{(5.31)}$$
The existence and uniqueness of the convex roof for $E_{\mathcal{T}}$ is guaranteed because it is a continuous function on the set of pure states \cite{uhl00}. This entanglement measure is expected to satisfy conditions (a), (b) and (c) given in section 5.4 and is expected to be (d) convex under discarding of information, i.e., $$\sum_{i}p_i E_{\mathcal{T}}(\rho_i)\geq E_{\mathcal{T}}(\sum_{i}p_i\rho_i). \eqno{(5.32)}$$

The criteria (a)-(d) above are considered to be the minimal set of requirements for any entanglement measure so that it is an entanglement monotone \cite{vida00}.

Evidently, criteria (a) and (b) are satisfied by $E_{\mathcal{T}}(\rho)$ defined via the convex roof as they are satisfied by
$E_{\mathcal{T}}$ for pure states. Condition (d) follows from the fact that every convex hull (roof) is a convex function \cite{uhl98}. We need to prove (c), which is summarized in proposition 5.5.1.\\
The proof follows from the monotonicity of $E_{\mathcal{T}}(|\psi\ran)$  for pure states, that is, propositions 5.4.5, 5.4.6 and 5.4.7.
Bennett {\it et al.} prove a version of proposition 5.5.1 in \cite{bdsw}, which applies to any measure satisfying propositions 5.4.5, 5.4.6 and 5.4.7. Thus the same proof applies to proposition 5.5.1. However, we give it here for the sake of completeness. \\
\noi\textbf{Proposition 5.5.1}: If  an $N$-qubit mixed state $\rho$ is subjected to a local operation on $i$th qubit giving outcomes $k_i$ with probabilities $p_{k_i}$ and leaving residual $N$-qubit mixed state $\rho_{k_i}$, then the expected entanglement $\sum_{k_i} p_{k_i} E_{\mathcal{T}}(\rho_{k_i})$ of the residual state is not greater than the entanglement $E_{\mathcal{T}}(\rho)$ of the original state, $$\sum_{k_i} p_{k_i} E_{\mathcal{T}}(\rho_{k_i}) \le E_{\mathcal{T}}(\rho).$$ (If the operation is simply throwing away part of the system, then there will be only one value of $k_i$, with unit probability.)

\noi\textbf{Proof :} Given mixed state $\rho$ there will exist some minimal-entanglement ensemble
$$\mathcal{E}=\{p_j,|\psi\rangle\}\eqno{(5.33)}$$
of pure states realizing $\rho$.

For any ensemble $\mathcal{E}'$ realizing $\rho$,
$$E(\rho)\le E_{\mathcal{T}}(\mathcal{E}').\eqno{(5.34)}$$
Applying the propositions 5.4.5, 5.4.6 and 5.4.7 to each pure state in the minimal-entanglement ensemble $\mathcal{E}$, we get, for each $j$,
$$\sum_{k_i} p_{k_i|j} E_{\mathcal{T}}(\rho_{jk_i})\le E_{\mathcal{T}}(|\psi_i\rangle),\eqno{(5.35)}$$
where $\rho_{jk_i}$ is the residual state if pure state $|\psi_j\rangle$ is subjected to $i$th partite's operation and yields result $k_i$, and $p_{k_i|j}$ is the conditional probability of obtaining this outcome when the initial state is $|\psi_j\rangle.$

Note that when the the outcome $k_i$ has occurred the residual mixed state is described by the density matrix
$$\rho_{k_i}=\sum_j p_{j|k_i} \rho_{jk_i}. \eqno{(5.36)}$$
Multiplying Eq.(5.35) by $p_j$ and summing over $j$ gives $$\sum_{j,k_i} p_j p_{k_i|j} E_{\mathcal{T}}(\rho_{jk_i}) \le \sum_j p_j E_{\mathcal{T}}(|\psi_j\rangle)=E_{\mathcal{T}}(\rho).\eqno{(5.37)}$$
By Bayes theorem $$p_{j,k_i}=p_j p_{k_i|j}=p_{k_i} p_{j|k_i},\eqno{(5.38)}$$
Equation (5.37) becomes
$$\sum_{j,k_i} p_{k_i} p_{j|k_i} E_{\mathcal{T}}(\rho_{jk_i}) \le  E_{\mathcal{T}}(\rho).\eqno{(5.39)}$$
Using the bound Eq.(5.34), we get
$$\sum_{k_i} p_{k_i} E_{\mathcal{T}}(\rho_{k_i})\le \sum_{k_i} p_{k_i} \sum_{j}p_{j|k_i} E_{\mathcal{T}}(\rho_{jk_i}) \le  E_{\mathcal{T}}(\rho).\eqno{(5.40)}$$
\hfill $\blacksquare$\\

Note that any sequence of local operations comprises local operations drawn from the set of basic local operations (i)-(iv) above, so that proposition 5.5.1 applies to any such sequence. Thus we can say that the expected entanglement of a $N$-qubit system, measured by $E_{\mathcal{T}}(\rho)$, does not increase under local operations. \\

\section{ A Related Entanglement Measure}

We consider the following entanglement measure. Consider $$\mathcal{E}_{\mathcal{T}}(|\psi\ran)=log_2||\mathcal{T}^{(N)}||=log_2[E_{\mathcal{T}}(|\psi\ran)+1],$$

where $\mathcal{T}^{(N)}$ is the $N$-way correlation tensor occuring in the Bloch representation of $\rho=|\psi\ran\lan\psi|.$

Proofs of propositions 5.4.1-5.4.7 and 5.5.1 easily go through for $\mathcal{E}_{\mathcal{T}}(|\psi\ran)$. We prove continuity as follows.

\noi\textit{Continuity of $\mathcal{E}_{\mathcal{T}}(|\psi\ran)$}. We have to show, for two $N$-qubit states $|\psi\ran$ and $|\phi\ran$, that $||(|\psi\ran\lan\psi|-|\phi\ran\lan\phi|)||\rightarrow 0 \Rightarrow \Big{|}\mathcal{E}_{\mathcal{T}}(|\psi\ran)-\mathcal{E}_{\mathcal{T}}(|\phi\ran)\Big{|}\rightarrow 0.$

We have  $||(|\psi\ran\lan\psi|-|\phi\ran\lan\phi|)||\rightarrow 0 $
$\Rightarrow ||\mathcal{T}^{(N)}_{|\psi\ran}-\mathcal{T}^{(N)}_{|\phi\ran}||\rightarrow 0. $

But  $ ||\mathcal{T}^{(N)}_{|\psi\ran}-\mathcal{T}^{(N)}_{|\phi\ran}|| \ge \big{|}||\mathcal{T}^{(N)}_{|\psi\ran}||- ||\mathcal{T}^{(N)}_{|\phi\ran}||\big{|}.$

Further, whenever $|||\mathcal{T}^{(N)}_{|\psi\ran}|| \ge 1$ and $|\mathcal{T}^{(N)}_{|\phi\ran}|| \ge 1$

we have $\big{|}||\mathcal{T}^{(N)}_{|\psi\ran}||- ||\mathcal{T}^{(N)}_{|\phi\ran}||\big{|} \ge \big{|}log_2(||\mathcal{T}^{(N)}_{|\psi\ran}||)- log_2(||\mathcal{T}^{(N)}_{|\phi\ran}||)\big{|}.$

Thus $ ||\mathcal{T}^{(N)}_{|\psi\ran}-\mathcal{T}^{(N)}_{|\phi\ran}||\rightarrow 0 \Rightarrow \big{|}||\mathcal{T}^{(N)}_{|\psi\ran}||- ||\mathcal{T}^{(N)}_{|\phi\ran}||\big{|} \rightarrow 0 \Rightarrow \big{|}log_2(||\mathcal{T}^{(N)}_{|\psi\ran}||)- log_2(||\mathcal{T}^{(N)}_{|\phi\ran}||)\big{|} \rightarrow 0 \Rightarrow \Big{|}\mathcal{E}_{\mathcal{T}}(|\psi\ran)-\mathcal{E}_{\mathcal{T}}(|\phi\ran)\Big{|}\rightarrow 0 .$

However, $\mathcal{E}_{\mathcal{T}}(|\psi\ran)$ has the added advantage that it is additive [ while $E_{\mathcal{T}}(|\psi\ran)$ is superadditive].
Indeed, from section 5.4.1 we see that for $k$ copies
$$\mathcal{E}_{\mathcal{T}}(|\psi\ran\otimes|\psi\ran\otimes\cdots \otimes |\psi\ran)=k \mathcal{E}_{\mathcal{T}}(|\psi\ran).$$
Similarly $\mathcal{E}_{\mathcal{T}}(|\psi\ran\otimes|\phi\ran)=\mathcal{E}_{\mathcal{T}}(|\psi\ran)+\mathcal{E}_{\mathcal{T}}(|\phi\ran).$

The extension of $\mathcal{E}_{\mathcal{T}}(|\psi\ran)$ to mixed states via convex roof construction is similar to that of $E_{\mathcal{T}}(|\psi\ran).$ Thus $\mathcal{E}_{\mathcal{T}}(|\psi\ran)$ has all the properties of $E_{\mathcal{T}}(|\psi\ran)$, with an additional property that $\mathcal{E}_{\mathcal{T}}(|\psi\ran)$ is additive, while $E_{\mathcal{T}}(|\psi\ran)$ is superadditive.\\

\section{ Conclusion}

In conclusion, we have developed an experimentally viable entanglement measure for $N$-qubit pure states, which passes almost all the tests for being a good entanglement measure. This is a global entanglement measure in the sense that it does not involve partitions or cuts of the system in its definition or calculation. This measure has quadratic computational complexity for symmetric or antisymmetric states. Computational tractability is not a serious problem if $N$ is not too large, and the measure can be easily computed for systems comprising small number of qubits, which can have many important applications such as teleportation of multiqubit states, quantum cryptography, dense coding, distributed evaluation of functions \cite{bras01} etc. However, finding other classes of states for which $E_{\mathcal{T}}$ can be computed polynomially will be useful. It will be very interesting to seek applications of this measure to situations
like quantum phase transitions \cite{on02}, transfer of entanglement along spin chains \cite{bb07}, NOON states in quantum lithography \cite{bkabwd} etc.
Finally, we have extended our measure to the mixed states and established its various properties, in particular, its monotonicity. We may also note that neither its definition nor its properties depends in an essential way on the fact that we are dealing with qubits, so that this measure can be defined and applied to a general $N$-partite quantum system.

\chapter{Summary and Future Directions}
\begin{center}
\scriptsize \textsc{This is not the end. Nor is this a beginning of the end.\\ This may at most be the end of a beginning.\\
 -{\it Sir Winston Churchill.}\footnote{From his last address to the British parliament as the prime minister of U.K. ( 3rd june 1946)}.}
\end{center}

In this thesis, I have tried to enhance understanding of the following two questions:\\
A-  Given a multipartite quantum state (possibly mixed), how to find out whether it is entangled or separable? (Detection of entanglement.)\\
B- Given an entangled state, how to decide how much entangled it is? (Measure of entanglement.)

 Answers to both these questions are known for bipartite pure states. For multipartite states, general answers to both these questions are not known. Many separability criteria are proposed. Example: Generalizations of Peres-Horodecki criterion.
The genuine entanglement of pure multipartite quantum state is established by checking whether it is entangled in all bipartite cuts, which can be tested using Peres-Horodecki criterion.

For mixed states this strategy does not work because there are mixed states which are separable in all bipartite cuts but are genuinely entangled. A direct and independent detection of genuine multipartite entanglement is lacking.
We have explored two approaches. \\
I- In the first approach, we assign a weighted graph with multipartite quantum state and address the question of separability  in terms of these graphs and various operations involving them (Chapters 2 and 3). \\
II- In the second approach, we use the so called Bloch representation of multipartite quantum states to establish new criteria for detection of multipartite entangled states (Chapter 4). We further give a new measure for entanglement in $N$-qubit entangled pure state and formally extend it to cover $N$-qubit mixed states (Chapter 5).

In the following, I give some of the key results obtained in the this thesis.\\
\noi\textbf{Chapter 2.}\\
1- We have given rules to associate a graph with a quantum state and a quantum state to a graph, with a positive semidefinite generalized Laplacian, for states in real as well as complex Hilbert space.\\
2- We have shown that projectors involving states in the standard basis are associated with the edges of the graph.\\
3- We have given graphical criteria for a state being pure. In particular, we have shown that a pure state must have a graph which is a clique plus isolated vertices.\\
4-  We have given an algorithm to construct graph corresponding to a convex combination of density matrices, in terms of the graphs of these matrices.\\
5- We have defined a modified tensor product of two graphs in terms of the graph operators $\cL, \eta, \cN, \Om$ and obtained the properties of these operators. We have shown that this product is associative and distributive with respect to the disjoint edge union  of graphs.\\
6- We have proved that the density matrix of the modified tensor product of two graphs is the tensor product of the density matrices of the factors. For density matrices,  we show that a convex combination of the products of density matrices has a graph which is the edge union of the modified tensor products of the graphs for these matrices. Thus we can code werner's definition of separability in terms of graphs.\\
7-  We have generalized the separability criterion given by  S. L. Braunstein {\it et al.} to the real density matrices having graphs without loops.\\
8- We have found the quantum superoperators corresponding to the basic operations on graphs, namely addition and deletion of edges and vertices. it is straightforward to see that all quantum operations on states result in the addition / deletion of edges and/ or vertices, or redistribution of weights. However, addition / deletion of edges / vertices correspond to quantum operations which are irreversible, in general. Hence it seems to be difficult to encode a unitary operator, which has to be reversible, in terms of the operations on graphs. Further, graphs do not offer much advantage for quantum operations which only redistribute the weights, without changing the topology of the graph, as in this case the graph is nothing more than a clumsy way of writing the density matrix.\\
10-  Finally, we have given several graphical criteria for the positive semidefiniteness of the generalized Laplacian associated with a graph. This characterizes a large class of graphs coding quantum states.\\
\noi\textbf{Chapter 3.}\\
1- We settle the so-called degree conjecture for the separability of multipartite quantum states, which are normalized graph Laplacians, first given by Braunstein {\it et al.} \cite{bgmsw}.
 The conjecture states that a multipartite quantum state is separable if and only if the degree matrix of the graph associated with the state is equal to the degree matrix of the partial transpose of this graph. We call this statement to be the strong form of the conjecture. In its weak version, the conjecture requires only the necessity, that is, if the state is separable, the corresponding degree matrices match. We prove the strong form of the conjecture for pure multipartite quantum states using the modified tensor product of graphs defined in chapter 2, as both necessary and
sufficient condition for separability.\\
2- Based on this proof, we give a polynomial-time algorithm for completely factorizing any pure multipartite quantum state. By polynomial-time algorithm, we mean that the execution time of this algorithm increases as a polynomial in $m$, where $m$ is the number of parts of the quantum system. \\
3- We give a counterexample to show that the conjecture fails, in general,
even in its weak form, for multipartite mixed states.\\
4- Finally, we prove this conjecture, in its weak form, for a class of multipartite mixed states, giving only a necessary condition for separability.\\
\noi\textbf{Chapter 4.}\\
1- We give a new separability criterion, a necessary condition for separability of $N$-partite quantum states. The criterion is based on the Bloch representation of a $N$-partite quantum state and makes use of multilinear algebra, in particular, the matrization of tensors. Our criterion applies to {\it arbitrary} $N$-partite quantum states in $\mathcal{H}=\mathcal{H}^{d_1}\otimes \mathcal{H}^{d_2} \otimes \cdots \otimes \mathcal{H}^{d_N}.$ The criterion can test whether a $N$-partite state is entangled and can be applied to different partitions of the $N$-partite system.\\
2-  We provide examples that show the ability of this criterion to detect entanglement. We show that this criterion can detect bound entangled states.\\
3- We prove a sufficiency condition for separability of a three-partite state, straightforwardly generalizable  to the case  $N > 3,$ under certain  condition.\\
4- We also give a necessary and sufficient condition for separability of a class of $N$-qubit states which includes $N$-qubit PPT states.\\
\noi\textbf{Chapter 5.}\\
1- We present a multipartite entanglement measure for $N$-qubit pure states, using the norm of the correlation tensor which occurs in the Bloch representation of the state.\\
2- We compute this measure for  several important classes of $N$-qubit pure states such as GHZ states, W states and their superpositions. We compute this measure for interesting applications like one dimensional Heisenberg antiferromagnet.  We use this measure to follow the entanglement dynamics of Grover's algorithm.\\
3- We prove that this measure possesses almost all the properties expected of a good entanglement measure, including monotonicity. \\
4- Finally, we extend this measure to $N$-qubit mixed states via convex roof construction  and establish its various properties, including its monotonicity. \\
5- We also introduce a related measure which has all properties of the above measure and is also additive.

Here are some of the interesting research problems emerging from our work.

(i) The principal achievement of first two chapters, apart from giving a new formulation is the proof of the degree criterion for separability of N-partite pure states and their factorization into entangled parts. One of the open problems of this new formulation is to find graphical criteria for the non-negativity of the generalized Laplacian associated with a graph. As we have seen, degree criterion fails, in general, for the mixed states. We could prove this criterion only for states with density matrices with real weighted graphs without loops. Such matrices have all elements real so there are no coherences \cite{ctdl} and elements in every row and every column sum up to zero \cite{wu06}. Can we then use graph topology to classify quantum states based on separability and seek (possibly different) criteria for separability of different classes of states?  Is it possible to code LOCC operation in terms of operations on graphs? It may be a good idea to use Jamiolkowaski isomorphism \cite{jam72} between states and quantum operations. If we combine these two questions, we can seek the classification of N-partite entanglement in terms of classes of states not inter-convertible via SLOCC \cite{pv07}. These are some of the interesting questions on the basis of chapters 2, 3, but we feel that its a long way to get there, if at all we can.

(ii) There is a variety of questions emerging from chapters 4 and 5. It is interesting to look for lower bound on say, concurrence \cite{gfw06} of three partite state to the violation of separability condition based on the criterion stated in chapter 5. Further, it will be interesting to seek a new PPT entangled state which is detected by the criterion in chapter 4 but not by any other criterion. It will be interesting to generalize the measure in chapter 5 to d-level systems instead of qubits. Using our measure, can we get tight upper or lower bounds on the entanglement of superposition of multipartite states \cite{ccaslc}? A very interesting question is whether we can obtain the entanglement dynamics of a multipartite system in terms of our entanglement measure? For this, we will have to get the effect on the correlation tensor $\mathcal{T}^{(N)}$ of the action of a SLOCC operator or a local $SL(N,\mathcal{C})$ group \cite{osterloh} on the state \cite{mwb95}. If this programme is successful, we hope to classify the multipartite entanglement into classes that are SLOCC inequivalent. All this may have applications in thermal entanglement of many particles systems \cite{fcgaww}. Finally, it will be very interesting to seek applications of our entanglement measure to situations like quantum phase transitions \cite{oromy}, transfer of entanglement across spin chains \cite{aosbb} etc.


\end{document}